\documentclass[11pt]{article}
\usepackage{latexsym,psfig}
\usepackage[dvips]{graphics}
\usepackage{amsmath} 
\usepackage{amscd}     \usepackage{amsxtra}
\usepackage{upref}     \usepackage{amsthm}
\usepackage{amssymb}   
\usepackage{amsfonts}

\begin{document}

\pagenumbering{roman}

\title{{\bf Extended Electrodynamics}:\\
A Brief Review}

\author{{\bf S.Donev} \thanks{e-mail: sdonev@inrne.bas.bg} \
and {\bf M. Tashkova} \\
Institute for Nuclear Research and Nuclear Energy,\\
Bulg.Acad.Sci., 1784 Sofia, blvd.Tzarigradsko chaussee 72\\
Bulgaria}

\date{}

\maketitle

\begin{abstract}
This paper presents a brief review of the newly developed {\it Extended
Electrodynamics}.  The relativistic and non-relativistic approaches to the
extension of Maxwell equations are considered briefly, and the further study
is carried out in relativistic terms in Minkowski space-time.  The non-linear
vacuum solutions are considered and fully described. It is specially pointed
out that solitary waves with various, in fact arbitrary, spatial structure
and photon-like propagation properties exist.  The {\it null} character of
all non-linear vacuum solutions is established and extensively used further.
Coordinate-free definitions are given to the important quantities {\it
amplitude} and {\it phase}. The new quantity, named {\it scale factor}, is
introduced and used as a criterion for availability of rotational component
of propagation of some of the nonlinear, i.e.  nonmaxwellian, vacuum
solutions.  The group structure properties of the nonlinear vacuum solutions
are analyzed in some detail, showing explicitly the connection of the vacuum
solutions with some complex valued functions.  Connection-curvature
interpretations are given and a special attention is paid to the curvature
interpretation of the intrinsic rotational (spin) properties of some of the
nonlinear solutions. Several approaches to coordinate-free local description
and computation of the integral spin momentum are considered.  Finally, a
larage family of nonvacuum spatial soliton-like solutions is explicitly
written down, and a procedure to get (3+1) versions of the known (1+1)
soliton solutions is obtained.

\end{abstract}

\newpage
\tableofcontents

\newpage

\pagenumbering{arabic}
\setcounter{page}{1}

\section{Introduction}
\subsection{The concept of Physical Object}
When we speak about free physical objects, e.g. classical particles, solid
bodies, elementary particles, etc., we always keep in mind that, although we
consider them as free, they can not in principle be absolutely free. What is
really understood under "free object" (or more precisely "relatively free
object") is, that some {\it definite properties (e.g. mass, velocity) of the
object under consideration do not change in time during its evolution under
the influence of the existing environment}.  The availability of such
time-stable features of any physical object guarantees its {\it
identification} during its existence in time.  Without such an availability
of constant in time properties (features), which are due to the object's
resistance abilities, we could not speak about objects and knowledge at all.
So, for example, a classical mass particle in external gravitational field is
free with respect to its mass, and it is not free with respect to its
behavior as a whole (its velocity changes), because in classical mechanics
formalism its mass does not change during the influence of the external field
on its accelerated way of motion.

The above view implies that three kinds of quantities will be necessary to
describe as fully as possible the existence and the evolution of a given
physical object:

	1. {\bf Proper (identifying) characteristics}, i.e. quantities which
do NOT change during the entire existence of the object. The availability of
such quantities allows to distinguish a physical object among the other ones.

	2. {\bf Kinematical characteristics}, i.e. quantities, which describe
the allowed space-time evolution, where "allowed" means {\it consistent with
the constancy of the identifying characteristics}.

	3. {\bf Dynamical characteristics}, i.e. quantities which are
functions (explicit or implicit) of the {\it proper} and of the {\it
kinematical} characteristics.

Some of the dynamical characteristics have the following two important
properties:  they are {\bf universal}, i.e. every physical object carries
nonzero value of them (e.g. energy-momentum), and they are {\bf
conservative}, i.e. they may just be transferred from one physical object to
another (in various forms) but no loss is allowed.

Hence, the evolution of a physical object subject to bearable/acceptable
exterior influence (perturbation), coming from the existing environment,
has three aspects:

	1. {\it constancy of the proper (identifying) characteristics},

	2. {\it allowed kinematical evolution},  and

	3. {\it exchange of dynamical quantities with the physical
environment}.

If the physical object under consideration is space-extended
(continuous) and is described by a many-component mathematical object we may
talk about {\it internal exchange} of some dynamical characteristic among the
various components of the object. For example, the electromagnetic field has
two vector components $(\mathbf{E},\mathbf{B})$ and internal energy-momentum
exchange between $\mathbf{E}$ and $\mathbf{B}$ should be considered as
possible.

The third feature suggests that the dynamical equations, describing locally
the evolution of the object, may come from giving an explicit form of the
quantities controlling the local internal and external exchange processes
i.e. writing down corresponding local balance equations.  Hence, denoting
the local quantities that describe the external exchange processes by $Q_i,
i=1,2,\dots$, the object should be considered to be $Q_i$-free,
$i=1,2,\dots$, if the corresponding integral values are time constant,  which
can be achieved only if $Q_i$ obey differential equations presenting
appropriately (implicitly or explicitly) corresponding local versions of the
conservation laws (continuity equations). In the case of absence of external
exchange these equations should describe corresponding {\it internal}
exchange processes.  The corresponding evolution in this latter case may be
called {\it proper evolution}.

\vskip 0.3cm
In trying to formalize these views we have to give some initial explicit
formulations of some most basic features (properties) of what we call
physical object, which features would  lead us to a, more or less,
adequate theoretical notion of our intuitive notion of a physical object.
Anyway, the following properties of the theoretical concept "physical object"
we consider as necessary:

      1. It can be created.

      2. It can be destroyed.

      3. It occupies finite 3-volumes at any moment of its existence, so it has structure.

      4. It has a definite stability to withstand some external disturbances.

      5. It has definite conservation properties.

      6. It necessarily carries energy-momentum, and, possibly,
other measurable (conservative or nonconservative) physical quantities.

      7. It exists in an appropriate environment (called usually vacuum), which
provides all necessary existence needs.

      8. It can be detected by the rest of the world through allowed
energy-momentum exchanges.

      9. It  may combine with other appropriate objects to form new
objects of higher level structure.

      10. Its death gives necessarily birth to new objects following definite
rules of conservation.

\vskip 0.3cm
\noindent {\bf Remark}.
The property {\it to be finite} we consider as a very essential one. So, the
above features do NOT allow the classical material points and the infinite
classical fields (e.g.  plane waves) to be considered as physical objects
since the former have no structure and cannot be destroyed, and the latter
carry infinite energy, so they cannot be created. Hence, the Born-Infeld
"principle of finiteness" stating that {\it a satisfactory theory should avoid
letting physical quantities become infinite}, declared in [9], may be
strengthened as follows: {\it all real physical objects are spatially finite
entities and NO infinite values of the physical quantities carried by them
are allowed}.

\vskip 0.2cm
Clearly, together with the purely qualitative
features physical objects carry important quantitatively described physical
properties, and any external interaction may be considered as an exchange of
such quantities provided both the object and the environment carry them.
Hence, the more universal is a physical quantity the more useful for us it
is, and this moment determines the exclusively important role of
energy-momentum, which modern physics considers as the most universal one,
i.e. every physical object necessarily carries energy-momentum.

From pure formal point of view a description of the evolution of a given
continuous physical object must include obligatory two mathematical objects:

	{\bf 1. First}, the mathematical object $\Psi$ which is meant to
represent as fully as possible the integrity of the object under
consideration, and when subject to appropriate operators, $\Psi$ must
reproduce explicitly all important information about the {\it structure} and
admissible {\it dynamical evolution} of the physical object;

	{\bf 2. Second}, the mathematical object $D\Psi$ which represents
the admissible changes, where $D$ is appropriately chosen differential
operator acting mainly on the kinematical and dynamical characteristics.  If
$D$ does not depend on $\Psi$ and its derivatives, and, so, on the
corresponding proper characteristics of $\Psi$, the relation $D\Psi=0$ then
would mean that those kinematical and dynamical properties of $\Psi$, which
feel the action of $D$, are constant with respect to $D$, so the evolution
prescribed by $D\Psi=0$, would have "constant" character and would say
nothing about possible changes of those characteristics, which do not feel
$D$.

From principle point of  view, it does not seem so important if the changes
$D\Psi$ are zero, or not zero, the {\it significantly important} point is
that the {\it changes $D\Psi$ are admissible}. The changes $D\Psi$ are
admissible in the following two cases: {\bf first}, when related to the very
object $\Psi$ through some "projection" $P$ upon $\Psi$, the "projections"
$P(D\Psi)$ vanish, and then the object is called {\it free} (with respect to
those characteristics which feel $D$); {\bf second}, when the projections
$P(D\Psi)$ do not vanish, but the object still survives, then the object is
called {\it not free} (with respect to the same characteristics).  In the
first case the corresponding admissible changes $D\Psi$ have to be considered
as having an {\it intrinsic} for the object nature, and they should be
generated by some {\it necessary} for the very existence of the object {\it
internal} energy-momentum redistribution during evolution (recall point 6 of
the above stated 10 properties of a physical object). In the second case the
admissible changes $D\Psi$, in addition to the intrinsic factors, depend also
on external factors, so, the corresponding projections $P(D\Psi)$ should
describe explicitly or implicitly some energy-momentum exchange with the
environment.

Following this line of considerations we come to a conclusion that every
description of a free physical object must include some mathematical
expression of the kind $\mathbb{F}(\Psi,D\Psi;S)=0$, specifying (through the
additional quantities $S$) what and how changes, and specifying also what is
projected and how it is projected.  If the object is not free but it survives
when subject to the external influence, then it is very important the
quantity $\mathbb{F}(\Psi,D\Psi;S)\neq 0$ to have, as much as possible,
universal character and to present a change of a conservative quantity, so
that this same quantity to be expressible through the characteristics
$\mathcal{F}$ of the external object(s).  Hence, specifying differentially
some conservation/balance properties of the object under consideration, and
specifying at every space-time point the corresponding admissible exchange
process with the environment through the equation $\mathbb{F}(\Psi,D\Psi;Q)=
\mathbb{G}(\mathcal{F},d\mathcal{F};\Psi,D\Psi,...)$
we obtain corresponding equations of motion being consistent with the
corresponding integral conservation properties.

In accordance with these considerations it seems reasonable to note also that
this two-sided {\it change-conservation} nature of a physical object could be
mathematically interpreted in terms of the {\it
integrability-nonintegrability} properties of a system of PDE. In fact,
according to the Frobenius integrability theorems (see Sec.9), integrability
means that every Lie bracket of two vector fields of the corresponding
differential system is linearly expressible through the vector fields
generating the very system, so, the corresponding curvature is zero and {\it
nothing flows out of the system}.  From physical point of view this would
mean that such physical objects are difficult to study, because, strictly
speaking, they seem to be strictly isolated, and, therefore, unobservable.
When subject to appropriate external perturbation they lose isolation, so
they are no more integrable, and have to transform to other physical objects,
which demonstrate {\it new integrability} properties.

On the contrary, nonintegrability means nonzero curvature, so there are {\it
outside directed Lie brackats} $[X,Y]$ (called usually {\it vertical
projections}) of the generators of the differential system, and the
correspondig flows of these Lie brackets carry the points of the expected
integral manifold out of it, so, {\it something is possible to flow out of the
system}.  These sticking out Lie brackets look like system's tools, or
"tentakles", which carry out the necessary for its very existence contacts
with the rest of the world. We can say that namely these "nonintegrability"
features of the physical system makes it accessible to be studied.
Figuratively speaking, "the nonintegrability properties protect and support
the integrability properties" of the system, so these both kind of properties
are important for the system's existence, and their
coavailability/coexistence demonstrates the dual nature of physical systems.

We note that this viewpoint assumes somehow, that for an accessible to study
time-stable physical system it is in principle expected both, appropriate
{\it integrable} and {\it nonintegrable } differential/Pfaff systems to be
associated. And this fits quite well with the structure of a fiber bundle:
as we know, {\it the vertical subbundle} of a fiber bundle {\it is always
integrable} with integral manifolds the corresponding fibers; {\it a
horizontal subbundle} chosen {\it may carry nonzero curvature}, and this
curvature characterizes the system's properties that make possible some
local studies to be performed.

We recall that this idea of {\it change-conservation} has been used firstly
by Newton in his momentum balance equation $\dot{\mathbf{p}}=\mathbf{F}$
which is the restriction of the nonlinear partial differential system
$\nabla_{\mathbf{p}}\mathbf{p}=\mathrm{m}\mathbf{F}$, or
$\mathbf{p}^i\nabla_i \mathbf{p}^j=\mathrm{m}\mathbf{F}^j$, on some
trajectory.  This Newton's system of equations just says that there are
physical objects in Nature which admit the "point-like" approximation, and
which can exchange energy-momentum with "the rest of the world" but keep
unchanged some other of their properties, and this allows these objects to be
identified in space-time and studied as a whole, i.e. as point-like ones. The
corresponding "tentakles" of these objects seem to carry out contacts with the
rest of the world by means of energy-momentum exchange.

In macrophysics, as a rule, the external influences are such that they do NOT
destroy the system, and in microphysics a full restructuring is allowed: the
old ingredients of the system may fully transform to new ones, (e.g.  the
electron-positron annihilation) provided the energy-momentum conservation
holds. The essential point is that {\it whatever the interaction is, it
always results in appearing of relatively stable objects, carrying
energy-momentum and some other particular physically measurable quantities}.
This conclusion emphasizes once again the importance of having an adequate
notion of what is called a physical object, and of its appropriate
mathematical representation.

Modern science requires a good adequacy between the real objects and the
corresponding mathematical model objects.  So, the mathematical model
objects $\Psi$ must necessarily be spatially finite, and even temporally
finite if the physical object considered has by its intrinsic nature finite
life-time.  This most probably means that $\Psi$ must satisfy nonlinear
partial differential equation(s), which should define in a consistent way the
{\it admissible} changes and the {\it conservation} properties of the object
under consideration. Hence, talking about physical objects we mean
time-stable spatially finite entities which have a well established balance
between change and conservation, and this balance is kept by a permanent and
strictly fixed interaction with the environment.

This notion of a (finite continuous) physical object sets the problem to try
to consider and represent its integral characteristics through its local
dynamical properties, i.e. to try to understand its nature and integral
appearance as determined and caused by its dynamical structure.  The integral
appearance of the local features may take various forms, in particular, it
might influence the spatial structure of the object. In view of this, the
propagational behavior of the object as a whole, considered in terms of its
local translational and rotational components of propagation, which, in turn,
should be related to the internal energy-momentum redistribution during
propagation, could be stably consistent only with some distinguished spatial
structures. We note that the local rotational components of propagation NOT
always produce integral rotation of the object, but if they are available,
time stable, and consistent with the translational components of propagation,
some specific conserved quantity should exist.  And if the rotational
component of propagation shows some consistent with the object's spatial
structure periodicity, clearly, the corresponding frequency may be used to
introduce such a quantity.

Finally we note that the rotational component of propagation of
a (continuous finite) physical object may be of two different origins: {\it
relative} and {\it intrinsic}. In the "relative" case the corresponding
physical quantity, called {\it angular momentum}, depends on the
choice of some {\it external} to the system factors, usually these are {\it
relative axis} and {\it relative point}. In the "intrinsic" case the
rotational component (if it is not zero) is meant to carry intrinsic
information about the system considered, so, the corresponding physical
quantity, usually called {\it spin}, should NOT depend on any external
factors.

As an idealized (mathematical) example of an object as outlined above, let's
consider a spatial region  $D$ of one-step piece of a helical cylinder with
some proper (or internal) diameter $r_o$, and let $D$ be winded around some
straightline axis $Z$. Let at some (initial) moment $t_o$ our mathematical
model-object $\Psi$ be different from zero only inside $D$.  Let now at
$t>t_o$ the object $\Psi$, i.e. the region $D$, begin moving as a whole along
the helical cylinder with some constant along $Z$ (translational) velocity
$c$ in such a way that every point of $D$ follows its own (helical)
trajectory around $Z$ and never crosses the (helical) trajectory of any other
point of $D$.  Obviously, the rotational component of propagation is
available, but the object does NOT rotate as a whole.  Moreover, since the
translational velocity $c$ along $Z$ is constant, the spatial periodicity
$\lambda$, i.e.  the height of $D$ along $Z$, should be proportional to the
time periodicity $T$, and for the corresponding frequency $\nu=1/T$ we obtain
$\nu=c/\lambda$. Clearly, this is an idealized example of an object with a
space-time consistent dynamical structure, but any physical interpretation
would require to have explicitly defined $D$, $\Psi$, corresponding dynamical
equations, local and integral conserved quantities, $\lambda$, $c$ and,
probably, some other parameters.  We shall show that in the frame of EED such
finite solutions of photon-like nature do exist and can be explicitly
written (see the pictures on pp.62-63).

\newpage

\subsection{Notes on Faraday-Maxwell Electrodynamics}
The 19th century physics, due mainly to Faraday and Maxwell, created the
theoretical concept of {\it classical field} as a model of a class of
spatially continuous (extended) physical objects having dynamical structure.
As a rule, the corresponding {\it free fields} are considered to satisfy
linear dynamical equations, and the corresponding time dependent solutions
are in most cases {\it infinite}. The most important examples seem to be the
free electromagnetic fields, which satisfy the charge-free Maxwell equations.
The concepts of {\it flux of a vector field through a 2-dimensional surface}
and {\it circulation of a vector field along a closed curve} were coined and
used extensively.  Maxwell's equations in their integral form establish where
the time-changes of the fluxes of the electric and magnetic fields go to, or
come from, in both cases of a closed 2-surface and a 2-surface with a
boundary. We note that these fluxes are specific to the continuous character
of the physical object under consideration and it is important also to note
that {\bf Maxwell's field equations have not the sense of direct
energy-momentum balance relations as the Newton's law $\dot{\mathbf
p}={\mathbf F}$ has}.  Nevertheless, they are consistent with energy-momentum
conservation in finite regions, as it is well known, but they impose much
stronger requirements on the field components, so,  the class of admissible
solutions is much more limited than the local conservation equations would
admit.  Moreover, from qualitative point of view, the electromagnetic
induction experiments show that a definite quantity of the field
energy-momentum is transformed to mechanical energy-momentum carried away by
the charged particles, but, on the contrary, the corresponding differential
equation does not allow such energy-momentum transfer (see further).  Hence,
the introduced by Newton basic approach to derive dynamics from local
energy-momentum balance equations had been left off, and this step is still
respected today.

Talking about description of {\it free} fields by means of differential
equations we mean that these equations must describe locally the {\it
intrinsic} dynamics of the field, so NO boundary conditions or any other
external influence should be present.  The pure field Maxwell equations
(although very useful for considerations in finite regions with boundary
conditions) have time-dependent vacuum solutions (i.e free-field solutions,
so, solutions in the whole space) that give an inadequate description of the
real free fields.  As a rule, if these solutions are time-stable, they occupy
the whole 3-space or an infinite subregion of it, and they do not go to zero
at infinity, hence, they carry {\it infinite} energy and momentum.  As an
example we recall the (transverse with zero invariants) plane wave solution,
given by the electric and magnetic fields of the form
\[
{\mathbf E}=\Bigl\{ u(ct+\varepsilon z),
p(ct+\varepsilon z), 0\Bigr\} ;\ {\mathbf B}=\Bigl\{\varepsilon
p(ct+\varepsilon z), -\varepsilon u(ct+\varepsilon z), 0\Bigr\} ,\
\varepsilon=\pm1,
\]
where $u$ and $p$ are arbitrary differentiable
functions. Even if $u$ and $p$ are soliton-like with respect to the
coordinate $z$, they do not depend on the other two spatial coordinates
$(x,y)$. Hence, the solution occupies the whole $\mathbb{R}^3$, or its
infinite subregion, and clearly it carries infinite integral energy
\[
E=\int_{\mathbb{R}^3}(u^2+p^2)dxdydz=\infty.
\]
In particular, the popular harmonic plane wave
$$
u=U_o{\rm cos}(\omega t\pm k_z.z),\quad
p=P_o {\rm sin}(\omega t\pm k_z.z),\quad k_z^2=\omega^2,\quad
U_o=const,\quad P_o=const,
$$
clearly occupies the whole 3-space, carries infinite energy
$$
E=\int_{\mathbb{R}^3}(U_o^2+P_o^2)dxdydz =\infty
$$
and, therefore, could hardly be a model of a really created field, i.e. of
really existing field objects.

We recall also that, according to Poisson's theorem for the D'Alembert wave
equation (which equation is necessarily satisfied by every component of
$\mathbf{E}$ and $\mathbf{B}$ in the pure field case), a spatially
finite and smooth enough initial field configuration
is strongly time-unstable [1]:  the initial condition blows up radially and
goes to infinity.  Hence, Maxwell's equations {\it cannot describe spatially
finite and time-stable time-dependent free field configurations with
soliton-like behavior}.  The contradictions between theory and experiment
that became clear at the end of the 19th century were a challenge to
theoretical physics.  Planck and Einstein created the notion of {\it
elementary field quanta}, named later by Lewis [2] the {\it photon}. The
concept of {\it photon} proved to be adequate enough and very seminal, and
has been widely used in the 20th century physics.  However, even now, after a
century, we still do not have a complete and satisfactory self-consistent
theory of single (or individual) photons.  It is worth recalling at this
point the Einstein's remarks of dissatisfaction concerning the linear
character of Maxwell theory which makes it not able to describe the
microstructure of radiation [3].  Along this line we may also note here some
other results, opinions and nonlinearizations [4]-[14].

According to the non-relativistic formulation of Classical Electrodynamics
(CED) the electromagnetic field has two aspects: {\it electric} and {\it
magnetic}. These two aspects of the field are described by two vector
fields, or by the corresponding through the Euclidean metric 1-forms, on
$\mathbb{R}^3$ :  the electric field ${\bf E}$ and the magnetic field ${\bf
B}$, and a parametric dependence of ${\bf E}$ and ${\bf B}$ on time is
admitted.  Maxwell's equations read (with $\mathbf{j}$ - the electric current
and in Gauss units)

\begin{equation}
\mathrm{rot}\,\mathbf{B}-\frac 1c \frac{\partial \mathbf{E}}{\partial t}=
\frac {4\pi}{c}{\bf j}, \quad \mathrm{div}\,\mathbf{E}=4\pi\rho, 
\end{equation}
\begin{equation}
\mathrm{rot}\,\mathbf{E}+\frac 1c \frac{\partial \mathbf{B}}{\partial t}=0,
\quad \mathrm{div}\,\mathbf{B}=0.                            
\end{equation}
The pure field equations are
\begin{equation}
\mathrm{rot}\,\mathbf{B}-\frac 1c \frac{\partial {\mathbf E}}{\partial t}=0,
\quad \mathrm{div}\,\mathbf{E}=0, 
\end{equation}
\begin{equation}
\mathrm{rot}\,\mathbf{E}+\frac 1c \frac{\partial {\bf B}}{\partial t}=0,
\quad \mathrm{div}\,\mathbf{B}=0.  
\end{equation}
The first of equations (2) is meant to represent mathematically the results
of electromagnetic induction experiments. But, it does NOT seem to have done
it quite adequately. In fact, as we mentioned earlier, these experiments and
studies show that {\it field energy-momentum is transformed to mechanical
energy-momentum}, so that the corresponding material bodies (magnet needles,
conductors, etc.) change their state of motion.  The vector equation (2),
which is assumed to hold inside all media, including the vacuum, manifests
different feature, namely, it gives NO information about any field
energy-momentum transformation to mechanical one in this way:  the time
change of $\mathbf{B}$ is compensated by the spatial nonhomogenity of
$\mathbf{E}$, in other words, this is an intrinsic dynamic property of the
field and it has nothing to do with any direct energy-momentum exchange with
other physical systems.  This is quite clearly seen in the relativistic
formulation of equations (2) and the divergence of the stress-energy-momentum
tensor (see further eqns. (8)-(9)).

This interpretation of the electromagnetic induction experiments has had a
very serious impact on the further development of field theory. The dual
vector equation (3), introduced by Maxwell through the so called
"displacement current", completes the time-dependent differential description
of the field inside the frame of CED. The two additional equations
$\mathrm{div}\,\mathbf{B}=0,\ \mathrm{div}\,\mathbf{E}=0$ impose restrictions
on the admissible spatial configurations (initial conditions, spatial shape)
in the pure field case.  The usual motivation for these last two equations is
that there are no magnetic charges, and NO electric charges are present
inside the region under consideration. This interpretation is also much
stronger than needed: a vector field may have non-zero divergence even if it
has no singularities, i.e. points where it is not defined, e.g.  the vector
field $(\mathrm{sech}(x), 0, 0)$ has NO singularities and is not
divergence-free.  The approximation for continuously distributed electric
charge, i.e.  $\mathrm{div}\,\mathbf{E}=4\pi\rho$, does not justify the
divergence-free case $\rho=0$, on the contrary, it is consistent with the
latter, since NO non-zero divergences are allowed outside of the continuous
charge distribution.

All these clearly unjustified by the experiment and additionally tacitly
imposed requirements lead, in our opinion, to an inadequate mathematical
formulation of the observational facts: as we mentioned above, a mathematical
consequence of (3)-(4) is that every component $\Phi$ of the field {\it
necessarily} satisfies the wave D'Alembert equation $\Box \Phi =0$, which is
highly undesirable in view of the required by this equation "blow up" of
the physically sensible initial conditions (see the comments in the next
section).

Another feature of the vacuum Maxwell system (3)-(4), which brings some
feeling of dissatisfaction (especially to the mathematically inclined
physicists) is the following.  Let's consider $\mathbf{E}$ and
$\mathbf{B}$ as 1-forms on the manifold $\mathbb{R}^4$ with coordinates
$(x,y,z,\xi=ct)$. It seems natural to expect that every solution to (3)-(4)
would define completely integrable Pfaff system $(\mathbf{E},\mathbf{B})$.
For example, the plane wave solution given above defines such a completely
integrable Pfaff system, which is easy to check through verifying the
relations
\[
\mathbf{d}\mathbf{E}\wedge\mathbf{E}\wedge\mathbf{B}=0,\quad
\mathbf{d}\mathbf{B}\wedge\mathbf{E}\wedge\mathbf{B}=0,
\]
with the plane wave $\mathbf{E}$ and $\mathbf{B}$. Now, if
we change slightly this plane wave $\mathbf{E}$ to
$\mathbf{E'}=\mathbf{E}+const.dz$, $const\neq 0$, and keep the same plane
wave $\mathbf{B}$, the new couple $(\mathbf{E'},\mathbf{B})$ will give again
vacuum solution, but will no more define completely integrable Pfaff system,
since we obtain
$$
\mathbf{d}\mathbf{E'}\wedge\mathbf{E'}\wedge\mathbf{B}=
-\frac12const.(u^2+p^2)_\xi dx\wedge dy\wedge dz\wedge d\xi,
$$
$$
\mathbf{d}\mathbf{B}\wedge\mathbf{E'}\wedge\mathbf{B}=
const.(pu_\xi-up_\xi)dx\wedge dy\wedge dz\wedge d\xi.
$$
These expressions are different from zero in the general case, i.e. for
arbitrary functions $u$ and $p$.

As for the charge-present case, i.e. equations (1)-(2), some
dissatisfaction of a quite different nature arises. In fact, every
mathematical equation $A=B$ means that $A$ and $B$ are different notations
for the same element, in other words, it means that on the two sides of "="
stays {\bf the same mathematical object}, but expressed in two different
forms. So, every physically meaningful equation {\it necessarily} implies
that {\it the physical nature of $A$ and $B$ {\bf must} be the same}. For
example, the Newton's law $\dot{\mathbf{p}}=\mathbf{F}$ equalizes {\bf the
same change of momentum} expressed in two ways:  through the characteristics
of the particle, and through the characteristics of the external field, and
this energy-momentum exchange does not destroy neither the particle nor the
field. Now we ask: which physical quantity, that is well defined as a local
characteristic of the field $F_{\mu\nu}$, as well as a local characteristic
of the charged particles, is represented simultaneously through the field's
derivatives as $\delta F_\mu$, and through the characteristics of
the charged particles as $j_\mu$? All known experiments show that the field
$F_{\mu\nu}$, considered as a physical object, can NOT carry electric charge,
only mass objects may be carriers of electric charge. So, from this
(physical) point of view $\delta F_\mu$ should NEVER be made equal to
$j_\mu$, these two quantities have different physical nature and can NOT be
equivalent. Equations (1) may be considered as acceptable only if they
formally, i.e.  without paying attention to their physical nature, present
some sufficient condition for another properly from physical point of view
justified relation to hold.  For example, the relativistic version of (1) is
given by $\delta F_\mu=4\pi j_\mu$, and it is sufficient for the relation
$F_{\mu\nu}\delta F^\nu=4\pi F_{\mu\nu}j^\nu$ to hold, but both sides of this
last equation have the same physical nature (namely, energy momentum change),
therefore, it seems better justified and more reliable.  Note, however, that
an admission of this last equation means that the very concept of
electromagnetic field, {\it considered as independent physical object}, has
been DRASTICALLY changed (extended), e.g.  it propagates NO MORE with the
velocity of light which is a {\it Lorentz-invariant property}. In our view, in
such a case, any solution $F$ represents just one aspect, the electromagnetic
one, of a very complicated physical system and should not be considered as
electromagnetic field in the proper sense, i.e. as a physical object.

The important observation at such a situation is that the adequacy of Maxwell
description is established experimentally NOT directly through the field
equations, BUT indirectly, mainly through the corresponding Lorentz force,
i.e. through the local energy-momentum consevation law.  Therefore, it seems
naturally to pass to a direct energy-momentum description, i.e to extend
Maxwell equations to energy-momentum exchange equations, and in this way to
overcome the above mentioned physically qualitative difference between
$\delta F_\mu$ and $j_\mu$.

In this paper we present a concise review of the newly developed {\it
Extended Electrodynamics} (EED) [15]-[20], which was built in trying to find
such a reasonable nonlinear extension of Maxwell equations together with
paying the corresponding respect to the {\bf soliton-like view} on the
structure of the electromagnetic field.  In the free field case EED follows
the rules:

	1. NO new objects should be introduced in the new equations.

	2. The Maxwell energy-momentum quantities should be kept the same,
and the local conservation relations together with their symmetry (see
further relation (7)) should hold.

\noindent

So, the first main goal of EED is, of course, to give a consistent field
description of {\it single time-stable and spatially finite free-field
configurations that show a consistent translational-rotational propagation as
their intrinsic property}. As for the electromagnetic fields interacting
continuously with some classical medium, i.e. a medium carrying continuous
{\it bounded} electric charges and magnetic moments, and having abilities to
{\it polarize} and {\it magnetize} when subject to external electromagnetic
field, as we mentioned above, our opinion is that the very concept of
electromagnetic field in such cases is quite different. The observed
macro-phenomena concerning properties and propagation may have very little
to do with those of the free fields. We are not even sure that such fields
may be considered as physical objects (in the sense considered in
Sec.1.1 to this paper), and to expect appearance of some kind of {\it
proper} dynamics. The very complicated nature of such systems seems hardly to
be fully described at elementary level, where at every space-time
micro-region so many "annihilation-creation" elementary acts are performed,
the exchanged energy-momentum acquires mechanical degrees of freedom, the
entropy change becomes significant, some reorganization of the medium may
take place and new, not electromagnetic, processes may start off, etc. So,
the nature of the field-object that survives during these continuous
"annihilation-creation" events is, in our view, very much different from the
one we know as electromagnetic free field; the corresponding solution $F$ may
be considered just as a seriously transformed image of the latter, and it is
just one component of a many-component system, and, in some cases, even not
the most important one.  Shortly speaking, the free field solutions are meant
to represent physical objects, while the other solutions represent just some
aspects of a much more complicated physical system.  In our view, only some
macro-picture of what really happens at micro-level we should try to obtain
through solving the corresponding equations. And this picture could be the
closer to reality, the more adequate to reality are our free field notions
and the energy-momentum balance equations. That's why the emphasis in our
approach is on the internal energy-momentum redistribution in the free field
case.

As for the 4-potential approach to Maxwell equations we recall that in the
vacuum case the two equations $\mathbf{d}F=0, \mathbf{d}*F=0$, considered on
topologically non-trival regions, admit solutions having NO 4-potentials. For
example, the field of a point source is given by (in spherical coordinates
originating at the source) $F=(q/r^2)dr\wedge d(ct),
*F=q\,sin\theta d\theta\wedge d\varphi$ and is defined on $S^2\times
\mathbb{R}^2$.  Now the new couple $\mathcal{F}=aF+b*F, \mathcal{*F}=-bF+a*F$
with $a,b=const.$ is a solution admitting NO 4-potential on
$S^2\times\mathbb{R}^2$. This remark suggests that the 4-potential does not
seem to be a fundamental quantity.

Turning back to the free field case we note that, in accordance with the
soliton-like view, which we more or less assume and shall try to follow,
the components of the field at every moment $t$ have to be represented by
{\bf smooth}, i.e.  nonsingular, functions, different from zero, or
concentrated, only inside a finite 3-dimensional region
$\Omega_t\subset\mathbb{R}^3$, and this must be achieved by the
nonlinearization assumed. Moreover, we have to take care for the various
spatial configurations to be admissible, so, it seems desirable the
nonlinearization to admit time-stable solutions with {\it arbitrary} initial
spatial shape. Finally, the observed space-time periodical nature of the
electromagnetic radiation (the Planck's formula $W=h\nu$ in case of photons)
requires the new nonlinear equations to be able to describe such a periodical
nature together with the corresponding frequency dependence during
propagation.

The above remarks go along with Einstein's view that "the whole theory must
be based on partial differential equations and their singularity-free
solutions" [21].

\vskip 1cm
\section {From Classical Electrodynamics to Extended Electrodynamics}
\vskip 0.5cm

The above mentioned idea, together with the assumption for a different
interpretation of the concept of field: first, considered as independent
physical object (field in vacuum), and second, considered in presence of
external charged matter, is strongly supported by the fact that Maxwell's
equations together with the Lorentz force give very good and widely accepted
expressions for the energy density $\bf w$, the momentum density (the
Poynting vector) ${\bf \vec{s}}$, and for the intra-field energy-momentum
exchange during propagation - the Poynting equation:
\begin{equation}
{\bf w}=\frac{1}{8\pi}({\bf E}^2+{\bf B}^2),\quad {\bf
\vec{s}}=\frac{c}{4\pi}({\bf E}\times {\bf B}),\quad                 
\frac{\partial {\bf w}}{\partial t}=-\mathrm{div}\,{\bf \vec{s}}.
\end{equation}

Let's consider now the D'Alembert wave equarion $\square \Phi=0$. A solution
of this equation, meant to describe a finite field configuration with
arbitrary spatial structure and moving as a whole along some spatial
direction, which we choose for $z$-coodinate, would look like as
$\Phi(x,y,z+\varepsilon ct), \varepsilon=\pm 1$, where $\Phi$ is a
bounded finite with respect to $(x,y,z)$ function. Substituting this $\Phi$
in the equation, clearly, the second derivatives of $\Phi$ with respect to
$z$ and $t$ will cancel each other, so, the equation requires $\Phi$ to be a
harmonic function with respect to $(x,y):  \Phi_{xx}+\Phi_{yy}=0$.  According
to the Liouville theorem in the theory of harmonic functions on
$\mathbb{R}^2$ if a harmonic on the whole $\mathbb{R}^2$ function is bounded
it must be constant. Therefore, $\Phi$ shall not depend on
$(x,y)$, i.e.  $\Phi=\Phi(z+\varepsilon ct)$.

Considering now the Poynting equation we'll find that it really {\bf admits}
spatially finite solutions with soliton-like propagation along the direction
chosen.  In fact, it is immediately verified that if in the above given plane
wave expressions for $\mathbf{E}$ and $\mathbf{B}$ we allow arbitrary
dependence of $u$ and $p$ on the three spatial coordinates $(x,y,z)$, i.e.
$u=u(x,y,\xi+\varepsilon z),\ p=p(x,y,\xi+\varepsilon z)$, then the
corresponding $\mathbf{E}$ and $\mathbf{B}$ will satisfy the Poynting
equation and will NOT satisfy Maxwell's vacuum equations. So, the Poynting
equation is able to describe collections of co-moving photons, which may be
considered classically as spatially finite radiation pulses with photon-like
behavior. Such finite pulses do not "blow up" radially, so, this suggests to
look for new dynamical equations which also must be consistent with the
Maxwell local conservation quantities and laws.  Hence, if the equations we
look for should be consistent with the photon structure of electromagnetic
radiation, i.e. with the view that any EM-pulse consists of a collection of
{\it real spatially finite objects, and each one propagates as a whole in
accordance with the energy-momentum that it has been endowed with in the
process of its creation}, we have to find new field equations.

We begin now the process of extension of Maxwell equations. First,
we specially note the well known invariance, usually called {\it
electric-magnetic duality}, of the pure field equations (3)-(4) with respect
to the transformation
\begin{equation}
({\bf E},{\bf B})\rightarrow ({\bf -B},{\bf E}):                    
\end{equation}
the first couple (3) is transformed into the second one (4), and vice versa.
Hence, instead of assuming equations (4) we could require invariance of (3)
with respect to the transformation (6).  Moreover, the energy density ${\bf
w}$, the Poynting vector ${\bf \vec{s}}$, as well as the Poynting relation
are also invariant with respect to this transformation. This symmetry
property of Maxwell equations we consider as a {\it structure one} and we'll
make a serious use of it.

The above symmetry transformation (6) is extended to the more general duality
transformation
\begin{equation}
(\mathbf{E},\mathbf{B})\rightarrow (\mathbf{E}',\mathbf{B}')          
=(\mathbf{E},\mathbf{B}).\alpha(a,b)
=(\mathbf{E},\mathbf{B})\begin{Vmatrix} a & b \\ -b & a \end{Vmatrix}
=(a\mathbf{E}-b\mathbf{B}, b\mathbf{E}+a\mathbf{B}), \
a=const, \ b=const.
\end{equation}
Note tnat we assume a "right action" of the matrix $\alpha$ on the
couple $(\mathbf{E},\mathbf{B})$. The couple $(\mathbf{E}',\mathbf{B}')$
gives again a solution to Maxwell vacuum equations, the new energy density
and Poynting vector are  equal to the old ones multiplied by $(a^2+b^2)$, and
the Poynting equation is satisfied of course.  Hence, the space of all vacuum
solutions factors over the action of the group of matrices of the kind
\[
\alpha(a,b)= \begin{Vmatrix} a & b \\ -b & a \end{Vmatrix},
\quad (a^2+b^2)\neq 0.
\]
All such matrices with nonzero determinant form a group with respect to the
usual matrix product and further this group will be denoted by $\mathbb{G}$.

This invariance is very important from the point of view of passing to
equations having direct energy-momentum exchange sense. In fact, in the
vacuum case the Lorentz force
$$
F=\rho\mathbf{E}+\frac1c(\mathbf{j}\times\mathbf{B})=
\frac{1}{4\pi}\left[\mathbf{E}\,\mathrm{div}\mathbf{E}+
\left(
\mathrm{rot}\,\mathbf{B}-\frac 1c \frac{\partial \mathbf{E}}{\partial t}
\right)\times\mathbf{B}\right]
$$
should be equal to zero for every solution, so, the transformed
$(\mathbf{E}',\mathbf{B}')$ must give also zero Lorentz force for every couple
$(a,b)$. This observation will bring us later to the new vacuum equations.

In the relativistic formulation of CED the difference between the electric
and magnetic components of the field is already quite conditional, and from
the invariant-theoretical point of view there is no any difference. However,
the 2-aspect (or the 2-component) character of the field is kept in a new
sense and manifests itself at a different level. To show this we go to
Minkowski space-time $M$, spanned by the standard coordinates $x^\mu=(x,y,
z,\xi=ct)$ with the pseudometric $\eta$ having components
$\eta_{\mu\mu}=(-1,-1,-1,1)$, and $\eta_{\mu\nu}=0$ for
$\mu\neq \nu$ in these
coordinates. We shall use also the Hodge star operator, defined by
$$
\alpha\wedge *\beta=-\eta (\alpha,\beta)\sqrt{|det(\eta_{\mu\nu})|}
dx\wedge dy\wedge dz\wedge d\xi,
$$
where $\alpha$ and $\beta$ are differential forms on $M$. With this
definition of $*$ we obtain for a 2-form $F$ in canonical coordinates
$(*F)_{\mu\nu}=-\frac12 \varepsilon_{\mu\nu\alpha\beta}F^{\alpha\beta}$.
The exterior derivative ${\bf d}$ and the Hodge $*$ combine to give the
coderivative (or divergence) operator $\delta=(-1)^{p}*^{-1}{\bf d}*$, which
in our case reduces to $\delta=*{\bf d}*$, and on 2-forms we have
$(\delta F)_\mu=-\nabla_\nu F^{\nu}\,_{\mu}$. Note that, reduced on 2-forms,
the Hodge $*$-operator induces a complex structure:
$**_2=-id_{\Lambda^2(M)}$, and the same property has transformation (6).
Explicitly, $F$ and $*F$ look as follows:
$$
F_{12}=\mathbf{B}_3,\ F_{13}=-\mathbf{B}_2,\ F_{23}=\mathbf{B}_1,
\ F_{14}=\mathbf{E}_1,\ F_{24}=\mathbf{E}_2,\ F_{34}=\mathbf{E}_3.
$$
$$
(*F)_{12}=\mathbf{E}_3,\ (*F)_{13}=-\mathbf{E}_2,\ (*F)_{23}=\mathbf{E}_1,
\ (*F)_{14}=-\mathbf{B}_1,\ (*F)_{24}=-\mathbf{B}_2,\
(*F)_{34}=-\mathbf{B}_3.
$$
So, if the non-relativistic vector couple $({\bf E},{\bf B})$ corresponds to
the relativistic 2-form $F$, then the transformed by transformation (6)
couple $({\bf -B},{\bf E})$ obviously corresponds to the relativistic 2-form
$*F$. Hence, the Hodge $*$ is the relativistic image of transformation (6).

Recall now the relativistic energy-momentum tensor, which can be
represented in the following two equivalent forms (see the end of
this section):
\begin{equation}
Q_\mu^\nu=\frac {1}{4\pi}\biggl[\frac 14 F_{\alpha\beta}F^
{\alpha\beta}\delta_\mu^\nu-F_{\mu\sigma}F^{\nu\sigma}\biggr]=
\frac {1}{8\pi}\biggl[-F_{\mu\sigma}F^{\nu\sigma}-                  
(*F)_{\mu\sigma}(*F)^{\nu\sigma}\biggr].
\end{equation}
It is quite clearly seen, that $F$ and $*F$ participate in the same way in
$Q_\mu^\nu$, so $Q_\mu^\nu$ is invariant with respect to $F\rightarrow *F$
and the full energy-momentum densities of the field are obtained
through summing up the energy-momentum densities, carried by $F$ and $*F$.
That is how the above mentioned invariance of the energy density and Poynting
vector look like in the relativistic formalism. Now, the Poynting relation
corresponds to the zero-divergence of $Q_\mu^\nu$:
\begin{equation}
\nabla_\nu Q_\mu^\nu=\frac {1}{4\pi}\biggl[F_{\mu\nu}(\delta        
F)^\nu+(*F)_{\mu\nu}(\delta *F)^\nu\biggr]=0.
\end{equation}
The obvious invariance of the Poynting relation with respect to (6)
corresponds here to the invariance of (9) with respect to the Hodge $*$.
Moreover, the above relation (9) obviously suggests that the field is
potentially able to exchange energy-momentum through $F$, as well as through
$*F$ independently, provided appropriate external field-object is present.

Now, the idea of the generalization CED$\rightarrow$EED could be formulated
as follows. The above mentioned momentum balance
interpretation of the second Newton's law $\dot{\bf p}={\bf F}$ says: the
momentum gained by the particle is lost by the external field, this is the
{\it true} sense of this relation. If there is no external field, then ${\bf
F}=0$, and giving the explicit form of the particle momentum ${\bf p}=m{\bf
v}=m\dot{\bf r}$ we obtain the equations of motion $\dot{\mathbf{p}}=m\dot
{\bf v}=0$ of a free particle, implying $\dot{m}=0$.  The same argument
should work with respect to the field:  if there is no particles then $m=0$,
$\mathbf{p}=0$ and $\dot{\bf p}=0$, so, we must obtain the equations of
motion (at least one of them) for the free field:  ${\bf F}=0$, after we give
the explicit form of the dependence of ${\bf F}$ on the field functions and
their derivatives.  In other words, {\bf we have to determine in terms of the
field functions and their derivatives how much energy-momentum the field is
potentially able to transfer to another physical system in case of the
presence of the latter and its ability to interact with the field, i.e. to
accept the corresponding energy-momentum}.

Following this idea in non-relativistic terms we have to compute the
expression
\[
\frac{d}{dt}\left[P_{mech}+\vec{\mathbf{s}}\right]
\]
making use of Maxwell equations (1)-(2), and to put the result equal to zero.
In the free field case $\frac{d}{dt}P_{mech}=0$.  The corresponding
relation may be found in the textbooks
(for example see [22], ch.6, Paragraph 9) and it reads
\begin{equation}
\left(\mathrm{rot}\,\mathbf{B}-\frac{\partial\,\mathbf{E}}{\partial
\xi}\right) \times\mathbf{B}+\mathbf{E}\,\mathrm{div}\,\mathbf{E}+      
\left(\mathrm{rot}\,\mathbf{E}+\frac{\partial\,\mathbf{B}}{\partial
\xi}\right) \times\mathbf{E}+\mathbf{B}\,\mathrm{div}\,\mathbf{B}=0.
\end{equation}
In the frame of Maxwell theory the first two terms on the left of (10)
give the Lorentz force in field terms, so, in the free field case we must
have
\begin{equation}
\left(\mathrm{rot}\,\mathbf{B}-
\frac{\partial \mathbf{E}}{\partial \xi}\right)
\times\mathbf{B}+\mathbf{E}\,\mathrm{div}\,\mathbf{E}=0.               
\end{equation}
Hence, it follows now from (10) and (11) that we must also have
\begin{equation}
\left(\mathrm{rot}\,\mathbf{E}+
\frac{\partial \mathbf{B}}{\partial \xi}\right)
\times\mathbf{E}+\mathbf{B}\,\mathrm{div}\,\mathbf{B}=0.               
\end{equation}
If we forget now about the sourceless Maxwell equations (3)-(4) we may
say that these two {\bf nonlinear} vector equations (11)-(12) describe
locally some {\it intrinsic} energy-momentum redistribution during the
propagation.

In order to have the complete local picture of the intrinsic
energy-momentum redistribution we must take also in view terms of the
following form:
\[
\frac1c\left(\frac{\partial \mathbf{E}}{\partial t}\times \mathbf{E} +
\mathbf{B}\times\frac{\partial \mathbf{B}}{\partial t}\right),\quad
\mathbf{E}\times\mathrm{rot}\mathbf{B},\quad
\mathbf{B}\times\mathrm{rot}\mathbf{E},\quad
\mathbf{E}\mathrm{div}\mathbf{B},\quad
\mathbf{B}\mathrm{div}\mathbf{E}.
\]
A reliable approach to obtain the right expression seems to be the same
invariance with respect to the transformation (7) to hold for (11)-(12). This
gives the additional vector equation:
\begin{equation}
\left(\mathrm{rot}\,\mathbf{E}+
\frac{\partial \mathbf{B}}{\partial \xi}\right)
\times\mathbf{B}-\mathbf{E}\,\mathrm{div}\,\mathbf{B}+                  
\left(\mathrm{rot}\,\mathbf{B}-
\frac{\partial \mathbf{E}}{\partial \xi}\right)
\times\mathbf{E}-\mathbf{B}\,\mathrm{div}\,\mathbf{E}=0.
\end{equation}

The above consideration may be put in the following way.  According to our
approach, the more important aspects of the field from the point of view to
write its proper dynamical equations are its abilities to exchange
energy-momentum with the rest of the world: after finding out the
corresponding quantities which describe quantitatively these abilities we put
them equal to zero.

The important moment in such an approach is to have an
adequate notion of the mathematical nature of the field. The dual symmetry of
Maxwell equations suggests that in the nonrelativistic formalism the
mathematical nature of the field is 1-form $\omega=\mathbf{E}\otimes
e_1+\mathbf{B}\otimes e_2$ on $\mathbb{R}^3$ with values in an appropriate
2-dimensional vector space $\mathcal{G}$ with the basis $(e_1,e_2)$, where
the group $\mathbb{G}$ acts as linear transformations. In fact, if
$\alpha(a,b)\in \mathbb{G}$, then consider the new basis
$(e'_1,e'_2)$ given by
\[
e'_1=\frac{1}{a^2+b^2}(ae_1-be_2),\quad e'_2=\frac{1}{a^2+b^2}(be_1+ae_2).
\]
Accordingly, $\alpha\in\mathbb{G}$ transforms the basis through right
action by means of $(\alpha^{-1})^*=\alpha/det(\alpha)$.  Then the
"new" solution $\omega'(\mathbf{E}',\mathbf{B}')$
is, in fact, the "old" solution $\omega(\mathbf{E},\mathbf{B})$:
\[
\omega'=\mathbf{E}'\otimes e'_1+\mathbf{B}'\otimes e'_2=
(a\mathbf{E}-b\mathbf{B})\otimes \frac{ae_1-be_2}{a^2+b^2}+
(b\mathbf{E}+a\mathbf{B})\otimes \frac{be_1+ae_2}{a^2+b^2}=
\mathbf{E}\otimes e_1+\mathbf{B}\otimes e_2=\omega,
\]
i.e., the "new" solution $\omega'(\mathbf{E}',\mathbf{B}')$, represented in
the new basis $(e'_1,e'_2)$ coincides with the "old" solution
$\omega(\mathbf{E},\mathbf{B})$, represented in the old basis $(e_1,e_2)$.

In view of this we may  consider transformations (7) as {\it nonessential},
i.e. we may consider $(\mathbf{E},\mathbf{B})$ and
$(\mathbf{E'},\mathbf{B'})$ as two different representations in corresponding
 bases of $\mathcal{G}$  of the same solution $\omega$, and the two
bases are connected with appropriate element of $\mathbb{G}$.

Such an interpretation is approporiate and useful if the field shows
some {\it invariant properties} with respect to this class of
transformations. We shall show further that the {\it nonlinear} solutions of
(11)-(13) do have such invariant properties. For example, if the field
has {\it zero invariants} (the so called "null field"):
$$
I_1=(\mathbf{B}^2-\mathbf{E}^2)=0, \quad I_2=2\mathbf{E}.\mathbf{B}=0,
$$
then all transformations (7) keep unchanged these zero-values of $I_1$ and
$I_2$.  In fact, under the transformation (7) the two invariants transform in
the following way:
\[
I_1'=(a^2-b^2)\,I_1+2ab\,I_2,\quad I_2'=-2ab\,I_1+(a^2-b^2)\,I_2,
\]
and the determinant of this transformation is $(a^2+b^2)^2\neq 0$. So, a null
field stays a null field under the dual transformations (7). Moreover, NO
non-null field can be transformed to a null field by means of
transformations (7), and, conversely, NO null field can be transformed to a
non-null field in this way. Further we shall show that the "null fields"
admit also other $\mathbb{G}$-invariant properties being closely connected
with availability of rotational component of propagation.

From this $\mathbb{G}$-{\it covariant} point of view, the well known Lorentz
force appears to be just one component of the corresponding covariant with
respect to transformations (7), or to the group $\mathbb{G}$, {\it
$\mathbb{G}$-covariant Lorentz force}.  So, in order to obtain this
$\mathbb{G}$-covariant Lorentz force we have to replace $\mathbf{E}$ and
$\mathbf{B}$ in the expression
$$
\left(\mathrm{rot}\,\mathbf{B}- \frac{\partial
\mathbf{E}}{\partial \xi}\right)
\times\mathbf{B}+\mathbf{E}\,\mathrm{div}\,\mathbf{E},
$$
by $(\mathbf{E'},\mathbf{B'})$ given by (7). In the vacuum case, i.e. when
the field keeps its integral energy-momentum unchanged, the
$\mathbb{G}$-covariant Lorentz force $\mathbb{F}$ obtained must be zero for
any values $(a,b)$ of the transformation parameters.

After the corresponding computation we obtain
\[
a^2\left[
\left(\mathrm{rot}\,\mathbf{B}- \frac{\partial \mathbf{E}}{\partial
\xi}\right) \times\mathbf{B}+\mathbf{E}\,\mathrm{div}\,\mathbf{E}\right]+
b^2\left[\left(\mathrm{rot}\,\mathbf{E}+
\frac{\partial \mathbf{B}}{\partial \xi}\right)
\times\mathbf{E}+\mathbf{B}\,\mathrm{div}\,\mathbf{B}\right]+
\]
\[
+ab\left[
\left(\mathrm{rot}\,\mathbf{E}+
\frac{\partial \mathbf{B}}{\partial \xi}\right)
\times\mathbf{B}-\mathbf{E}\,\mathrm{div}\,\mathbf{B}+
\left(\mathrm{rot}\,\mathbf{B}-
\frac{\partial \mathbf{E}}{\partial \xi}\right)
\times\mathbf{E}-\mathbf{B}\,\mathrm{div}\,\mathbf{E}\right]=0.
\]
Note that the $\mathbb{G}-covariant$ Lorentz force $\mathbb{F}$ has three
vector components given by the quantities in the brackets. Since the
constants $(a,b)$ are arbitrary the equations (11)-(13) follow.

Another formal way to come to equations (11)-(13) is the following. Consider
a 2-dimensional real vector space $\mathcal{G}$ with basis
$(e_1,e_2)$, and the objects
\[
\Phi=\left(\mathrm{rot}\,\mathbf{E}+\frac{\partial\,\mathbf{B}}{\partial
\xi}\right)\otimes e_1+
\left(\mathrm{rot}\,\mathbf{B}-\frac{\partial\,\mathbf{E}}{\partial
\xi}\right)\otimes e_2,\quad
\omega_1=\mathbf{E}\otimes e_1+\mathbf{B}\otimes e_2
\]
\[
\omega_2=-\mathbf{B}\otimes e_1+\mathbf{E}\otimes e_2,\quad
\theta=-(\mathrm{div}\,\mathbf{B})\otimes e_1+
(\mathrm{div}\,\mathbf{E})\otimes e_2.
\]
Now define a new object by means of the bilinear maps "vector product
$\times$", "symmetrized tensor product $\vee$" and usual product of vector
fields and functions (denoted by ".") as follows:
\[
(\times,\,\vee)\left[
\left(\mathrm{rot}\,\mathbf{E}+\frac{\partial\,\mathbf{B}}{\partial
\xi}\right)\otimes e_1+
\left(\mathrm{rot}\,\mathbf{B}-\frac{\partial\,\mathbf{E}}{\partial
\xi}\right)\otimes e_2,\,\mathbf{E}\otimes e_1+\mathbf{B}\otimes e_2\right]+
\]
\[
(.,\vee)\left[-\mathbf{B}\otimes e_1+\mathbf{E}\otimes e_2,\,
-(\mathrm{div}\,\mathbf{B})\otimes e_1+
(\mathrm{div}\,\mathbf{E})\otimes e_2\right]=
\]
\[
=\left[\left(\mathrm{rot}\,\mathbf{E}+\frac{\partial\,\mathbf{B}}{\partial
\xi}\right)\times\mathbf{E}+\mathbf{B}\mathrm{div}\,\mathbf{B}\right]
\otimes e_1\vee e_1+
\left[\left(\mathrm{rot}\,\mathbf{B}-\frac{\partial\,\mathbf{E}}{\partial
\xi}\right)\times\mathbf{B}+\mathbf{E}\mathrm{div}\,\mathbf{E}\right]
\otimes e_2\vee e_2+
\]
\[\left[
\left(\mathrm{rot}\,\mathbf{E}+
\frac{\partial \mathbf{B}}{\partial \xi}\right)
\times\mathbf{B}-\mathbf{E}\,\mathrm{div}\,\mathbf{B}+
\left(\mathrm{rot}\,\mathbf{B}-
\frac{\partial \mathbf{E}}{\partial \xi}\right)
\times\mathbf{E}-\mathbf{B}\,\mathrm{div}\,\mathbf{E}\right]
\otimes e_1\vee e_2 =\mathbb{F}.
\]
Now, putting this product equal to zero, we obtain the required equations.

The left hand sides of equations (11)-(13), i.e. the components of the
3-dimensional (nonrelativistic) $\mathbb{G}$-covariant Lorentz
force $\mathbb{F}$, define three vector quantities of energy-momentum change,
which appear to be the field's tools controlling the energy-momentum exchange
process during interaction. We can also say that these three terms define
locally the energy-momentum which the field is potentially able to give to
some other physical system, so, the field demonstrates three different
abilities of possible energy-momentum exchange.

Obviously, equations (11)-(13) contain all Maxwell vacuum
solutions, and they also give new, non-Maxwellean
nonlinear solutions satisfying the following inequalities:
\begin{equation}
\left(\mathrm{rot}\,\mathbf{E}+
\frac{\partial \mathbf{B}}{\partial \xi}\right)\neq 0,\quad
\left(\mathrm{rot}\,\mathbf{E}-                                     
\frac{\partial \mathbf{B}}{\partial \xi}\right)\neq 0,\quad
\mathrm{div}\,\mathbf{E}\neq 0,\quad \mathrm{div}\,\mathbf{B}\neq 0.
\end{equation}
It is obvious that in the nonlinear case we obtain immediately from
(11)-(13) the following relations:
\begin{equation}
\left(\mathrm{rot}\,\mathbf{B}-
\frac{\partial \mathbf{E}}{\partial \xi}\right).\mathbf{E}=0,\quad
\left(\mathrm{rot}\,\mathbf{E}+
\frac{\partial \mathbf{B}}{\partial \xi}\right).\mathbf{B}=0,\quad
\mathbf{E}.\mathbf{B}=0,                                              
\end{equation}
and combining with (13) we obtain additionally (in view of
$\mathbf{E}.\mathbf{B}=0$)
\begin{equation}
{\mathbf B}.\left(\mathrm{rot}\mathbf{B}-
\frac{\partial \mathbf{E}}{\partial \xi}\right)-
\mathbf{E}.\left(\mathrm{rot}\mathbf{E}
+\frac{\partial {\mathbf B}}{\partial \xi}\right)=          
\mathbf{B}.\mathrm{rot}\mathbf{B}-
\mathbf{E}.\mathrm{rot}\mathbf{E}=0.
\end{equation}
So, in the nonlinear case, from (15) we obtain readily the Poynting relation,
 and (in hydrodynamic terms) we may say that relation (16) requires the
local helicities $\mathbf{B}.\mathrm{rot}\mathbf{B}$ and
$\mathbf{E}.\mathrm{rot}\mathbf{E}$ of $\mathbf{E}$ and $\mathbf{B}$
[23,24] to be always equal.

In relativistic terms we first recall Maxwell's equations in presence of
charge distribution:
\begin{equation}
\delta *F=0,\quad \delta F=4\pi j .                       
\end{equation}
In the pure field case we have
\[
\delta *F=0,\quad \delta F=0,
\]
or, in terms of {\bf d}
\[
{\bf d}F=0,\quad {\bf d}*F=0.
\]
Recall now from relations (8)-(9) the explicit forms of the
stress-energy-momentum tensor and its divergence.  From Maxwell equations
(17) it follows that field energy-momentum is transferred to another physical
system (represented through the electric current $j$) ONLY through the term
$4\pi F_{\mu\nu}j^\nu=F_{\mu\nu}(\delta F)^{\nu}$ (the relativistic Lorentz
force), since the term $(*F)_{\mu\nu}(\delta *F)^\nu$ is always equal to zero
because of the first of equations (17) which is meant to represent the
Faraday's induction law.

Following the same line of consideration as in the nonrelativistic case
we extend Maxwell's free field equations $\delta F=0,\ \delta *F=0$
as follows. In the field expression of the relativistic Lorentz force
$F_{\mu\nu}\delta F^\nu$ we replace $F$ by $aF+b*F$, so we obtain the
relativistic components of the $\mathbb{G}-covariant$ Lorentz force
$\mathbb{F}$:
\[
a^2 F_{\mu\nu}\delta F^\nu +b^2 (*F)_{\mu\nu}(\delta *F)^\nu
+ab\big[F_{\mu\nu}(\delta *F)^\nu+(*F)_{\mu\nu}\delta F^\nu\big].
\]
In view of the arbitrariness of the parameters $(a,b)$ after putting this
expression equal to zero we obtain equations (11)-(13) in relativistic
notation. The equation
\begin{equation}
F_{\mu\nu}\delta F^\nu=0                                        
\end{equation}
extends the relativistic Maxwell's equation $\delta F=0$ and in
standard coordinates (18) gives equations (11) and the first of (15).
Equations (12) and the second of (15) are presented by the equation
\begin{equation}
(*F)_{\mu\nu}(\delta *F)^\nu=0.                                   
\end{equation}
Looking back to the divergence relation (9) for the energy-momentum tensor
$Q_\mu^\nu$ we see that (18) and (19) require zero values for the two
naturally arising distinguished parts of $\nabla_\nu Q^\nu_\mu$, and so the
whole divergence is zero. This allows $Q_\mu^\nu$ to be assumed as
energy-momentum tensor for the nonlinear solutions.

The relativistic version of (13) and (16) looks as follows:
\begin{equation}
F_{\mu\nu}(\delta *F)^\nu +(*F)_{\mu\nu}\delta F^\nu=0.       
\end{equation}

This relativistic form of the equations (11)-(13) and (15)-(16) goes along
with the above interpretation: equations (18)-(19), together with equation
(9), mean that no field energy-momentum is transferred to any other physical
system through any of the two components $F$ and $*F$, so, only intra-field
energy-momentum redistribution should take place; equation (20) characterizes
this intra-field energy momentum exchange in the following sense:  the
energy-momentum quantity, transferred locally from $F$ to $*F$ and given by
$F_{\mu\nu}(\delta *F)^\nu$,
is always equal to that transferred locally from $*F$ to $F$ which, in turn,
is given by $(*F)_{\mu\nu}(\delta F)^\nu$.

In terms of the "change-conservation" concept considered in Sec.1.1
, we could say that the self-projections
$F_{\mu\nu}(\delta F)^\nu$ and $(*F)_{\mu\nu}(\delta *F)^\nu$
of the admissible changes $\delta F$ and $\delta *F$ on $F$ and $*F$,
respectively, vanish, and the cross-projections
$(*F)_{\mu\nu}(\delta F)^\nu$ and $(F)_{\mu\nu}(\delta *F)^\nu$
have the same absolute value.

So, we can think of the electromagnetic field as a 2-vector component field
$\Omega=F\otimes e_1+*F\otimes e_2$, where $(e_1,e_2)$ is an appropriately
chosen basis of the same real 2-dimensional vector space $\mathcal{G}$,
and later the nature of $\mathcal{G}$ will be specially considered.

In presence of external fields (media), which exchange energy-momentum with
our field $F$, the right hand sides of (18)-(20) will not be zero in general.
The energy-momentum local quantities that flow out of our field through its
two components $(F,*F)$ and are given by the left hand sides of (18)-(20),
in accordance with the local energy-momentum conservation law have to
be absorbed by the external field. Hence, these same quantities have to be
expressed in terms of the field functions of the external field.

Maxwell equations give in the most general case only one such expression,
namely, when the external field is represented by continuously distributed
free and bounded charged particles. This expression reads:
$$
4\pi F_{\mu\nu}j^\nu,\quad j^\nu=j^\nu_{free}+j^\nu_{bound},
$$
where $j^\nu_{free}=\rho\,u^\nu$ is the standard 4-current for freely moving
charged particles, $j^\nu_{bound}$ is expressed through
the polarization $\mathbf{P}$ and magnetization $\mathbf{M}$ vectors of
the (dielectric) medium by
$$
j^{\nu}_{bound}=(\mathbf{j}_{bound}, \rho_{bound})=
\left\{c\,\mathrm{rot}\,\mathbf{M}+\frac{\partial \mathbf{P}}
{\partial t},\ \ -\mathrm{div}\,\mathbf{P}\right\}.
$$
So, Maxwell equations allow NO energy-momentum exchanges through $*F$. The
usual justification for this is the absence of magnetic charges. In our view
this motivation is insufficient and has to be left off.

In the frame of EED, in accordance with the concept of
$\mathbb{G}-${\it covariant Lorentz force}, we reject this limitation, in
general. Energy-momentum exchanges through $F$, as well as, through $*F$ are
allowed.  Moreover, EED does not forbid some media to influence the
energy-momentum transfers between $F$ and $*F$, favoring exchanges with $F$
or $*F$ in correspondence with medium's own structure.  Formally, this means
that the right hand side of (20), in general, may also be different from
zero. Hence, in the most general case, EED assumes the following equations:
\begin{align}
F_{\mu\nu}\delta F^\nu
& =F_{\mu\nu}(\alpha^1)^\nu,\\
(*F)_{\mu\nu}(\delta *F)^\nu
&=(*F)_{\mu\nu}(\alpha^4)^\nu,\\
F_{\mu\nu}(\delta *F)^\nu +(*F)_{\mu\nu}\delta F^\nu
& =F_{\mu\nu}(\alpha^2)^\nu +(*F)_{\mu\nu}(\alpha^3)^\nu.       
\end{align}

The new objects $\alpha^i, i=1,2,3,4$ are four 1-forms, and they represent the
abilities of the corresponding medium for energy-momentum exchange with the
field. Clearly, their number corresponds to the four different abilities of
the field to exchange energy-momentum, which are represented by
$F_{\mu\nu}\delta F^\nu,\ (*F)_{\mu\nu}(\delta *F)^\nu,\
(*F)_{\mu\nu}\delta F^\nu$ and $F_{\mu\nu}(\delta *F)^\nu$.
These four 1-forms shall be expressed by means of the external field functions
and their derivatives. This is a difficult problem and its appropriate
resolution requires a serious knowledge of the external field considered. In
most practical cases however people are interested only in how much
energy-momentum has been transferred, and the transferred energy-momentum is
given by some bilinear functions on $F$ and $\alpha^i, i=1,2,3,4$ as it is in
the Maxwell case.

Here is the 3-dimensional form of the above equations (the bold
$\mathbf{a^i}$ denotes the spatial part of the corresponding $\alpha^i$):
\[
\left({\rm rot}{\bf B}-\frac{\partial {\bf E}}{\partial \xi}\right)\times
{\bf B}+{\bf E}{\rm div}{\bf E}=
{\bf a}^1\times {\bf B}+{\bf E}(\alpha^1)^4,
\]
\[
{\bf E}.\left({\rm rot}{\bf B}-\frac{\partial {\bf E}}{\partial
\xi}\right)={\bf E}.{\bf a}^1,
\]
\[
\left({\rm rot}{\bf E}+\frac{\partial {\bf B}}{\partial \xi}\right)\times
{\bf E}+{\bf B}{\rm div}{\bf B}=
{\bf a}^4\times {\bf E}-{\bf B}(\alpha^4)^4,
\]
\[
{\bf B}.\left({\rm rot}{\bf E}+\frac{\partial {\bf B}}{\partial \xi}\right)=
{\bf B}.{\bf a}^4,
\]
\[
\left({\rm rot}{\bf E}+\frac{\partial {\bf B}}{\partial \xi}\right)\times
{\bf B}+ \left({\rm rot}{\bf B}-\frac{\partial {\bf E}}{\partial
\xi}\right)\times {\bf E}- {\bf B}{\rm div}{\bf E}-{\bf E}{\rm div}{\bf B}=
\]
\[
={\bf a}^2\times {\bf B}+{\bf E}(\alpha^2)^4+{\bf a}^3\times {\bf E}-
{\bf B}(\alpha^3)^4,
\]
\[
{\bf B}.\left({\rm rot}{\bf B}-\frac{\partial {\bf E}}{\partial \xi}\right)-
{\bf E}.\left({\rm rot}{\bf E}+\frac{\partial {\bf B}}{\partial
\xi}\right)={\bf B}.{\bf a}^3-{\bf E}.{\bf a}^2.
\]

As an example we recall the case of availability of magnetic charges with
density $\rho_m$ and magnetic current $\mathbf{j}_m$. Then the corresponding
four 1-forms are
\[
\alpha^1=\left[\frac{4\pi}{c}\mathbf{j}_e, 4\pi \rho_e\right], \ \
\alpha^3=\left[\frac{4\pi}{c}\mathbf{j}_e, -4\pi \rho_e\right], \ \
\alpha^2=\alpha^4=\left[-\frac{4\pi}{c}\mathbf{j}_m, -4\pi \rho_m\right].
\]
The corresponding three vector equations (we omit the scalar ones) look like
\[
\left({\rm rot}{\bf E}+\frac{\partial
{\bf B}}{\partial \xi}\right)\times
{\bf E}+{\bf B}{\rm div}{\bf B}=
\frac{4\pi}{c}\left(-{\bf j}_m\times{\bf E}\right)
+4\pi\rho_m{\bf B},
\]
\[
\left({\rm rot}{\bf B}-\frac{\partial
{\bf E}}{\partial \xi}\right)\times
{\bf B}+{\bf E}{\rm div}{\bf E}=
\frac{4\pi}{c}\left({\bf j}_e\times{\bf B}\right)+
4\pi\rho_e{\bf E},
\]
\[
\begin{split}
&\left({\rm rot}{\bf B}-\frac{\partial
{\bf E}}{\partial \xi}\right)\times{\bf E}+
\left({\rm rot}{\bf E}+\frac{\partial
{\bf B}}{\partial \xi}\right)\times{\bf B}-
{\bf B}{\rm div}{\bf E}-{\bf E}{\rm div}{\bf B}=
\\
&=\frac{4\pi}{c}\left(-{\bf j}_m\times{\bf B}+{\bf j}_e\times{\bf E}\right)-
4\pi(\rho_m{\bf E}+\rho_e{\bf B}).
\end{split}
\]

Returning to the general case we note that the important moment is that
although the nature of the field may significantly change,
{\it the interaction, i.e. the energy-momentum exchange, must NOT destroy the
medium}.  So, {\it definite integrability properties of the medium MUST
be available}, and these integrability properties should be expressible
through the four 1-forms $\alpha^i, i=1,2,3,4$.  In the Maxwell case,
where the charged particles represent any medium, this property implicitly
presents through the implied stability of the charged particles, and it is
mathematically represented by the local integrability of the electric current
vector field $j^\mu$: the corresponding system of ordinary differential
equations $\dot{x^\mu}=j^\mu$ has always solution at given initial
conditions.

If the physical system "electromagnetic field+medium" is energy-momentum
isolated, i.e. no energy-momentum flows out of it, and the system does NOT
destroy itself during interaction, EED assumes, in addition to equations
(21)-(23), the following:

\vskip 0.3cm
{\bf Every couple $(\alpha^i, \alpha^j),
i\neq j$, defines a completely integrable 2-dimensional Pfaff system}.
\vskip 0.3cm
This assumption means that the following equations hold:
\begin{equation}
{\bf d}\alpha^i\wedge \alpha^i\wedge \alpha^j=0, \ \ i,j=1,2,3,4.  
\end{equation}
Equations (21)-(24) constitute the basic system of equations of EED. Of
course, the various special cases can be characterized by adding some new
consistent with (21)-(24) equations and relations.

We are going now to express equations (21)-(23) as one relation, making use
of the earlier introduced object $\Omega=F\otimes e_1+*F\otimes e_2$ and
combining the 1-forms $\alpha^i, i=1,2,3,4$ in two $\mathcal{G}$-valued
1-forms:
\[
\Phi=\alpha^1\otimes e_1+\alpha^2\otimes e_2,\ \ \ \
\Psi=\alpha^3\otimes e_1+\alpha^4\otimes e_2 .
\]
The basis $(e_1,e_2)$ defines two projections $\pi_1$ and $\pi_2$:
\[
\pi_1\Omega=F\otimes e_1,\ \ \ \pi_2\Omega=*F\otimes e_2.
\]
Every bilinear map $\varphi:\mathcal{G}\times\mathcal{G}\rightarrow W$,
where $W$ is some linear space, defines corresponding product in the
$\mathcal{G}$-valued differential forms by means of the relation
\[
\varphi(\Omega_1^i\otimes e_i,\Omega_2^j\otimes e_j)=
\Omega_1^i\wedge\Omega_2^j\otimes\varphi(e_i,e_j).
\]
Now, let $\varphi=\vee$, where "$\vee$" is the symmetrized tensor product.
Recalling $*_\circ *_2=-id$ we obtain
\[
\vee(\Phi,*\pi_1\Omega)+\vee(\Psi,*\pi_2\Omega)=
\alpha^1\wedge *F\otimes e_1\vee e_1-
\alpha^4\wedge F\otimes e_2\vee e_2+
(-\alpha^3\wedge F+\alpha^2\wedge *F)\otimes e_1\vee e_2.
\]
Now, equations (21)-(23) are equivalent to
\begin{equation}
\vee(\delta\Omega,*\Omega)=
\vee(\Phi,*\pi_1\Omega)+\vee(\Psi,*\pi_2\Omega).               
\end{equation}
The Maxwell case (with zero magnetic charges)
corresponds to $\alpha^2=\alpha^4=0, \ \alpha^1=\alpha^3=
4\pi(\delta S+j)$, where the 2-form $S$ is defined by the polarization
3-vector $\mathbf{P}$ and the magnetization 3-vector $\mathbf{M}$ in the
similar
way as $F$ is defined by $(\mathbf{E},\mathbf{B})$. Explicitly,
$\pi_2\delta\Omega=0$, $\pi_1\delta\Omega=4\pi(\delta S+j)\otimes e_1$.

In vacuum (25) reduces to
\begin{align*}
\mathbb{F}=&\vee(\delta\Omega,*\Omega)=
F_{\mu\nu}\delta F^\nu dx^\mu\otimes e_1\vee e_1+
(*F)_{\mu\nu}(\delta *F)^\nu dx^\mu\otimes e_2\vee e_2+\\
&+\Big[F_{\mu\nu}(\delta *F)^\nu +
(*F)_{\mu\nu}\delta F^\nu\Big] dx^\mu\otimes e_1\vee e_2=0,
\end{align*}
which says that the relativistic $\mathbb{G}$-covariant Lorentz force,
denoted also by $\mathbb{F}$, is equal to zero.

As for the energy-momentum tensor $Q_{\mu\nu}$ of the vacuum solutions,
considered as a symmetric 2-form on $M$, it is defined in terms of $\Omega$
as follows:
\[
Q(X,Y)=\frac12 *g\big[i(X)\Omega,*i(Y)\Omega\big],
\]
where
$(X,Y)$ are two arbitrary vector fields on $M$, $g$ is the metric in
$\mathcal{G}$ defined by $g(\alpha,\beta)=\frac12 tr(\alpha.\beta^*)$, and
$\beta^*$ is the transposed to $\beta$.

Further we are going to consider the pure field (vacuum) case $\alpha^i=0$
fully. As for the non-vacuum case, we shall show explicitly
some (3+1)-soliton solutions with well defined integral conserved quantities.
We note the following useful relations in Minkowski space-time.
Let $\alpha$ be 1-form, $F,G$ be two 2-forms and $H$ be a 3-form. Then we
have the identities:
\begin{align}
&*(\alpha\wedge *F)=-\alpha^\sigma F_{\sigma\mu}dx^\mu;\\
&*(F\wedge *H)=\frac12 F^{\sigma\nu}H_{\sigma\nu\mu}dx^\mu;\\   
&\frac12 F_{\alpha\beta}G^{\alpha\beta}\delta_\mu^\nu=
F_{\mu\sigma}G^{\nu\sigma}-(*G)_{\mu\sigma}(*F)^{\nu\sigma}.
\end{align}
Making use of the identities (26)-(28) equations (21)-(23) can be represented
equivalently and \linebreak respectively as follows:
\begin{align}
(*F)^{\sigma\tau}(\mathbf{d}*F)_{\sigma\tau\mu}&=
F_{\mu\nu}(\alpha^1)^\nu,\quad \sigma<\tau \\
F^{\sigma\tau}(\mathbf{d}F)_{\sigma\tau\mu}&=
(*F)_{\mu\nu}(\alpha^4)^\nu,\quad \sigma<\tau\\                
-(*F)^{\sigma\tau}(\mathbf{d}F)_{\sigma\tau\mu}
-F^{\sigma\tau}(\mathbf{d}*F)_{\sigma\tau\mu}&=
F_{\mu\nu}(\alpha^2)^\nu +(*F)_{\mu\nu}(\alpha^3)^\nu,\quad \sigma<\tau.
\end{align}

Finally we note the following. The identity (28) directly leads to relation
(8). In fact, we put in (28) $G=F$, after that we  multiply by $\frac12$, now
we represent the so obtained first term on the right as
$F_{\mu\sigma}F^{\nu\sigma}- \frac12F_{\mu\sigma}F^{\nu\sigma}$, and finally
we transfer $F_{\mu\sigma}F^{\nu\sigma}$ to the left side.

Another suggestion comes if we put in (28) $G=*F$. Then multiplying by
$\frac12$, taking in view that $**F=-F$, and transferring everything to the
left we obtain the identity
\[
\frac14 F_{\alpha\beta}(*F)^{\alpha\beta}\delta_\mu^\nu-
F_{\mu\sigma}(*F)^{\nu\sigma}=0.
\]
This last relation suggests that NO interaction energy-momentum exists
between $F$ and $*F$, which is in correspondence with our equation (20)
which states that even if some energy-momentum is transferred locally from
$F$ to $*F$, the same quantity of energy-momentum is simultaneously locally
transferred from $*F$ to $F$.  We note also that these last remarks are of
pure algebraic nature and they do NOT depend on whether $F$ satisfies or
does not satisfy any equations or additional conditions.

\section{EED: General properties of the vacuum non-linear solutions}

First we note the following coordinate free form of equations (18)-(20)
in terms of $\delta$ and $\mathbf{d}$:
\begin{align}
(*F)\wedge*\mathbf{d}*F & \equiv\delta F\wedge *F=0,\\
F\wedge *\mathbf{d}F & \equiv -F\wedge \delta *F=0,\\
F\wedge *\mathbf{d}*F+(*F)\wedge*\mathbf{d}F             
& \equiv \delta F\wedge F-\delta *F\wedge *F=0,
\end{align}
We begin with establishing some elementary symmetry properties of these
equations.
\vskip 0.3cm
\noindent
{\bf Proposition 1.}\ \ Equations (32)-(34) are invariant with respect to a
conformal change of the metric.
\vskip 0.3cm
\noindent
{\bf Proof.} Let $\eta\rightarrow g=f^2 \eta$, where $f^{2}(m)>0, m\in M$. It
is seen from the coordinate-free $\mathbf{d}$-form of the vacuum equations
that only the restrictions $*_2$ and $*_3$ of the Hodge $*$ participate,
moreover, $*_3$ is not differentiated. Since $*_2$ is conformally invariant,
the general conformal invariance of the equations will depend on $*_3$.  But,
for this restriction of $*$ we have $(*_g)_3= f^{-2}(*_\eta)_3$, so the
 left-hand sides of the equations are multiplied by $f\neq 0$ of a given
degree, which does not change the equations.
\vskip 0.3cm
\noindent
{\bf Proposition 2.} Equations (29)-(31) are invariant with respect to the
transformation:
\[
F\rightarrow *F,\ \ \alpha^1\rightarrow\alpha^4,\ \
\alpha^2\rightarrow (-\alpha^3),\ \ \alpha^3\rightarrow \alpha^2,\ \
\alpha^4\rightarrow (-\alpha^1).
\]
{\bf Proof}. Obvious.
\vskip 0.3cm
\noindent
{\bf Proposition 3.} The vacuum equations (32)-(34) are invariant with respect
to the transformation
\[
F\rightarrow \mathcal{F}=aF-b*F,\ \ *F\rightarrow*\mathcal{F}=bF+a*F,\ \
a,b\in \mathbb{R}.
\]
{\bf Proof}. We substitute and obtain:
\[
\delta \mathcal{F}\wedge *\mathcal{F}=a^2(\delta F\wedge *F)-
b^2(\delta *F\wedge F)+ab(\delta F\wedge F-\delta *F\wedge *F)
\]
\[
\delta *\mathcal{F}\wedge \mathcal{F}=a^2(\delta *F\wedge F)-
b^2(\delta F\wedge *F)+ab(\delta F\wedge F-\delta *F\wedge *F)
\]
\[
\delta\mathcal{F}\wedge\mathcal{F}-\delta *\mathcal{F}\wedge *\mathcal{F}=
(a^2-b^2)(\delta F\wedge F-\delta *F\wedge *F)-
2ab(\delta F\wedge *F+\delta *F\wedge F).
\]
It is seen that if $F$ defines a solution then $\mathcal{F}$ also defines a
solution. Conversely, if $\mathcal{F}$ defines a solution then subtracting the
second equation from the first and taking in view that $a^2+b^2\neq 0$ we
obtain $\delta F\wedge *F=(\delta *F)\wedge F$. So, if
$\delta\mathcal{F}\wedge *\mathcal{F}=
\delta *\mathcal{F}\wedge \mathcal{F}=0$ from the first two equations we
obtain $\delta F\wedge F-\delta *F\wedge *F=
-(\delta F\wedge F-\delta *F\wedge *F)=0$.
The proposition follows.
\vskip 0.3cm
We recall the following property of vectors in Minkowski space-time (further
referred to as {\bf BP1}):
\vskip 0.2cm
\noindent
{\bf Basic property 1.}: There are NO mutually orthogonal time-like
vectors in Minkowski space-time.
\vskip 0.2cm

\noindent
We recall also [27] the following relation between the eigen
properties of $F_{\mu}^{\nu}$ and $Q_{\mu}^{\nu}$.
\vskip 0.3cm
\noindent
{\bf Basic property 2}: All eigen vectors of $F$ and $*F$ are eigen vectors
of $Q$ too.  \newline (Further referred to as {\bf BP2}).
\vskip 0.3cm
\noindent
It is quite clear that the solutions of our non-linear equations
are naturally divided into three subclasses:
\vskip 0.3cm
1. {\bf Linear
(Maxwellean)}, i.e. those, satisfying $\delta F=0,\ \delta *F=0$. This
subclass will not be of interest since it is well known.
\vskip 0.2cm
2. {\bf Semilinear}, i.e.  those, satisfying $\delta *F=0,\ \delta F\neq 0$.
(or $\delta *F\neq 0,\ \delta F=0$).
This subclass of solutions, as it will soon become clear, does NOT admit
rotational component of propagation, so it is appropriate to describe just
running wave kind behavior.
\vskip 0.2cm
3. {\bf Nonlinear}, i.e. those, satisfying $\delta F\neq 0,\ \delta *F\neq 0$.
\vskip 0.3cm

Further we shall consider only the nonlinear solutions, and the semilinear
solutions will be characterized as having the additional property $\delta
*F=0, (\text{or}\ \  \delta F=0)$.

The following Proposition is of crucial importance for all further studies of
the nonlinear solutions.

\vskip 0.5cm
{\bf Proposition 4.}\  All nonlinear solutions have zero invariants:
\newline
$$
I_1=\frac 12 F_{\mu\nu}F^{\mu\nu}=
\pm\sqrt{\mathrm{det}[(F\pm*F)_{\mu\nu}]}=0,
$$
$$
\ I_2=\frac 12 (*F)_{\mu\nu}F^{\mu\nu}=
\pm 2\sqrt{\mathrm{det}(F_{\mu\nu})}=0.
$$
\indent{\bf Proof}.\ Recall the field equations in the form:
\[
F_{\mu\nu}(\delta F)^\nu=0,\ (*F)_{\mu\nu}(\delta*F)^\nu=0,\
F_{\mu\nu}(\delta*F)^\nu+(*F)_{\mu\nu}(\delta F)^\nu=0.
\]
It is clearly seen that the first two groups of these equations may be
considered as two linear homogeneous systems with respect to $\delta F^\mu $
and $\delta *F^\mu $ respectively. These
homogeneous systems have non-zero solutions, which is possible only if
$\mathrm{det}(F_{\mu\nu})=\mathrm{det}((*F)_{\mu\nu})=0$, i.e. if
$I_2=2{\bf E}.{\bf B}=0$.
Further, summing up these three systems of equations, we obtain
\[
(F+*F)_{\mu\nu}(\delta F+\delta *F)^\nu=0.
\]
If now $(\delta F+\delta *F)^\nu\neq 0$, then
\[
0=\mathrm{det}||(F+*F)_{\mu\nu}||=\left [\frac 14
(F+*F)_{\mu\nu}(*F-F)^{\mu\nu}\right ]^2= \left [-\frac 12
F_{\mu\nu}F^{\mu\nu}\right ]^2=(I_1)^2.
\]
If $\delta F^\nu=-(\delta *F)^\nu\neq 0$, we sum up the first two systems and
obtain $(*F-F)_{\mu\nu}(\delta *F)^\nu=0$. Consequently,
\[
0=\mathrm{det}||(*F-F)_{\mu\nu}||=
\left [\frac 14 (*F-F)_{\mu\nu}(-F-*F)^{\mu\nu}\right ]^2=
\left [\frac 12 F_{\mu\nu}F^{\mu\nu}\right ]^2=(I_1)^2.
\]
This completes the proof.

{\bf Corollary}. All nonlinear
solutions are {\it null fields}, so $(*F)_{\mu\sigma}F^{\nu\sigma}=0$ and
$F_{\mu\sigma}F^{\nu\sigma}=(*F)_{\mu\sigma}(*F)^{\nu\sigma}$.

{\bf Corollary}. The vector $\delta F^\mu$ is an eigen vector of $F_\mu^\nu$;
the vector $(\delta *F)^\mu$ is an eigen vector of $(*F)_\mu^\nu$.

{\bf Corollary}. The vectors $\delta F^\mu$ and $(\delta *F)^\mu$ are eigen
vectors of the energy tensor $Q^\mu_\nu$.

 \vskip 0.3cm {\bf Basic Property 3.}[27] In the {\it null field}
case $Q_\mu^\nu$ has just one isotropic eigen direction, defined by the
isotropic vector $\zeta$, and all of its other eigen directions are
space-like. (Further referred to as {\bf BP3}).
\vskip 0.5cm
{\bf Proposition 5.}\  All nonlinear solutions satisfy the conditions
\begin{equation}
(\delta F)_\mu (\delta *F)^\mu =0,\
\left|\delta F\right|=\left|\delta *F\right|             
\end{equation}
\indent{\bf Proof}.\  We form the inner product
$i(\delta *F)(\delta F\wedge *F)=0$  and get
\[
(\delta *F)^\mu(\delta F)_\mu(*F)-\delta F\wedge (\delta
*F)^\mu(*F)_{\mu\nu}dx^\nu=0.
\]
Because of the obvious nullification of the second term the first term will be
equal to zero (at non-zero $*F$) only if
$(\delta F)_\mu (\delta *F)^\mu=0$.

Further we form the inner product
$i(\delta *F)(\delta F\wedge F-\delta *F\wedge *F)=0$ and obtain

\[
(\delta *F)^\mu(\delta F)_\mu F-
\delta F\wedge (\delta *F)^\mu F_{\mu\nu}dx^\nu-
\]
\[
-(\delta *F)^2(*F)+\delta *F\wedge(\delta *F)^\mu (*F)_{\mu\nu}dx^\nu=0.
\]
Clearly, the first and the last terms are equal to zero. So, the inner
product by $\delta F$ gives
\[
(\delta F)^2 (\delta *F)^\mu F_{\mu\nu}dx^\nu-
\left[(\delta F)^\mu (\delta *F)^\nu F_{\mu\nu}\right]\delta F
+(\delta *F)^2(\delta F)^\mu (*F)_{\mu\nu}dx^\nu=0.
\]
The second term of this equality is zero. Besides,
$(\delta *F)^\mu F_{\mu\nu}dx^\nu=-(\delta F)^\mu (*F)_{\mu\nu}dx^\nu$.
So,
\[
\left[(\delta F)^2-(\delta *F)^2\right](\delta F)^\mu(*F)_{\mu\nu}dx^\nu=0.
\]
Now, if $(\delta F)^\mu(*F)_{\mu\nu}dx^\nu\neq 0$, then the relation
$\left|\delta F\right|=\left|\delta *F\right|$ follows immediately.
If \linebreak
$(\delta F)^\mu(*F)_{\mu\nu}dx^\nu=0=-(\delta *F)^\mu F_{\mu\nu}dx^\nu$
according to the third equation of (22), we shall show that
 $(\delta F)^2=(\delta *F)^2=0$. In fact, forming the inner product
$i(\delta F)(\delta F\wedge *F)=0$ , we get
\[
(\delta F)^2*F-\delta F\wedge (\delta F)^\mu (*F)_{\mu\nu}dx^\nu=
(\delta F)^2*F=0.
\]
In a similar way, forming the inner product
$i(\delta *F)(\delta *F\wedge F)=0$ we have
\[
(\delta *F)^2 F-\delta (*F)\wedge (\delta *F)^\mu F_{\mu\nu}dx^\nu=
(\delta *F)^2 F=0.
\]
This completes the proof. It follows from this Proposition and from {\bf BP1}
that

1. $\delta F$ and  $\delta *F$ can NOT be time-like,

2. $\delta F$ and $\delta *F$ are {\bf simultaneously space-like}, i.e.
$(\delta F)^2=(\delta *F)^2<0$, or {\bf simultaneously isotropic}, i.e.
$|\delta F|=|\delta *F|=0$.  We note that in this last case the isotropic
vectors $\delta F$ and $\delta *F$ are also eigen vectors of $Q_\mu^\nu$, and
since $Q_\mu^\nu$ has just one isotropic eigen direction, which we denoted by
 $\zeta$, we conclude that $\delta F, \delta*F$ and $\zeta$ are collinear.

In our further study of the nonlinear solutions we shall make use of the
following. As it is shown in [27] at zero invariants $I_1=I_2=0$ the
following representation holds:
\[
F=A\wedge \hat \zeta,\ *F=A^*\wedge \hat \zeta,
\]
where $A$ and $A^*$ are 1-forms,
$\hat \zeta$ is the corresponding to $\zeta$ through the pseudometric $\eta$
1-form: $\hat \zeta_\mu=\eta_{\mu\nu}\zeta^\nu$. It follows that
\[
F\wedge\hat\zeta=*F\wedge\hat\zeta=0.
\]
\vskip 0.3cm
\noindent
{\bf Remark}. Further we are going to skip
the "hat" over $\zeta$, and from the context it will be clear the meaning of
$\zeta$: one-form, or vector field.
\vskip 0.3cm
We establish now some useful properties of these quantities.
\vskip 0.3cm
{\bf Proposition 6.} The following relations hold:

	$1^o. A_\mu\zeta^\mu=A^*_\mu\zeta^\mu=0$

	$2^o. A^\sigma(*F)_{\sigma\mu}=0,\ \ (A^*)^\sigma F_{\sigma\mu}=0$

	$3^o. A^\sigma A^*_\sigma=0$,\ \ $A^2=(A^*)^2<0$.

{\bf Proof.} In order to prove $1^o$ we note
\begin{align*}
& 0=I_1=\frac12 F_{\mu\nu}F^{\mu\nu}=\frac12(A_\mu\zeta_\nu-A_\nu\zeta_\mu)
(A^\mu\zeta^\nu-A^\nu\zeta^\mu)  \\
&=\frac12(2A_\mu A^\mu.\zeta_\nu\zeta^\nu-2(A_\mu\zeta^\mu)^2)=
-(A_\mu\zeta^\mu)^2.
\end{align*}
The second of $1^o$ is proved in the same way just replacing $F$ with $*F$
and $A$ with $A^*$.

To prove $2^o$ we make use of relation (26).
\[
0=*(A^*\wedge A^*\wedge\zeta)=*(A^*\wedge *F)=
-(A^*)^\sigma F_{\sigma\mu}dx^\mu
\]
Similarly
\[
0=*(A\wedge A\wedge\zeta)=*(A\wedge F)=
A^\sigma (*F)_{\sigma\mu}dx^\mu.
\]
Hence, $A^*$ is an eigen vector of $F$ and $A$ is an eigen vector of $*F$.
Therefore, $A^2<0,\ \ (A^*)^2<0$, i.e. these two vectors (or 1-forms) are
space-like. The case $|A|=0$ is not considered since then $A$ is collinear to
$\zeta$ and $F=A\wedge \zeta=0$, so $*F=0$ too.

Now, $3^o$ follows from $2^o$ because
\[
0=-(A^*)^\sigma F_{\sigma\mu}=-(A^*)^\sigma(A_\sigma\zeta_\mu-
A_\mu\zeta_\sigma)=-(A^*.A)\zeta_\mu+A_\mu(A^*.\zeta)=-(A^*.A)\zeta_\mu.
\]

Finally we express $Q_\mu^\nu$ in terms of $A$, or $A^*$, and $\zeta$. Since
$I_1=0$ and $Q_\mu^\nu(F)=Q_\mu^\nu(*F)$ we have
\[
Q_\mu^\nu=-\frac{1}{4\pi}\Big[F_{\mu\sigma}F^{\nu\sigma}\Big]=
-\frac{1}{4\pi}(A_\mu\zeta_\sigma-
A_\sigma\zeta_\mu)(A^\nu\zeta^\sigma-A^\sigma\zeta^\nu)=
-\frac{1}{4\pi}A^2\zeta_\mu\zeta^\nu=
-\frac{1}{4\pi}(A^*)^2\zeta_\mu\zeta^\nu.
\]
Hence, since $Q_4^4>0$ and $\zeta_4\zeta^4>0$ we obtain again that
$A^2=(A^*)^2<0$.

We normalize $\zeta$, i.e. we divide $\zeta_\mu$ by $\zeta_4\neq 0$,
so, we assume further that $\zeta_\mu=(\zeta_1,\zeta_2,\zeta_3,1)$. Now, from
the local conservation law $\nabla_\nu Q^{\mu\nu}=0$ it follows that the
isotropic eigen direction defined by $\zeta$ defines geodesic lines:
$\zeta^\nu\nabla_\nu \zeta^\mu=0$. In fact,
\[
\nabla_\nu Q^{\mu\nu}=
\nabla_\nu\left(-\frac{1}{4\pi}A^2\zeta^\mu\zeta^\nu\right)=
-\frac{1}{4\pi}\Big[\zeta^\mu(\nabla_\nu A^2\zeta^\nu)+
A^2\zeta^\nu\nabla_\nu\zeta^\mu\Big]=0.
\]
This relation holds for any $\mu=1,2,3,4$. We consider it for $\mu=4$ and
recall that for $\mu=4$ we have $\zeta^4=1$. In our coordinates
$\nabla_\nu=\partial_\nu$, and we obtain that the second term becomes zero,
so $\nabla_\nu (A^2\zeta^\nu)=0$.  Therefore,
$\zeta^\nu\nabla_\nu\zeta^\mu=0$, which means that {\bf all the trajectories
of} $\zeta$ {\bf are parallel straight isotropic} lines. Hence, with every
nonlinear solution $F$ we are allowed to introduce $F$-{\it adapted}
coordinate system by the requirement that the trajectories of $\zeta$ to be
parallel to the plane $(z,\xi)$. In such a coordinate system we obtain
$\zeta_\mu=(0,0,\varepsilon,1),\ \varepsilon=\pm 1$. From 3-dimensional point
of view this means that the field propagates along the coordinate $z$, and
$\varepsilon =-1$ implies propagation along $z$ from $-\infty$ to $+\infty$,
while $\varepsilon =+1$ implies propagation from $\infty$ to $-\infty$.

\vskip 0.4cm
{\bf Corollary}. The translational direction of propagation of any
null-field is determined {\it intrinsically}.

\vskip 0.4cm
We note that in Maxwell theory, as well as in Born-Infeld theory [9] (the
latter reduces to Maxwell theory in the null-field case) all time
stable null-field solutions are {\it spatially infinite}, otherwise they have
to blow-up radially according to Poisson theorem, i.e. $\zeta$ shall not be
unique. We shall see that EED has no problems in this respect, i.e. spatially
finite and time stable null field solutions are allowed.

We express now $F, *F, A$ and $A^*$ in the corresponding $F$-{\it adapted}
coordinate system making use of the relations $F=A\wedge\zeta,\
*F=A^*\wedge\zeta$, where $\zeta=\varepsilon dz+d\xi$.
\begin{equation}
\begin{split}
&F_{12}=F_{34}=0,\ \ F_{13}=\varepsilon F_{14},
\ \ F_{23}=\varepsilon F_{24},\\
&(*F)_{12}=(*F)_{34}=0,\ \ (*F)_{13}=\varepsilon (*F)_{14}=-F_{24},  
\ \ (*F)_{23}=\varepsilon (*F)_{24}=F_{14}.\\
\end{split}
\end{equation}
Moreover, from $A.\zeta=A^*.\zeta=0$ it follows that in an $F$-adapted
coordinate system we obtain
\begin{equation}
\begin{split}
&A=A_1 dx+A_2 dy +f.\zeta=(F_{14})dx+(F_{24})dy +f.\zeta \\
&A^*=A^*_1 dx+A^*_2 dy +f^*.\zeta=(-F_{23})dx+(F_{13})dy +f^*.\zeta=
-(\varepsilon A_2)dx+(\varepsilon A_1)dy+f^*.\zeta,                 
\end{split}
\end{equation}
where $f$ and $f^*$ are two arbitrary functions.
\vskip 0.3cm
Having this in mind we prove the following
\vskip 0.2cm
\noindent
{\bf Proposition 6.}\  All nonlinear solutions satisfy the relations:
\begin{equation}
\zeta^\mu(\delta F)_\mu=0,\ \zeta^\mu(\delta *F)_\mu=0.       
\end{equation}
\vskip 0.2cm
\noindent
{\bf Proof}.\ We form the inner product
 $i(\zeta)(\delta F\wedge *F)=0$ :
\[
\left[\zeta^\mu
(\delta F)_\mu\right]*F-\delta F\wedge(\zeta)^\mu(*F)_{\mu\nu}dx^\nu=
\]
\[
=\left[\zeta^\mu (\delta F)_\mu\right]A^*\wedge\zeta-
(\delta F\wedge\zeta)\zeta^\mu (A^*)_\mu
+(\delta F\wedge A^*)\zeta^\mu \zeta_\mu=0.
\]
Since the second and the third terms are equal to zero and $*F\neq 0$, then
$\zeta^\mu(\delta F)_\mu=0$. Similarly, from the equation
$(\delta *F)\wedge F=0$ we get
$\zeta^\mu(\delta *F)_\mu=0$. The proposition is proved.
\vskip 0.3cm
\noindent
{\bf Proposition 7.}\ If $F\neq 0$ is a nonlinear solution with
$|\delta F|\neq 0$
then $\delta F\wedge F=\delta *F\wedge *F\neq 0$.
\vskip 0.2cm
\noindent
{\bf Proof}. From $F\neq 0$ we have $(\delta F)^2\neq 0$.
Recall that $*(\delta F\wedge F)=
i(\delta F)*F=(\delta F)^\mu (*F)_{\mu\nu}dx^\nu$. Assume now the opposite,
i.e. that $\delta F\wedge F=0$. We compute
\[
0=i(\delta F)(\delta F\wedge *F)=
(\delta F)^2 *F-\delta F\wedge i(\delta F)*F=(\delta F)^2 *F,
\]
i.e. $(\delta F)^2=0$, or $*F=0$, which contradicts the assumption $F\neq 0$.

Conversely, if $\delta F\wedge F\neq 0$ then surely $\delta F\neq 0$. So, we
may have $|\delta F|\neq 0$, or $|\delta F|=0$. If $|\delta F|=0$ then
$\delta F$ must be colinear to $\zeta$, i.e. $\zeta\wedge F\neq 0$, which is
not possible since $F=A\wedge \zeta$.
\vskip 0.3cm
We also note that the 3-form $\delta F\wedge F$ is isotropic: $(\delta
F\wedge F)^2=0$. In fact
\[
(\delta F\wedge F)^{\mu\nu\sigma}(\delta F\wedge F)_{\mu\nu\sigma}=
F^{\mu\nu}\big[(\delta F)^2 F_{\mu\nu}\big]-
F^{\mu\nu}\big[\delta F\wedge i(\delta F)F\big]_{\mu\nu}=
(\delta F)^2 F^{\mu\nu}F_{\mu\nu}=0, \
 \mu<\nu<\sigma.
\]

Finally we note that there are no nonlinear spherically symmetric solutions,
i.e. if $F$ is a spherically symmetric solution then it is a solution
of Maxwell's equations. In fact, the most general spherically symmetric
2-form in spherical coordinates, originating at the symmetry center, is
\[
F=f(r,\xi)\,dr\wedge d\xi+
h(r,\xi)\,\mathrm{sin}\theta\, d\theta\wedge d\varphi.
\]
Now the equation $F\wedge *\mathbf{d}F=0$ requires
$h(r,\xi)=const$, and the equation $\delta F\wedge *F=0$ requires
$f(r,\xi)=const/r^2$. It follows: $\mathbf{d}F=0,\ \delta F=0$.


\vskip 0.5cm
\section{EED: Further properties of the nonlinear solutions}
Let's introduce the notations
$F_{14}\equiv u,\ F_{24}\equiv p$, so, in an $F$-adapted coordinate system
we can write
\begin{equation}
\begin{split}
&F=\varepsilon udx\wedge dz + \varepsilon pdy\wedge dz + udx\wedge d\xi +
pdy\wedge d\xi \\
&*F=-pdx\wedge dz + udy\wedge dz - \varepsilon pdx\wedge d\xi  +   
\varepsilon udy\wedge d\xi.
\end{split}
\end{equation}
This form of $F$ shows that every nonlinear solution defines a map
$$
\alpha_F: M\rightarrow \mathbb{R}^2,\quad
\alpha_F(x,y,z,\xi)=[u(x,y,z,\xi), p(x,y,z,\xi)].
$$
We endow now $\mathbb{R}^2$ with the canonical complex
structure $J$ and identify $(\mathbb{R}^2,J)$ with the complex numbers
$\mathbb{C}$. Hence, our nonlinear solution $F$ defines the complex valued
function $\alpha_F=u+ip$. This correspondence and its inverse will be studied
further.

In the $F$-adapted coordinate system only 4 of the components $Q_\mu^\nu$ are
different from zero, namely:
$$
Q_4^4=-Q_3^3=\varepsilon Q_3^4= -\varepsilon
Q_4^3=\left|A^2\right|=u^2+p^2.
$$

The complex valued function $\alpha_F$ defined above by the nonlinear
solution $F$ has module $|\alpha_F|=\sqrt{u^2+p^2}$ and phase
$\psi=\mathrm{arctg}(p/u)=\mathrm{arccos}(u/\sqrt{u^2+p^2})$. We shall come
now to these two quantities in a coordinate free way.  We show first how the
nonlinear solution $F$ defines at every point a pseudoorthonormal basis in
the corresponding tangent and cotangent spaces.  The nonzero 1-forms
$\tilde{A}=A-f\zeta$ and $\tilde{A^*}=A^*-f^*\zeta$ are normed to
$\tilde{\mathbf{A}}= \tilde{A}/|\tilde{A}|$ and
$\tilde{\mathbf{A^*}}=\tilde{A^*}/|\tilde{A^*}|$.  We note that
$|A|=|\tilde{A}|$ and $|A^*|=|\tilde{A^*}|$.  Two new unit 1-forms
$\mathbf{R}$ and $\mathbf{S}$ are introduced through the equations:
\[
{\bf R}^2=-1,\ {\bf \tilde{A}}^\nu {\bf R}_\nu=0,\
({\bf \tilde{A^*}})^\nu {\bf R}_\nu=0,\
\zeta^\nu{\bf R}_\nu=\varepsilon,\ {\bf S}=\zeta+\varepsilon{\bf R}.
\]
The only solution of the first 4 equations in the $F$-adapted coordinate
system is $\mathbf{R}_\mu=(0,0,-1,0)$, so, $(\mathbf{R})^2=-1$. Then for
$\mathbf{S}$ we obtain $\mathbf{S}_\mu=(0,0,0,1),\  (\mathbf{S})^2=1$. This
pseudoorthonormal co-tangent basis $({\bf \tilde{A}, \tilde{A^*}, R, S})$ is
carried over to a tangent pseudoorthonormal basis by means of $\eta$.

In this way at every point, where the field is different from zero, we have
three frames: the pseudoorthonormal ($F$-adapted) coordinate frame
$(dx,dy,dz,d\xi)$, the pseudoorthonormal frame $\chi^0=({\bf
\tilde{A}},\varepsilon{\bf \tilde{A^*}},{\bf R}, {\bf S})$ and the
pseudoorthogonal frame $\chi=(\tilde{A},\varepsilon \tilde{A}^*,{\bf R}, {\bf
S})$.  The matrix $\chi_{\mu\nu}$ of  $\chi$ with respect to the coordinate
frame is
\[
\chi_{\mu\nu}=\begin{Vmatrix}
u  &-p  & 0  &0 \\
p  & u  & 0  &0 \\
0  & 0  &-1  &0 \\
0  & 0  & 0  &1
\end{Vmatrix}.
\]
We define now the amplitude $\phi>0$ of the solution $F$ by
\begin{equation}
\phi=\sqrt{|det(\chi_{\mu\nu})|}.                              
\end{equation}
Clearly, in an $F$-adapted coordinate system
$\phi=\sqrt{u^2+p^2}=\sqrt{Q_4^4}=|\alpha_F|=|A|$.

Each of these 3 frames defines its own volume form:
\[
\omega=dx\wedge dy\wedge dz\wedge d\xi;\quad
\omega_{\chi^{o}}=\mathbf{A}\wedge\mathbf{\varepsilon A}\wedge
\mathbf{R}\wedge\mathbf{S}=-\omega;\quad
\omega_{\chi}=A\wedge A^{*}\wedge\mathbf{R}\wedge\mathbf{S}=
-(u^2+p^2)\omega.
\]
We proceed further to define the {\it phase} of the nonlinear solution $F$.
We shall need the matrix $\chi^0_{\mu\nu}$ of the frame $\chi^0$ with respect
to the coordinate basis. We obtain
\[
\chi^0_{\mu\nu}=\begin{Vmatrix}
\frac{u}{\sqrt{u^2+p^2}}  &\frac{-p}{\sqrt{u^2+p^2}}  & 0  &0 \\
\frac{p}{\sqrt{u^2+p^2}}  &\frac{u}{\sqrt{u^2+p^2}}   & 0  &0 \\
0                         & 0                         &-1  &0 \\
0                         & 0                         & 0  &1
\end{Vmatrix}.
\]
The {\it trace} of this matrix is
$$
tr(\chi^0_{\mu\nu})=\frac{2u}{\sqrt{u^2+p^2}}.
$$
Obviously, the inequality $|\frac 12 tr(\chi^0_{\mu\nu})|\leq 1$ is
fulfilled. Now, by definition, the quantities $\varphi$ and $\psi$
defined by
\begin{equation}
\varphi=\frac 12 tr(\chi^0_{\mu\nu}),\
\psi=\mathrm{arccos}(\varphi)=
\mathrm{arccos}\left(\frac 12 tr(\chi^0_{\mu\nu})\right)
\end{equation}                                                 
will be called {\it the phase function} and {\it the phase} of the solution,
respectively. Clearly, in the $F$-adapted coordinate system
$\varphi=\mathrm{cos}[arg(\alpha_F)]$.

Making use of the amplitude function $\phi$, of the phase function $\varphi $
and of the phase $\psi$, we can write
\begin{equation}
u=\phi.\varphi=\phi\,\mathrm{cos}\psi,\
\ p=\phi\,(\pm\sqrt{1-\varphi^2})=\phi\,\mathrm{sin}\psi.  
\end{equation}

We consider now the pair of 1-forms $(\tilde{A},\tilde{A^*})$ on a region
$U\subset M$, where $\tilde{A}\neq 0, \ \tilde{A^*}\neq 0$, and
since $\tilde{A}.\tilde{A^*}=0$, we have a 2-dimensional Pfaff system on
$U$.
\vskip 0.3cm
\noindent
{\bf Proposition 8.} The above 2-dimensional Pfaff system
$(\tilde{A},\tilde{A^*})$ is completely integrable, i.e. the following
equations hold:
\[
{\bf d}\tilde{A}\wedge \tilde{A}\wedge \tilde{A^*} =0,
\ {\bf d}\tilde{A^*}\wedge \tilde{A}\wedge \tilde{A^*}=0.
\]
\vskip 0.2cm
\noindent
{\bf Proof.}
In fact, $\tilde{A}\wedge \tilde{A^*}=(u^2+p^2)dx\wedge dy$, and in every
term of ${\bf d}\tilde{A}$ and ${\bf d}\tilde{A^*}$ at least one of the basis
vectors $dx$ and $dy$ will participate, so the above exterior products will
vanish.  The proposition is proved.
\vskip 0.3cm
\noindent
{\bf Remark}.
As we shall see further, this property of Frobenius integrability  is dually
invariant for the nonlinear solutions.
\vskip 0.3cm
\noindent
{\bf Remark}. These considerations stay in force also
for those time stable linear solutions, which have zero invariants
$I_1=I_2=0$.  But Maxwell's equations require $u$ and $p$ to be {\it infinite
running plane waves} in this case, so the corresponding amplitudes will NOT
depend on two of the spatial coordinates and the phase functions will be also
running waves.  As we'll see further, the phase functions for the nonlinear
solutions are arbitrary bounded functions.

We proceed further to define the new and important concept of {\it scale
factor} $\mathcal{L}$ for a given nonlinear solution $F$ with $|\delta
F|\neq 0$. First we recall a theorem from vector bundle theory, which
establishes some important properties of those vector bundles which admit
pseudoriemannian structure [30]. The theorem says that if a vector bundle
$\Sigma$ with a base manifild $B$ and standard fiber $V$
admits pseudoriemannian structure $g$ of signature $(p,q), p+q=dimV$, then it
is always possible to introduce in this bundle a riemannian structure $h$ and
a linear automorphism $\varphi$ of the bundle, such that two subbundles
$\Sigma^+$ and $\Sigma^-$ may be defined with the following properties:

\begin{enumerate}
\item $g(\Sigma^+,\Sigma^+)=h(\Sigma^+,\Sigma^+)$,
\item $g(\Sigma^-,\Sigma^-)=-h(\Sigma^-,\Sigma^-)$,
\item $g(\Sigma^+,\Sigma^-)=h(\Sigma^+,\Sigma^-)=0$.
\end{enumerate}
The automorphism $\varphi$ is defined by
\[
g_x(u_x,v_x)=h_x(\varphi(u_x),v_x),\quad u_x,v_x\in T_xB, \ \ x\in B.
\]
In components we have
\[
g_{ij}=h_{ik}\varphi^k_j \rightarrow \ \varphi^k_i=g_{im}h^{mk}.
\]
In the tangent bundle case this theorem allows to separate a subbundle of the
tangent bundle if the manifold admits pseudoriemannian metric. In the simple
case of Minkowski space $(M,\eta),\  sign(\eta)=(-,-,-,+)$, introducing
the standard Euclidean metric $h$ on $M$ we may separate 1-dimensional
subbundle, i.e. a couple of vector fields $\pm X_o$, being eigen vectors
of $\varphi$, and to require $\eta(X_o,X_o)=1$. In our canonical coordinates
we obtain $X_o=\partial/\partial \xi$, so, the coordinate free relativistic
definitions of the electric $\mathbf{E}$ and magnetic $\mathbf{B}$ $1$-forms
through $F$ and $*F$ are:
\[
\mathbf{E}=-i(X_o)F,\quad \mathbf{B}=i(X_o)(*F).
\]
Now, the scale factor $\mathcal{L}$ for a nonlinear solution $F$ with
$|\delta F|\neq 0$ is defined by
\begin{equation}
\mathcal{L}=\frac{|i(X_o)F|}{|\delta F|}= \frac{|A|}{|\delta     
F|}=\frac{|i(X_o)(*F)|}{|\delta *F|}=\frac{|A^*|}{|\delta*F|}.
\end{equation}
It is interesting to see how $\mathcal{L}$ is defined in non-relativistic
terms.  Consider the vector fields
\begin{equation}
\vec{\mathcal{F}}=\mathrm{rot}\,\mathbf{E}+
\frac{\partial \mathbf{B}}{\partial \xi}+
\frac{\mathbf{E}\times\mathbf{B}}
{|\mathbf{E}\times\mathbf{B}|}\mathrm{div}\,\mathbf{B},   
\end{equation}
\begin{equation}
\vec{\mathcal{M}}=\mathrm{rot}\,\mathbf{B}-
\frac{\partial {\mathbf E}}{\partial \xi}-                   
\frac{\mathbf{E}\times\mathbf{B}}
{|\mathbf{E}\times\mathbf{B}|}\mathrm{div}\,\mathbf{E}.
\end{equation}

It is obvious that on the solutions of Maxwell's vacuum equations
$\vec{\mathcal{F}}$ and $\vec{\mathcal{M}}$ are equal to zero. Note also that
under the transformation (6) we get
$\vec{\mathcal{F}}\rightarrow
-\vec{\mathcal{M}}$ and $\vec{\mathcal{M}}\rightarrow \vec{\mathcal{F}}$.

We shall consider now the relation between  $\vec{\cal F}$ and $\mathbf{E}$,
and between $\vec{\cal M}$ and $\mathbf{B}$ on the nonlinear solutions of our
equations when $\vec{\cal F}\neq 0$ and $\vec{\cal M}\neq 0$.

Recalling eqns.(11)-(16) we obtain
$$
({\mathbf E}\times {\mathbf
B})\times {\mathbf E}= -{\mathbf E}\times ({\mathbf E}\times {\mathbf B})=
-[{\mathbf E}({\mathbf E}.{\mathbf B})-
{\mathbf B}({\mathbf E}.{\mathbf E})]={\mathbf B}({\mathbf E}^2),
$$
and since $|{\mathbf E}\times {\mathbf B}|={\mathbf E}^2={\mathbf B}^2$,
we get
$$
\vec{\cal F}\times {\mathbf E}=
\left(\mathrm{rot}{\mathbf E}+\frac{\partial {\mathbf B}}{\partial \xi}\right)
\times {\mathbf E}+{\mathbf B}\mathrm{div}{\mathbf B}=0.
$$
In the same way we get $\vec{\cal M}\times {\mathbf B}=0$. In other words, on
the nonlinear solutions we obtain that $\vec{\cal F}$ is co-linear to
${\mathbf E}$ and $\vec{\cal M}$ is co-linear to ${\mathbf B}$.  Hence, we
can write the relations
\begin{equation}
\vec{\cal F}=f_1.{\mathbf E},\ \ \vec{\cal M}=f_2.{\mathbf B},        
\end{equation}
where $f_1$ and $f_2$ are two functions, and of course, the
interesting cases are $f_1\neq 0,\infty;\ f_2\neq 0,\infty$.
Note that the physical dimension of $f_1$ and $f_2$ is the
reciprocal to the dimension of coordinates, i.e.
$[f_1]=[f_2]= [length]^{-1}$.

We shall prove now that $f_1=f_2$. In fact, making use of the same formula
for the double vector product, used above, we easily obtain
\[
\vec{\cal F}\times {\mathbf B}+\vec{\cal M}\times {\mathbf E}=
\]
\[
=\left(\mathrm{rot}{\mathbf E}+\frac{\partial {\mathbf B}}{\partial
\xi}\right)
\times {\mathbf B}+
\left(\mathrm{rot}{\mathbf B}-\frac{\partial {\mathbf E}}{\partial \xi}\right)
\times {\mathbf E}-{\mathbf E}\mathrm{div}{\mathbf B}
-{\mathbf B}\mathrm{div}{\mathbf E}=0 .
\]
Therefore,
\[
\vec{\cal F}\times {\mathbf B}+\vec{\cal M}\times {\mathbf E}=
\]
\[
=f_1{\mathbf E}\times {\mathbf B}+f_2{\mathbf B}\times {\mathbf E}=
(f_1-f_2){\mathbf E}\times {\mathbf B}=0.
\]
The assertion follows. The relation $|\vec{\cal F}|=|\vec{\cal M}|$ also
holds. It is readily verified now that the scale factor $\mathcal{L}$ may be
defined by
\begin{equation}
\mathcal{L}(\mathbf{E},\mathbf{B})=
\frac{1}{|f_1|}=\frac{1}{|f_2|}=
\frac{|\mathbf{E}|}{|\vec{\mathcal{F}}|}=                      
\frac{|\mathbf{B}|}{|\vec{\mathcal{M}}|}.
\end{equation}

We go back to the 4-dimensional consideration.
Obviously, for all nonlinear solutions $F$ with $|\delta F|\neq 0$ we have
$0<\mathcal{L}(F)<\infty$, and for all other cases we readily obtain
$\mathcal{L}\rightarrow \infty$.  From the definition (43) of $\mathcal{L}$
is seen that $\mathcal{L}$ depends on the point in general.  Note that if the
scale factor $\mathcal{L}$, defined by the nonlinear solution $F$, is a {\it
finite} and {\it constant} quantity, we can introduce a {\it characteristic}
finite time-interval $T(F)$ by the relation $$ cT(F)=\mathcal{L}(F), $$ and
also, in some cases, the corresponding {\it characteristic frequency} by
$$
\nu(F)=1/T(F).
$$
In these wave terms the modified scale factor $2\pi\mathcal{L}$ acquires the
meaning of wave length.

When $\mathcal{L}=const\neq 0$ we can say that the inverse scale factor
$\mathcal{L}^{-1}$ is a measure of "how far" a nonlinear solution $F$ (with
$|\delta F|\neq 0$) is from the linear Maxwell solutions.

We modify now the volume form $\omega_{\chi}$ to $(-\frac 1c\omega_{\chi})$.
The integral of $(-\frac 1c\omega_{\chi})$ over the 4-region \newline
$\mathbb{R}^3\times\mathcal{L}, \mathcal{L}=const$, is well defined for all
3d-finite solutions, and its value is an integral invariant. Making use of an
$F$-adapted coordinate system we readily obtain
\[
-\frac1c\int_{\mathbb{R}^3\times\mathcal{L}}\omega_{\chi}=
\frac 1c\int_{\mathbb{R}^3\times\mathcal{L}}
(u^2+p^2)dx\wedge dy\wedge dz\wedge d\xi
=\frac 1c E\mathcal{L}=ET,
\]
where $E$ is the integral energy of the solution in this frame. Note that
this integral invariant depends only on the 3d-finite nature of the solution
and on the finite constant value of the scale factor $\mathcal{L}$.

\section{EED: Explicit non-linear vacuum solutions}
As it was shown with every nonlinear solution $F$ of our nonlinear equations
a class of $F$-adapted coordinate systems is associated, such that $F$ and
$*F$ acquire the form (39):
\[
\begin{split}
&F=\varepsilon udx\wedge dz + udx\wedge d\xi + \varepsilon pdy\wedge dz +
pdy\wedge d\xi \\
&*F=-pdx\wedge dz - \varepsilon pdx\wedge d\xi + udy\wedge dz +
\varepsilon udy\wedge d\xi.
\end{split}
\]
Since we look for non-linear solutions of (18)-(20), we substitute these $F$
and $*F$ and after some elementary calculations we obtain
\vskip 0.3cm
{\bf Proposition 9.} Every $F$ of the above kind satisfies the equation
$\delta F\wedge F-\delta *F\wedge *F=0$.
\vskip 0.2cm
{\bf Proof.} It is directly verified.
\vskip 0.2cm
\noindent
Further we obtain
\[
\delta F=(u_\xi-\varepsilon u_z)dx +(p_\xi-\varepsilon p_z)dy +
\varepsilon(u_x + p_y)dz +(u_x + p_y)d\xi,
\]
\[
\delta *F=-\varepsilon(p_\xi -\varepsilon p_z)dx
+\varepsilon(u_\xi-\varepsilon p_z)dy - (p_x - u_y)dz -
\varepsilon(p_x - u_y)d\xi,
\]
\[
F_{\mu\nu}(\delta F)^\nu
dx^\nu= (*F)_{\mu\nu}(\delta *F)^\nu dx^\nu=
\]
\[
=\varepsilon\left[p(p_\xi-\varepsilon p_z)+u(u_\xi-\varepsilon u_z)\right]dz+
\left[p(p_\xi-\varepsilon p_z)+u(u_\xi-\varepsilon u_z)\right]d\xi,
\]
\[
(\delta F)^2=(\delta *F)^2=
-(u_\xi-\varepsilon u_z)^2-(p_\xi-\varepsilon p_z)^2
=-\phi^2(\psi_\xi-\varepsilon \psi_z)^2.
\]
We infer that our equations reduce to only one equation, namely
\begin{equation}
p(p_\xi-\varepsilon p_z)+u(u_\xi-\varepsilon u_z)=
\frac 12\left[(u^2+p^2)_\xi-\varepsilon(u^2+p^2)_z\right]=0.   
\end{equation}
The obvious solution to this equation is
\begin{equation}
u^2+p^2=\phi^2 (x,y,\xi+\varepsilon z),                         
\end{equation}
where $\phi$ is an arbitrary differentiable function of its arguments.
The solution obtained shows that the equations impose some limitations only
on the amplitude function $\phi=|\alpha_F|$ and that the phase function
$\varphi$ is arbitrary except that it is bounded: $|\varphi|\leq 1$. The
amplitude $\phi$ is a running wave along the specially chosen coordinate $z$,
which is common for all $F$-adapted coordinate systems. Considered as a
function of the spatial coordinates, the amplitude $\phi$ is {\it arbitrary},
so it can be chosen {\it spatially finite}. The time-evolution does not
affect the initial form of $\phi$, so it will stay the same in time, but the
whole solution may change its form due to $\varphi$ . Since $|\varphi|\leq 1$
and the two independent field components are given by
$F_{14}=u=\phi\,\varphi$, $F_{24}=p =\pm\phi\,\sqrt{1-\varphi^2}$ this shows,
that {\it among the nonlinear solutions of our equations there are (3+1)
soliton-like solutions}. The spatial structure of $\phi$ can be determined by
initial condition, and the phase function $\varphi$ can be used to describe
additional structure features and {\it internal dynamics} of the solution.

We compute $\delta F\wedge F=\delta *F\wedge *F$ and obtain
\[
\delta F\wedge F=\Big[\varepsilon p(u_\xi-\varepsilon u_z)-
\varepsilon u(p_\xi-\varepsilon p_z)\Big]dx\wedge dy\wedge dz+
\Big[p(u_\xi-\varepsilon u_z)-
u(p_\xi-\varepsilon p_z)\Big]dx\wedge dy\wedge d\xi.
\]
In terms of $\phi$ and $\psi=\mathrm{arccos}(\varphi)$ we obtain
\[
p(u_\xi-\varepsilon u_z)-u(p_\xi-\varepsilon p_z)=
-\phi^2(\psi_\xi-\varepsilon \psi_z),
\]
and so
\[
\delta F\wedge F=-\varepsilon\phi^2(\psi_\xi-
\varepsilon \psi_z)dx\wedge dy\wedge dz-
\phi^2(\psi_\xi-\varepsilon \psi_z)dx\wedge dy\wedge d\xi.
\]
Applying $*$ from the left we get
\[
*(\delta F \wedge F)=-\phi^2(\psi_\xi-\varepsilon \psi_z)dz-
\varepsilon\phi^2(\psi_\xi-\varepsilon \psi_z)d\xi=
-\varepsilon\phi^2(\psi_\xi-\varepsilon\psi_z)\zeta.
\]
\vskip 0.3cm
{\bf Corollary.} $\delta F\wedge F$ is isotropic, and it is equal to zero
iff $\psi$ is a running wave along $z$.
\vskip 0.3cm
{\bf Corollary.} $F$ is a running wave along $z$ iff $\psi$ is a running wave
along $z$, i.e. iff $\delta F\wedge F=0$.
\vskip 0.3cm
Computing the 4-forms
$\mathbf{d}\tilde{A}\wedge\tilde{A}\wedge\zeta$ and
$\mathbf{d}\tilde{A}^*\wedge\tilde{A}^*\wedge\zeta$ we obtain
\[
\begin{split}
\mathbf{d}\tilde{A}\wedge\tilde{A}\wedge\zeta=
 \mathbf{d}\tilde{A}^*\wedge\tilde{A}^*\wedge\zeta &=
\varepsilon\big[u(p_\xi-\varepsilon p_z)-p(u_\xi-\varepsilon u_z)\big]
dx\wedge dy\wedge dz\wedge d\xi \\
&=\varepsilon\phi^2(\psi_\xi-
\varepsilon \psi_z)dx\wedge dy\wedge dz\wedge d\xi.
\end{split}
\]
Hence, since $(\delta F)^2=-\phi^2(\psi_\xi-\varepsilon \psi_z)^2$,
and $\phi\neq0$ is a running wave, the nonlinear solutions $F$ satisfying
$|\delta F|=|\delta *F|=0$, imply $\psi_\xi-\varepsilon \psi_z=0$, i.e.
{\it absence of rotational component of propagation}. The following
relations are equivalent:
\vskip 0.1cm
\hspace{2cm}1. $\delta F\wedge F=0$.

\hspace{2cm}2. $|\delta F|=|\delta *F|=0$.

\hspace{2cm}3. $\psi$ is a running wave along $z$: $L_\zeta \psi=0$.

\hspace{2cm}4. $\mathbf{d}\tilde{A}\wedge\tilde{A}\wedge\zeta=
\mathbf{d}\tilde{A}^*\wedge\tilde{A}^*\wedge\zeta=0.$
\vskip 0.1cm
We give now two other relations that are equivalent to the above four.
First, consider a nonlinear solution $F$, and the corresponding $(1,1)$ tensor
$F_\mu^\nu=\eta^{\nu\sigma}F_{\sigma\mu}$. We want to compute the
corresponding Fr\"olicher-Nijenhuis tensor $S_F=\big[F,F\big]$, which is
a 2-form on $M$ with values in the vector fields on $M$. The components of
$S_F$ in a coordinate frame are given by
\[
(S_F)_{\mu \nu }^\sigma =2\left[ F_\mu ^\alpha \frac{\partial F_\nu
^\sigma} {\partial x^\alpha }-F_\nu ^\alpha \frac{\partial F_\mu ^\sigma
}{\partial x^\alpha }-F_\alpha ^\sigma \frac{\partial F_\nu ^\alpha
}{\partial x^\mu } +F_\alpha ^\sigma \frac{\partial F_\mu ^\alpha }{\partial
x^\nu }\right].
\]
We recall the two unit vector fields $\tilde{\mathbf{A}}$ and
$\varepsilon\mathbf{\tilde{A^*}}$, given by (in a $F$-adapted coordinate
system)
\[
\mathbf{\tilde{A}}=-\varphi \frac{\partial} {\partial x}-\sqrt{1-\varphi ^2}
\frac{\partial}{\partial y},\quad
\varepsilon\mathbf{\tilde{A^*}} =
\sqrt{1-\varphi^2}\frac{\partial}{\partial x}-
\varphi \frac{\partial} {\partial y},
\]
and we compute $S_F(\mathbf{\tilde{A}},\varepsilon\mathbf{\tilde{A^*}})$.
\[
(S_F)_{\mu \nu }^\sigma
{\bf\tilde{A}}^\mu {\bf \varepsilon\tilde{A^*}}^\nu
=(S_F)_{12}^\sigma ({\bf\tilde{A}}^1{\bf\varepsilon\tilde{A^*}}^2-
{\bf\tilde{A}}^2{\bf\varepsilon\tilde{A^*}}^1).
\]
For $(S_F)_{12}^\sigma $ we get
\[
(S_F)_{12}^1=(S_F)_{12}^2=0,\quad
(S_F)_{12}^3=-\varepsilon (S_F)_{12}^4=2\varepsilon \{p(u_\xi -\varepsilon
u_z)-u(p_\xi -\varepsilon p_z)\}.
\]
It is easily seen that
${\bf A}^1{\bf \varepsilon A^*}^2-{\bf A}^2{\bf \varepsilon A^*}^1=1$, so,
the relation $S_F(\mathbf{\tilde{A}},\varepsilon\mathbf{\tilde{A^*}})=0$ is
equivalent to the above four.

Second, recall that if $\big[\mathcal{A}, (+,.)\big]$ is an algebra (may
graded), then the (anti)derivations
$\mathcal{D}:\mathcal{A}\rightarrow\mathcal{A}$ satisfy:
$\mathcal{D}(a.b)=\mathcal{D}a.b+\varepsilon_a a.\mathcal{D}b$, where $a,b\in
\mathcal{A}$ and $\varepsilon_a$ is the parity of $a\in\mathcal{A}$.  So,
the derivations are not morphisms of $\mathcal{A}$, and satisfy the
generalized Leibniz rule. The difference
$$
\Delta_D(a,b)=\Big[Da.b+\varepsilon_a a.Db\Big]-D(a.b)\quad,
a,b\in\mathcal{A}
$$
is called the Leibniz bracket of the operator $D$, and
$D$ is a (anti)derivation if its Leibniz bracket vanishes. If $\mathcal{A}$
is the exterior algebra of differential forms on a manifold $M$ and $D$ is
the coderivative $\delta$ with respect to a given metric, then the
corresponding Leibniz bracket (sometimes called the Schouten bracket) is
denoted by $\{,\}$. So, if $F,G$ are two forms on $M$ then $$ \{F,G\}=\delta
F\wedge G+(-1)^p F\wedge \delta G -\delta(F\wedge G).  $$ Note that the
brackets $\{F,F\}$ do not vanish in general.

Now, if the 2-form $F$ on the Minkowski space is a nonlinear solution, then
$F\wedge F=0$ and
\[
\{F,F\}=\delta F\wedge F+(-1)^2F\wedge \delta F-\delta(F\wedge F)
=2\delta F\wedge F.
\]
So, the above relations are equivalent to the requirement that the
Leibniz/Schouten bracket $\{F,F\}$ vanishes.

Finally, all these conditions are equivalent to $\mathcal{L}=\infty$.

We note the very different nature of these seven conditions.  The complete
integrability of any of the two Pfaff 2-dimensional systems
$(\tilde{A},\zeta)$ and $(\tilde{A}^*,\zeta)$ is equivalent to zero value of
$|\delta F|$ on the one hand, and to the zero value of the quantity
$S_F(\mathbf{\tilde{A}},\varepsilon\mathbf{\tilde{A^*}})$ on the other hand,
and both are equivalent to the vanishing of the Leibniz/Schouten bracket and
to the infinite value of $\mathcal{L}$. This could hardly be occasional, so, a
physical interpretation of these quantities in the nonzero case, i.e. when
$\psi$ is not a running wave, is strongly suggested. In view of the above
conclusion that the condition $|\delta F|=0$ implies absence of rotational
component of propagation, our interpretation is the following:
\vskip 0.4cm
{\bf A nonlinear solution will carry rotational component of propagation,
i.e. {\it intrinsic angular (spin) momentum}, only if $\psi$ is NOT a running
wave along the direction of translational propagation}.
\vskip 0.4cm
Natural measures of this spin momentum appear to be $\delta
F\wedge F\neq 0$, or $|\delta F|\neq 0$, or $S_F(\mathbf{\tilde{A}},
\varepsilon\mathbf{\tilde{A^*}})$. The most attractive seems to be $\delta
F\wedge F$, because it is a 3-form, and imposing the requirement
$\mathbf{d}(\delta F\wedge F)=0$ we obtain both: the equation for $\psi$ and
the corresponding conserved (through the Stokes' theorem) quantity
$\mathcal{H}=\int{i^*(\delta F\wedge F)}dx\wedge dy\wedge dz$, where
$i^*(\delta F\wedge F)$ is the restriction of $\delta F\wedge F$ to
$\mathbb{R}^3$.

Making use of the relations $u=\phi\,\mathrm{cos}\psi$ and
$p=\phi\,\mathrm{sin}\psi$, we get
\begin{equation}
\begin{split}
&A=\phi\,\mathrm{cos}\,\psi\,dx+\phi\,\mathrm{sin}\,\psi\,dy +f\zeta,
\quad A^*=-\varepsilon\phi\,\mathrm{sin}\,\psi\,dx+
\varepsilon\phi\,\mathrm{cos}\,\psi\,dy + f^*\zeta, \\
&\delta F=-\phi\,\mathrm{sin}\,\psi\,(\psi_\xi-\varepsilon \psi_z)\,dx+
\phi\,\mathrm{cos}\,\psi\,(\psi_\xi-\varepsilon \psi_z)\,dy+
(u_x+p_y)\zeta \\
&=-\varepsilon(L_\zeta \psi)A^*+(\varepsilon f^*L_\zeta \psi+u_x+p_y)\zeta \\
&\delta *F=-\varepsilon\phi\,\mathrm{cos}\,\psi\,                     
(\psi_\xi-\varepsilon \psi_z)\,dx-
\varepsilon\phi\,\mathrm{sin}\,\psi\,(\psi_\xi-\varepsilon \psi_z)\,dy-
\varepsilon (p_x-u_y)\zeta \\
&=\varepsilon(L_\zeta \psi)A-
\varepsilon(fL_\zeta \psi+p_x-u_y)\zeta,
\end{split}
\end{equation}
where $L_\zeta$ is the Lie derivative with respect to $\zeta$.
We obtain also
\begin{equation}
\begin{split}
|\delta F| & =|\delta *F|=
\frac{\phi|\varphi_\xi-\varepsilon\varphi_z|}{\sqrt{1-\varphi^2}}=
\phi|\psi_\xi-\varepsilon \psi_z|=\phi|L_\zeta \psi|,\\       
\mathcal{L} & =\frac{|A|}{|\delta F|}=
\frac{\sqrt{1-\varphi^2}}{|\varphi_\xi-\varepsilon\varphi_z|}=
\frac{1}{|\psi_\xi-\varepsilon \psi_z|}=|L_\zeta \psi|^{-1}.
\end{split}
\end{equation}
Now we can write
\begin{equation}
\delta F=\pm\varepsilon\frac{A^*}{\mathcal{L}}+
\left(\mp \varepsilon\frac{f^*}{\mathcal{L}}+(u_x+p_y)\right)\zeta,\
\quad                                                                 
\delta *F=\mp\varepsilon\frac{A}{\mathcal{L}}+ \left(\pm
\varepsilon\frac{f}{\mathcal{L}}-\varepsilon (p_x-u_y)\right)\zeta.
\end{equation}
\vskip 0.3cm
{\bf Corollary.} Obviously, the following relations hold:
$$
A_\mu(\delta F)^\mu=0,\ \
(A^*)_\mu(\delta *F)^\mu=0,\ \
A_\mu(\delta *F)^\mu=\pm \varepsilon\frac{\phi^2}{\mathcal{L}}
=-(A^*)_\mu(\delta F)^\mu.
$$

Finally we note that since the propagating along the given $\zeta$ nonlinear
solutions in canonical coordinates are parametrized by one function $\phi$ of
3 independent variables and one {\it bounded} function $\varphi$ of 4
independent variables, the separation of various subclasses of nonlinear
solutions is made by imposing additional conditions on these two functions.

\section{Homological Properties of the Nonlinear Solutions}
From pure algebraic point of view we speak about homology (or, cohomology)
every time when we meet a linear map $D$ in a vector space $\mathbb{V}$ over
a field (e.g. $\mathbb{R}$, or $\mathbb{C}$), or in a module $\mathbb{W}$
over some ring, having the property $D\circ D=0$. Then we have two related
subspaces, $Ker(D)=\{x\in \mathbb{V}: D(x)=0\}$ and
$Im(D)=D(\mathbb{V})$. Since $Im(D)$ is a subspace of $Ker(D)$, we can
factorize, and the corresponding factor space $H(D,\mathbb{V})=Ker(D)/Im(D)$
is called the {\it homology space} for $D$. The dual linear map $D^*$ in the
dual space $\mathbb{V}^*$ has also the property $D^*\circ D^*=0$, so we
obtain the corresponding {\it cohomology space} $H^*(D^*,\mathbb{V}^*)$. In
such a situation the map $D$ (resp. $D^*$) is called {\bf boundary operator}
(resp {\bf coboundary operator}). The elements of $Ker(D)$ (resp. $Ker(D^*)$)
are called {\it cycles} (resp. {\it cocycles}), and the elements of $Im(D)$
(resp. $Im(D^*)$) are called {\it boundaries} (resp. {\it coboundaries}).

The basic property of a boundary operator $D$ is that every linear map
$\mathcal{B}:\mathbb{V}\rightarrow \mathbb{V}$ which commutes with $D$:
$D\circ \mathcal{B}=\mathcal{B}\circ D$, induces a linear map
$\mathcal{B}_*: H(D)\rightarrow H(D)$. So, a boundary operator realizes the
general idea of distinguishing some properties of a class of objects which
properties are important from a definite point of view, and to find those
transformations which keep invariant these properties.

The basic example for boundary operator used in theoretical physics is the
exterior derivative $\mathbf{d}$ (the so called de Rham cohomology).
This operator acts in the space of differential forms over a manifold, e.g.
the Euclidean space $(\mathbb{R}^3,g)$, or the Minkowski space-time
$M=(\mathbb{R}^4,\eta)$. The above mentioned basic property of every boundary
operator $D$ appears here as a commutation of $\mathbf{d}$ with the smooth
maps $\mathfrak{f}$ of the manifold considered:
$\mathbf{d}\circ\mathfrak{f}^*=\mathfrak{f}^*\circ \mathbf{d}$.  A well known
physical example for cocycles of $\mathbf{d}$ (which are called here {\it
closed differential forms}) comes when we consider a spherically symmetric
gravitational or electrostatic field generated by a point source.  Since the
field is defined only outside of the point-source, i.e. on the space
$N=\mathbb{R}^3-\{0\}$, then a natural object representing the field is a
closed differential 2-form $\omega, \mathbf{d}\omega=0$. In view of the
spherical symmetry such a (spherically symmetric or $SO(3)$-invariant) closed
2-form is defined up to a constant coefficient $q$, and in standard spherical
coordinates originating at the point-source we obtain
$\omega=q\,\mathrm{sin}\,\theta\, d\theta\wedge d\varphi$. The Hodge star on
$N$ gives $*\omega=(q/r^2)dr$, which is usually called electric field
generated by the point source $q$.  Now, the Stokes' theorem establishes the
charge $q$ as a topological invariant, characterizing the nontrivial topology
of the space $N$.

The above mentioned boundary operator $\mathbf{d}$ is a
differential operator. Pure algebraic boundary operators in a linear space
$V$ also exist and one way to introduce such operators is as
follows.  Let $V^*$ be the dual to $V$ space and $\langle,\rangle$ denotes
the canonical conjugation:  $(x^*,x)\rightarrow\langle x^*,x\rangle$, where
$x^*\in V^*$ and $x\in V$.  Now fix $x\in V$ and let $x^*$ be such that
$\langle x^*,x\rangle =0$.  Consider now the decomposable element $x^*\otimes
x \in V^*\otimes V$. The isomorphism between $V^*\otimes V$ and $L(V,V)$
allows to consider $x^*\otimes x$ as a linear map
$\varphi_{(x^*,x)}:V\rightarrow V$ as follows:
$$
\varphi_{(x^*,x)}(y)=(x^*\otimes x)(y)=\langle x^*,y\rangle x, \ \ y\in V.
$$
Hence, the linear map $\varphi_{(x^*,x)}$ sends all elements of $V$ to the
1-dimensional space determined by the non-zero element $x\in V$. Clearly,
$\varphi_{(x^*,x)}$ is a boundary operator since
\[
\varphi_{(x^*,x)}\circ\varphi_{(x^*,x)}(y)=
\varphi_{(x^*,x)}(\langle x^*,y\rangle x)=
\langle x^*,y\rangle\varphi_{(x^*,x)}(x)=
\langle x^*,y\rangle\langle x^*,x\rangle x=0,\ \ y\in V.
\]
So, if $\{e_i\}$ and $\{\varepsilon^j\}$ are two dual bases in the
$n$-dimensional linear spaces $V$ and $V^*$ respectively, then every couple
$(\varepsilon^j,e_i), i\neq j$, determines the linear boundary operators
$\varphi_{(\varepsilon^j,e_i)}=\varepsilon^j\otimes e_i, i\neq j$.

If $g$ is an inner product in $V$ and the two elements $(x,y)$ are
$g$-orthogonal: $g(x,y)=0$, then denoting by $\tilde{g}$ the linear
isomorphism $\tilde{g}: V\rightarrow V^*$ (lowering indices) we can define
the linear map $\tilde{g}(x)\otimes y$, which is obviously a boundary
operator in $V$:  $$ (\tilde{g}(x)\otimes y)\circ(\tilde{g}(x)\otimes y)(z)=
\langle \tilde{g}(x),z\rangle(\tilde{g}(x)\otimes y)(y)=
\langle \tilde{g}(x),z\rangle g(x,y)(y)=0, \ \ z\in V.
$$
In particular, if $(V,\eta)$ is the Minkowski space, then every isotropic
vector $\zeta: \eta(\zeta,\zeta)=\zeta^2=0$ defines a boundary operator
$\tilde{\eta}(\zeta)\otimes\zeta=\tilde{\zeta}\otimes\zeta$. In fact,
\[
(\tilde{\eta}(\zeta)\otimes\zeta)\circ
(\tilde{\eta}(\zeta)\otimes\zeta)(x)=
(\tilde{\zeta}\otimes\zeta)\circ(\tilde{\zeta}\otimes\zeta)(x)=
\langle \tilde{\zeta},x\rangle (\tilde{\zeta}\otimes\zeta)(\zeta)=
\langle \tilde{\zeta},x\rangle \langle \tilde{\zeta},\zeta\rangle(\zeta)=0,
\ \ x\in V.
\]
We note that the duality between $V$ and
$V^*$ allows to consider the element $x^*\otimes x$ as a linear map in $V^*$
as follows $(x^*\otimes x)(y^*)=\langle y^*,x\rangle x^*$, and to build the
corresponding boundary operators in $V^*$. Also, in the finite dimensional
case we always get $dim(V)-2=dim[Ker(x^*\otimes x)/Im(x^*\otimes x)]$, where
$<x^*,x>=0$.

The above mentioned property that the image space of any such boundary
operator $x^*\otimes x$, where $\langle x^*, x\rangle =0$, is 1-dimensional,
implies that the natural extensions of these boundary operators to
derivations in the graded exterior algebras $\Lambda(V)$ and $\Lambda(V^*)$
[28] define boundary operators of degree zero in these graded algebras (see
further).

Further in this section we show that the energy-momentum tensors of the
nonlinear vacuum solutions in EED give examples of such boundary operators,
and we also present some initial study of the corresponding homology.
We have the following property of the electromagnetic
stress-energy-momentum tensor $Q(F)_\mu^\nu$ established by Rainich [29]

\begin{equation}
Q(F)_\mu^\sigma Q(F)_\sigma^\nu=\frac14(I_1^2+I_2^2)\delta
_\mu^\nu,                                                      
\end{equation}
where $I_1=\frac12 F_{\mu\nu}F^{\mu\nu}$ and
$I_2=\frac12 F_{\mu\nu}(*F)^{\mu\nu}$ are the two invariants. Considering
$Q(F)_\mu^\nu$ as a linear map in the module of vector fields, or 1-forms,
over the Minkowski space-time, we see that if the two invariants are equal to
zero, {\it as it is in the case of nonlinear solutions of our equations}, we
obtain a boundary operator at all points of $M$ where $F\neq 0$.
Our purpose now is to consider how the corresponding homology is connected
with the structure of the nonlinear solutions $F$ of the vacuum EED
equations.
\vskip 0.3 cm
{\bf Proposition 10.} The image space $Im(Q_F)$ coincides with the only
isotropic eigen direction of $Q_F$.
\vskip 0.2cm
{\bf Proof.} The linear map $Q_F$ is given in an $F$-adapted
coordinate system by
\begin{equation}
Q_F=-\phi^2 dz\otimes \frac{\partial}{\partial z}+
\varepsilon \phi^2 dz\otimes\frac{\partial}{\partial\xi}
- \varepsilon \phi^2 d\xi\otimes\frac{\partial}{\partial z} +
\phi^2 d\xi\otimes \frac{\partial}{\partial \xi}.
\end{equation}
Clearly, $Q_F=\phi^2\tilde{\zeta}\otimes\zeta$, where
$\tilde\zeta=\varepsilon dz+d\xi$ in the $F$-adapted coordinate system
and $\zeta=-\varepsilon \partial_z+\partial_\xi$
defines the only isotropic eigen direction of $Q_F$.  Let in this coordinate
system the arbitrary vector field $X$ be presented by its components
$(X^\mu),\ \mu=1,\dots,4$. We obtain

\[
\begin{split}
Q_F(X)&=-\phi^2 dz(X)\frac{\partial}{\partial z}+
\varepsilon \phi^2 dz(X)\frac{\partial}{\partial\xi}
- \varepsilon \phi^2 d\xi(X)\frac{\partial}{\partial z} +
\phi^2 d\xi(X)\frac{\partial}{\partial \xi} \\
&=-\phi^2 X^3\frac{\partial}{\partial z}+
\varepsilon \phi^2 X^3\frac{\partial}{\partial\xi}
- \varepsilon \phi^2 X^4\frac{\partial}{\partial z} +
\phi^2 X^4\frac{\partial}{\partial \xi} \\
&= \varepsilon\phi^2X^3\zeta+\phi^2 X^4\zeta=
\phi^2(\varepsilon X^3+X^4)\zeta.
\end{split}
\]
If $\alpha=\alpha_{\mu}dx^\mu$ is a 1-form then in the same way we obtain
\[
Q_F(\alpha)=(Q_F)^\nu_\mu\alpha_\nu dx^\mu=
\phi^2(-\varepsilon\alpha_3+\alpha_4)\zeta.
\]
The proposition is proved.
\vskip 0.3cm
{\bf Proposition 11.} The kernel space $Ker(Q_F)$ coincides with the 3-space
spanned by the vectors $A, A^*$ and $\zeta$.
\vskip 0.2cm
{\bf Proof.}
We know that $Q_F(A)=Q_F(A^*)=Q_F(\zeta)=0$. Now, if $X$ is an arbitrary
vector, then from the above Prop.10 follows that $Q_F(X)=\phi^2(\varepsilon
X^3+X^4)\zeta$, so we
conclude that $Q_F(X)$ will be equal to zero only if $X$ is a linear
combination of $A, A^*$ and $\zeta$.
\noindent
The proposition is proved,

Hence, we may write $Ker(Q_F)=\{A\}\oplus\{A^*\}\oplus\{\zeta\}$. The
corresponding factor space
$$
H(Q_F)=Ker(Q_F)/Im(Q_F)
$$
is isomorphic to $\{A\}\oplus\{A^*\}$.  The classes defined by $A$ and $A^*$
are given
by $[A]=A+f\zeta$ and $[A^*]=A^*+f^*\zeta$, where $f$ and $f^*$ are
functions.  Recall now that $Q_F(\delta F)=Q_F(\delta *F)=0$, so $[\delta
F]=\delta F +h\zeta$ and $[\delta *F]=\delta *F+h^*\zeta$, where $h$ and
$h^*$ are functions, give the corresponding homology classes.

According to the above mentioned property, every symmetry of a boundary
operator induces a linear map inside the homology space. Therefore,
the homology spaces are invariant with respect to
the linear isomorphisms which commute with the boundary operators. In our
case we have to find those linear maps $\Phi$ in the module of vector fields
over $M$, which commute with $Q_F$, i.e.  $\Phi\circ Q_F=Q_F\circ\Phi$. It is
readily obtained that in the $F$-adapted coordinate system every such $\Phi$
is given by a matrix of the following kind:
\[
\Phi=\begin{Vmatrix}a & b & c & -\varepsilon c\\
		    m & n & q & -\varepsilon q\\
		    r & s & w &  0\\
	 \varepsilon r & \varepsilon s & 0 & w\end{Vmatrix},
\]
where all nine independent entries of this matrix are functions of the
coordinates. It follows that the $Q_F$- homology spaces are invariant with
respect to all diffeomorphisms $\varphi: M\rightarrow M$ which generate
isomorphisms $d\varphi: TM\rightarrow TM$ of the tangent bundle of $M$ given
in the $F$-adapted coordinate system by a nondegenerate matrix of the above
kind.

An important property of the boundary operator $Q_F$ is that its image space
$Im(Q_F)$ is 1-dimensional. As it was mentioned earlier, this allows to
extend $Q_F$ as a boundary operator in the graded exterior algebra of
differential forms over $M$. In fact, recall that a linear map $\varphi$ in a
linear space $V$ induces derivation $\varphi^{\wedge}$ in
the exterior algebra $\Lambda(V)$ according to the rule
\[
\varphi^{\wedge}(x_1\wedge x_2\wedge \dots \wedge
x_p)= \varphi(x_1)\wedge x_2\wedge\dots \wedge x_p +x_1\wedge
\varphi(x_2)\wedge\dots \wedge x_p+ \dots +x_1\wedge x_2\wedge\dots
\wedge\varphi(x_p).
\]
\vskip 0.3cm
\noindent
{\bf Remark}.\ If we try to extend $Q_F$ to antiderivation with respect to the
usual involution \linebreak
$\omega(\alpha)=(-1)^p \alpha,\  \alpha\in\Lambda^p(M)$ we'll
find that this is not possible since the necessary condition for this, given
by $Q_F(\alpha)\wedge\alpha+\omega(\alpha)\wedge Q_F(\alpha)=0$, does not
hold for every $\alpha\in\Lambda^1(M)$.
\vskip 0.3cm
Hence, if $Im(\varphi)=\varphi(V)$ is 1-dimensional, then
every summond of $\varphi^{\wedge}\circ\varphi^{\wedge}(x_1\wedge x_2\wedge
\dots \wedge x_p)$
will contain two elements of the kind $\varphi(x_i)$ and
$\varphi(x_j)$, and if these two elements are collinear, their exterior
product is zero and the corresponding summond is zero. In our case
$\varphi=Q_F$ and $Im(Q_F)=\{\zeta\}$ is 1-dimensional, so we shall have
$Q_{F}^{\wedge}\circ Q_{F}^{\wedge}(\alpha)=0, \alpha\in \Lambda(M)$.
\vskip 0.3cm
{\bf Corollary}. The extension $Q_F^{\wedge}$ defines a boundary operator
of degree zero in $\Lambda(M)$.
\vskip 0.3cm
\noindent
{\bf Remark}. Further the extension $Q_F^{\wedge}$ will be denoted just by
$Q_F$.
\vskip 0.3cm
{\bf Corollary}. The extension of $Q_F$ to derivation
in $\Lambda (M)$ introduces in $\Lambda(M)$ some structure of graded
{\it differential} algebra with corresponding {\it graded homology algebra}
$H(Q_F)(\Lambda(M))$.
\vskip 0.3cm
The following relations are readily verified:
\[
Q_F(F)=0,\ \ Q_F(*F)=0,\ \
Q_F(\delta F\wedge F)=0.
\]
For example,
$Q_F(F)=Q_F(A\wedge\zeta)=Q_F(A)\wedge\zeta+A\wedge Q_F(\zeta)=0$.  So, $F,
*F$ and $\delta F\wedge F$ are $Q_F$-cycles.

From relations (50) we obtain
\[
[\delta F]=-\varepsilon (L_\zeta\psi)[A^*],\quad
[\delta *F]=\varepsilon (L_\zeta\psi)[A],
\]
i.e. $\delta F$ and $A^*$ define the same $Q_F$-homology classes, and
$\delta *F$ and $A$ define the same $Q_F$-homology classes in $\Lambda^1(M)$.
\vskip 0.3 cm
{\bf Proposition 12.} The scale factor $\mathcal{L}=|A|/|\delta F|$
depends only on the classes of $A$ and $\delta F$.
\vskip 0.2cm
{\bf Proof.}
Since $|[A]|=|A+f\zeta|=|A|$ and $|[\delta F]|=|\delta F+h\zeta|$ we obtain
$\mathcal{L}=|A|/|\delta F|=|[A]|/|[\delta F]|$.  \vskip 0.3cm Clearly,
$([A],[A^*])$ and $([\delta F],[\delta *F])$ represent two bases of
$H(Q_F)$ in $\Lambda^1(M)$.
\vskip 0.3cm
{\bf Corollary.} The transformation $([A],[A^*])\rightarrow
([\delta F],[\delta *F])$ is given by
\begin{equation}
([A],[A^*])\begin{Vmatrix}0 & \varepsilon L_\zeta\psi \\   
-\varepsilon L_\zeta\psi & 0\end{Vmatrix}=
(-\varepsilon L_\zeta \psi[A^*], \varepsilon L_\zeta\psi [A])=
([\delta F],[\delta *F]).
\end{equation}
The above formula (55) shows that the transformation matrix, further denoted
by $\mathcal{M}$, between these two bases is
$(\pm\varepsilon\mathcal{L})^{-1}J$,
where $J$ is the canonical complex structure in a real 2-dimensional space.
This fact may give another look on the duality symmetry (Recall Prop.3),
because of the invariance of $J$ with respect to the transformation
$J\rightarrow S.J.S^{-1}$, where $S$ is given on p.7  :
\[
\begin{Vmatrix}a & b \\-b & a\end{Vmatrix}
\begin{Vmatrix}0 & 1 \\-1 & 0\end{Vmatrix}
\begin{Vmatrix}a &-b \\ b & a\end{Vmatrix}\frac{1}{a^2+b^2}=
\begin{Vmatrix}0 & 1 \\-1 & 0\end{Vmatrix}.
\]
One could say that the duality symmetry of the nonlinear solutions is a
consequence of the null-field homology presented. In other words, every
initial null-field configuration given by $(A,A^*,\zeta)$, with $|\delta
F|\neq 0, |\delta *F|\neq 0, \zeta^2=0$, compulsory has
rotational-translational {\it dynamical} nature, so, it is {\it intrinsically
forced} to propagate with rotational component of propagation in space-time,
because, according to relation (55), the nonzero $F_{\mu\nu}$, i.e. the
nonzero $(A,A^*)$, imply nonzero values of the derivatives of $F_{\mu\nu}$
including the nonzero value of $L_\zeta \psi$ even if $\psi$ is
time-independent, and the basis $([A],[A^*)]$ is continuously forced to
rotate. In fact, the running-wave character of $\phi=|A|$ drags the solution
along the coordinate $z$ and the nonzero $L_\zeta \psi$ implies
$\mathrm{cos}\,\psi\neq 0, \mathrm{sin}\,\psi\neq 0$. The evolution obtained
is strongly connected with the nonzero {\it finite} value of the scale factor
$\mathcal{L}=|L_\zeta \psi|^{-1}$, which, in turn, determines rotation in the
homology space $H_F(Q)$. This rotation is determined entirely by the Lie
derivative of the phase $\psi$ with respect to $\zeta$, and it is
intrinsically consistent with the running wave translational propagation of
the energy-density $\phi^2$. It is seen that the field configuration has a
rotational component of propagation, while the energy density has just
translational component of propagation.

It is interesting to see the action of $Q_F$ as {\it derivation}
in $\Lambda(M)$. We shall do this in a
$F$-adapted coordinate system. Let $F$ define a nonlinear solution and
$Q_F: \Lambda(M)\rightarrow \Lambda(M)$ be the corresponding derivation
with $\phi^2=u^2+p^2$ the corresponding energy-density.  We give first the
action of $Q_F$ as {\it derivation} on the bases elements.
\begin{align*}
&Q_F(dx)=Q_F(dy)=0,\quad
Q_F(dz)=-\varepsilon\phi^2 \zeta,\quad
Q_F(d\xi)=\phi^2 \zeta
\end{align*}
\begin{alignat*}{3}
 Q_F(dx\wedge dy)&=0, & \quad
 Q_F(dx\wedge dz) &=-\varepsilon\phi^2 dx\wedge\zeta, & \quad
 Q_F(dy\wedge dz) &=-\varepsilon\phi^2 dy\wedge\zeta,  \\
 Q_F(dx\wedge d\xi) &=\phi^2 dx\wedge\zeta, & \quad
 Q_F(dy\wedge d\xi) &=\phi^2 dy\wedge\zeta, & \quad
 Q_F(dz\wedge d\xi) &=0.
\end{alignat*}
\begin{alignat*}{2}
Q_F(dx\wedge dy\wedge dz)&=-\varepsilon\phi^2 dx\wedge dy\wedge\zeta, &\qquad
Q_F(dx\wedge dy\wedge d\xi)&=\phi^2 dx\wedge dy\wedge\zeta, \\
Q_F(dx\wedge dz\wedge d\xi)&=0, & \qquad
Q_F(dy\wedge dz\wedge d\xi)&=0,
\end{alignat*}
\[
Q_F(dx\wedge dy\wedge dz\wedge d\xi)=0.
\]

Let now the arbitrary 2-form $G$ be represented in this coordinate system by
$G=G_{\mu\nu}dx^\mu\wedge dx^\nu,\ \ \mu<\nu $. Making use of the above given
explicit form for the action of $Q_F$ on the basis elements as {\it
derivation} we obtain
\begin{equation}
\begin{split}
Q_F(G) & =\varepsilon\phi^2(-\varepsilon G_{13}+G_{14})dx\wedge dz+
\varepsilon\phi^2(-\varepsilon G_{23}+G_{24})dy\wedge dz+\\
& \phi^2(-\varepsilon G_{13}+G_{14})dx\wedge d\xi+          
\phi^2(-\varepsilon G_{23}+G_{24})dy\wedge d\xi.
\end{split}
\end{equation}
This result makes possible the following conclusions concerning 2-forms:

\hspace{1.5cm}1. The space $Im(Q_F)$ consists of null fields,
i.e. every nonlinear solution $F$ determines a subspace
$Im(Q_F)\subset \Lambda^2(M)$ of null-fields .

\hspace{1.5cm}2. The space $Ker(Q_F)$ consists of 2-forms, which in this
coordinate system satisfy:
$\varepsilon G_{13}=G_{14},\ \varepsilon G_{23}=G_{24}$,
and $(G_{12},G_{34})$-arbitrary.

\hspace{1.5cm}3. The eigen spaces of $Q_{F}(G)$ coincide with the eigen
spaces of $F$ for every (nonzero) $G\in \Lambda^2(M)$.

If $G$ is a 3-form with components $G_{123}, G_{124}, G_{134}, G_{234}$ in
the same $F$-adapted coordinate system, we obtain
\[
Q_F(G)=\varepsilon \phi^2(-\varepsilon G_{123}+G_{124})dx\wedge dy\wedge dz+
\phi^2(-\varepsilon G_{123}+G_{124})dx\wedge dy\wedge d\xi.
\]
So, $Q_F(G)$ is isotropic, and a 3-form $G$ is in $Ker(Q_F)$ only if
$\varepsilon G_{123}=G_{124}$ in this coordinate system. Moreover, since
$Q_F(G)$ does not depend on $G_{134}dx\wedge dz\wedge d\xi$ and
$G_{234}dy\wedge dz\wedge d\xi$ we conclude that the kernel of $Q_F$ in this
case consists of time-like 3-forms.

Finally, if $G$ is a 4-form, then $Q_F(G)=0$.
\vskip 0.3cm
{\bf Corollary.} If $G\in\Lambda(M)$ lives in $Im(Q_F)$ where $F$ is a
nonlinear solution, then $Q_F(G)$ is isotropic.
\vskip 0.3cm
This may be extended to the smooth functions $f\in C^{\infty}(M)$ if we assume
$Q_F(f)=0$.

Let's resume. Every space-like (straight-line) direction may be chosen for
$z$-coordinate on $M$, and the 1-form $\zeta=\varepsilon dz+d\xi$ determines
an isotropic direction along which a class of null-fields $F$ are defined by
the formula (39). The corresponding linear map $Q_F$ satisfies
$Q_F\circ Q_F=0$ and defines homology in the spaces of 1-forms and of vector
fields.

Since the image space $Q_F(\Lambda^1 M)$ is 1-dimensional, $Q_F$ extends to a
boundary operator in the whole exterior algebras over the 1-forms and vector
fields.  The image space of the extended $Q_F$ consists of isotropic (null)
objects.  If $G$ is a 2-form then $Q_F(G)$ has, in general, the same eigen
properties as $F$.  Hence, every 2-form $F$ with zero invariants lives in
just one such subclass and the whole set of these 2-forms divides to such
nonintersecting subclasses. Moreover, every 2-form $G$ has its (null-field)
image in every such subclass.

For every nonlinear solution $F(u,p)$, ($\delta F\neq 0$) the corresponding
$\phi^2=u^2+p^2$ propagates translationally, i.e. is a running
wave, along the space-like direction chosen (considered as the coordinate
$z$). The 4-dimensional versions of the corresponding electric and magnetic
fields are presented by the nonisotropic parts of the homology classes
defined by the mutually orthogonal space-like 1-forms $A$ and $(-A^*)$. The
nonlinear solutions with $|\delta F|=|\delta *F|=0$ propagate only
translationally, i.e. without rotation.  Rotational components of
propagation, or spin-momentum, may have just those nonlinear solutions having
nonzero finite scale factor $\mathcal{L}=|A|/|\delta F|$, or equivalently,
satisfying one of the conditions given above (pp.27-28).  The isotropic
3-form $\delta F\wedge F$ defines a $Q_F$-homology class since
$Q_F(\delta F\wedge F)=0$, and it appears as a natural candidate representing
locally the spin-momentum if we assume the additional equation
$\mathbf{d}(\delta F\wedge F)=0$, which should reduce to an equation for the
phase $\psi$.  The two mutually orthogonal space-like 1-forms $\delta F$ and
$\delta *F$ define the same homology classes as $A^*$ and $A$ respectively.
The transformation matrix $\mathcal{M}$ between the two bases $(A,A^*)$ and
$(\delta *F,\delta F)$ defines a complex structure in the 2-dimensional
homology space through the scale factor:
$\mathcal{M}[A]=\mathcal{M}(A+f\zeta)=\mathcal{M}(A)+f\zeta=\delta
*F+f\zeta=[\delta *F]$
and $\mathcal{M}=\pm\varepsilon\mathcal{L}^{-1}J$, where $J$ is the canonical
complex structure in a 2-dimensional space. The 2-parameter duality symmetry
coincides with the symmetries of $\mathcal{M}$ and transforms solutions to
solutions inside the subclass of solutions propagating along the spatial
direction chosen.

\section{Structure of the Nonlinear Solutions}
We return now to the correspondence $F\rightarrow \alpha_F$.

\subsection{Properties of the duality matrices}
We consider the set $\mathbb{G}$ of matrices $\alpha$ of the kind
\begin{equation}
\alpha=\begin{Vmatrix} a & b \\-b & a                              
\end{Vmatrix}, \quad \text{where}\quad a,b\in\mathbb{R}.
\end{equation}
As we mentioned earlier, the nonzero matrices of this kind form a group
$\mathbb{G}$ with respect to the usual matrix multiplication. Together with
the zero $2\times 2$ matrix $I_o$ they also form a 2-dimensional linear space
$\mathcal{G}=\mathbb{G}\bigcup\{I_o\}$ over $\mathbb{R}$ with respect to the
usual addition of matrices. As is well known $\mathcal{G}$ gives the real
representation of the field of complex numbers.

A natural basis of the linear space $\mathcal{G}$ is given by the two
matrices
\[
I=\begin{Vmatrix}1 & 0 \\ 0 & 1\end{Vmatrix},\quad
J=\begin{Vmatrix}0 & 1 \\-1 & 0\end{Vmatrix}.
\]
The group $\mathbb{G}$ is commutative, in fact
\[
\begin{Vmatrix}a & b \\-b & a\end{Vmatrix}.  \begin{Vmatrix}m & n \\-n &
m\end{Vmatrix}= \begin{Vmatrix}m & n \\-n & m\end{Vmatrix}.
\begin{Vmatrix}a & b \\-b & a\end{Vmatrix}=
\begin{Vmatrix}am-bn & (an+bm) \\-(an+bm) & am-bn\end{Vmatrix} .
\]
Every element $\alpha\in\mathcal{G}$ can be
represented as $\alpha=aI+bJ,\ a,b\in \mathbb{R}$. Recall the  natural
representation of $\mathbb{G}$ in $\mathcal{G}$ given by
$$
\rho(\alpha)(aI+bJ)=a\,(\alpha^{-1})^*(I)+b\,(\alpha^{-1})^*(J)
=\frac{1}{a^2+b^2}\Big[a\,\alpha(I)+b\,\alpha(J)\Big].
$$
From now on we shall consider
$\alpha(I)$ and $\alpha(J)$  just as matrix product, so we have
\[
\alpha(I)=\alpha.I=\alpha=aI+bJ,\quad
\alpha(J)=\alpha.J=-bI+aJ.
\]
Since $(J^{-1})^*=J$, $J$ generates a complex structure in $\mathcal{G}$:
$J\circ J(x)=-x, \ x\in \mathcal{G}$.

The product of two matrices $\alpha=aI+bJ$ and $\beta=mI+nJ$ looks like
$\alpha.\beta=(am-bn)I+(an+bm)J$. The commutativity of $\mathbb{G}$ means
symmetry, in particular, every $\alpha\in \mathbb{G}$ is a symmetry of $J$:
$\alpha.J=J.\alpha$.

Finally we note, that
the inner product $g_e$ in $\mathcal{G}=T_e(\mathbb{G})$, where $e$ is the
identity of $\mathbb{G}$, given by
\[
g_e(\alpha,\beta)=\frac12 tr(\alpha\circ\beta^*),
\quad \alpha,\beta\in\mathcal{G},
\]
generates a (left invariant) riemannian metric on $\mathbb{G}$ by means
of the (left) group multiplication:
\[
g_{\sigma}=(L_{\sigma^{-1}})^*g_{e},\quad \sigma\in\mathbb{G}.
\]

\subsection{The action of $\mathbb{G}$ in the space of 2-forms on $M$}
We consider now the space
$\Lambda^2(M)$ - the space of 2-forms on
$M$ with its natural basis:
\[
dx\wedge dy,\ \ dx\wedge dz,\ \ dy\wedge dz,\ \ dx\wedge d\xi,\ \
dy\wedge d\xi,\ \ dz\wedge d\xi.
\]
We recall that the Hodge $*$ acts in $\Lambda^2(M)$ as a
complex structure $*=\mathcal{J}$ and on
the above basis its action is given by:
\begin{align*}
\mathcal{J}(dx\wedge dy) &=-dz\wedge d\xi &
\mathcal{J}(dx\wedge dz) &=dy\wedge d\xi  &
\mathcal{J}(dy\wedge dz) &=-dx\wedge d\xi \\
\mathcal{J}(dx\wedge d\xi) &=dy\wedge dz &
\mathcal{J}(dy\wedge d\xi) &=-dx\wedge dz &
\mathcal{J}(dz\wedge d\xi) &=dx\wedge dy.
\end{align*}
Hence, in this basis the matrix of $\mathcal{J}$ is off-diagonal with
entries $(1,-1,1,-1,1,-1)$.

Let now $\mathcal{I}$ be the identity map in $\Lambda^2(M)$. We
define a representation $\rho$ of $\mathbb{G}$ in $\Lambda^2(M)$ as
follows:
\begin{equation}                                                   
\rho(\alpha)=
\rho(aI+bJ)=a\mathcal{I}+ b\mathcal{J}, \quad \alpha\in\mathbb{G}.
\end{equation}
The map $\rho$ may be considered as a restriction of a linear map
$\mathcal{G}\rightarrow L_{\Lambda^2(M)}$ to the nonzero elements of
$\mathcal{G}$, i.e. to the elements of $\mathbb{G}$. Every $\rho(\alpha)$ is
a linear isomorphism, in fact, its determinant $\mathrm{det}||\rho(\alpha)||$
is equal to $(a^2+b^2)^3$.  The identity $I$ of $\mathbb{G}$ is sent to the
identity transformation $\mathcal{I}$ of $\Lambda^2(M)$, and the complex
structure $J$ of the vector space $\mathcal{G}$ is sent to the complex
structure $\mathcal{J}$ of $\Lambda^2(M)$. This map is surely a
representation, because
$\rho(\alpha.\beta)=\rho(\alpha).\rho(\beta),\ \ \alpha,\beta\in\mathbb{G}$.
In fact,
\[
\rho(\alpha.\beta)=
\rho\big[(aI+bJ).(mI+ nJ)\big]=
\]
\[
\rho\big[(am-bn)I+(a\varepsilon_2 n+bm)J\big]=
(am-bn)\mathcal{I}+
(an+bm)\mathcal{J}.
\]
On the other hand
\[
\rho(\alpha).\rho(\beta)
=\rho(aI+bJ).\rho(mI+nJ)
\]
\[
=(a\mathcal{I}+b\mathcal{J}).
(m\mathcal{I}+ n\mathcal{J})=
(am-bn)\mathcal{I}+
(an+bm)\mathcal{J}.
\]

We consider now the space $\Lambda^2(M,\mathcal{G})$ of $\mathcal{G}$-valued
2-forms on $M$.  Every such 2-form $\Omega$ can be represented as
$\Omega=F_1\otimes I+F_2\otimes J$, where $F_1$ and $F_2$ are 2-forms. We
have the joint action of $\mathbb{G}$ in $\Lambda^2(M,\mathcal{G})$ as
follows:
\[
[\rho(\alpha)\times \alpha](\Omega)=
\rho(\alpha).F_1\otimes (\alpha^{-1})^*(I)+
\rho(\alpha).F_2\otimes (\alpha^{-1})^*(J).
\]
We obtain
\[
det(\alpha).[\rho(\alpha)\times \alpha](\Omega)=
\]
\[
\big[(a^2\mathcal{I}+ab\mathcal{J})F_1-
(b^2\mathcal{J}+ab\mathcal{I})F_2\big]\otimes I+
\big[(b^2\mathcal{J}+ ab\mathcal{I})F_1+
(a^2\mathcal{I}+ab\mathcal{J})F_2\big]\otimes J.
\]
In the special case $\Omega=F\otimes I+\mathcal{J}.F\otimes J$ it readily
follows that
\begin{equation}
\left[\rho(\alpha)\times\alpha\right](\Omega)=\Omega.       
\end{equation}
In this sense the forms $\Omega=F\otimes I+\mathcal{J}.F\otimes J$ are
{\it equivariant} with respect to the joint action of $\mathbb{G}$.

Explicitly for a general 2-form $F$ we have
\begin{equation*}
\begin{split}
\rho(\alpha).F & =a\mathcal{I}.F+b\mathcal{J}.F \\
& = (aF_{12}+bF_{34})dx\wedge dy+
(aF_{13}- bF_{24})dx\wedge dz+
(aF_{23}+\ bF_{14})dy\wedge dz \\
& +(aF_{14}-bF_{23})dx\wedge d\xi+
(aF_{24}+ bF_{13})dy\wedge d\xi
+(aF_{34}-bF_{12})dz\wedge d\xi.\\
\end{split}
\end{equation*}
If $F_{\varepsilon}$ is of the kind (39) we
modify correspondingly the representation as follows:
$\rho(\alpha)=a\mathcal{I}+\varepsilon b\mathcal{J}$, and obtain
\begin{equation}
\begin{split}
\rho(\alpha).F_{\varepsilon} & =\varepsilon(au-bp)dx\wedge dz+
\varepsilon(ap+bu)dy\wedge dz \\
&+(au-bp)dx\wedge d\xi+               
(ap+bu)dy\wedge d\xi.
\end{split}
\end{equation}
Recalling now equation (48) the above formula (60) shows that if
$F_{\varepsilon}$ is a nonlinear solution then $\rho(\alpha).F_{\varepsilon}$
will be a nonlinear solution if the quantity
$$
\big[(\rho(\alpha).F_\varepsilon)_{14}\big]^2+
\big[(\rho(\alpha).F_\varepsilon)_{24}\big]^2=
(au-bp)^2+(ap+bu)^2
$$
is a running wave along $z$. But this quantity is equal to
$(a^2+b^2)(u^2+p^2)$ and since $(a^2+b^2)=const$, we see that
$\rho(\alpha).F_{\varepsilon}$ is again a nonlinear solution for any
$\alpha\in\mathbb{G}$. In other words, the group $\mathbb{G}$ acts as group of
symmetries of our nonlinear equations.  Moreover, in view of the conclusions
at the end of the preceding section, $\mathbb{G}$ acts inside every subclass
of solutions defined by the chosen space-like direction (the coordinate $z$).
Hence, if $F$ is a nonlinear solution, we may write $\mathbb{G}.F\subset
Q_F(\Lambda^2(M))$, i.e. any orbit $\mathbb{G}.F$ lives entirely and always
inside the subclass $Q_F(\Lambda^2(M))$.

As for the phase $\psi$ of the solution (60) we obtain (in this coordinate
system)
\[
\psi=\mathrm{arccos}\left[\frac{\varepsilon(au-bp)}
{\sqrt{(a^2+b^2)(u^2+p^2)}}\right]=
 \psi(F_{\varepsilon}(u,p))+\psi(\alpha(a,b)),
\]
where $\psi(F_{\varepsilon}(u,p))$ is the phase of the solution
$F_{\varepsilon}(u,p)$ and
$\psi(\alpha(a,b))=\mathrm{arccos}\big[a/\sqrt{a^2+b^2}\big]$ is the phase
of the complex number $\alpha=aI+bJ$. Now, since $\psi(\alpha(a,b))=const$
we obtain the

{\bf Corollary}. The 1-form $\mathbf{d}\psi$
and the scale factor $\mathcal{L}=|L_\zeta\psi|^{-1}$  are
 $\mathbb{G}$-invariants:
$\mathbf{d}(\psi(F))=\mathbf{d}(\psi(\rho(\alpha).F)), \quad
\mathcal{L}(F)=\mathcal{L}(\rho(\alpha).F)$.

We are going to show now that the action
$\rho(\alpha):F_\varepsilon\rightarrow \rho(\alpha).F_\varepsilon$ can be
generated by an appropriate action $T: \mathbb{G}\times M\rightarrow M$ of
$\mathbb{G}$ on the space-time $M$. The term appropriate here means that the
action $T$ takes in view some isotropic straight-line direction, which will
be considered to lie inside the planes $(z,\xi)$. The action $T$ is defined
by
\[
(\alpha(a,b);x,y,z,\xi)\rightarrow (x',y',z, \xi)=
\left(\frac{ax-by}{a^2+b^2},\, \frac{bx+ay}{a^2+b^2},\, z,\, \xi\right),
\quad \alpha(a,b)\in\mathbb{G}.
\]
Since $x=ax'+by',\  y=-bx'+ay'$ and
writing again $(x,y)$ instead of $(x',y')$) we obtain:
\[
T^*F_\varepsilon=\varepsilon (au-bp)dx\wedge dz+
\varepsilon (ap+bu)dy\wedge dz+(au-bp)dx\wedge d\xi +(ap+bu)dy\wedge d\xi
\]
We obtain also
\[
T^*\zeta=\zeta,\ \ T^*A=(au-bp)dx+(ap+bu)dy,\ \
T^*(A^*)=-\varepsilon(ap+bu)dx+\varepsilon(au-bp)dy,\ \
\]
\[
\delta(T^*F)=\big[a(u_\xi-\varepsilon u_z)-b(p_\xi-\varepsilon p_z)\big]dx+
\big[b(u_\xi-\varepsilon u_z)+a(p_\xi-\varepsilon p_z)\big]dy+
\]
\[
+\varepsilon\big[a(u_x+p_y)-b(p_x-u_y)\big]dz+
\big[a(u_x+p_y)-b(p_x-u_y)\big]d\xi,
\]
\[
|T^*A|^2=|T^*(A^*)|^2=(a^2+b^2)(u^2+p^2),\ \
|\delta(T^*F)|^2=(a^2+b^2)|\delta F|^2.
\]
The invariance of the scale factor follows directly now from the
above relations.

The corresponding fundamental vector fields $Z_I$ and $Z_J$, generated by the
 basis $(I,J)$ of $\mathcal{G}$ are
\[
Z_I=-\left(x\frac{\partial}{\partial x}+
y\frac{\partial}{\partial y}\right),
\quad Z_J=-y\frac{\partial}{\partial x}+
x\frac{\partial}{\partial y}.
\]
From the above corollary and from the commutation relation
$\mathbf{d}\circ L_X=L_X\circ\mathbf{d}$ it
follows now that the phase $\psi$ satisfies the equations
\[
L_{Z_I}(\mathbf{d}\psi)=L_{Z_J}(\mathbf{d}\psi)=0.
\]
The $\mathbb{G}$-invariance of the scale factor $\mathcal{L}$,
considered in this coordinate system, implies the relations
$Z_I(\mathcal{L})=Z_J(\mathcal{L})=0$, which lead to the conclusion that
$\mathcal{L}$ may depend only on $(z,\xi)$ in this coordinate system. In
fact, we obtain the linear system
\[
x\mathcal{L}_x+y\mathcal{L}_y=0, \quad
-y\mathcal{L}_x+x\mathcal{L}_y=0,
\]
and since $x^2+y^2\neq 0$, the only solution is
$\mathcal{L}_x=\mathcal{L}_y=0$, i.e. $\mathcal{L}=\mathcal{L}(z,\xi)$.

Now we have the following result:

{\bf Corollary}:
The scale factor $\mathcal{L}$ is a constant quantity:
$\mathcal{L}=const$, only if it is a Lorentz-invariant quantity.

In fact, every {\it constant} is Lorentz invariant. On the other hand,
the invariance of $\mathcal{L}$ with respect to rotations in the plane
$(x,y)$ is obvious in this coordinate system:
\[
y\frac{\partial \mathcal{L}(z,\xi)}{\partial x}-
x\frac{\partial \mathcal{L}(z,\xi)}{\partial y}=0.
\]
Further, the generators of the
proper Lorentz rotations are
\[
L_1=x\frac{\partial}{\partial \xi}+\xi\frac{\partial}{\partial x},\ \
L_2=y\frac{\partial}{\partial \xi}+\xi\frac{\partial}{\partial y},\ \
L_3=z\frac{\partial}{\partial \xi}+\xi\frac{\partial}{\partial z}.
\]
So, any of the relations $L_1(\mathcal{L})= 0,\ \ L_2(\mathcal{L})=0$ implies
that $\mathcal{L}$ does not depend on $\xi$, and, finally, the relation
$L_3(\mathcal{L})=0$ requires not-dependence on $z$.
\vskip 0.3cm
{\bf Corollary}. If the scale factor is Lorentz invariant:
$\mathcal{L}=const$, then the 3-form $F\wedge\delta F$ is closed:
$\mathbf{d}(F\wedge\delta F)=0$.  \
\vskip 0.3cm
In fact, from relation (51) it follows that if $\mathcal{L}=const$ then
$\psi_\xi-\varepsilon\psi_z=const$ too. We recall that
\[
\delta F\wedge F=
-\varepsilon\phi^2(\psi_\xi-\varepsilon\psi_z)dx\wedge dy\wedge dz-
\phi^2(\psi_\xi-\varepsilon\psi_z)dx\wedge dy\wedge d\xi.
\]
The assertion now follows from the notice that $\Phi^2$ is a running wave
along $z$. Hence, when the scale factor $\mathcal{L}$ is Lorentz-invariant we
obtain another conservative quantity, namely, the integral of the restriction
of $\delta F\wedge F$ on $\mathbb{R}^3$ over the whole 3-space will not
depend on time.

We also note that under the action of $\alpha(a,b)$ we have
\[
A\rightarrow A'=[au-bp,ap+bu,0,0],\quad
A^*\rightarrow (A^*)'=[ap+bu,-(au-bp),0,0],
\]
and this is equivalent to
\[
A'=aA-bA^*,\quad (A^*)'=bA+aA^*.
\]
Hence, in the $F$-adapted coordinate systems the dual transformation, as
given above, restricts to transformations in the $(x,y)$-plane, so we have
the
\vskip 0.3cm
{\bf Corollary}. The Frobenius integrability of the 2-dimensional Pfaff
system $(A,A^*)$ is a $\mathbb{G}$-invariant property.
\vskip 0.3cm
{\bf Corollary}. The Frobenius nonintegrability of the 2-dimensional Pfaff
systems $(A,\zeta)$ and $(A^*,\zeta)$ is a $\mathbb{G}$-invariant property.
\vskip 0.3cm
\noindent
{\bf Remark}. For a possible connection of $\delta F\wedge F$ to the
Godbillon-Vey closed 3-form $\Gamma=\mathbf{d}\theta\wedge\theta$ see further
in sec.11.
\vskip 0.3cm
Finally we note that inside an orbit $\mathbb{G}.F$ a limited
superposition rule holds: if $F$ is a solution and if
$[\alpha(a,b)+\beta(m,n)]\in\mathbb{G}$ then, obviosly,
$\alpha(a,b).F+\beta(m,n).F$ is also a solution.

\subsection{Point dependent group parameters}
We are going now to see what happens if the group parameters become functions
of the coordinates: $\alpha=a(x,y,z,\xi)I+b(x,y,z,\xi)J$. Consider the
nonlinear solution $F$ given by
\[
F=\varepsilon udx\wedge dz + \varepsilon p dy\wedge dz+
udx\wedge d\xi + p dy\wedge d\xi,
\]
and the matrix
\[
\beta_F=\beta(u,p)=\begin{Vmatrix}u & p \\-p & u\end{Vmatrix}.
\]
Since $F$ is a nonlinear solution then $u$ and $p$
satisfy the equation
\begin{equation}
L_\zeta\Big[\mathrm{det}||\beta(u,p)||\Big]=         
(u^2+p^2)_\xi -\varepsilon(u^2+p^2)_z=0,
\end{equation}
where $L_\zeta$ is the
Lie derivative with respect to the intrinsically defined by $F$ isotropic
vector field $\zeta$.

Consider now the 2-form (in the $F$-adapted coordinate system)
\[
F_o=dx\wedge \zeta=\varepsilon dx\wedge dz + dx\wedge d\xi,
\]
which, obviously, defines a (linear constant)
solution, and define a map $\beta:M\rightarrow \mathcal{G}$ as follows:
\[
\beta(x,y,z,\xi)=u(x,y,z,\xi)I+\varepsilon p(x,y,z,\xi)J,
\]
The action of
$$
\rho(\beta(x,y,z,\xi))
=u(x,y,z,\xi)\mathcal{I}+\varepsilon p(x,y,z,\xi)\mathcal{J}
$$
on $F_o$ gives exactly $F$, i.e. the (linear) solution $F_o$ is transformed
to a nonlinear solution:
\[
\rho(\beta(x,y,z,\xi))(F_o)=
F=\varepsilon udx\wedge dz + \varepsilon p dy\wedge dz+
udx\wedge d\xi + p dy\wedge d\xi.
\]
This suggests to check if this is
true in general, i.e. if we have a solution $F$ defined by the two functions
$u$ and $p$, and we consider a map
$\alpha:M\rightarrow\mathbb{G}\subset{\mathcal{G}}$,
such that the components $a(x,y,z,\xi)$ and
$b(x,y,z,\xi)$ of $\alpha=a(x,y,z,\xi)I + b(x,y,z,\xi)J$ determine a
solution, then whether the 2-form $\tilde F=\alpha(x,y,z,\xi).F=
\big[a(x,y,z,\xi)\mathcal{I} +\varepsilon b(x,y,z,\xi)\mathcal{J}\big].F$
will be a solution?

For $\tilde F$ we obtain
\[
\tilde F=\rho(\alpha).F=(a\mathcal{I}+\varepsilon b\mathcal{J}).F=
\]
\[
\varepsilon (au-\varepsilon bp)dx\wedge dz +
\varepsilon (ap+\varepsilon bu)dy\wedge dz +
(au-\varepsilon bp)dx\wedge d\xi +
(ap+\varepsilon bu)dy\wedge d\xi,
\]
which looks the same as given by (60), but here $a$ and $b$ are functions of
the coordinates.

Now, $\tilde F$ will be a solution iff
\begin{equation}
\big[(au-\varepsilon bp)^2 +
(ap+\varepsilon bu)^2\big]_\xi -                   
\varepsilon\big[(au-\varepsilon bp)^2 +
(ap+\varepsilon bu)^2\big]_z=0.
\end{equation}
This relation is equivalent to
\[
\big[(a^2+b^2)_\xi-\varepsilon(a^2+b^2)_z\big](u^2+p^2)+
\big[(u^2+p^2)_\xi-\varepsilon(u^2+p^2)_z\big](a^2+b^2)=0.
\]
This shows that if $F(u,p)$ is a solution, then
$\tilde F(u,p;a,b)=\rho(\alpha(a,b)).F(u,p)$ will be a solution
iff  $F(a,b)$ is a solution, i.e. iff $\rho(\alpha(a,b)).F_o$ is a solution.
So we have the
\vskip 0.3cm
{\bf Proposition 14.}
Every nonlinear solution
$F(a,b)=\big[a\mathcal{I}+\varepsilon b\mathcal{J}\big]F_o$, defines a map
$\Phi(a,b):F(u,p)\rightarrow (\Phi F)(u,p;a,b)$, such that if $F(u,p)$ is
a solution then $(\Phi F)(u,p;a,b)$ is also a solution.
\vskip 0.3cm
Further in this section we shall denote by the same letter $F$ the solution
$F$ and the corresponding map $\Phi(F):F_1\rightarrow F.F_1$.

The above {\bf Prop.14} says that the set $\Sigma(F_o)$ of nonlinear
solutions, defined by the chosen $F_o$, has a {\it commutative group
structure} with group multiplication
\[
F(a,b).F(u,p)=\rho(\alpha(a,b))F_o.\rho(\beta(u,p))F_o
=\rho(\alpha.\beta)F_o=\rho(\beta.\alpha)F_o=F(u,p).F(a,b).
\]
Clearly, $F_o$ should be identified with $\rho(I).F_o$ because
$\rho(1.I+0.\varepsilon J)(F_o)=\mathcal{I}(F_o)=F_o$.

The identification of a nonlinear solution $F(u,p)$ with the corresponding
map $\Phi(u,p)=\rho(\alpha(u,p))=u\mathcal{I}+\varepsilon p\mathcal{J}$
according to the rule
$F(u,p)=\Phi(u,p).F_o=(u\mathcal{I}+\varepsilon p\mathcal{J}).F_o$ gives
the possibility to introduce in an obviously invariant way the concept of
amplitude of a nonlinear solution. Together with the solution $F(u,p)$ we
have its conjugate solution $F(u,-p)$. We identify these two solutions with
the corresponding linear maps $\Phi(u,p)$ and $\Phi(u,-p)$  in
$\Lambda^2(M)$ and compute the quantity
$\frac 16 tr\big[\Phi(u,p)\circ\Phi(u,-p)\big]$.
\[
\frac 16 tr\big[\Phi(u,p)\circ \Phi(u,-p)\big]=
\frac 16 tr\left[\big(u\mathcal{I}+ \varepsilon p\mathcal{J}\big)\circ
\big(u\mathcal{I}- \varepsilon p\mathcal{J}\big)\right]=
\frac 16 tr\big[(u^2+p^2)\mathcal{I}\big]=u^2+p^2=\phi^2(u,p).
\]

In the general case
$\alpha=\alpha_p.\alpha_{p-1}\dots\alpha_1$ we readily obtain
$F(\alpha)=F(\alpha_p).F(\alpha_{p-1})\dots F(\alpha_1)$, and
for the corresponding amplitude
$\phi(F_p.F_{p-1}\dots F_1)$ and phase
$\psi(F_p.F_{p-1}\dots F_1)$ we obtain
\begin{equation}
\phi(F_p.F_{p-1}\dots F_1))=\phi(F_p).\phi(F_{p-1})\dots\phi(F_1),
\quad \psi(F_p.F_{p-1}\dots F_1)=                     
\sum_{i=1}^p(\psi_i).
\end{equation}
The corresponding scale factor is
\[
\mathcal{L}(F_p.F_{p-1}\dots F_1)=\frac{1}{|L_\zeta \psi|}=
\frac{1}{|\sum_{i=1}^p (\psi_i)_\xi-
\varepsilon\sum_{i=1}^p(\psi_i)_z|}.
\]
Denoting $\mathcal{L}_k=\mathcal{L}(F_k)$ we obtain the nonequality
\[
\mathcal{L}(F_p.F_{p-1}...F_1)\geq\frac
{\mathcal{L}_1.\mathcal{L}_2...\mathcal{L}_p}
{\mathcal{L}_2.\mathcal{L}_3...\mathcal{L}_p+
\mathcal{L}_1.\mathcal{L}_3...\mathcal{L}_p+...+
\mathcal{L}_1.\mathcal{L}_2...\mathcal{L}_{p-1}}=
\frac{1}{\mathcal{L}_1^{-1}+\mathcal{L}_2^{-1}+...+\mathcal{L}_p^{-1}}.
\]
If $\alpha(a,b)\neq 0$ at some point
of $M$ then we have $F^{-1}=\rho(\alpha^{-1}_\varepsilon).F_o$ given by
$$
(F)^{-1}=\left(\frac{u}{u^2+p^2}\mathcal{I}
-\frac{\varepsilon p}{u^2+p^2}\mathcal{J}\right).F_o .
$$
Clearly $\psi(F_o)=0$, which corresponds to $\mathcal{L}(F_o)=\infty$.
\vskip 0.3cm
Finally, we note that, in general, there is no superposition inside the
subclass of solutions defined by a given solution $F(u,p)$. The verification
shows that if $\alpha(a,b)$ generates the solution $F(a,b)$ and
$\alpha(m,n)$ generates the solution $F(m,n)$ inside the same
$\zeta$-subclass, then the sum $F(a,b).F(u,p)+F(m,n).F(u,p)$ will be a
solution only if the phase difference $\psi_{F(a,b)}-\psi_{F(m,n)}$ is a
running wave along the corresponding $z$-coordinate. In particular, if this
phase difference is a constant, then the superposition will work, which
recalls the interference rule in Maxwell vacuum electrodynamics, that two
waves may interfere only if their phase difference is a constant quantity.

\vskip 0.5cm

The considerations in this section allow the following conclusions and
interpretations.  The whole set of nonlinear solutions divides to subclasses
$S(F_o)$ of the kind (39) (every such subclass is determined by the spatial
direction along which the solution propagates translationally, it is the
coordinate $z$ in our consideration, or by the corresponding $\zeta$).  Every
solution $F$ of a given subclass is obtained by means of the action of a
solution $\alpha(u,p)$ of equation (61) on the corresponding $F_o$ through the
representation $\rho$, namely, $F=\rho(\alpha).F_o$. We also note that every
such subclass divides to subsubclasses $\mathbb{G}.F$, determined by a given
solution $F$ and the action of $\mathbb{G}$ with {\it constant} coefficients.
Every such subsubclass may be considered as one solution represented in
different bases of $\mathcal{G}$.

Further, the relation $\rho(\alpha(u,p))F_o=F$ suggests to consider $F_o$, as
a "vacuum state along $\zeta$" of the subclass considered, and the action of
$\rho(\alpha)$ on $F_o$ as a "creation operator along $\zeta$", provided
$a(x,y,z,\xi)\neq 0, b(x,y,z,\xi)\neq 0, (x,y,z,\xi)\in D\subset M$, and
$\alpha=aI+b\mathcal{J}$ satisfies (61) in $D$.  Since
$\rho(\alpha).\rho(\alpha^{-1}).F_o= F_o$ we see that every such "creation
operator along $\zeta$" in $D$ may be interpreted also as an "annihilation
operator along $\zeta$" in $D$, i.e. in the same subclass of solutions. A
given $\alpha$ may generate a creation operator with respect to some subclass
of solutions only if $|\alpha|$ is a running wave along the corresponding
isotropic vector $\zeta$. Finally, the only universal "annihilation" operator
is the $\rho$-image of the zero element of $\mathcal{G}$.

Clearly, every solution $F=\rho(\alpha).F_o$ of a given subclass may be
represented in various ways in terms of other solutions to (61) in the same
domain $D$, i.e.  we have an example of a {\it nonlinear} "superposition"
inside a given subclass, represented by the above mentioned commutative group
structure.

The interpretation of a given subsubclass determined by
a given solution $F$ and the action of $\mathbb{G}$ with {\it constant}
coefficients as one solution, suggests also the extension of
this interpretation to the point dependent case. In
fact, at every point $m=(x,y,z,\xi)\in M$ we may consider an isomorphic image
$\mathcal{G}_m$ of $\mathcal{G}$ with basis $\{e_1(m),e_2(m)\}$, where
\[
e_1(m)=\begin{Vmatrix}u(m) & 0 \\ 0 & u(m)\end{Vmatrix},\quad
e_2(m)=\begin{Vmatrix}0 & p(m) \\ -p(m) & 0\end{Vmatrix}.
\]
Now, the element $[e_1(m)+e_2(m)]\in\mathcal{G}_m$ is sent through the
representation $\rho_m:\mathcal{G}_m\rightarrow L_{\Lambda^2_m(M)}$ to the
linear map $\Phi(m)=u(m)\mathcal{I}+p(m)\mathcal{J}$, and when $(u^2+p^2)$
satisfies equation (61) we obtain the solution $\Phi(F_o)$. Hence, all
solutions $F(u,p)$ of a given subclass may be considered as one solution
represented in different point dependant bases $\{e_1(m),e_2(m)\}$ of
$\mathcal{G}_m,\, m\in M$. In fiber bundle terms we could say that the
nonlinear solutions are determined by sections of the bundle
$M\times\mathcal{G}$ which satisfy (61) and their $\rho$-actions on
corresponding "vacuums" $F_o$.

Looking back we see that in the frame of linear solutions (Maxwell
solutions) we have {\it additive} commutative group structure (any linear
combination with constant coefficients of linear solutions is again a
solution), while in the frame of the nonlinear solutions we have a {\it
multiplicative} commutative group structure: a product (as given above) of
any number of nonlinear solutions along $\zeta$ is again a solution along
$\zeta$.  The important difference between these two structures is that the
additive structure of the linear solutions holds for all solutions, and the
multiplicative group structure holds just inside a given $\zeta$-subclass.
So, the whole set of nonlinear solutions consists of orbits of the
(multiplicative) group of those complex valued functions $\alpha(x,y,z,\xi)$
(the product is point-wise), the module $|\alpha|=\sqrt{a^2+b^2}$ of which is
a running wave along some fixed isotropic direction $\zeta$.

We see also that the amplitude and the
phase of a solution in a natural way acquire the interpretations of the
{\it amplitude} $\phi=|\alpha|=\sqrt{a^2+b^2}$ and {\it
phase} $\psi=\mathrm{arccos}(\varphi)$ (see relation (41)) of the
corresponding complex fields $\alpha(x,y,z,\xi)$. Clearly, through the well
known Moivre formulae, every $F=\rho(\alpha).F_o$ generates
$(F)^n=\rho(\alpha ^n).F_o$ and $\sqrt[n]{F}=\rho(\sqrt[n]{\alpha}).F_o$, and
in this way we obtain new solutions of the same subclass.

\section{A Heuristic Approach to the Vacuum Nonlinear Equations}
In this section we give a heuristic, in our view, approach as for how to come
to the nonlinear vacuum equations of EED. Only two things turn out to be
needed:  first, accepting that the time dependent EM fields propagate along
straight lines, second, some experience with differential geometry concepts.
\vskip 0.4cm
Our basic assumption reads: {\bf The translational component of propagation
of time dependent vacuum EM fields is along isotropic straight lines in
Minkowski space-time}.

\vskip 0.4cm
\noindent
We shall denote by $M$ the Minkowski space-time, endowed with the canonical
coordinates $(x,y,z,\xi=ct)$. From the above assumption it follows that at
any particular case we may find isotropic vector field the integral
(isotropic) lines of which can be described by the vector field
$\zeta=-\varepsilon\partial_z+\partial_\xi$ or by the corresponding through
the (flat) Lorentz metric $\eta$, $sign(\eta)=(-,-,-,+)$, 1-form
$\zeta=\varepsilon dz+d\xi$. This 1-form $\zeta$ is closed, so, it defines
1-dimensional completely integrable Pfaff system. The corresponding
3-dimensional differential system (or distribution) on $M$ is spanned by the
 vector fields
\[
\bar A=-u\partial_x-p\partial_y,\ \ \bar A'=-m\partial_x-n\partial_y,\ \
\bar \zeta=-\varepsilon\partial_z+\partial_\xi,
\]
which are $\eta$-images of the space-like 1-forms $(A,A')$,
given by
\[
A=udx+pdy,\ \ A'=mdx+ndy,
\]
and of $\zeta=\varepsilon dz+d\xi$, where
$(u,p,m,n)$ are four independent functions on $M$.  We form now the tensor
$Q=\phi^2\,\zeta\otimes\bar\zeta$, where the function $\phi$ is still to be
determined.  This tensor field defines linear map in the spaces of vector
fields and 1-forms on $M$. Clearly, $Im\,Q$ is spanned by $\zeta$, and
$Ker\,Q$ is spanned by $(A,A',\zeta)$, where $A$ and $A'$ are linearly
independent. The factor space $Ker\,Q/Im\,Q$ is isomorphic to the plane
spanned by $(A,A')$, and further we choose $\eta(A,A')=0$, i.e.
$A'=-pdx+udy$, and we shall denote this choice of $A'$ by $A^*$. It is easily
verified that the 2-dimensional Pfaff system $(A,A^*)$ is completely
integrable, and that the relations $A.\zeta=A^*.\zeta=0$ hold.

We define now the 2-forms
\[
F=A\wedge\zeta,\ \ F^*=A^*\wedge\zeta.
\]
The following relations hold:
\[
i(\bar A)F^*=0,\ \ i(\bar A^*)F=0,\ \ i(\zeta)F=i(\zeta)F^*=0.
\]
Moreover, if $*F$ is the Hodge-$\eta$ dual to $F$ we have $F^*=*F$.

Obviously, $F\wedge F=0$ and $F\wedge *F=0$, so $F$ is a null field.
Now, the tensor field
$$
-\frac12[F_{\mu\sigma}F^{\nu\sigma}+(*F)_{\mu\sigma}(*F)^{\nu\sigma}]
$$
is equal to $(u^2+p^2)\zeta_\mu\bar\zeta^\nu$, so we define
$\phi^2=u^2+p^2=-A^2=-(A^*)^2$.
We compute $\mathbf{d}F$ and $\mathbf{d}*F$:
\[
\mathbf{d}F=\varepsilon(p_x-u_y)dx\wedge dy\wedge dz+(p_x-u_y)dx\wedge
dy\wedge d\xi-
\]
\[
\varepsilon(u_\xi-\varepsilon u_z)dx\wedge dz\wedge d\xi-
\varepsilon(p_\xi-\varepsilon p_z)dy\wedge dz\wedge d\xi,
\]
\[
\mathbf{d}*F=(p_y+u_x)dx\wedge dy\wedge dz+\varepsilon(p_y+u_x)dx\wedge
dy\wedge d\xi-
\]
\[
(p_\xi-\varepsilon p_z)dx\wedge dz\wedge d\xi+
(u_\xi-\varepsilon u_z)dy\wedge dz\wedge d\xi.
\]
The relation $F\wedge *\mathbf{d}*F+(*F)\wedge *\mathbf{d}F=0$ is
now immediately verified. We compute the quantities
$i(\bar F)\mathbf{d}F=i(\bar A\wedge\bar \zeta)\mathbf{d}F=
i(\bar\zeta)\circ i(\bar A)\mathbf{d}F$
and
$i(\overline{*F})\mathbf{d}*F=i(\bar A^*\wedge\bar \zeta)\mathbf{d}*F=
i(\bar\zeta)\circ i(\bar A^*)\mathbf{d}*F$,
and obtain
\[
i(\bar A\wedge\bar \zeta)\mathbf{d}F=
i(\bar A^*\wedge\bar \zeta)\mathbf{d}*F=
\frac12\big[(u^2+p^2)_\xi-\varepsilon(u^2+p^2)_z\big]\zeta.
\]
The equations
\[
F\wedge *\mathbf{d}*F+(*F)\wedge *\mathbf{d}F=0,\ \
i(\bar F)\mathbf{d}F=0,\ \ i(\bar{*F})\mathbf{d}*F=0
\]
now give $\nabla_\nu Q_{\mu}^{\nu}=0$ and that the amplitude function $\phi$
is a running wave along the $z$-coordinate. From these relations we can
derive everything that we already know.

\vskip 0.5cm
The above considerations in this section made use, more or less, of the
metric structure of Minkowski space and of the corresponding Hodge
$*$-operator. A more careful look into the details, however, shows that we
can come to the important relations of EED without using any metric. We
proceed now to consider this possibility.

Instead of Minkowski space $M$ we consider the manifold $\mathbb{R}^4$
together with its tangent tensor and exterior algebra, and standard
coordinates $(x,y,z,\xi)$.  The main tool to be used is the Poincar\'e
isomorphism between the $p$-vectors and $(4-p)$-forms on $\mathbb{R}^4$ [28].
For convenience we briefly recall how it is introduced.  Let $\Lambda^p(V)$
be the space of $p$-tensors, i.e. all antisymmetric (contravariant) tensors,
and $\Lambda^{(n-p)}(V^*)$ be the space of $(n-p)$-forms, i.e. all
antisymmetric covariant tensors, over the pair of dual $n$-dimensional real
vector spaces $(V,V^*)$, and $p=1,\dots,n$.  If $\{e_i\}$ and
$\{\varepsilon^{i}\}$ are two dual bases we have the $n$-tensor
$\omega=e_1\wedge \dots \wedge e_n$ and the $n$-form
$\omega^*=\varepsilon^1\wedge \dots \wedge \varepsilon^n$. The duality
requires $<\varepsilon^i,e_j>=\delta^i_j$ and
$<\omega^*,\omega>=1$.  If $x\in V$ and $\alpha\in
\Lambda^p(V^*)$ then by means of the insertion operator $i(x)$ we obtain the
$(p-1)$-form $i(x)\alpha$. If $\alpha$ is decomposable:
$\alpha=\alpha^1\wedge \alpha^2\wedge \dots \wedge\alpha^p$, where
$\alpha^1,\dots,\alpha^p$ are 1-forms, then
\begin{align*}
\begin{split}
i(x)\alpha=&<\alpha^1,x>\alpha^2\wedge \dots \wedge \alpha^p
-<\alpha^2,x>\alpha^1\wedge \alpha^3\wedge \dots \wedge \alpha^p
+\dots  \\
&+(-1)^{p-1}<\alpha^p,x>\alpha^1\wedge \dots\wedge \alpha^{p-1}.
\end{split}
\end{align*}
Now let $x_1\wedge x_2\wedge \dots \wedge x_p \in \Lambda^p(V)$. The
Poincar\'e isomorphism
$\mathfrak{P}:\Lambda^p(V)\rightarrow\Lambda^{n-p}(V^*)$ acts as follows:
\[
\mathfrak{P}_p(x_1\wedge x_2\wedge\dots\wedge x_p)=
i(x_p)_\circ i(x_{p-1})_\circ \dots _\circ i(x_1)\omega^*.
\]
In the same way
\[
\mathfrak{P}^p(\alpha^1\wedge \dots \wedge\alpha^p)=
i(\alpha^p)_\circ \dots _\circ i(\alpha^1)\omega.
\]
For nondecomposable tensors $\mathfrak{P}$ is extended by linearity. We note
also the relations:
\[
\mathfrak{P}_{n-p}\mathfrak{P}^p=(-1)^{p(n-p)}id,\quad
\mathfrak{P}^{n-p}\mathfrak{P}_p=(-1)^{p(n-p)}id,
\]
\[
<\mathfrak{P}^p(\alpha^1\wedge \dots \wedge \alpha^p),
\mathfrak{P}_p(x_1\wedge \dots \wedge x_p)>=
<\alpha^1\wedge \dots \wedge \alpha^p,x_1\wedge \dots \wedge x_p>.
\]
On an arbitrary basis element $\mathfrak{P}$ acts in the following way:
\[
\mathfrak{P}^p(\varepsilon^{i_1}\wedge \dots \wedge \varepsilon^{i_p})=
(-1)^{\mbox{\small$\displaystyle\sum_{k=1}^p (i_k -k)$}}e_{i_{p+1}}
\wedge \dots \wedge e_{i_n},
\]
where $i_1<\dots<i_p$ and $i_{p+1}<\dots<i_n$ are complementary
$p-$ and $(n-p)-$ tuples.

On the manifold $\mathbb{R}^4$ we have the dual bases $\{dx^i\}$ and
$\left\{\frac{\partial}{\partial x^i}\right\}$ of the two modules of 1-forms
and vector fields. The corresponding $\omega$ and $\omega^*$ are
\[
\omega=\frac{\partial}{\partial x}\wedge\frac{\partial}{\partial y}\wedge
\frac{\partial}{\partial z}\wedge\frac{\partial}{\partial \xi},\quad
\omega^*=dx\wedge dy\wedge dz\wedge d\xi.
\]

Our basic initial objects of interest are the 1-form $\zeta$ and the vector
field $\bar\zeta$ defined by
\[
\zeta=\varepsilon dz+d\xi,\quad
\bar\zeta=-\varepsilon\frac{\partial}{\partial z}+
\frac{\partial}{\partial \xi}.
\]
Note that no unknown functions are introduced with these two objects. We
introduce now the tensor field $Q:=\phi^2\zeta\otimes\bar\zeta$, where the
function $\phi$ is to be determined. The tensor $Q$ defines linear maps in
the two spaces of vector fields and 1-forms on $\mathbb{R}^4$ by the
relations
\[
Q(X)=\phi^2<\zeta,X>\bar\zeta,\quad
Q^*(\alpha)=\phi^2<\alpha,\bar\zeta>\zeta,
\]
where $X$ is an arbitrary vector field, $\alpha$ is an arbitrary 1-form, and
$<\alpha,X>$ is the canonical conjugation between $T\mathbb{R}^4$ and
$T^*\mathbb{R}^4$. Clearly, the image space $ImQ$ of $Q$ is generated by
$\bar\zeta$, and the image space $ImQ^*$ of $Q^*$ is generated by $\zeta$.
The kernel space $KerQ$ is 3-dimensional and looks like:
$KerQ=\{\mathbf{a}\}\oplus\{\mathbf{b}\}\oplus\{\bar\zeta\}$, where the two
vectors $(\mathbf{a},\mathbf{b})$ are linearly independent:
$\mathbf{a}\wedge\mathbf{b}\neq 0$. The kernel space $KerQ^*$ is also
3-dimensional and looks like:
$KerQ^*=\{\alpha\}\oplus\{\beta\}\oplus\{\zeta\}$, where the two
1-forms $(\alpha,\beta)$ are linearly independent:
$\alpha\wedge\beta\neq 0$. The corresponding 2-dimensional factor spaces
$KerQ/ImQ$ and $KerQ^*/ImQ^*$ naturally arise, and further we shall consider
mainly the second one. The elements of $KerQ^*/ImQ^*$ have the form
$\alpha+f\zeta$ and $\beta+f\zeta$. Inside each of these two classes
there is only one element that has no component along $\zeta$, we denote
these two elements by $A$ and $A'$ correspondingly. These two 1-forms
live in the planes $(x,y)$. So, we have $A=udx+pdy,\ \ A'=mdx+ndy$, with the
relation $un-pm\neq 0$.

It is now readily verified the following
\vskip 0,4cm
{\bf Corollary}. The 2-dimensional
Pfaff system $(A,A')$ is completely integrable, and the two Pfaff systems
$(A,\zeta)$ and $(A,'\zeta)$ are not, in general, completely integrable.
\vskip 0.4cm
Hence,  the following relations hold:
\[
\mathbf{d}A\wedge A\wedge A'=0,\quad \mathbf{d}A'\wedge A\wedge A'=0,\quad
\mathbf{d}A\wedge A\wedge \zeta\neq 0,\quad
\mathbf{d}A'\wedge A'\wedge \zeta\neq 0.
\]

Intuitively, the {\it integrable} Pfaff system $(A,A')$ carries information
about those properties of the physical system under consideration which do
not change during the evolution, and the other two {\it nonintegrable} Pfaff
systems carry information about the {\it admissible} interaction of the
physical system with the rest of the world. Therefore, namely these two Pfaff
systems, $(A,\zeta)$ and $(A',\zeta)$, should be of primary interest for us,
(from the point of view of studying this physical system)
because studying them, we may get some knowledge about the {\it proper}, or
{\it identifying}, properties of the system (Recall Sec.1). The objects
$A\otimes\bar\zeta$ and $A'\otimes\bar\zeta$ are also expected to be of use.

Since $A$ and $A'$ are linearly independent we choose them as follows:
\[
A=udx+pdy, \quad A'=-pdx+udy,\quad \text{so} \quad A\wedge A'\neq 0.
\]

According to the above remark we turn our attention to the two 2-forms
$A\wedge\zeta$ and $A'\wedge\zeta$, and our first observation is that they
are defined on the classes determined by $A$ and $A'$ respectively. Our
second observation is that the corresponding curvatures, given by
\[
\mathbf{d}A\wedge A\wedge \zeta=
\mathbf{d}A'\wedge A'\wedge \zeta=
-\varepsilon\big[p(u_\xi-\varepsilon u_z)-
u(p_\xi-\varepsilon p_z)\big]\omega^*,
\]
give the same coefficient (Sec.5) which introduces rotational
component of propagation. As we shall see further (Sec.9) this is not
occasional.

We compute now $\mathfrak{P}^2(A\wedge\zeta)$ and
$\mathfrak{P}^2(A'\wedge\zeta)$ and obtain
\[
\mathfrak{P}^2(A\wedge\zeta)=-\varepsilon u\frac{\partial}{\partial y}\wedge
\frac{\partial}{\partial \xi}+\varepsilon p\frac{\partial}{\partial x}\wedge
\frac{\partial}{\partial \xi}+u\frac{\partial}{\partial y}\wedge
\frac{\partial}{\partial z}-p\frac{\partial}{\partial x}\wedge
\frac{\partial}{\partial z},
\]
\[
\mathfrak{P}^2(A'\wedge\zeta)=-u\frac{\partial}{\partial x}\wedge
\frac{\partial}{\partial z}-p\frac{\partial}{\partial y}\wedge
\frac{\partial}{\partial z}+\varepsilon u\frac{\partial}{\partial x}\wedge
\frac{\partial}{\partial \xi}+\varepsilon p\frac{\partial}{\partial y}\wedge
\frac{\partial}{\partial \xi}.
\]
The following relations are readily verified:
\[
i(A)\big[\mathfrak{P}^2(A\wedge\zeta)\big]=0, \ \
i(A')\big[\mathfrak{P}^2(A'\wedge\zeta)\big]=0, \ \
\]
\[
i(A\wedge\zeta)\big[\mathfrak{P}^2(A\wedge\zeta)\big]=0, \ \
i(A'\wedge\zeta)\big[\mathfrak{P}^2(A\wedge\zeta)\big]=0.
\]
We compute $\mathfrak{P}^3\big[\mathbf{d}(A\wedge\zeta)\big]$:
\[
\mathfrak{P}^3\big[\mathbf{d}(A\wedge\zeta)\big]=
\varepsilon(p_x-u_y)\bar\zeta +\varepsilon\left[(u_\xi-\varepsilon u_z)\frac
{\partial}{\partial y}-(p_\xi-\varepsilon p_z)\frac{\partial}{\partial
x}\right].
\]
The two objects $A\otimes\bar\zeta$ and $A'\otimes\bar\zeta$ may be
considered as 1-forms with values in the tangent bundle, so we can use the
insertion operator $i(X)$ as follows:
$i(X)(A\otimes\bar\zeta)=<A,X>\bar\zeta$. Denoting the two vector fields
$\mathfrak{P}^3(\mathbf{d}(A\wedge\zeta))$ and
$\mathfrak{P}^3(\mathbf{d}(A'\wedge\zeta))$ by $Z$ and $Z'$ respectively,
we obtain
\[
i(Z)(A\otimes\bar\zeta)=i(Z')(A'\otimes\bar\zeta)=
\varepsilon\big[p(u_\xi-\varepsilon u_z)-
u(p_\xi-\varepsilon p_z)\big]\bar\zeta,
\]
and
\[
i(Z)(A'\otimes\bar\zeta)=-i(Z')(A\otimes\bar\zeta)=
\varepsilon\big[u(u_\xi-\varepsilon u_z)+
p(p_\xi-\varepsilon p_z)\big]\bar\zeta.
\]
Recalling now from Sec.5 the explicit forms of $\delta F\wedge F=(\delta *F)
\wedge *F, \ \ \delta F\wedge *F$ and $(\delta *F)\wedge F$, and the equations
they satisfy, the same equations may be written as follows: the two equations
$\delta F\wedge *F=0$ and $(\delta *F)\wedge F=0$ are equivalent to
\[
i(Z)(A'\otimes\bar\zeta)=-i(Z')(A\otimes\bar\zeta)=0,
\]
and the equation $\delta F\wedge F=(\delta *F)\wedge *F$ is equivalent to
\[
i(Z)(A\otimes\bar\zeta)=i(Z')(A'\otimes\bar\zeta).
\]
The additional equation $\mathbf{d}(\delta F\wedge F)=0$ is equivalent to
\[
\mathbf{d}\Big[\mathfrak{P}_1\big[i(Z)(A\otimes\bar\zeta)\big]\Big]=0.
\]
Finally, choosing the two vector fields $\mathbf{a}$ and $\mathbf{b}$ in the
form $\mathbf{a}=u\partial_x + p\partial_y$ and
$\mathbf{b}=-p\partial_x + u\partial_y$, the function $\phi$ in
$Q=\phi^2\zeta\otimes\bar\zeta$ is obtained through the relation
\[
Q=(< , >,\otimes)(A\wedge\zeta,\mathbf{a}\wedge\bar\zeta)=
<A,\mathbf{a}>\zeta\otimes\bar\zeta=
(u^2+p^2)\zeta\otimes\bar\zeta,
\]
or through the relation
\[
Q=(< , >,\otimes)(A'\wedge\zeta,\mathbf{b}\wedge\bar\zeta)=
<A',\mathbf{b}>\zeta\otimes\bar\zeta=
(u^2+p^2)\zeta\otimes\bar\zeta,
\]

The considerations made so far were limited, more or less, inside a given
subclass of (nonlinear) solutions, which propagate translationally along the
same isotropic 4-direction in $M$, or along a given spatial direction which
we choose for $z$-coordinate. A natural question arises: is it possible to
write down equations which would simultaneously describe a set of $N$ such
non-interacting solutions, which propagate
translationally along {\it different} spatial directions. The answer to this
question is positive, and the equations look like:
\[
\sum_{k=1}^{N}\left(\delta F^{2k-1}\wedge *F^{2k-1}\right)\otimes e_{2k-1}\vee
e_{2k-1}-
\sum_{k=1}^{N}\left(\delta *F^{2k-1}\wedge F^{2k-1}\right)\otimes e_{2k}\vee
e_{2k}+
\]
\[
\sum_{k=1}^{N}\left(-\delta F^{2k-1}\wedge F^{2k-1}+
\delta *F^{2k-1}\wedge *F^{2k-1}\right)\otimes e_{2k-1}\vee e_{2k}=0,
\]
where the index $k$ enumerates the 2-space $(\mathbb{R}^2)^k$ for the
corresponding couple $(F^k,*F^k)$. So, for every $k=1,2,\dots,N$ we obtain
the system (32)-(34), i.e. the corresponding couple $(e_{2k-1},e_{2k})$
defines, in fact, the direction of translational propagation of the solution
$(F^k,*F^k)$.  The case of possible interaction of nonlinear solutions (e.g.
through "interference") will be considered elsewhere.

\section{Connection-Curvature Interpretations}
The concepts of connection and curvature have proved to be very useful in
field theory, for example, General Relativity and the later Yang-Mills theory
were built entirely in terms of these (geometric in nature) concepts. So, it
seems interesting to try to find corresponding interpretation of EED.

The concept of connection manifests itself in three main forms: {\it general
connection} on a manifold, {\it principal connection} in a principal
bundle, and {\it linear connection} in a vector bundle. In the first case the
general connection gives an appropriate geometric picture of a system of
partial differential equations, and the corresponding (general) curvature
shows when this system of equations is integrable and when it is not
integrable: the nonzero curvature forbids integrability. The principal
connection adapts appropriately the concept of general connection to the
special case of principal bundles making use of the rich geometric structure
of Lie groups and free Lie group actions on manifolds. The linear connection
allows to differentiate sections of vector bundles along vector
fields, obtaining new sections of the same vector bundle.

First we are going to introduce the necessary concepts in view of their use
for corresponding interpretation of EED.

\subsection{Integrability and Connection}
The problem for integration of a system of partial differential equations of
the kind
\begin{equation}
\frac{\partial y^a}{\partial x^i}=f^a_i(x^k,y^b),       
\ i,k=1,...,p;\ a,b=1,...,q,
\end{equation}
where $f^a_i(x^k,y^b)$ are given functions, obeying some definite smoothness
conditions, has contributed to the formulation of a number of concepts,
which in turn have become generators of ideas and research directions, and
most of them have shown an wide applicability in many branches of
mathematics and mathematical physics.  A particular case of the above system
(nonlinear in general) of equations is when there is only one independent
variable, i.e.  when all $x^i$ are reduced to $x^1$, which is usually denoted
by $t$ and the system acquires the form
\begin{equation}
\frac{dy^a}{dt}=f^a (y^b,t),\ a,b=1,...,q.                 
\end{equation}
We recall now some of the concepts used in considering the integrability
problems for these equations, making use of the geometric language of
manifold theory.

Let $X$ be a vector field on the $q$-dimensional manifold $M$ and the map
$c:I\rightarrow M$, where $I$ is an open interval in ${\mathbb{R}}$, defines a
smooth curve in $M$. Then if $X^a, a=1,2,\dots,q$, are the components of $X$
with respect to the local coordinates $(y^1, ..., y^q)$ and the equality
$c'(t)=X(c(t))$ holds for every $t\in I$, or in local coordinates,
\[
\frac{dy^a}{dt}=X^a (y^b),
\]
$c(t)$ is the {\it integral curve} of the vector field $X$, i.e. $c(t)$
defines a 1-dimensional manifold such that $X$ is tangent to it at every
point $c(t)$.  As it is seen, the difference with (65) is in the
additional dependence of the right side of (65) on the independent variable
$t$.  Mathematics approaches these situations in an unified way as follows.
The product ${\mathbb{R}}\times M$ is considered and the important theorem for
uniqueness and existence of a solution is proved: For every point $p\in M$
and point $\tau \in {\mathbb{R}}$ there exist a vicinity $U$ of $p$, a positive
number $\varepsilon $ and a smooth map $\Phi:(\tau-\varepsilon,\tau
+\varepsilon)\times U\rightarrow M$, $\Phi:(t,y)\rightarrow \varphi_t (y)$,
such that for every point $y\in U$ the following conditions are met:
$\varphi_\tau (y)=y,\ t\rightarrow \varphi_t (y)$ is an integral curve of
$X$, passing through the point $y\in M$; besides, if two such integral curves
of $X$ have at least one common point, they coincide. Moreover, if $(t',y),\
(t+t',y)$ and $(t,\varphi(y))$  are points of a vicinity $U'$ of $\{0\}\times
{\mathbb{R}}$ in ${\mathbb{R}}\times M$, we have $\varphi_{t+t'}(y)=\varphi_t
(\varphi_{t'}(y))$. This last relation gives the local group action: for
every $t\in I$ we have the local diffeomorphism $\varphi_t :U\rightarrow
\varphi_t (U)$. So, through every point of $M$ there passes only one
trajectory of $X$ and in this way the manifold $M$ is foliated to
non-crossing trajectories - 1-dimensional manifolds, and these 1-dimensional
manifolds define all trajectories of the defined by the vector field $X$
system of ODE. This fibering of $M$ to nonintersecting submanifolds, the
union of which gives the whole manifold $M$, together with considering $t$
and $y(t)$ as 1-d submanifolds of the same manifold, is the leading idea in
treating the system of partial differential equations (64), where the number
of the independent variables is more than 1, but finite. For example, if we
consider two vector fields on $M$, then through every point of $M$ two
trajectories will pass and the question: when a 2-dimensional surface,
passing through a given point can be built, and such that the representatives
of the two vector fields at every point of this 2-surface to be tangent to
the surface, naturally arises. The answer to this problem is given by the
following {\it Frobenius theorems}.

For simplicity, further we consider regions of the space
${\mathbb{R}}^p \times {\mathbb{R}}^q$, but this is not essentially important since
the Frobenius theorems are local statements, so the results will hold for any
$(p+q)$-dimensional manifold.

Let $U$ be a region in ${\mathbb{R}}^p\times{\mathbb{R}}^q$, and
$(x^1,...,x^p,y^1=x^{p+1},...,y^q =x^{p+q})$ are the canonical coordinates.
We set the question: for which points $(x_0,y_0)$ of $U$ the system of
equations (64) has a solution $y^a=\varphi ^a(x^i)$, defined for points $x$,
sufficiently close to $x_0$ and satisfying the initial condition
$\varphi (x_0)=y_0$? The answer to this question is: for this to happen it is
necessary and sufficient the functions $f_i^a$ on the right hand side of
(64) to satisfy the following conditions:
\begin{equation}
\frac{\partial f^a_i}{\partial x^j}(x,y)+
\frac{\partial f^a_i}{\partial y^b}(x,y).f^b_j(x,y)=
\frac{\partial f^a_j}{\partial x^i}(x,y)+                           
\frac{\partial f^a_j}{\partial y^b}(x,y).f^b_i(x,y).
\end{equation}
This relation is obtained as a consequence of two basic steps: first,
equalizing the mixed partial derivatives of $y^a$ with respect to $x^i$ and
$x^j$, second, replacing the obtained first derivatives of $y^a$ with
respect to $x^i$ on the right hand side of (64) again from the system (64).
If the functions $f^a_i$ satisfy the equations (66), the system (64) is called
{\it completely integrable}. In order to give a coordinate free formulation
of (66) and to introduce the {\it curvature, as a measure for
non-integrability} of (64), we shall first sketch the necessary
terminology [23].

Let $M$ be an arbitrary $n=p+q$ dimensional manifold. At every point $x\in M$
the tangent space $T_x(M)$ is defined. The union of all these spaces with
respect to the points of $M$ defines the {\it tangent bundle}. On the other
hand, the union of the co-tangent spaces $T^*_x(M)$ defines the {\it
co-tangent bundle}. At every point now of $M$ we separate a $p$ dimensional
subspace $\Delta_x(M)$ of $T_x(M)$ in a smooth way, i.e. the map
$x\rightarrow \Delta_x$ ix smooth. If this is done we say that a
$p$-dimensional {\it distribution} $\Delta$ on $M$ is defined.  From the
elementary linear algebra we know that every $p$-dimensional subspace
$\Delta_x$ of $T_x(M)$ defines unique $(n-p)=q$ dimensional subspace
$\Delta^*_x$ of the dual to $T_x(M)$ space $T^*_x(M)$, such that all elements
of $\Delta^*_x$ annihilate (i.e. send to zero) all elements of $\Delta_x$. In
this way we get a $q$-dimensional {\it co-distribution} $\Delta^*$ on $M$.
We consider those vector fields, the representatives of which at every point
are elements of the distribution $\Delta$, and those 1-forms, the
representatives of which at every point are elements of the co-distribution
$\Delta^*$. Clearly, every system of $p$ independent vector fields, belonging
to $\Delta$, defines $\Delta$ equally well, and in this case we call such a
system a {\it differential $p$-system} ${\cal P}$ on $M$.  The corresponding
system ${\cal P}^*$ of $q$ independent 1-forms is called $q$-dimensional {\it
Pfaff system}. Clearly, if $\alpha \in {\cal P}^*$ and $X\in {\cal P}$, then
$\alpha (X)=0$.

Similarly to the integral curves of vector fields, the concept of {\it
integral manifold} of a $p$-dimensional differential system is introduced.
Namely, a $p$-dimensional submanifold $V^p$ of $M$ is called {\it integral
manifold} for the $p$-dimensional differential system ${\cal P}$, or for the
$p$-dimensional distribution $\Delta$, to which ${\cal P}$ belongs, if the
tangent spaces of $V^p$ at every point coincide with the subspaces of the
distribution $\Delta$ at this point. In this case $V^p$ is called also
integral manifold for the $q$-dimensional Pfaff system ${\cal P}^*$. If
through every point of $M$ there passes an integral manifold for ${\cal P}$,
then ${\cal P}$ and ${\cal P}^*$ are called {\it completely integrable}.

Now we shall formulate the Frobenius theorems for integrability [25,26].
\vskip 0.5cm
{\it A differential system ${\cal P}$ is completely integrable if and only if
the Lie bracket of any two vector fields, belonging to ${\cal P}$, also
belongs to ${\cal P}$}.
\vskip 0.5cm
So, if $(X_1, ..., X_p)$ generate the completely integrable
differential system ${\cal P}$, then
\begin{equation}
\left[X_i,X_j\right]=C_{ij}^k X_k,                     
\end{equation}
where the coefficients $C_{ij}^k$ depend on the point. If we form the
$p$-vector $X_1\wedge X_2\wedge\dots\wedge X_p$ then the above relatin (67) is
equivalent to
\[
[X_i,X_j]\wedge X_1\wedge X_2\wedge\dots\wedge X_p=0,
\quad i,j=1,2,\dots,p.
\]

We turn now to the integrability criterion for the corresponding Pfaff
systems. If
the 1-forms $(\alpha^1, ..., \alpha^q)$ define the $q$-dimensional Pfaff
system ${\cal P}^*$ the following integrability criterion holds (the dual
Frobenius theorem):

\vskip 0.5cm
{\it The Pfaff system ${\cal P}^*$ is completely integrable if and only if}
\begin{equation}
{\bf d}\alpha^a=K^a_{bc}\alpha^b\wedge \alpha^c, \ b<c.    
\end{equation}
\vskip 0.5cm
Now, it is easily shown, that the above equations (68) are equivalent to the
following equations:
\begin{equation}
({\bf d}\alpha^a)\wedge \alpha^1\wedge\ .\ .\ .\wedge\alpha^a
\wedge\ .\ .\ .\wedge\alpha^q=0,\ a=1,...,q.                   
\end{equation}
We note that these criteria do not depend on the special choice of the vector
fields $X_i, i=1,\dots,p$, or 1-forms $\alpha^a, a=1,\dots,q$, where $p+q=dim
M$, which represent the distribution $\mathcal{P}$ and the codistribution
$\mathcal{P}^*$.

When a given Pfaff system ${\cal P}^*$, or the corresponding differential
system ${\cal P}$, are not integrable, then the relations (67)-(69) {\it are
not fulfilled}.  From formal point of view this means that there is at least
one couple of vector fields $(X,Y)$, belonging to ${\cal P}$, such that the
Lie bracket $\left[X,Y\right]$ {\it does not belong} to ${\cal P}$.
Therefore, if at the corresponding point $x\in M$ we choose a basis of
$T_x(M)$ such, that the first $p$ basis vectors form a basis of
$\Delta_x$, then $\left[X,Y\right]_x$ will have some nonzero components with
respect to those basis vectors, which belong to some complimentary to
$\Delta_x$ subspace $\Gamma_x:\ T_x(M)=\Delta_x \oplus \Gamma_x$.
As a rule, the distribution $\Gamma$ is chosen to be integrable, so
the choice of a projection operator $V_x:T_x(M)\rightarrow \Gamma_x(M), x\in
M$ with constant rank defines a new distribution $\Delta=Ker(V)$, such that
$KerV_x(M)\oplus\Gamma_x(M)=T_x(M)$. Then the distribution $\Gamma$ is called
{\it vertical} and the distribution $\Delta=Ker(V)$ is called {\it
horizontal}.

Now, the {\it curvature} ${\cal K}$ of the horizontal distribution $\Delta$
is defined by
\begin{equation}
{\cal K}(X,Y)=V([X,Y]), \ X,Y\in \Delta.                     
\end{equation}
It is clear, that the curvature ${\cal K}$ is a 2-form on $M$ with
values in the vertical distribution $\Gamma$. In fact, if $f$ is a
smooth function and $X,Y$ are two horizontal vector fields, then
\[
{\cal K}(X,fY)=V([X,fY])=V(f[X,Y]+X(f)Y)=
\]
\[
=fV([X,Y])+X(f)V(Y)=f{\cal K}(X,Y)
\]
because $V(Y)=0,\ Y-horizontal$.

The corresponding $q$-dimensional co-distribution $\Delta^*$ (or Pfaff system
${\cal P}^*)$ is defined locally by the 1-forms $(\theta^1,...,\theta^q)$,
such that $\theta^a(X)=0$ for all horizontal vector fields $X$. In this case
the 1-forms $\theta^a$ are called {\it vertical}, and clearly, they depend on
the choice of the horizontal distribution $\Delta$. In these terms the
non-integrability of $\Delta$ means
\[
{\bf d}\theta^a \neq K^a_{bc}\theta^b\wedge \theta^c, \ b<c.
\]
This non-equality means that at least one of the 2-forms ${\bf d}\theta^a$ is
not vertical, i.e. it has a nonzero horizontal projection
$H^*{\bf d}\theta^a$, which means that it does not annihilate all horizontal
vectors. In these terms it is naturally to define the curvature by
\[
H^*{\bf d}\theta ^a={\bf d}\theta ^a-K^a_{bc}\theta^b\wedge \theta^c, \ b<c.
\]
Now we see how this picture is defined by the equations (64).

Consider now the manifold $M=\mathbb{R}^p\times\mathbb{R}^q$ with coordinates
$(x^1,\dots,x^p,y^1,\dots,y^q)$. The vertical distribution is defined by
\[
\frac{\partial}{\partial y^a},\ a=p+1,...,p+q=n.
\]
Now, making use the equations of the system (64), i.e. the functions
$f^a_i$, we have to define the horizontal spaces at every point $(x,y)$,
i.e. local linearly independent vector fields $X_i, i=1,...,p$. The
definition is:
\begin{equation}
X_i=\frac{\partial}{\partial x^i}+f^a_i\frac{\partial}{\partial y^a}. 
\end{equation}
The corresponding Pfaff system shall consist of 1-forms $\theta^a,
a=p+1,...,p+q$ and is defined by
\begin{equation}
\theta^a=dy^a-f^a_i dx^i.                            
\end{equation}
In fact,
\[
\theta^a (X_i)=dy^a\left(\frac{\partial}{\partial x^i}\right)
+dy^a\left(f^b_i \frac{\partial}{\partial y^b}\right)-
f^a_j dx^j \left(\frac{\partial}{\partial x^i}\right)-
f^a_j dx^j\left(f^b_i\frac{\partial}{\partial y^b}\right)=
\]
\[
=0+f^b_i\delta_b^a -f^a_j\delta_i^j-0=0.
\]
\vskip 0.3cm
{\bf Remark}. The coordinate 1-forms $dy^a$ are not vertical and the
local vector fields $\frac{\partial}{\partial x^i}$ are not horizontal with
respect to the so defined horizontal distribution $(X_1,\dots,X_p)$.
\vskip 0.3cm
In this way the system (64) defines unique horizontal distribution. On the
other hand, if a horizontal distribution is given and the corresponding
vertical Pfaff system admits at least 1 basis, then this basis may be chosen
of the kind (72) always.  Moreover, in this kind it is unique. In fact, let
$(\theta'^1,...,\theta'^q)$ be any local basis of $\Delta^*$. Then in the
adapted coordinates we'll have
\[
\theta'^a=A^a_b dy^b+B^a_i dx^i,
\]
where $A^a_b, B^a_i$ are functions on $M$. We shall show that the matrix
$A^a_b$ has non-zero determinant, i.e. it is non-degenerate. Assuming the
opposite, we could find scalars $\lambda_a$, not all of which are equal to
zero, and such that the equality $\lambda_aA^a_b=0$  holds. We multiply now
the above equality by $\lambda^a$ and sum up with respect to $a$. We get
\[
\lambda_a \theta'^a=\lambda_a B^a_idx^i.
\]
Note now that on the right hand side of this last relation we have a
horizontal 1-form, while on the left hand side we have a vertical 1-form.
This is impossible by construction, so our assumption is not true, i.e. the
inverse matrix $(A^a_b)^{-1}$ exists, so multiplying on the left
$\theta'^a$ by $(A^a_b)^{-1}$ and putting $(A^a_b)^{-1} \theta'^b=\theta^a$
we obtain
\[
\theta^a=dy^a+(A^a_b)^{-1}B^b_i dx^i.
\]
We denote now $(A^a_b)^{-1}B^b_i=-f^a_i$ and get what we need. The uniqueness
part of the assertion is proved as follows. Assume there is another basis
$(\theta^a)''$ of the same kind. So, there must be a non-degenerate matrix
$C^a_b$, such that $(\theta^a)''=C^a_b\theta^b$. We get
\[
dy^a-(f^a_i)''dx^i=C^a_b dy^b-C^a_b f^b_idx^i,
\]
and from this relation it follows that the matrix $C^a_b$ is the identity.

Let's see now the explicit relation between the integrability condition (66)
of the system (64) and the curvature  $H^*{\bf d}\theta^a$ of the defined by
this system horizontal distribution. We obtain
\[
{\bf d}\theta^a=-\frac{\partial f^a_i}{\partial x^j} dx^j\wedge dx^i
-\frac{\partial f^a_i}{\partial y^b}dy^b\wedge dx^i.
\]
In order to define the horizontal projection of ${\bf d}\theta^a$
we note that $H^*(dx^i)=dx^i$, so, $H^*(dy^a)=H^*(\theta^a+f^a_i dx^i)=
f^a_i dx^i$, because $\theta^a$ are vertical. That's why for the curvature
$\Omega=H^*{\bf d}\theta^a\otimes\frac{\partial}{\partial y^a}$ we get
\begin{equation}
\Omega=(H^*{\bf d}\theta^a)\otimes\frac{\partial}{\partial y^a}
=\left(\frac{\partial f^a_j}{\partial x^i}-
\frac{\partial f^a_i}{\partial x^j}+                                   
\frac{\partial f^a_j}{\partial y^b}f^b_i-
\frac{\partial f^a_i}{\partial y^b}f^b_j\right)dx^i\wedge dx^j\otimes
\frac{\partial}{\partial y^a},\hspace{1cm} i<j.
\end{equation}
It is clearly seen that the integrability condition (66) coincides with the
requirement for zero curvature. The replacement of $dy^a$, making use of the
system (64), in order to obtain (66), acquires now the status of "horizontal
projection".

We verify now that the curvature, defined by $V(\left[X_i,X_j\right])$ gives
the same result.
\[
V([X_i,X_j])=
V\left(\left[\frac{\partial}{\partial x^i}+
f^a_i\frac{\partial}{\partial y^a},\frac{\partial}{\partial x^j}+
f^b_j\frac{\partial}{\partial y^b}\right]\right)=
\]
\[
=V\left(\frac{\partial f^b_j}{\partial x^i}\frac{\partial}{\partial
y^b}-\frac{\partial f^a_i}{\partial x^j}\frac{\partial}{\partial y^a}+
f^a_i\frac{\partial f^b_j}{\partial y^a}\frac{\partial}{\partial y^b}-
f^b_i\frac{\partial f^a_i}{\partial y^b}\frac{\partial}{\partial y^a}\right)=
\]
\[
=\left(\frac{\partial f^a_j}{\partial x^i}-\frac{\partial f^a_i}{\partial x^j}+
\frac{\partial f^a_j}{\partial y^b}f^b_i -
\frac{\partial f^a_i}{\partial y^b}f^b_j\right)\frac{\partial}{\partial y^a}.
\]

The basic moment in these considerations is that the system (64) defines a
vertical projection operator $V$ and a horizontal
projection operator $H=id-V$ in the tangent bundle $T(M)$. Explicitly, $H$
is defined in the basis
$(\frac{\partial}{\partial x^i},\frac{\partial}{\partial y^a})$ by
\[
H\left(\frac{\partial}{\partial x^i},\frac{\partial}{\partial y^a}\right)
=\left(\frac{\partial}{\partial x^i},\frac{\partial}{\partial y^a}\right)
\begin{Vmatrix}I & 0 \\ f^a_i & 0\end{Vmatrix}=
\left(\frac{\partial}{\partial x^i}+
f^a_i \frac{\partial}{\partial y^a},0,\dots,0\right),
\]
where $I$ denotes the identity $(p\times p)$-matrix and $0$ denotes the zero
$(p\times p)$ and the zero $(q\times q)$ matrices. So, $H_x$
projects the basis
$(\frac{\partial}{\partial x^i},\frac{\partial}{\partial y^a})$ at $x\in M$
on the horizontal subspace basis $(X_1(x),\dots,X_p(x))$. The projection
$V_x=(id_{T_xM}-H_x)$ projects every tangent space $T_xM$ to the corresponding
vertical subspace $VT_x$, so, the transposed matrix
$(V_x)^*=(id_{T_xM}-H_x)^*$ projects every cotangent space to the vertical
cotangent subspace. Explicitly we have (we omit to write the point $x\in M$)
\[
(id-H)^*(dx^i,dy^a)=
(dx^i,dy^a)\begin{Vmatrix}0 & -f^a_i \\ 0 & I\end{Vmatrix}
=(0,\dots,0,-f^a_idx^i+dy^a).
\]
Hence, a general connection on a manifold $M$ is defined by a projection
operator $H$ (or $V=(id_{T(M)}-H)$) in $T(M)$ with constant rank.
As a rule, if $H$ denotes the horizontal projection as it is in our case, the
image space of the corresponding vertical projection $V=(id_{T(M)}-H)$ is
considered to define an integrable distribution, so the kernel subspace
$Ker(id_{T(M)}-H)=KerV$ defines the horizontal distribution, and the
corresponding curvature carries information about the integrability of
$KerV$.

\subsection{General Connection Interpretation of EED}
We go back now to EED. According to {\bf Proposition 8.} the 1-forms
$(\tilde{A},\tilde{A^*})$, which live in the planes $(x,y)$, define a
completely integrable 2-dimensional Pfaff system. We choose the vertical
subspaces to live inside also in the $(x,y)$-planes, so, accordingly, the
coordinates are given in a different order, namely: $(z,\xi,x,y)$. The
horizontal spaces are defined by
\[
X_1=\frac{\partial}{\partial z}+\varepsilon u\frac{\partial}{\partial x}+
\varepsilon p\frac{\partial}{\partial y},\quad
X_2=\frac{\partial}{\partial \xi}+u\frac{\partial}{\partial x}+
 p\frac{\partial}{\partial y},
\]
which corresponds to the partial differential system
\[
\frac{\partial x}{\partial z}=\varepsilon u,\quad
\frac{\partial x}{\partial \xi}=u,\quad
\frac{\partial y}{\partial z}=\varepsilon p,\quad
\frac{\partial y}{\partial \xi}=p.
\]
Recalling from Sec.4 the unit 1-forms $\mathbf{R}$ and $\mathbf{S}$, and
denoting by prime (') the corresponding (through the metric $\eta$)
vector fields, we have the
\vskip 0.3cm
	{\bf Corollary}: The following relations hold:
\[
X_1=\mathbf{R'}+(i(\mathbf{R'})F)',\quad
X_2=\mathbf{S'}+(i(\mathbf{S'})F)'.
\]
\vskip 0.3cm
The horizontal projection $H_*$ in $TM$ is defined as follows (in the
$F$-adapted coordinate system)
\[
H_*
\left(\frac{\partial}{\partial z},\frac{\partial}{\partial \xi},
\frac{\partial}{\partial x},\frac{\partial}{\partial y}\right)=
\left(\frac{\partial}{\partial z},\frac{\partial}{\partial \xi},
\frac{\partial}{\partial x},\frac{\partial}{\partial y}\right)
\begin{Vmatrix}1 & 0 & 0 & 0\\
	       0 & 1 & 0 & 0\\
	       \varepsilon u & u & 0 & 0\\
	       \varepsilon p & p & 0 & 0
\end{Vmatrix}
=(X_1,X_2,0,0).
\]
In coordinate free form, i.e. as a section of $T^*(M)\otimes T(M)$, this
projection is given by
\[
H_*=-\mathbf{R}\otimes \big[\mathbf{R'}+(i(\mathbf{R'})F)'\big]+
\mathbf{S}\otimes \big[\mathbf{S'}+(i(\mathbf{S'})F)'\big].
\]

The corresponding vertical 1-forms are
\[
\theta^1=dx-\varepsilon udz-ud\xi,\quad
\theta^2=dy-\varepsilon pdz-pd\xi,
\]
which corresponds to the vertical projection $V^*=(id-H_*)^*$:
\[
(id-H_*)^*(dz,d\xi,dx,dy)=(dz,d\xi,dx,dy)
\begin{Vmatrix}
	       0 & 0 & -\varepsilon u & -\varepsilon p\\
	       0 & 0 & -u & -p\\
	       0 & 0 & 1 & 0\\
	       0 & 0 & 0 & 1
\end{Vmatrix}=(0,0,\theta^1,\theta^2).
\]
Respectively, for the horizontal projection of the coordinate 1-forms we
obtain
\[
H^*(dz,\,d\xi,\,dx,\,dy)=
\begin{Vmatrix}1 & 0 & 0 & 0\\
	       0 & 1 & 0 & 0\\
	       \varepsilon u & u & 0 & 0\\
	       \varepsilon p & p & 0 & 0
\end{Vmatrix}\begin{pmatrix}dz\\d\xi\\dx\\dy\end{pmatrix}
=(dz,\,d\xi,\,\varepsilon udz+ud\xi,\,\varepsilon pdz+pd\xi).
\]
Recalling our isotropic 1-form $\zeta=\varepsilon dz+d\xi$, we may write
\[
H^*(dz)=dz,\quad H^*(d\xi)=d\xi,
\quad H^*(dx)=dx-\theta^1=u\,\zeta, \quad H^*(dy)=dy-\theta^2=p\,\zeta.
\]
Making use of these relations we obtain
\[
H^*\mathbf{d}\theta^1=\varepsilon(u_\xi-\varepsilon u_z)dz\wedge d\xi,\ \
H^*\mathbf{d}\theta^2=\varepsilon(p_\xi-\varepsilon p_z)dz\wedge d\xi.
\]
Hence, for the curvature $\Omega$, determined by the nonlinear solution
 $F(u,p)$ we obtain
\[
 \Omega(u,p)=\varepsilon(u_\xi-\varepsilon u_z)dz\wedge d\xi\otimes
 \frac{\partial}{\partial x}+
 \varepsilon(p_\xi-\varepsilon p_z)dz\wedge d\xi\otimes
 \frac{\partial}{\partial y}.
\]
\vskip 0.3cm
{\bf Corollary}. The nonlinear solution $F(u,p)$ is a running wave only if
the curvature $\Omega(u,p)$ is equal to zero.

Denoting by $\eta(X_1)$ and $\eta(X_2)$ the $\eta$-corresponding 1-forms to
$X_1$ and $X_2$, we compute the expressions $<\eta(X_1),\Omega(u,p)>$ and
$<\eta(X_2),\Omega(u,p)>$:
\[
<\eta(X_1),\Omega(u,p)>=
\big[\varepsilon(u_\xi-\varepsilon u_z)<\eta(X_1),\frac{\partial}{\partial
x}>+ \varepsilon(p_\xi-\varepsilon p_z)<\eta(X_1),\frac{\partial}{\partial
y}>\big]dz\wedge d\xi=
\]
\[
=-\frac12\big[(u^2+p^2)_\xi-\varepsilon (u^2+p^2)_z\big]dz\wedge d\xi=
\varepsilon<\eta(X_2),\Omega(u,p)>.
\]
So, relations
$<\eta(X_1),\Omega(u,p)>=<\eta(X_2),\Omega(u,p)>=0$ give equation (48).

For the square of $\Omega$ we obtain
\[
\Omega_{\mu\nu}^\sigma\Omega^{\mu\nu}_\sigma=
(u_\xi-\varepsilon u_z)^2+(p_\xi-\varepsilon p_z)^2
=\phi^2(\psi_\xi-\varepsilon \psi_z)^2,\quad \mu<\nu.
\]
Recalling now the formulas in Sec.5 we have
\vskip 0.4cm
{\bf Corollary}: The square of the tensor field $\Omega$ is equal to
$-(\delta F)^2$, so $|\Omega|=|\delta F|$.
\vskip 0.4cm
{\bf Corollary}: A nonlinear solution may have rotational component of
propagation only if the curvature form $\Omega$ is time-like, i.e. if
$\Omega^2>0$.
\vskip 0.3cm
In order to define the scale factor $\mathcal{L}$ in these terms we observe
that the vertical components of the two coordinate vector fields
$\partial_z$ and $\partial_\xi$, given by
\[
V_*\left(\partial_z\right)=
\partial_z-H_*\left(\partial_z\right)=
\partial_z-X_1=-\varepsilon u\,\partial_x-\varepsilon p\,\partial_y,
\]
and
\[
V_*\left(\partial_\xi\right)=
\partial_\xi-H_*\left(\partial_\xi\right)=
\partial_\xi-X_2 =-u\,\partial_x-p\,\partial_y,
\]
have the same modules equal to $\sqrt{u^2+p^2}$. So, the scale factor
$\mathcal{L}$ is given by
\[
\mathcal{L}=\frac{\left|V_*\left(\partial_z\right)\right|}
{|\Omega|}=\frac{\left|V_*\left(\partial_\xi\right)\right|}
{|\Omega|}.
\]

Another way to introduce the scale factor only in terms of $H_*$ and $\Omega$
is the following. Recall that the space $L_W$ of linear maps in a linear
space $W$ is isomorphic to the tensor product $W^*\otimes W$, so, the space
$L_{W^*}$ of linear maps in $W^*$ is isomorphic to $W\otimes W^*$.
We shall define a map $\mathcal{F}: L_W\times L_{W^*}\rightarrow L_W$.
Let $\Phi\in L_W$ and $\Psi\in L_{W^*}$ and $\{e_i\}$ and $\{\varepsilon^j\}$
be two dual bases in $W$ and $W^*$ respectively. Then we have the
representations
\[
\Phi=\Phi_i^j\varepsilon^i\otimes e_j,\quad
\Psi=\Psi_n^m e_m\otimes\varepsilon^n.
\]
Now we define $\mathcal{F}(\Phi,\Psi)$ as follows:
\[
\mathcal{F}(\Phi,\Psi)=
\mathcal{F}(\Phi_i^j\varepsilon^i\otimes e_j,
\Psi_n^m e_m\otimes\varepsilon^n)=
\Phi_i^j\Psi_n^m<\varepsilon^i,e_m>\varepsilon^n\otimes e_j=
\Phi_i^j\Psi_n^i\varepsilon^n\otimes e_j
\]
For the composition $tr\circ\mathcal{F}(\Phi,\Psi)$ we obtain
\[
tr\circ\mathcal{F}(\Phi,\Psi)=\Phi_i^j\Psi_j^i.
\]
We compute now the matrix $\mathcal{A}$ of $\mathcal{F}(H_*,(id-H_*)^*)$
making use of the corresponding matrix representations of these two linear
maps given above.
\[
\mathcal{A}\big(\mathcal{F}(H_*,(id-H_*)^*)\big)=
\begin{Vmatrix} 0 & 0 & -\varepsilon u & -\varepsilon p\\
		0 & 0 & -u & -p\\
		0 & 0 & -2u^2 & -2up\\
		0 & 0 & -2up  & -2p^2
\end{Vmatrix},\quad tr(\mathcal{A})=-2(u^2+p^2).
\]
Hence,
the relation $L_\zeta(tr(\mathcal{A}))=0$ gives equation (48):
$(u^2+p^2)_\xi-\varepsilon(u^2+p^2)_z=0$, and we can define the scale factor
$\mathcal{L}$ as
\[
\mathcal{L}=\frac{\sqrt{\frac12|tr\circ\mathcal{F}(H_*,(id-H_*)^*)|}}
{|\Omega|}.
\]
On the other hand since $\mathcal{L}=|L_\zeta \psi|^{-1}$ (cf. relation
(51)), where $\psi$ is the phase of the solution, we obtain that $\psi$
satisfies the differential equation
\[
|L_\zeta \psi|=
\frac{|\Omega|}{\sqrt{\frac12|tr\circ\mathcal{F}(H_*,(id-H_*)^*)|}}.
\]

\subsection{Principal Connection Interpretation}
The principal connections are general connections on principal bundles [30]
appropriately adapted to the available bundle structure of the manifold.
A principal bundle $(\mathcal{P},\pi,B,G)$ consists of two manifolds:
the total space $\mathcal{P}$ and the base space $B$, a Lie group $G$, and a
surjective smooth map $\pi:\mathcal{P}\rightarrow B$. There is a {\it
free} smooth right action $R:\mathcal{P}\times
G\rightarrow \mathcal{P}$ of $G$ on $\mathcal{P}$, such, that the fibers
$\pi^{-1}(x),x\in B$, are orbits $G_x$ of $G$ through $x\in B$, and every
fiber $G_x$ is diffeomorphic to $G$. Since locally $\mathcal{P}$ is
diffeomorphic to $U_\alpha\times G$, where $U_\alpha\subset B$ is an open
subset of $B$, the right action of $G$ on $\mathcal{P}$ is made appropriately
consistent with the group product $G\times G\rightarrow G$, namely, there is
a coordinate representation $(U_\alpha,\psi_\alpha)$, where
$\psi_\alpha:U_\alpha\times G\rightarrow\pi^{-1}(U_\alpha)$ is a
diffeomorphism such, that $R(\psi_\alpha(x,a),b)=\psi_\alpha(x,a).b=
\psi_\alpha(x,ab); a,b\in G $.

If $\mathcal{G}$ is the Lie algebra of $G$ then the right action $R$ carries
the elements $h\in\mathcal{G}$ to fundamental (vertical) vector fields $Z_h$
on $\mathcal{P}$, and $Z_h, h\in \mathcal{G}$, define integrable distribution
$V_{\mathcal{G}}$ on $\mathcal{P}$. So, choosing additional to
$V_{\mathcal{G}}$ distribution $H$, we obtain a general connection on
$\mathcal{P}$. To make this general connection a principal connection we make
it consistent with the right action $R$ of $G$ on $\mathcal{P}$ in a natural
way: the subspace $H_{z.a}$ at the point $z.a\in\mathcal{P}$ coincides with
the image of $H_z$ through $(dR_a)_z$, i.e. $H_{z.a}=(dR_a)_z(H_z)$. The
projection on the horizontal distribution is denoted by $H_*$.

The horizontal distribution can be defined by a connection form $\omega$ on
$\mathcal{P}$, which is $\mathcal{G}$-valued 1-form and satisfies the
conditions: $\omega(Z_h)=h, h\in \mathcal{G}$;\  $\omega\circ H_*=0$
and $R_a^*\omega=Ad(a^{-1})\circ \omega,\
a\in G$.  Then the curvature $\Omega$ of the connection $\omega$ is given by
$\Omega=\mathbf{d}\omega\circ H_*=\mathbf{d}\omega+\frac12[\omega,\omega]$.
If the group is abelian, as it is our case, then $\Omega=\mathbf{d}\omega$.

If the bundle is trivial, i.e. $\mathcal{P}=M\times G$ then the projection
$\pi$ is the projection on the first member: $\pi(x,a)=x, x\in M, a\in G$.
In this case with every connection form $\omega$ can be associated a
$\mathcal{G}$-valued 1-form $\theta$ on the base space such, that
$\omega(x,e;X,Z_h)=h+\theta(x;X)$, where $x\in M, X\in T_xM, h\in\mathcal{G}$
and $e$ is the identity of $G$ [30, p.290]. For the curvature in the abelian
case we obtain $\Omega=\pi^*\mathbf{d}\theta$.

We go back now to EED. The base manifold of our trivial principal bundle
$\mathcal{P}=M\times G$ is, of course, the Minkowski space-time $M$, and the
abelian Lie group $\mathbb{G}$ is given by the $(2\times 2)$-real matrices
\[
\alpha(u,p) =\begin{Vmatrix}u & p\\-p & u\end{Vmatrix},\quad
u^2+p^2\neq 0.
\]
The corresponding Lie algebra $\mathcal{G}$, considered as a vecttor space,
has the natural basis $(I,J)$ and, as a set, it differs from $\mathbb{G}$
just by adding the zero $(2\times 2)$-matrix.  According to the above, in
order to introduce a connection on $\mathcal{P}=M\times G$ it is sufficient
to have a $\mathcal{G}$-valued 1-form on $M$. We recall now that a nonlinear
solution $F$ in the $F$-adapted coordinate system defines the 1-form
 $\zeta=\varepsilon dz+d\xi$ and the two functions $(u,p)$.  So, we
define $\theta$ as
\[
\theta=u\zeta\otimes I+p\zeta\otimes J.
\]
For the curvature we obtain
\[
\Omega=\Omega^1\otimes I+\Omega^2\otimes J=\mathbf{d}\theta=
\big[u_xdx\wedge\zeta+u_ydy\wedge\zeta-
\varepsilon(u_\xi-\varepsilon u_z)dz\wedge d\xi\big]\otimes I
\]
\[
+\big[p_xdx\wedge\zeta+p_ydy\wedge\zeta-
\varepsilon(p_\xi-\varepsilon p_z)dz\wedge d\xi\big]\otimes J.
\]
Since $\Omega$ is horizontal and, in our case, it is also
$\mathbb{G}$-invariant, there is just one $\mathcal{G}$-valued 2-form $\Phi$
on M such, that $\pi^*\Phi=\Omega$. Further we identify $\Phi$ and $\Omega$
and write just $\Omega$ since in our coordinates they look the same. So, we
can find $*\Omega$ with respect to the Minkowski metric in $M$:
\[
*\Omega=(*\Omega^1)\otimes I+(*\Omega^2)\otimes J=
\varepsilon\big[u_xdy\wedge\zeta-u_ydx\wedge\zeta-
\varepsilon(u_\xi-\varepsilon u_z)dx\wedge dy\big]\otimes I
\]
\[
+\varepsilon\big[p_xdy\wedge\zeta-p_ydx\wedge\zeta-
\varepsilon(p_\xi-\varepsilon p_z)dx\wedge dy\big]\otimes J.
\]

We recall that the canonical conjugation $\alpha\rightarrow\alpha^*$ in
$\mathcal{G}$, given by $(I,J)\rightarrow (I,-J)$, defines the inner
product in $\mathcal{G}$ by $<\alpha,\beta>=\frac12
tr(\alpha\circ\beta^*)$.  We have $<I,I>=1, <J,J>=1, <I,J>=0$.  We compute
the expressions $*<\Omega,*\Omega>$, $<\theta,*\Omega>$,
$\wedge(\theta,*\Omega)$, and obtain respectively:
\[
*<\Omega,*\Omega>=
*(\Omega^1\wedge *\Omega^1)<I,I>+ *(\Omega^2\wedge *\Omega^2)<J,J>
\]
\[
=(u_\xi-\varepsilon u_z)^2+(p_\xi-\varepsilon u_z)^2=-(\delta F)^2.
\]
\[
<\theta,*\Omega>=
-\big[u(u_\xi-\varepsilon u_z)+
p(p_\xi-\varepsilon p_z)\big]dx\wedge dy\wedge dz
\]
\[
-\varepsilon\big[u(u_\xi-\varepsilon u_z)
+p(p_\xi-\varepsilon p_z)\big]dx\wedge dy\wedge d\xi ,
\]
\[
\wedge(\theta,*\Omega)=
\Big\{\big[p(u_\xi-\varepsilon u_z)-
u(p_\xi-\varepsilon p_z)\big]dx\wedge dy\wedge dz
\]
\[
+\varepsilon\big[p(u_\xi-\varepsilon u_z)
-u(p_\xi-\varepsilon p_z)\big]dx\wedge dy\wedge d\xi\Big\}\otimes I\wedge J
=\delta F\wedge F\otimes I\wedge J.
\]
\vskip 0.4cm
{\bf Corollary}. Equations (18)-(20) are equivalent to the equation
$<\theta,*\Omega>=0$; a non-linear solution may have rotational component of
propagation only if $<\Omega,*\Omega>\neq 0$; the equation $\mathbf{d}(\delta
F\wedge F)=0$ is equivalent to $\mathbf{d}\big[\wedge(\theta,
*\Omega)\big]=0$.
\vskip 0.3cm
Since our principal bundle is trivial, the fundamental vector fields $Z_I$
and $Z_J$ on $M\times G$, generated by the basis vectors $(I,J)$ in
$\mathcal{G}$ coincide with the corresponding left invariant vector fields on
$\mathbb{G}$. Since we have a linear group action, a left invariant vector
field $Z_\sigma$ on $\mathbb{G}$, at the point $\alpha\in \mathbb{G}$,
generated by $\sigma\in \mathcal{G}$ is given by
$Z_\sigma=(\alpha,\alpha\circ \sigma)$.  So, in our case, at the point
$\big(x,\beta=(u,p)\big)\in M\times\mathbb{G}$ we obtain
$Z_I=uI+pJ;\ \ Z_J=-pI+uJ$. Hence, the metric on $\mathbb{G}$ gives
$|Z_I|=|Z_J|=\sqrt{u^2+p^2}$.  Therefore, for the scale factor $\mathcal{L}$
we readily obtain
\[
\mathcal{L}=\frac{|Z_I|}{|\Omega|}=\frac{|Z_J|}{|\Omega|}.
\]
\vskip 0.5cm
	{\bf Corollary}: The 1-form $\theta$, the curvature 2-form $\Omega$
and the scale factor $\mathcal{L}$ are invariant with respect to the group
action $(M\times\mathbb{G},\mathbb{G})\rightarrow (M\times\mathbb{G})$.
\vskip 0.5cm

Finally we note that we could write the 1-form $\theta$ in the form
\[
\theta=u\zeta\otimes I+p\zeta\otimes J=
\zeta\otimes uI+\zeta\otimes pJ=\zeta\otimes Z_I.
\]
If we start with the new 1-form
\[
\theta'=\zeta\otimes Z_J=\zeta\otimes (-pI)+\zeta\otimes (uJ)=
-p\zeta\otimes I+u\zeta\otimes J
\]
then, denoting $\mathbf{d}\theta'=\Omega'$, we obtain
\[
*<\Omega',*\Omega'>=*<\Omega,*\Omega>,\ \
<\theta',*\Omega'>=-<\theta,*\Omega>,\ \
\wedge(\theta',*\Omega')=-\wedge(\theta,*\Omega).
\]
Hence the last corollary and the definition of $\mathcal{L}$ stay in force
with respect to $\theta'$ too.
\vskip 0.5cm
	{\bf Remark}. The group $\mathbb{G}$ acts on the right on the basis
$(I,J)$ of $\mathcal{G}$. So, the transformed basis with
$\beta(u,\varepsilon p)\in\mathbb{G}$ is
$R_{\beta}(I,J)=(uI-\varepsilon pJ,\,\varepsilon pI+uJ)$.
Now let $\omega$ be 1-form on $M$ such that $\omega^2<0$ and $\omega\wedge
*\zeta=0$. Then the corresponding generalized field is given by
\[
F_\omega\otimes I+*F_\omega\otimes J=
=(\omega\wedge\zeta)\otimes (uI-\varepsilon pJ)+
*(\omega\wedge\zeta)\otimes (\varepsilon pI+uJ).
\]
In particular, $\omega=dx$ defines the field in an $F$-adapted coordinate
system.

\subsection{Linear connection interpretation}
Linear connections $\nabla$ are 1st-order differential operators in vector
bundles.
If such a connection $\nabla$ is given and $\sigma$ is a section of
the bundle, then $\nabla \sigma$ is 1-form on the base space valued in the
space of sections of the vector bundle, so if $X$ is a vector field on the
base space then $i(X)\nabla \sigma=\nabla_X \sigma$ is a new section of the
same bundle. If $f$ is a smooth function on the base space then
$\nabla(f\sigma)=df\otimes\sigma+f\nabla\sigma$, which justifies the
differential operator nature of $\nabla$: the components of $\sigma$ are
differentiated and the basis vectors are linearly transformed.

Let $e_a$ and $\varepsilon^b, a,b=1,2,\dots,r$ be two dual local bases of the
corresponding spaces of sections: $<\varepsilon^b,e_a>=\delta_a^b$, then we
can write
\[
\sigma=\sigma^a e_a,\quad \nabla=\mathbf{d}\otimes id+
\Gamma_{\mu a}^b dx^\mu\otimes(\varepsilon^a\otimes e_b),\quad
\nabla(e_a)=\Gamma_{\mu a}^b dx^\mu\otimes e_b,
\]
so
\[
\nabla(\sigma^m e_m)=\mathbf{d}\sigma^m\otimes e_m+
\sigma^m\Gamma_{\mu a}^b dx^\mu<\varepsilon^a,e_m>\otimes\, e_b=
\left[\mathbf{d}\sigma^b+\sigma^a\Gamma_{\mu a}^b dx^\mu\right]\otimes e_b,
\]
and $\Gamma_{\mu a}^b$ are the components of $\nabla$ with respect to
the coordinates $\{x^\mu\}$ on the base space and with respect to the bases
$\{e_a\}$ and $\{\varepsilon^b\}$. Since  the elements
$(\varepsilon^a\otimes e_b)$ define a basis of the space of (local) linear
maps of the local sections, it becomes clear that in order to define locally
a linear connection it is sufficient to have some 1-form $\theta$ on the base
space and a linear map $\phi=\phi_a^b\varepsilon^a\otimes e_b$ in the space
of sections. Then
$$
\nabla(\sigma)=\mathbf{d}\sigma^a\otimes e_a+\theta\otimes \phi(\sigma)
$$
defines a linear connection with components
$\Gamma_{\mu a}^b=\theta_\mu\phi_a^b$ in these bases.
So, locally, a linear connection $\nabla$ may be written as
$$
\nabla=\mathbf{d}\otimes(\varepsilon^a\otimes e_a)
+\Psi_{\mu a}^{b}dx^\mu\otimes(\varepsilon^a\otimes e_b).
$$
If $\Psi_1$ and $\Psi_2$ are two such 1-forms then a map
$(\Psi_1,\Psi_2)\rightarrow
(\wedge,\circledcirc)(\Psi_1,\Psi_2)$ is defined by (we
shall write just $\circledcirc$ for $(\wedge,\circledcirc)$ and the usual
$\circ$ will mean just composition)
\[
\circledcirc(\Psi_1,\Psi_2)=
(\Psi_1)_{\mu a}^{b}(\Psi_2)_{\nu m}^{n}dx^\mu\wedge dx^\nu\otimes
\big[\circ(\varepsilon^a\otimes e_b,\varepsilon^m\otimes e_n)\big]
\]
\[
=(\Psi_1)_{\mu a}^{b}(\Psi_2)_{\nu m}^{n}dx^\mu\wedge dx^\nu\,
\otimes\big[<\varepsilon^a,e_n>(\varepsilon^m\otimes e_b)\big]
=(\Psi_1)_{\mu a}^b(\Psi_2)_{\nu m}^a
dx^\mu\wedge dx^\nu\otimes(\varepsilon^m\otimes e_b),
\ \ \mu<\nu.
\]
Now, in the case of trivial vector bundles, the curvature of
$\nabla$ is given by [30, p.328]
$$
\left[\mathbf{d}(\Psi_{\mu a}^{b}dx^\mu)\right]\otimes(\varepsilon^a\otimes
e_a) +\circledcirc(\Psi,\Psi).
$$

We go back now to EED. The vector bundle under consideration is the
(trivial) bundle $\Lambda^2(M)$ of 2-forms on the Minkowski space-time $M$.
From Section 7 we know that a representation $\rho'$ of $\mathcal{G}$ is
defined by relation (58). So, if $\alpha(u,p)\in\mathcal{G}$ then
$\rho'(\alpha)$ is a linear map in $\Lambda^2(M)$, and recalling our 1-form
$\zeta=\varepsilon dz+d\xi$ we define a linear connection $\nabla$ in
$\Lambda^2(M)$ by
\[
\nabla=\mathbf{d}\otimes id_{\Lambda^2(M)}+\zeta\otimes\rho'(\alpha(u,p))=
\mathbf{d}\otimes id_{\Lambda^2(M)}
+\zeta\otimes(u\mathcal{I}+p\mathcal{J}) ,\quad
\alpha\in\mathcal{G}.
\]
Two other connections $\bar{\nabla}$ and $\nabla^*$ are defined by
$$
\rho'(\bar{\alpha}(u,p))=
\rho'(\alpha(u,-p))=u\mathcal{I}-p\mathcal{J},\ \  \text{and}\ \
(\rho')^*(\alpha(u,p))=\rho'(\alpha.J)=\rho'(\alpha(-p,u))=
-p\mathcal{I}+u\mathcal{J},
$$
and we shall denote:
$$
\chi=u\mathcal{I}+p\mathcal{J},\
\bar\chi=u\mathcal{I}-p\mathcal{J},\
\chi^*=-p\mathcal{I}+u\mathcal{J}.
$$
So, we may say that $\chi$ represents $F$ and $\chi^*$ represents $*F$.

Denoting $\Psi=\zeta\otimes\chi,\ \bar\Psi=\zeta\otimes\bar\chi,\
\Psi^*=\zeta\otimes\chi^*$,
we note that (since $\zeta\wedge\zeta=0$)
\[
\circledcirc(\Psi,\Psi)=
\circledcirc(\Psi,\bar\Psi)=
\circledcirc(\Psi,\Psi^*)=0.
\]

Now, since
$\Psi=u\zeta\otimes\mathcal{I}+p\zeta\otimes\mathcal{J},
\ \bar\Psi=u\zeta\otimes\mathcal{I}-p\zeta\otimes\mathcal{J},\
\Psi^*=-p\zeta\otimes\mathcal{I}+u\zeta\otimes\mathcal{J}$
for the corresponding curvatures we obtain
\[
\mathcal{R}=
\mathbf{d}(u\zeta)\otimes\mathcal{I}+
\mathbf{d}(p\zeta)\otimes\mathcal{J},\ \
\bar{\mathcal{R}}=
\mathbf{d}(u\zeta)\otimes\mathcal{I}-
\mathbf{d}(p\zeta)\otimes\mathcal{J},\ \
\mathcal{R}^*=
\mathbf{d}(-p\zeta)\otimes\mathcal{I}+
\mathbf{d}(u\zeta)\otimes\mathcal{J}.
\]
\noindent
{\bf Remark}. We have omitted here $\varepsilon$ in front of $p\mathcal{J}$,
but this is not essential since, putting $p\rightarrow\varepsilon p$ in the
expressions obtained, we easily restore the desired generality.
\vskip 0.3cm

By direct calculation we obtain:
\[
*\frac16 Tr\left[ \circledcirc(\bar\Psi,*\mathbf{d}\Psi)\right]=-
\varepsilon\big[u(u_\xi-\varepsilon u_z)+
p(p_\xi-\varepsilon p_z)\big]dz-
\big[u(u_\xi-\varepsilon u_z)+
p(p_\xi-\varepsilon p_z)\big]d\xi;
\]
\[
\frac16Tr\left[\circledcirc(\Psi^*,*\mathbf{d}\Psi)\right]=\varepsilon
\Big[p(u_\xi-\varepsilon u_z)-
 u(p_\xi-\varepsilon p_z)\Big]dx\wedge dy\wedge dz+
\]
\[
\Big[p(u_\xi-\varepsilon u_z)-
u(p_\xi-\varepsilon p_z)\Big]dx\wedge dy\wedge d\xi
=\delta F\wedge F;
\]
Denoting by $|\mathcal{R}|^2$ the quantity
$\frac16|*Tr\left[\circledcirc(\mathcal{R}\wedge*\bar{\mathcal{R}})\right]|$
we obtain
\[
|\mathcal{R}|^2=
\frac16|*Tr\left[\circledcirc(\mathbf{d}\Psi,
*\mathbf{d}\bar{\Psi})\right]|=
(u_\xi-\varepsilon u_z)^2+(p_\xi-\varepsilon p_z)^2=|\delta F|^2.
\]
Finally, since in our coordinates
\[
\frac16 tr(\chi)=\frac16 tr(u\mathcal{I}+p\mathcal{J})=u,\quad
\text{and}\quad
\frac16 tr\big[(\chi\circ\bar\chi)\big]=\frac16
tr\big[(u\mathcal{I}+p\mathcal{J})\circ(u\mathcal{I}-p\mathcal{J})\big]
=u^2+p^2,
\]
for the phase $\psi$ and for the scale factor
$\mathcal{L}$ we obtain respectively
\[
\psi=\mathrm{arccos}\frac{\frac16 tr\chi}
{\sqrt{\frac16 tr(\chi\circ\bar\chi)}},
\quad
\mathcal{L}=\frac{\sqrt{\frac16
tr(\chi\circ\bar\chi)}}{\sqrt{\frac16}|\mathcal{R}|}=
\frac{\sqrt{tr(\chi\circ\bar\chi)}}{|\mathcal{R}|}.
\]
These results allow to say that choosing such a linear connection in
$\Lambda^2(M)$ then our nonlinear equations  are given by
$Tr\left[\circledcirc(\bar\Psi,*\mathbf{d}\Psi)\right]=0$, and that the
non-zero value of the squared curvature invariant $|\mathcal{R}|^2$
guarantees availability of rotational component of propagation.

\vskip 0.5cm
As a brief comment to the three connection-curvature interpretations of the
basic relations of EED we would like to especially note the basic role of the
isotropic 1-form $\zeta$: in the general connection approach $\zeta$ defines
the horizontal projection of the coordinate 1-forms $dx,dy$: $H^*(dx)=u\zeta$
and $H^*(dy)=p\zeta$; in the principal connection approach $\zeta$ defines
the 1-form $\theta=\zeta\otimes Z_I$, or $\theta'=\zeta\otimes Z_J$; in the
linear connection approach it defines the components of $\nabla$ through
$\zeta\otimes \rho'(\alpha(u,p))$. It also participates in defining
the 2-form $F_o=dx\otimes \zeta$, which gives the possibility to identify a
nonlinear solution $F(u,p)$ with an appropriately defined linear map
$\rho'(\alpha(u,p))=u\mathcal{I}+p\mathcal{J}$ in $\Lambda^2(M)$.

This special importance of $\zeta$  is based on the fact that it defines the
direction of translational propagation of the solution, and its uniqueness
is determined by the EED equations (18)-(20): {\it zero invariants
$F_{\mu\nu}F^{\mu\nu}=F_{\mu\nu}(*F)^{\mu\nu}=0$ are required for all
nonlinear solutions}.

For all nonlinear solutions we have $\delta F\neq 0$, and all finite
nonlinear solutions have finite amplitude $\phi$:
$0<\phi^2=\frac16 tr(F\circ\bar{F})=
\frac16 tr(\chi\circ\bar\chi)<\infty$.
The scale factor $\mathcal{L}$ separates the finite nonlinear
solutions to two subclasses: if $\mathcal{L}\rightarrow \infty$, i.e.
$|\delta F|=|\Omega|=|\mathcal{R}|=0$, the solution has no
spin properties; if $\mathcal{L}<\infty$, i.e.
$|\delta F|=|\Omega|=|\mathcal{R}|\neq 0$, the solution
carries spin momentum.

Hence, in terms of curvature we can say that {\it the nonzero curvature
invariant $|\Omega|$, or $|\mathcal{R}|$,  is responsible for availability of
rotational component of propagation}, in other words, {\it the spin
properties of a nonlinear solution require non-zero curvature}.

\section{Nonlinear Solutions with Intrinsic Rotation}
Before to start with spin-carrying solutions we briefly comment the nonlinear
solutions with running wave character. These solutions require $|\delta
F|=|\delta *F|=0$, so, for the two spatially finite functions $u$ and $p$ we
get in an $F$-adapted coordinate system $u=u(x,y,\xi+\varepsilon z)$ and
$p=p(x,y,\xi+\varepsilon z)$. Whatever the spatial shape and spatial
structure of these two finite functions could be the whole solution will
propagate {\it only translationally} along the coordinate $z$ with the
velocity of light $c$ without changing its shape and structure. In this sense
this class of nonlinear solutions show soliton-like behavior: finite 3d
spatial formations propagate translationally in vacuum. If we forget about
the spin properties of electromagnetic radiation, we can consider such
solutions as mathematical models of classical finite electromagnetic
macro-formations of any shape and structure, radiated by ideal parabolic
antennas. Maxwell equations can NOT give such solutions.

Now we turn to spin-carrying solutions. The crucial moment here is to find
reasonable additional conditions for the phase function $\varphi$, or
for the phase $\psi=\mathrm{arccos}\,\varphi$.

\subsection{The Basic Example}
The reasoning here follows the idea that
these additional conditions {\it have to express some internal consistency
among the various characteristics of the solution}. A suggestion what kind of
internal consistency to use comes from the observation that {\it the
amplitude function $\phi$ is a first integral of the vector field $\zeta$},
i.e.
\[
\zeta(\phi)=\left(-\varepsilon \frac{\partial}{\partial z}+\frac
{\partial} {\partial \xi}\right)(\phi)=-\varepsilon \frac{\partial}{\partial
z}\phi(x,y,\xi+\varepsilon z) +\frac {\partial}{\partial
\xi}\phi(x,y,\xi+\varepsilon z)=0.
\]
In order to extend this consistency between $\zeta$ and $\phi$ we require the
phase function $\varphi$ to be first integral of some of the available
$F$-generated vector fields.  Explicitly, we require the following:
\vskip 0.5cm
\noindent
{\it The phase function}\ $\varphi$\
{\it is a first integral of the three vector fields}
${\bf A,A^*}$ and ${\bf S}$:
$$
{\bf A}(\varphi)=
{\bf A^*}(\varphi)={\bf S}(\varphi)=0,
$$
{\it and the scale factor $\mathcal{L}$ is a first integral of} $\mathbf{R}:
\mathbf{R}(\mathcal{L})=0$.
\vskip 0.5cm
The first two requirements ${\bf A}(\varphi)={\bf A^*}(\varphi)=0$
define the following system of differential equations for $\varphi$:
\[
-\varphi \frac{\partial \varphi}{\partial
x}-\sqrt{1-\varphi^2}\frac{\partial \varphi}{\partial y}=0,\
\sqrt{1-\varphi^2}\frac{\partial \varphi}{\partial x}-
\varphi\frac{\partial \varphi}{\partial y}=0.
\]
Noticing that the matrix
\[
\begin{Vmatrix}
-\varphi              &-\sqrt{1-\varphi^2}\\
\sqrt{1-\varphi^2}    &-\varphi
\end{Vmatrix}
\]
has non-zero determinant, we conclude that the only solution of the above
system is the zero-solution:
\[
\frac{\partial \varphi}{\partial x}=\frac{\partial \varphi}{\partial y}=0.
\]
We conclude that in the coordinates used the phase function $\varphi$ may
depend only on $(z,\xi)$. The third equation $\mathbf{S}(\varphi)=0$ requires
$\varphi$ not to depend on $\xi$ in this coordinate system, so,
$\varphi=\varphi(z)$. For $\mathcal{L}$ we get
$$
\mathcal{L}=\frac{\sqrt{1-\varphi^2}}{|\varphi_z|}.
$$
Now, the last requirement, which in these coordinates reads
\[
\mathbf{R}(\mathcal{L})=\frac{\partial \mathcal{L}}{\partial z}=
\frac{\partial}{\partial z}\frac{\sqrt{1-\varphi^2}}{|\varphi_z|}=0,
\]
means that the scale factor $\mathcal{L}$ is a pure constant:
$\mathcal{L}=const$.  In this way the defining relation for $\mathcal{L}$
turns into a differential equation for $\varphi$:
\begin{equation}
\mathcal{L}=\frac{\sqrt{1-\varphi^2}}{|\varphi_z|}\  \rightarrow
\frac{\partial \varphi}{\partial z}=                               
\mp \frac{1}{\mathcal{L}}\sqrt{1-\varphi^2}.
\end{equation}
The obvious solution to this equation is
\begin{equation}
\varphi(z)=\mathrm{cos}\left(\kappa\frac{z}{\mathcal{L}}+const\right),
\end{equation}                                    
where $\kappa=\pm 1$. We note that the naturally arising in this case spatial
periodicity $2\pi\mathcal{L}$ and {\it characteristic frequency}
$\nu=c/2\pi\mathcal{L}$ have nothing to do with the corresponding concepts
in CED.  In fact, {\it the quantity $\mathcal{L}$ can not be defined in
Maxwell's theory}.

The above considerations may be slightly extended and put in terms of the
phase $\psi$, and in these terms they look simpler. In fact, we have the
equation
\[
\psi_\xi-\varepsilon \psi_z=\kappa\frac{1}{\mathcal{L}}, \ \ \kappa=\pm 1,
\]
where $\mathcal{L}=const$. So, we get the two basic solutions
\[
\psi_1=\kappa\frac{z}{\mathcal{L}}+const, \quad
\psi_2=\kappa\frac{\xi}{\mathcal{L}}+const.
\]

We get two kinds of periodicity: spatial periodicity along the coordinate $z$
and time-periodicity along the time coordinate $\xi$. The two values of
$\kappa=\pm 1$ determine the two possible rotational structures: left-handed
(left polarized), and right-handed (right polarized).
Further we are going to
concentrate on the spatial periodicity because it is strongly connected with
the spatial shape of the solution. In particular, it suggests to localize the
amplitude function $\phi$ inside a helical cylinder of height
$2\pi\mathcal{L}$, so, the solution will propagate along the prolongation of
this finite initial helical cylinder in such a way that all points of the
spatial support shall follow their own helical trajectories without
crossings. For such solutions with $|\delta F|\neq 0$ we are going to
consider various ways for quantitative description of the available intrinsic
rotational momentum, or the {\it spin momentum}, of these solutions. We call
it {\it spin-momentum} by obvious reasons: it is of intrinsic nature and does
not depend on any {\it external} point or axis as it is the case of angular
momentum. And it seems quite natural to follow the suggestions coming from
the corresponding considerations in Sec.5.

\subsection{The $\mathcal{G}$-Approach}
In this first approach we make use of the corresponding
scale factor $\mathcal{L}=const$, of the isotropic 1-form $\zeta$ and
of the two objects $Z_I=uI+pJ$ and $Z_J=-pI+uJ$, considered as
$\mathcal{G}$-valued functions on $M$. By these quantities we build the
following $\mathcal{G}\wedge\mathcal{G}$-valued 1-form $H$:
\begin{equation}
H=2\pi\kappa \frac{\mathcal{L}}{c} \zeta \otimes(Z_I\wedge Z_J).
\end{equation}
In components we have
\[
H_\mu^{ab}=2\pi\kappa\frac{\mathcal{L}}{c} \zeta_\mu(
Z_I^aZ_J^b-Z_I^bZ_J^a).
\]
In our system of coordinates we get
\[
H=2\pi\kappa\frac{\mathcal{L}}{c}\phi^2\
(\varepsilon dz+d\xi)\otimes I\wedge J,
\]
hence, the only non-zero components are
\[
H_3^{12}=2\pi\kappa\varepsilon\frac{\mathcal{L}}{c} \phi^2,
\ H_4^{12}=2\pi\kappa\frac{\mathcal{L}}{c} \phi^2.
\]
It is easily seen that $*H$ is closed: $\mathbf{d}*H=0$. In fact,
\[
\mathbf{d}*H=2\pi\kappa\frac{\mathcal{L}}{c}\big[\big(\phi^2\big)_\xi-
\varepsilon\big(\phi^2\big)_z\big](dx\wedge dy\wedge d\wedge d\xi)\otimes
(I\wedge J)=0
\]
because $\phi^2$ is a running wave along the coordinate $z$. We restrict now
$*H$ to $\mathbb{R}^3$ and obtain
\[
(*H)_{\mathbb{R}^3}=
2\pi\kappa\frac{\mathcal{L}}{c}
\phi^2(dx\wedge dy\wedge dz)\otimes(I\wedge J).
\]
According to Stokes theorem, for finite solutions, we obtain the finite
(conserved) quantity
\[
{\bf H}=\int_{\mathbb{R}^3}*H_{\mathbb{R}^3}
=2\pi\kappa\frac{\mathcal{L}}{c} E=\kappa ET\,I\wedge J,
\]
which is a volume form in $\mathcal{G}$, $T=2\pi\mathcal{L}/c$, and $E$ is
the integral energy of the solution. The module $|{\bf H}|$ of {\bf H} is
$|{\bf H}|=ET$.

We see the basic role of the two features of the solutions: their spatially
finite/concentrated nature, giving finite value of all spatial integrals,
and their translational-rotational dynamical nature with $|\delta F|\neq 0$,
allowing finite value of the scale factor $\mathcal{L}$.

\subsection{The FN-Bracket Approach}
We proceed to the second approach to introduce spin-momentum. We recall the
Fr\"oliher-Nijenhuis bracket $S_F$ of the finite nonlinear solution $F$,
evaluated on the two unit vector fields $\mathbf{A}$ and
$\varepsilon\mathbf{A}^*$:
\[
(S_F)_{\mu \nu }^\sigma{\bf A}^\mu {\bf \varepsilon A^*}^\nu =
(S_F)_{12}^\sigma ({\bf A}^1{\bf \varepsilon A^*}^2-
{\bf A}^2{\bf \varepsilon A^*}^1).
\]
For $(S_F)_{12}^\sigma $ we get
\[
(S_F)_{12}^1=(S_F)_{12}^2=0,\quad
(S_F)_{12}^3=-\varepsilon (S_F)_{12}^4=2\varepsilon \{p(u_\xi -\varepsilon
u_z)-u(p_\xi -\varepsilon p_z)\}.
\]
It is easily seen that the following relation holds:
${\bf A}^1{\bf \varepsilon A^*}^2-{\bf A}^2{\bf \varepsilon A^*}^1=1.$
Now, for the above obtained solution for $\varphi$ we have
\[
u=\phi (x,y,\xi +\varepsilon z)
\cos\left(\kappa \frac{z}{\mathcal{L}} +const\right),\quad
p=\phi (x,y,\xi +\varepsilon z)
\sin \left(\kappa\frac{z}{\mathcal{L}} +const\right).
\]
We obtain
\[
(S_F)_{12}^3=-\varepsilon (S_F)_{12}^4=
-2\varepsilon \frac{\kappa}{\mathcal{L}} \phi ^2,
\]
\[
(S_F)_{\mu\nu }^\sigma {\bf A}^\mu {\bf \varepsilon A^*}^\nu
=\left[0,0,-2\varepsilon
\frac{\kappa}{\mathcal{L}}\phi ^2,
2\frac{\kappa}{\mathcal{L}}\phi ^2\right].
\]
Since $\phi^2$ is a running wave along the $z$-coordinate, the vector
field $S_F({\bf A,\varepsilon A^*})$ has zero divergence:
$\nabla_\nu \left[S_F({\bf A,\varepsilon A^*})\right]^\nu=0$.
Now, defining the {\it helicity vector} of the solution $F$ by
\[
\Sigma_F=\frac{\mathcal{L} ^2}{2c}S_F({\bf A,\varepsilon A^*}),
\]
 then $\Sigma_F$ has also zero divergence, and the
integral quantity
\[
\int{\left(\Sigma_F\right)_4}dxdydz
\]
does not depend on time and is equal to $\kappa ET$.

\subsection{The $\mathbf{d}(F\wedge\delta F)=0$ Approach}
Here we make use of the equation $\mathbf{d}(F\wedge\delta F)=0$ and see what
restrictions this equation imposes on $\psi$. In our system of coordinates
this equation is reduced to
\begin{equation}
\mathbf{d}(F\wedge \delta F)=\varepsilon\Phi^2\left(\psi_{\xi\xi}+\psi_{zz}-
2\varepsilon\psi_{z\xi}\right)dx\wedge dy\wedge dz\wedge d\xi=0,   
\end{equation}
i.e.
\begin{equation}
\psi_{\xi\xi}+\psi_{zz}-2\varepsilon\psi_{z\xi}=
\left(\psi_\xi-\varepsilon\psi_z\right)_\xi -
\varepsilon\left(\psi_\xi-\varepsilon\psi_z\right)_z =0.           
\end{equation}
Equation (78) has the following solutions:

1$^o $. Running wave solutions $\psi=\psi(x,y,\xi+\varepsilon z)$,

2$^o $. $\psi = \xi.g(x,y,\xi+\varepsilon z)+b(x,y)$,

3$^o $. $\psi = z.g(x,y,\xi+\varepsilon z)+b(x.y)$,

4$^o $. Any linear combination of the above solutions with coefficients which
are allowed to depend on $(x,y)$.

\noindent The functions $g(x,y,\xi+\varepsilon z)$ and $b(x,y)$ are arbitrary
in the above expressions.

The running wave solutions $\psi_1$, defined by $1^o$, lead to $F\wedge \delta
F=0$ and to $|\delta F|=0$, and by this reason they have to be ignored. The
solutions $\psi_2$ and $\psi_3$, defined respectively by 2$^o$ and 3$^o$,
give the scale factors $\mathcal{L}=1/|g|$, and since $\mathcal{L}$ is
dually invariant it should not depend on $(x,y)$ in this coordinate system.
Hence, we obtain $g=g(\xi+\varepsilon z)$, so, the most natural choice seems
$g=const$, which implies also $\mathcal{L}=const$. A possible dependence of
$\psi$ on $(x,y)$ may come only through $b(x,y)$.  Note that the physical
dimension of $\mathcal{L}$ is {\it length} and $b(x,y)$ is
dimensionless.

We turn now to the integral spin-momentum computation.
In this approach its density is given by the correspondingly
normalized 3-form $F\wedge \delta F$. So we normalize it as follows:
\begin{equation}
\beta=2\pi\frac{\mathcal{L} ^2}{c}F\wedge\delta F=
2\pi\frac{\mathcal{L} ^2}{c}\left[-\varepsilon                    
\phi^2(\psi_\xi-\varepsilon \psi_z)dx\wedge dy\wedge
dz-\phi^2(\psi_\xi-\varepsilon\psi_z)dx\wedge dy\wedge d\xi\right].
\end{equation}
The physical dimension of $\beta$ is "energy-density $\times $ time".
We see that $\beta$ is closed: $\mathbf{d}\beta=0$,
so we may use the Stokes' theorem.  The restriction of
$\beta$ to $\mathbb{R}^3$ is:
\[
\beta_{\mathbb{R}^3}=2\pi\frac{\mathcal{L} ^2}{c}\left[-\varepsilon
\phi^2(\psi_\xi-\varepsilon \psi_z)dx\wedge dy\wedge dz\right].
\]
Let's consider first the solutions $3^o$ of equation (78) with
$\mathcal{L}=const$.  The corresponding phase
$\psi=\kappa\frac{z}{\mathcal{L}}+b(x,y), \kappa=\pm1$, requires {\bf
spatial} periodicity of $2\pi\mathcal{L}$ along the coordinate $z$. So, if we
restrict the spatial extension of the solution along $z$  to one such period
$l_o=2\pi\mathcal{L}$, our solution will occupy at every moment a one-step
part of a helical (screw) cylinder. Its time evolution will be a
translational-rotational propagation along this helical cylinder. So, we have
an example of an object with helical spatial structure and with intrinsical
rotational component of propagation, and this rotational component of
propagation does NOT come from a rotation of the object as a whole around
some axis.

On the contrary, the solutions defined by $2^o$, are NOT obliged to have
spatial periodicity. Their evolution includes $z$-translation and rotation
around the $z$-axis {\it as a whole} together with some rotation around
themselves.

For the case $3^o$ with $\mathcal{L}=const$ we can integrate
$$
\beta_{\mathbb{R}^3}=
\frac{2\pi \mathcal{L}}{c}\kappa\phi^2 dx\wedge dy\wedge dz
$$
over the 3-space and obtain
\begin{equation}
\int_{\mathbb{R}^3}{\beta}=                                       
\kappa E\frac{2\pi \mathcal{L}}{c}=
\kappa ET=\pm ET,
\end{equation}
where $E$ is the integral energy of the solution, $T=2\pi\mathcal{L}/c$ is
the intrinsically defined time-period, and $\kappa=\pm 1$ accounts for the
two polarizations.  According to our interpretation this is the integral
spin-momentum of the solution for one period $T$.

\subsection{The Nonintegrability Approach}
Here we make use of the observation that the two Pfaff systems $(A,\zeta)$
and $(A^*,\zeta)$ are nonintegrable when $\mathcal{L}\neq 0$. From Sec.5 we
have
\[
\begin{split}
\mathbf{d}\tilde{A}\wedge\tilde{A}\wedge\zeta=
 \mathbf{d}\tilde{A}^*\wedge\tilde{A}^*\wedge\zeta &=
\varepsilon\big[u(p_\xi-\varepsilon p_z)-p(u_\xi-\varepsilon u_z)\big]
dx\wedge dy\wedge dz\wedge d\xi \\
&=\varepsilon\phi^2(\psi_\xi-
\varepsilon \psi_z)dx\wedge dy\wedge dz\wedge d\xi.
\end{split}
\]

Integrating the 4-form
\[
\frac{2\pi\varepsilon\mathcal{L}}{c}\mathbf{d}\tilde{A}\wedge\tilde{A}\wedge\zeta
\]
on the 4-volume $\mathbb{R}^3\times \mathcal{L}$ we obtain $\kappa ET$.
\vskip 0.5cm

We recall also that for finite solutions the electromagnetic volume form
$\omega_\chi=-\frac1c
\tilde{A}\wedge\tilde{\varepsilon A^*}\wedge\mathbf{R}\wedge\mathbf{S}$
gives the same quantity $ET$ when integrated over the 4-volume
$\mathbb{R}^3\times \mathcal{L},\  \mathcal{L}=const$.
\vskip 0.5cm

\subsection{The Godbillon-Vey 3-form as a Conservative Quantity}
According to the Frobenius integrability theorems having a completely
integrable $(n-1)$-dimensional differential system on a $n$-manifold $M$ is
equivalent to having a suitable completely integrable 1-dimensional Pfaff
system on the same manifold. This Pfaff system is determined by a suitable
1-form $\omega$, defined up to a nonvanishing function:
$f\omega, f(x)\neq 0, x\in M$, and $\omega$ satisfies
the equation $\mathbf{d}\omega\wedge\omega=0$ (so obviously, $f\omega$ also
satisfies $\mathbf{d}(f\omega)\wedge f\omega=0$). From this last equation it
follows that there is 1-form $\theta$ such, that
$\mathbf{d}\omega=\theta\wedge\omega$.  Now, the Godbillon-Vey theorem [31]
says that the 3-form $\beta=\mathbf{d}\theta\wedge\theta$ is closed:
$\mathbf{d}\beta=\mathbf{d}(\mathbf{d}\theta\wedge\theta)=0$, and, varying
$\theta$ and $\omega$ in an admissible way: $\theta\rightarrow
(\theta+g\omega); \ \omega\rightarrow f\omega$, leads to adding an
exact 3-form to $\beta$, so we have a cohomological class $\Gamma$ defined
entirely by the integrable 1-dimensional Pfaff system. From physical point
of view the conclusion is that each completely integrable 1-dimensional Pfaff
system on Minkowski space generates a conservation law through the
restriction of $\beta$ on $\mathbb{R}^3$.

Recall now the following objects on our Minkowski space-time: $A$, $A^*$ and
$\zeta$. These are 1-forms. We form the corresponding vector fields through
the Lorentz-metric and denote them by $\vec{A}, \vec{A^*}, \vec{\zeta}$.
Let's consider the 1-form $\omega=f\zeta=\varepsilon fdz+fd\xi$, where $f$ is
a nonvanishing function on $M$. We have the relations:
\[
\omega(\vec{A})=0,\ \ \ \omega(\vec{A^*})=0,\ \ \ \omega(\vec{\zeta})=0.
\]
Moreover, since $\zeta$ is closed, $\omega=f\zeta$ satisfies the Frobenius
integrability condition:
\[
\mathbf{d}\omega\wedge\omega=f\mathbf{d}f\wedge\zeta\wedge\zeta=0.
\]
Therefore, the corresponding 1-dimensional Pfaff system, defined by
$\omega=f\zeta$, is completely integrable, and there exists a new 1-form
$\theta$, such that $\mathbf{d}\omega=\theta\wedge\omega$, and
$\mathbf{d}(\mathbf{d}\theta\wedge\theta)=0$.

From the point of view of generating a conservative quantity through
integrating the restriction $i^*\beta$ of
$\beta=\mathbf{d}\theta\wedge\theta$ to $\mathbb{R}^3$, through the
imbedding $i: (x,y,z)\rightarrow(x,y,z,0)$ it is not so important whether
$\Gamma$ is trivial or nontrivial.  The important point is the 3-form $\beta$
to have appropriate component $\beta_{123}$ in front of the basis element
$dx\wedge dy\wedge dz$, because only this component survives after the
restriction considered is performed, which formally means that we put
$d\xi=0$ in $\beta$.  The value of the corresponding conservative quantity
will be found provided the integration can be carried out successfully, i.e.
when $(i^*\beta)_{123}$ has no singularities and $(i^*\beta)_{123}$ is
concentrated in a finite 3d subregion of $\mathbb{R}^3$.

In order to find appropriate $\theta$ in our case we are going to take
advantage of the freedom we have when choosing $\theta$: the 1-form $\theta$
is defined up to adding to it an 1-form $\gamma=g\,\omega$, where $g$
is an arbitrary function on $M$, because $\theta$ is defined by the relation
$\mathbf{d}\omega=\theta\wedge\omega$, and
$(\theta+g\,\omega)\wedge\omega=\theta\wedge\omega$
always. The freedom in choosing $\omega$ consists in choosing the
function $f$, and we shall show that $f$ may be chosen in such a way:
$\omega=f\,\zeta$, that the corresponding integral of $i^*\beta$ to present a
finite conservative quantity.

Recalling that $\mathbf{d}\zeta=0$, we have
\[
\mathbf{d}\omega=\mathbf{d}(f\zeta)=\mathbf{d}f\wedge\zeta+f\mathbf{d}\zeta=
\mathbf{d}f\wedge\zeta.
\]
Since $\mathbf{d}\omega$ must be equal to $\theta\wedge\omega$ we obtain
\[
\mathbf{d}\omega=\mathbf{d}f\wedge\zeta=\theta\wedge\omega=
\theta\wedge(f\zeta)=f\theta\wedge\zeta.
\]
It follows
\[
\theta\wedge\zeta=
\frac{1}{f}\mathbf{d}f\wedge\zeta=\mathbf{d}(ln\,f)\wedge\zeta=
\Big[\mathbf{d}(ln\,f)+h\zeta\Big]\wedge\zeta,
\]
where $h$ is an arbitrary function. Hence, in general, we obtain
$\theta=\mathbf{d}(ln\,f)+h\zeta$. Therefore, since
$\mathbf{d}\theta=\mathbf{d}h\wedge\zeta$ for $\mathbf{d}\theta\wedge\theta$
we obtain
\[
\mathbf{d}\theta\wedge\theta=\mathbf{d}(ln\,f)\wedge\mathbf{d}h\wedge\zeta.
\]
Denoting for convenience $(ln\,f)=\varphi$ for the restriction $i^*\beta$ we
obtain
\[
i^*\beta=\varepsilon(\varphi_x h_y-\varphi_y h_x)dx\wedge dy\wedge dz.
\]

In order to find appropriate interpretation of $i^*\beta$ we recall that
$*(\delta F\wedge F)=-\varepsilon\phi^2(\psi_\xi-\varepsilon \psi_z)\zeta$,
so, $*(\delta F\wedge F)$ is of the kind $f\,\zeta$, and it
defines the same 1-dimensional Pfaff system as $f\,\zeta$ does. We recall also
that if the scale factor $\mathcal{L}=1/|\psi_\xi-\varepsilon \psi_z|$ is a
nonzero constant then $\phi^2(\psi_\xi-\varepsilon \psi_z)$ is a running
wave and the 3-form $\delta F\wedge F$ is closed. Hence, the
interpretation of $i^*\beta$ as
$i^*(\delta F\wedge F)=
-\frac{1}{\mathcal{L}}\varepsilon\kappa\phi^2 dx\wedge dy\wedge dz$
requires appropriate definition of the two functions $f$ and $h$. So we must
have
\[
\varphi_x h_y -h_x\varphi_y=-\frac{\kappa}{\mathcal{L}}\phi^2.
\]
If we choose
\[
f=exp(\varphi)=exp\left[\int{\phi^2}dx\right],\ \
h=-\frac{\kappa}{\mathcal{L}}y+const
\]
all requirements will be fulfilled, in particular,
$\varphi_{xy}=\varphi_{yx}=(\phi^2)_y$ and $h_{xy}=h_{yx}=0$.

Hence, the above choice of $f$ and $h$ allows
 the spatial restriction of the Godbillon-Vey 3-form $\beta$ to be
interpreted as the spatial restriction of $F\wedge \delta F$. So,
the curvature expressions found in the previous sections, as well as the
corresponding {\it spin}-properties of the nonlinear solutions being
available when $\delta F\wedge F\neq 0$, are being connected with the
integrability of the Pfaff system $\omega=f\,\zeta$.

\vskip 0.5cm
On the two figures below are given two theoretical examples with $\kappa=-1$
and $\kappa=1$ respectively, amplitude function $\phi$ located inside a
one-step helical cylinder with height of $2\pi \mathcal{L}$, and phase
function $\varphi=\mathrm{cos}(\kappa z/\mathcal{L})$. The solutions
propagate left-to-right along the coordinate $z$.
\begin{center}
\begin{figure}[ht!]
\centerline{
{\mbox{\psfig{figure=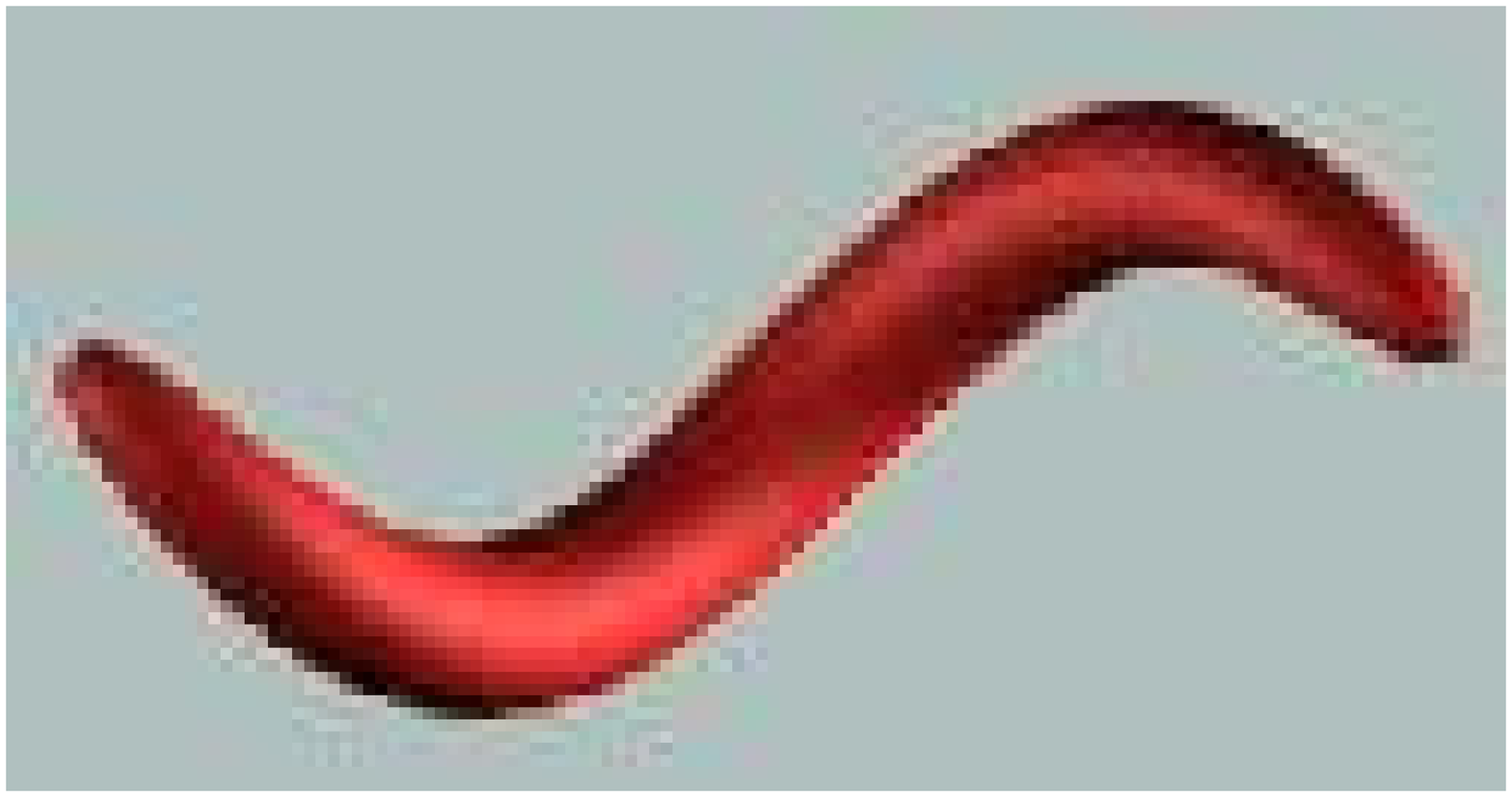,height=1.8cm,width=3.5cm}}
\mbox{\psfig{figure=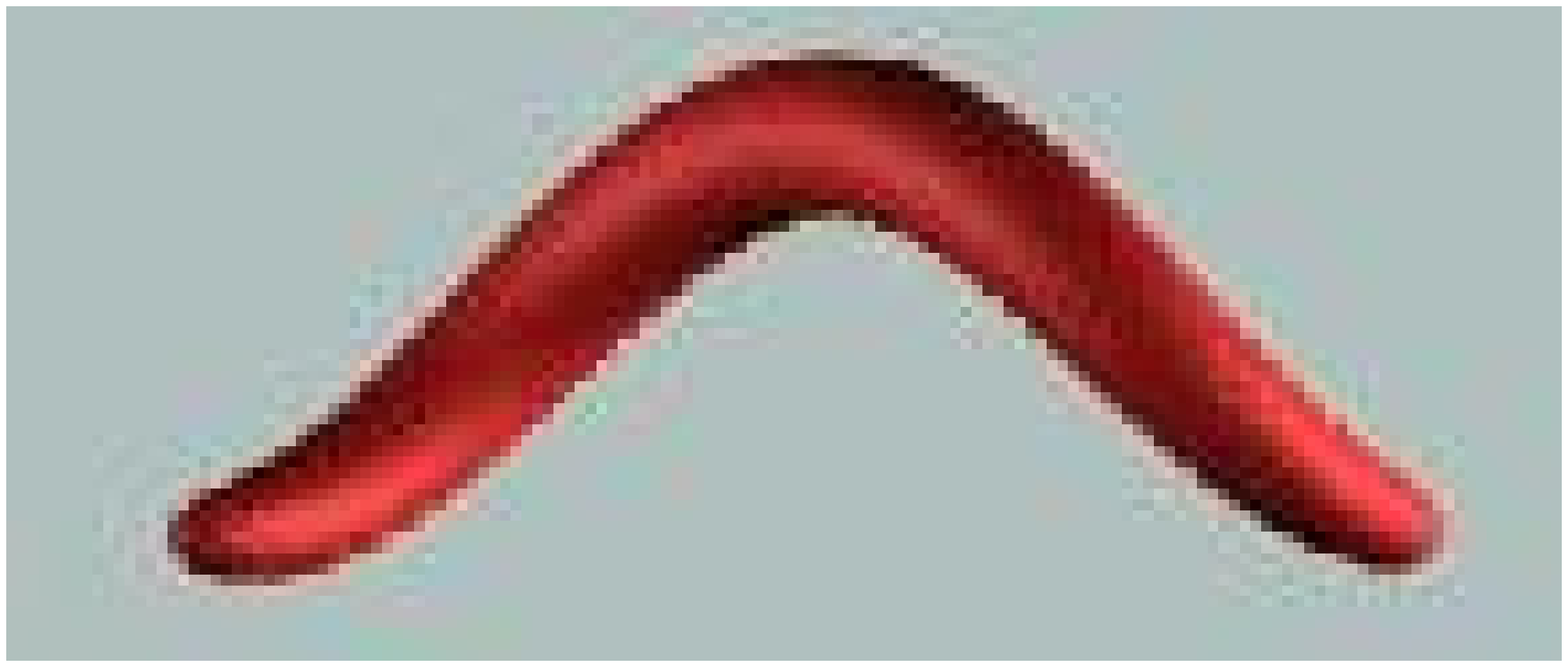,height=1.8cm,width=4.2cm}}
\mbox{\psfig{figure=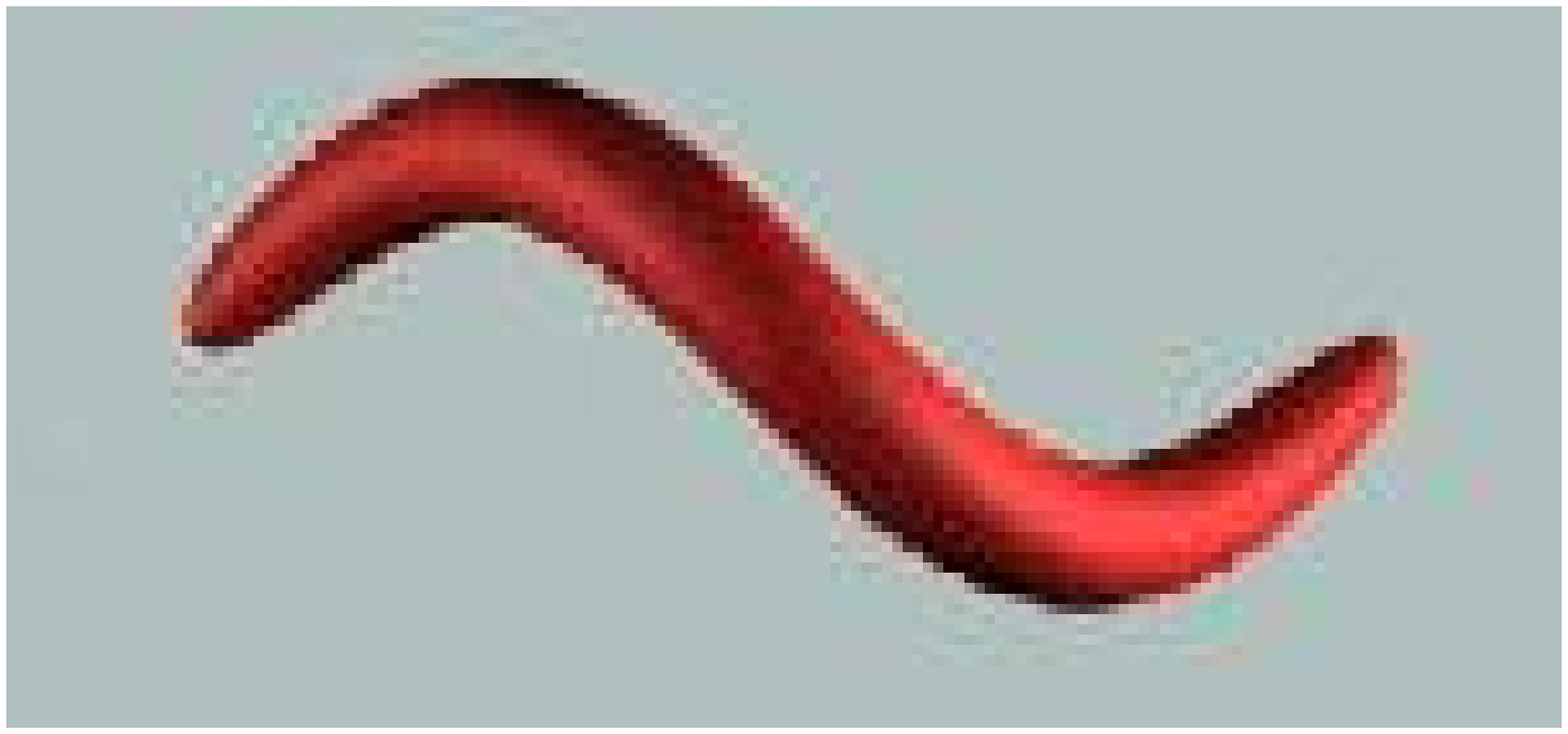,height=1.8cm,width=4.2cm}}}}
\caption{Theoretical example with $\kappa=-1$. The Poynting vector is
directed left-to-right.}
\end{figure}
\end{center}
\begin{center}
\begin{figure}[ht!]
\centerline{
{\mbox{\psfig{figure=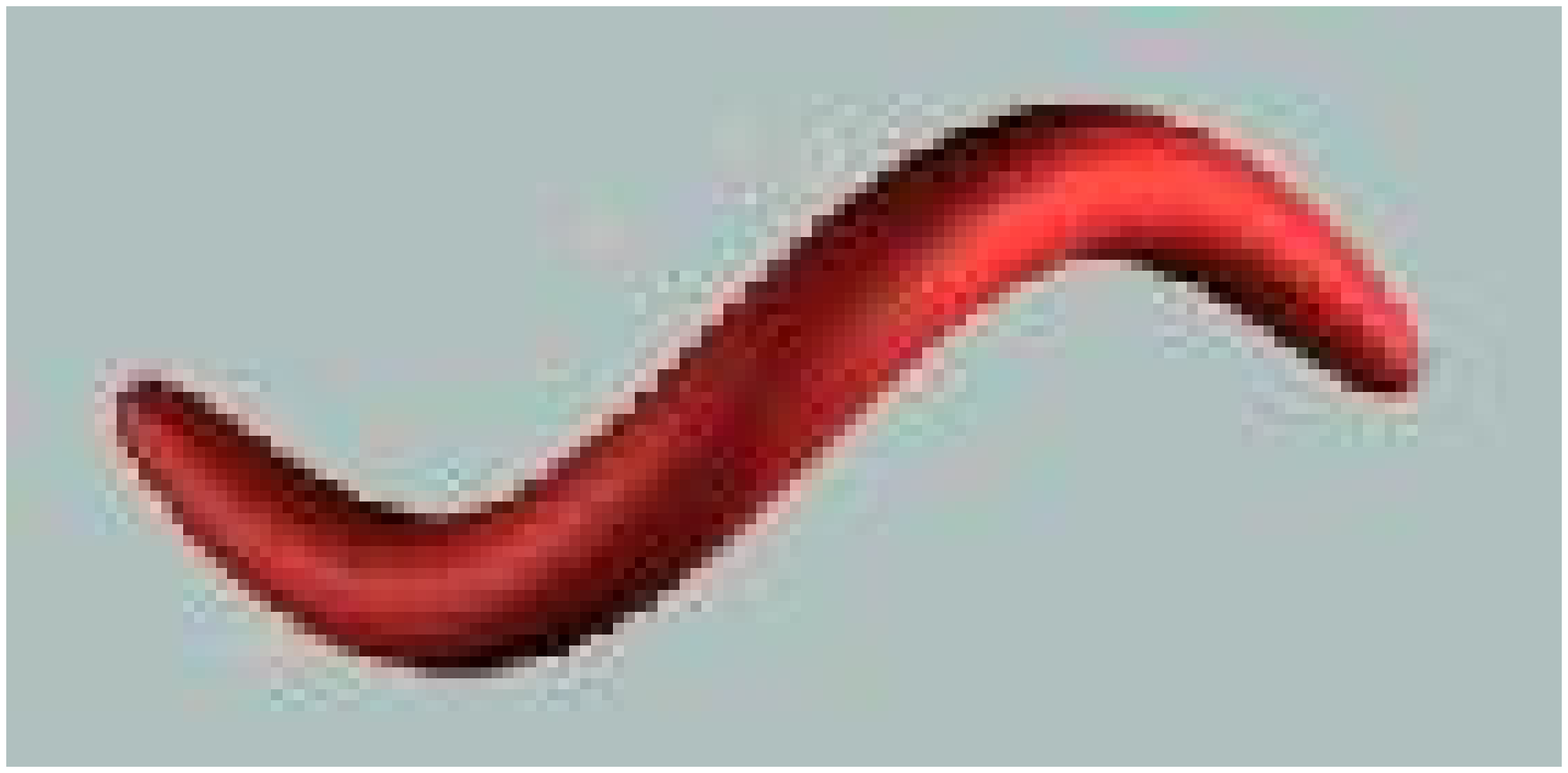,height=1.8cm,width=3.5cm}}
\mbox{\psfig{figure=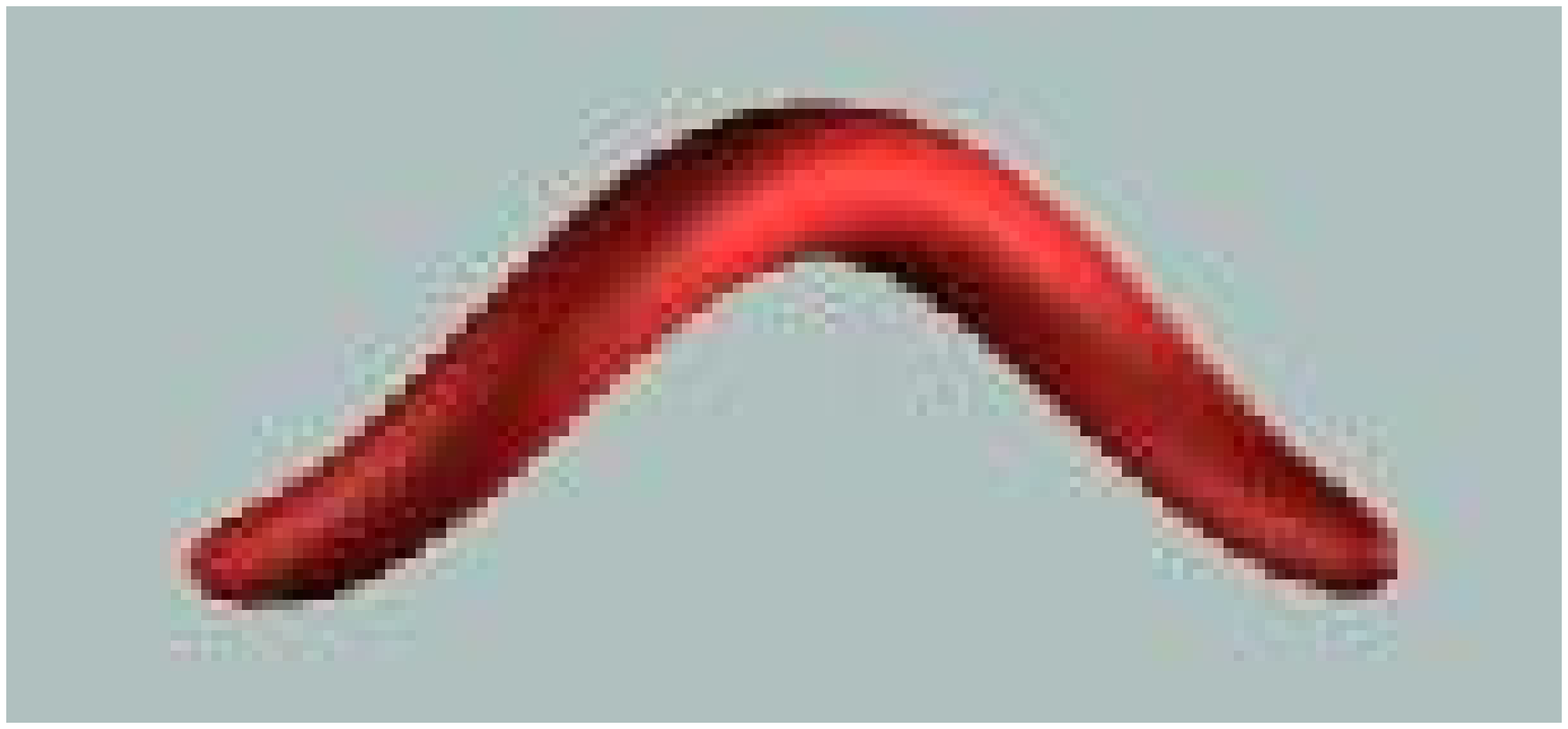,height=1.8cm,width=4.2cm}}
\mbox{\psfig{figure=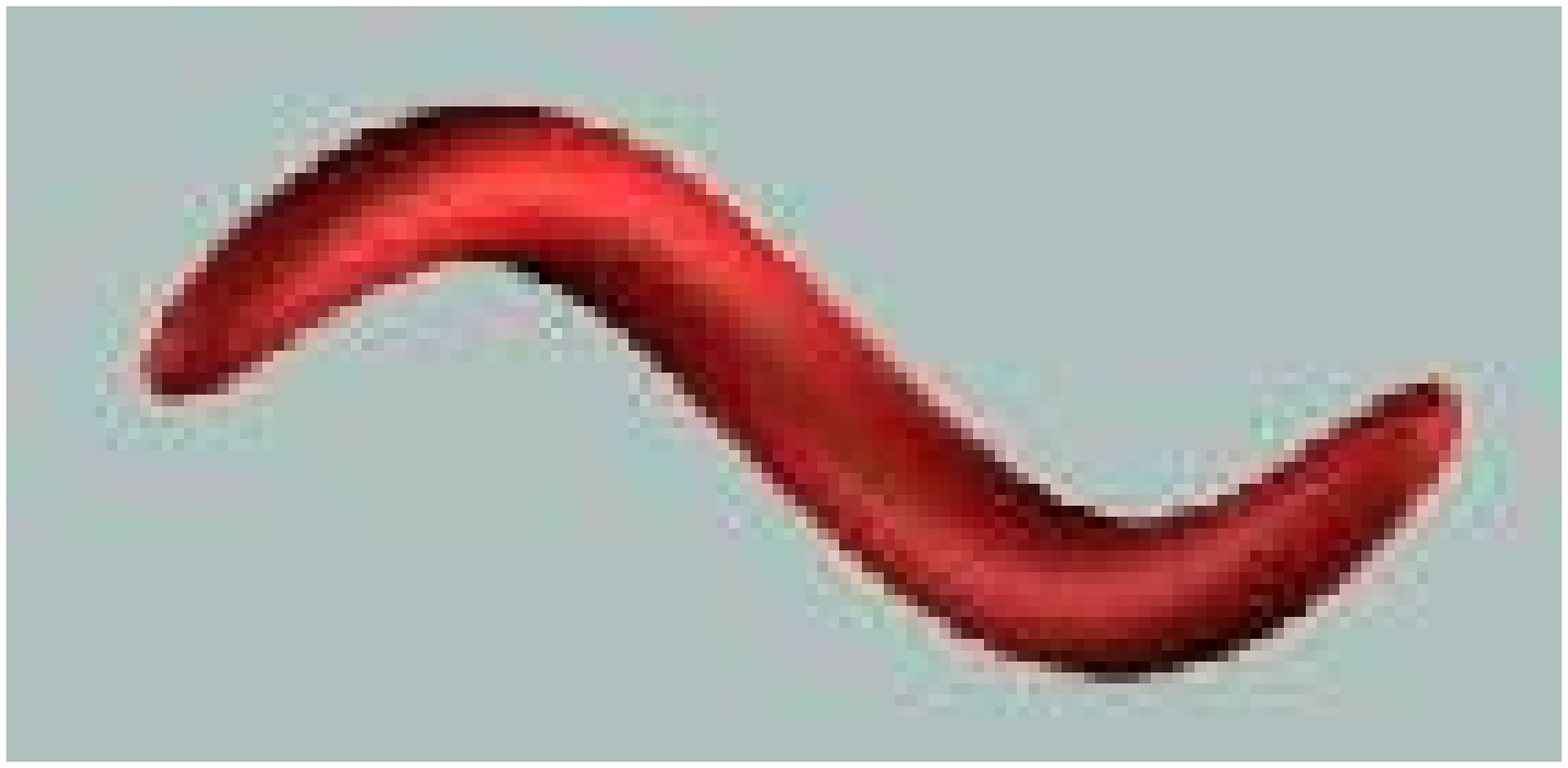,height=1.8cm,width=4.2cm}}}}
\caption{Theoretical example with $\kappa=1$. The Poynting vector is
directed left-to-right.}
\end{figure}
\end{center}

In the case $\kappa=-1$ the magnetic vector is always directed to the
rotation axis, i.e. it is normal to the rotation, and the electric vector
is always tangent to the rotation, so, looking from behind (i.e. along the
Poynting vector) we find clock-wise rotation. In the case $\kappa=1$ the
two vectors exchange their roles and, looking from behind again, we find
anti-clock-wise rotation. From structural point of view the case $\kappa=-1$
is obtained from the case $\kappa=1$ through rotating the couple
$(\mathbf{E},\mathbf{B})$ anti-clock-wise to the angle of $\pi/2$, hence we
get the (dual) transformation
$(\mathbf{E},\mathbf{B})\rightarrow(\mathbf{-B},\mathbf{E})$.
The dynamical roles of the two vectors are exchanged: now the electric vector
drags the points of the object towards rotation axis, the magnetic vector
generates rotation. In both cases the Poynting vector "pushes" the object
along the rotation axis.

The above pictures suggest the interpretation  that the rotation axis
directed vectors keep the object from falling apart, and such kind of
dynamical structure seems to be the only possible time-stable one.

If we project the object on the plane orthogonal to the Poynting vector we
shall obtain a sector between two circles with the same center, and this
sector has nontrivial topology. One of the two vectors is always directed to
the center of the circles and this stabalizes the solution, and the
other is tangent to the circles and correspondingly oriented.  The stability
of the construction is in accordance also with the fact that, when $z$ runs
from zero to $2\pi\mathcal{L}$ any of the two vectors performs {\it just one}
full rotation.  Any other rotaional evolution would make the center-directed
vector leave its directional behavior and this would bring to structural
changes and, most probably, to falling apart of the structure.  Choosing
orientation and computing the corresponding {\it rotation numbers} of
$\mathbf{E}$ and $\mathbf{B}$ we shall obtain, depending on the orientation
chosen, $(+1)$, or $(-1)$, and in definite sense these values guarantee from
mathematical viewpoint the dynamical stability of the solution-object.


\section{\bf Explicit Non-vacuum Solutions}
Before to start searching solutions with nonzero $\alpha^i, i=1,2,3,4$, we
must keep in mind that the field $F$ now will certainly be of {\it quite
different nature compare to the vacuum case}, and its interpretation as
electromagnetic field is {\it much conditional}: no more propagation with the
Lorentz invariant speed, no more photon-likeness, etc. The solution is meant
to represent a field interacting {\it continuously} with other physical
system, so the situation is quite different and, correspondingly, the
properties of the solution may differ drastically from the vacuum solutions
properties.  For example, contrary to the nonlinear vacuum case where the
solutions propagate translationally with the velocity of light, here it is
not excluded to find solutions which do not propagate with respect to an
appropriate Lorentz frame.  In {\it integrability-nonintegrability} terms
this would mean that at least some of the proper integrability properties of
the vacuum solutions have been lost. On the other hand, the proper
integrability properties of the external field are, at least partially,
guaranteed to hold through the requirement for the Frobenius integrability of
the 2-dimensional Pfaff systems $(\alpha^i,\alpha^j)$.

As a first step we expect possible physical interpretation just in the frame
of macro phenomena, where the continuity of the energy-momentum exchange is
easily understood, but we do not exclude interpretations in the frame of
microworld in view of the large quantity of variables we have (in fact, 22
functions), and in view of the universality of the physical process
described, namely, energy-momentum exchange.

From purely formal point of view finding a solution, whatever it is,
legitimizes the equations considered as a consistent system. Our purpose in
looking for solutions in the nonvacuum case,
however, is not purely formal, we'd like to consider the corresponding
solutions as physically meaningful, in other words, we are interested in
solutions, which can be, more or less, {\it physically interpretable}, i.e.
presenting more or less approximate models of real objects and processes.
That's why we'll try to meet the following.  First, the solutions must be
somehow {\it physically clear}, which means that the anzatz assumed should be
comparatively simple and its choice should be made on the base of a
preliminary analysis of the physical situation in view of the mathematical
model used.  Second, {\it it is absolutely obligatory the solutions to have
well defined local and integral energy and momentum}. Third, {\it the
solutions are desirable to be in the spirit of the soliton-like consideration
of the real natural objects} when a corresponding physical interpretation is
meant to be done.  Finally, it would be nice, the solutions found to be
comparatively simple and to have something to do with "popular" and well
known solutions of "well liked" equations.

\subsection{Choice of the anzatz and finding the solutions}

Let's now get started. We shall be interested in time-dependent solutions,
therefore, the "electric" and the "magnetic" components should present. The
simplest $(F,*F)$, meeting this requirement, look as follows (we use the
above assumed notations):
\begin{equation}
F=-udy\wedge dz -vdy\wedge d\xi,\ *F=vdx\wedge dz + udx\wedge d\xi,   
\end{equation}
where $u$ and $v$ are two functions on $M$.
Second, when choosing the 1-forms $\alpha^i$ we shall follow the requirement,
that the "medium" {\it does not influence} the $F\leftrightarrow *F$
exchange, so we put
\begin{equation}                                           
\alpha^2=\alpha^3=0.
\end{equation}
Third, since the rest two 1-forms , $\alpha^1$ and $\alpha^4$ must be {\it
linearly independent} (they must define a 2-dimensional Pfaff system),
we choose them to be mutually orthogonal and let $\alpha^4$
be {\it time-like}, and the other to be {\it space-like}. The simplest
time-like 1-form looks like $\alpha=A(x,y,z,\xi)d\xi$, moreover,
any such 1-form defines {\it integrable} 1-dimensional Pfaff system:
\[
{\bf d}\alpha\wedge \alpha=\left(A_x dx\wedge d\xi +A_y dy\wedge d\xi +
A_zdz\wedge d\xi\right)\wedge Ad\xi=0.
\]
So, we choose $\alpha^4=A(x,y,z,\xi)d\xi$. Finally, for $\alpha^1$,
denoted further by $\beta$, we obtain in general,
$\beta=\beta_1dx+\beta_2dy+\beta_3dz$. Clearly, $\beta^2<0 $, and
$\alpha.\beta=0$, so they are linearly independent. Now, $\beta$ participates
in the equations through the expression $\beta\wedge *F$, and from the
explicit form of $*F$ is seen, that the coefficient $\beta_1$ in front of
$dx$ does not take part in $\beta\wedge *F$, therefore, we put $\beta_1=0$.
So, putting $\beta_2=b$ and $\beta_3=-B$ for convenience, we get
\begin{equation}
\alpha^1\equiv \beta=bdy-Bdz,\quad \alpha^4\equiv \alpha=Ad\xi,\quad
\alpha^2=\alpha^3=0.
\end{equation}
We note, that the so chosen $\beta$ does {\it not} define in general an
integrable 1-dimensional Pfaff system, so, accordingly:
 ${\bf d}\alpha\wedge \alpha=0,\
{\bf d}\beta\wedge \beta\neq 0$.

At these conditions our equations
\[
\delta *F\wedge F=\alpha\wedge F,\quad
\delta F\wedge* F=\beta\wedge * F,\quad
\delta *F\wedge *F-\delta F\wedge F=0,
\]
\[
{\bf d}\alpha\wedge \alpha\wedge \beta=0,\quad
{\bf d}\beta\wedge \alpha\wedge \beta=0
\]
take the form: $\delta *F\wedge *F-\delta F\wedge F=0$ is reduced to
\[
-vu_y+uv_y=0,\quad  -uv_x +vu_x=0,
\]
the Frobenius equations  ${\bf d}\alpha\wedge \alpha\wedge \beta=0,\
{\bf d}\beta\wedge \alpha\wedge \beta=0$ reduce to
\[
\left(-b_xB+B_xb\right).A=0,
\]
$\delta *F\wedge F=\alpha\wedge F$ is reduced to
\[
u\left(u_\xi-v_z\right)=0,\quad
v\left(u_\xi-v_z\right)=0,\quad
uu_x-vv_x=Au,
\]
finally, $\delta F\wedge* F=\beta\wedge *F$ reduces to
\[
v\left(v_\xi-u_z\right)=-bv,\quad
u\left(v_\xi-u_z\right)=-bu,\quad
uu_y-vv_y=Bu.
\]
In this way we obtain 7 equations for 5 unknown functions $u,v,A,B,b$.

The equations $-vu_y+uv_y=0,\  -uv_x +vu_x=0$ have the following solution:
\[
u(x,y,z,\xi)=f(x,y)U(z,\xi),\quad v(x,y,z,\xi)=f(x,y)V(z,\xi).
\]
That's why
\[
AU=f_x\left(U^2-V^2\right),\quad BU=f_y\left(U^2-V^2\right),\quad
f\,\left(V_\xi-U_z\right)=-b,\quad U_\xi-V_z=0.
\]
It follows that $b$ should be of the kind $b(x,y,z,\xi)=f(x,y)b^o(z,\xi)$,
so the equation $B_xb-Bb_x=0$ takes the form
\[
ff_{xy}=f_xf_y.
\]
The general solution of this last equation is $f(x,y)=g(x)h(y)$.
The equation $gh\left(V_\xi-U_z\right)=-b$ reduces to
\[
V_\xi-U_z=-b^o.
\]

The relations obtained show how to build a solution of this class. Namely,
first, we choose the function $V(z,\xi)$, then we determine the function
$U(z,\xi)$ by
\[
U(z,\xi)=\int{V_z d\xi} +l(z),
\]
where $l(z)$ is an arbitrary function, which may be assumed equal to $0$.
After that we define $b^{o}=U_z-V_\xi$. The functions $g(x)$ and $h(y)$ are
arbitrary, and for $A$ and $B$ we find
\[
A(x,y,z,\xi)=g'(x)h(y)\frac{U^2-V^2}{U},\quad
B(x,y,z,\xi)=g(x)h'(y)\frac{U^2-V^2}{U}.
\]
In this way we obtain a family of solutions, which is parametrized by one
function $V$ of the two variables $(z,\xi)$ and two functions $g(x),\ h(y)$,
each depending on one variable.

In order to find corresponding conserved quantities we sum up the nonzero
right-hand sides of the equations and obtain
$*(\alpha\wedge F+\beta\wedge *F)$. We represent this expression in a
divergence form as follows:
\[
*(\alpha\wedge F+\beta\wedge *F)=Audx-Budy-budz-bvd\xi=
\]
\[
=\frac12 (U^2-V^2)\left[(gh)^2\right]_x dx-
\frac12 (U^2-V^2)\left[(gh)^2\right]_y dy-
\]
\[
-(gh)^2\left(\int{Ub^{o}dz}\right)_{z}dz-(gh)^2\left(\int{Vb^{o}d\xi}\right)_{\xi}d\xi=
=-\left\{\frac{\partial}{\partial x^\nu}P_\mu^\nu\right\}dx^\nu,
\]
where the interaction energy-momentum tensor is defined by the matrix
\[
P_\mu^\nu=\begin{Vmatrix}
-\frac12(gh)^2 Z  &0                &0                     &0                      \cr
0                 &\frac12(gh)^2 Z  &0                     &0                      \cr
0                 &0                &(gh)^2\int{Ub^{o}dz}  &0                      \cr
0                 &0                &0                     &(gh)^2\int{Vb^{o}d\xi}
\end{Vmatrix},
\]
and the notation $Z\equiv U^2-V^2$ is used. For the components of the
full energy tensor $T_\mu^\nu=Q_\mu^\nu +P_\mu^\nu$ we obtain
\[
T_3^3=(gh)^2\left[\int{Ub^{o}dz} - \frac12(U^2+V^2)\right],
\]
\[
T_3^4=-T_4^3=(gh)^2 UV,
\]
\[
T_4^4=(gh)^2\left[\int{Vb^{o}d\xi}+\frac12(U^2+V^2)\right],
\]
and all other components are zero.

\vskip 0.5cm
\subsection{Examples}

In this subsection we consider some of the well known and well studied
(1+1)-dimensional soliton equations as generating procedures for choosing the
function $V(z,\xi)$, and only the 1-soliton solutions will be explicitly
elaborated. Of course, there is no anything standing in our way to consider
other (e.g. multisoliton) solutions. We do not give the corresponding
formulas just for the sake of simplicity.

We turn to the soliton equations mainly because of two reasons. First, most
of the solutions have a clear physical sense in a definite part of physics
and, according to our opinion, they are sufficiently attractive for models of
real physical objects with internal structure. Second, the soliton solutions
describe free and interacting objects with {\it no dissipation} of energy and
momentum, which corresponds to our understanding of the Frobenius
integrability equations as criteria for availability of integral properties of
the medium considered.

\vskip 0.5cm
\noindent{\it 1. Nonlinear (1+1) Klein-Gordon Equation}.
In this example we define our functions $U$ and $V$ through the derivatives
of the function $k(z,\xi)$ in the following way: $U=k_z,\ V=k_\xi$. Then the
equation $U_\xi-V_z=k_{z\xi}-k_{\xi z}=0$ is satisfied automatically, and
the equation $U_z-V_\xi=b^{o}$ takes the form $k_{zz}-k_{\xi\xi}=b^{o}$. Since
$b^o$ is unknown, we may assume $b^o=b^o(k)$, which reduces the whole
problem to solving the general nonlinear Klein-Gordon equation when $b^o$
depends nonlinearly on $k$. Since in this case $V=k_\xi$ we have
\[
\int{Vb^o(k)d\xi}=\int{k_\xi b^o(k)d\xi}=
\int{\left[\frac{\partial}{\partial \xi}\int{b^o(k)dk}\right]d\xi}=
\int{b^o(k)dk}.
\]
For the full energy density we get
\[
T_4^4=\frac12(gh)^2\left\{k_z^2+k_\xi^2+2\int{b^o(k)dk}\right\}.
\]
Choosing $b^o(k)=m^2 sin(k)$, $m=const$, we get the well known and widely
used in physics Sine-Gordon equation, and accordingly, we can use {\it
all} solutions of this (1+1)-dimensional nonlinear equation to generate
(3+1)-dimensional solutions of our equations following the above described
procedure.  When we consider the (3+1) extension of the soliton solutions of
this equation, the functions $g(x)$ and $h(y)$ should be localized too. The
determination of the all 5 functions in our approach is straightforward, so
we obtain a (3+1)-dimensional version of the soliton solution chosen. As it
is seen from the above given formulas, the integral energy of the solution
differs from the energy of the corresponding (1+1)-dimensional solution just
by the $(x,y)$-localizing factor $[g(x)h(y)]^2$.

For the 1-soliton solution (kink) we have:
\[
k(z,\xi)=4arctg\left\{exp\left[\pm \frac{m}{\gamma}(z-\frac wc
\xi)\right]\right\},\quad \gamma=\sqrt{1-\frac{w^2}{c^2}}
\]
\[
U(z,\xi)=k_z=\frac{\pm 2m}{\gamma
ch\left[\pm\frac{m}{\gamma}\left(z-\frac{w}{c} \xi \right)\right]},\quad
V(z,\xi)=k_\xi=\frac{\pm 2mw}{c\gamma
ch\left[\pm\frac{m}{\gamma}\left(z-\frac{w}{c} \xi \right)\right]},
\]
\[
A=g'(x)h(y)\frac{\pm 2m\gamma}{ch\left
[\pm\frac{m}{\gamma}\left(z-\frac{w}{c} \xi \right)\right]},\quad
B=g(x)h'(y)\frac{\pm 2m\gamma}{ch\left[\pm\frac{m}{\gamma}\left(z-\frac{w}{c} \xi \right)\right]},\
\]
\[
b^o=U_z-V_\xi=
\frac{-2m^2 sh\left[\pm\frac{m}{\gamma}
\left(z-\frac{w}{c} \xi \right)\right]}
{ch\left[\pm\frac{m}{\gamma}\left(z-\frac{w}{c} \xi \right)\right]},\quad
T_4^4=\frac{(gh)^2 4m^2}{\gamma^2 ch^2\left[\pm\frac{m}{\gamma}
\left(z-\frac{w}{c} \xi \right)\right]}
\]
and for the 2-form $F$ we get
\[
F=-\frac{\pm2mg(x)h(y)}{\gamma ch
\left[\pm\frac{m}{\gamma}(z-\frac wc \xi)\right]}dy\wedge dz+
\frac wc \frac{\pm2mg(x)h(y)}{\gamma ch\left[\pm\frac{m}{\gamma}
(z-\frac wc \xi)\right]}dy\wedge d\xi.
\]
In its own frame of reference this soliton looks like
\[
F=-\frac{\pm2mg(x)h(y)}{ ch(\pm mz)}dy\wedge dz.
\]
From this last expression and from symmetry considerations, i.e. at
homogeneous and isotropic medium, we come to the most natural (but not
necessary) choice of the functions $g(x)$ and $h(y)$:

$$
g(x)=\frac{1}{ch(mx)},\quad h(y)=\frac{1}{ch(my)}.
$$

\vskip 0.5cm
2.{\it Korteweg-de Vries equation.} This nonlinear equation has the following
general form:
\[
f_\xi+a_1ff_z+a_2f_{zzz}=0,
\]
where $a_1$ and $a_2$ are 2 constants. The well known 1-soliton solution is
\[
f(z,\xi)=\frac{a_o}{ch^2\left[\frac zL -\frac{w}{cL}\xi\right]},\quad
L=2\sqrt{\frac{3a_2}{a_o a_1}},\quad w=\frac{ca_oa_1}{3},
\]
where $a_o$ is a constant. We choose $V(z,\xi)=f(z,\xi)$ and get
\[
U=-\frac{a_oc}{w}\frac{1}{ch^2\left[\frac zL
-\frac{w}{cL}\xi\right]},\quad
b^o=U_z-V_\xi=\left(\frac{c}{Lw}-
\frac wc\right)\frac{2a_o}{ch^3\left[\frac zL -\frac{w}{cL}\xi\right]},
\]
\[
T_4^4=(gh)^2\frac{a_o^2c^2(1+L)}{2w^2Lch^4\left[\frac zL -
\frac{w}{cL}\xi\right]}.
\]
\vskip 0.5cm

3. {\it Nonlinear Schr\"odinger equation}. In this case we have an equation
for a complex-valued function, i.e. for two real valued functions. The
equation reads
\[
if_\xi+f_{zz}+|f|^2 f=0,
\]
and its 1-soliton solution, having oscillatory character, is
\[
f(z,\xi)=2\beta^2\frac{exp\left[-i\left(2\alpha z+
4(\alpha^2-\beta^2)\xi -\theta\right)\right]}
{ch\left(2\beta z+8\alpha\beta \xi -\delta\right)},
\]
where $\alpha, \beta, \delta$ and $\theta$ are constants.
The natural substitution $f(z,\xi)=\sqrt{\rho}.exp(i\varphi)$ brings this
equation to the following two equations
\[
\rho_\xi + (2\rho \varphi_z)_z=0,\quad
4\rho + \frac{2\rho\rho_{zz}-\rho_z^2}{\rho^2}=4(\varphi_\xi+\varphi_z^2).
\]
For the 1-soliton solution we get
\[
\rho=\frac{4\beta^4}{ch^2\left(2\beta z+
8\alpha\beta \xi -\delta\right)},\quad
\varphi=-\left[2\alpha z+4(\alpha^2-\beta^2)\xi-\theta\right].
\]
We put $U=\rho, \ V=-2\rho\varphi_z$ and obtain
\[
U=\rho=\frac{4\beta^4}{ch^2\left(2\beta z+8\alpha\beta \xi -
\delta\right)},\quad
V=-2\rho\varphi_z=
\frac{16\alpha\beta^4}{ch^2\left(2\beta z+8\alpha\beta \xi -\delta\right)},
\]
\[
A=g'(x)h(y)\frac{4\beta^4(1-16\alpha^2)}{ch^2\left(2\beta z+
8\alpha\beta \xi -\delta\right)},\quad
B=g(x)h'(y)\frac{4\beta^4(1-16\alpha^2)}{ch^2\left(2\beta z+
8\alpha\beta \xi -\delta\right)},
\]
\[
b^o=\frac{16\beta^3(16\alpha^2\beta^2-1)sh(2\beta z+
8\alpha\beta \xi -\delta)}
{ch^3(2\beta z+8\alpha\beta \xi -\delta)},
\]
\[
T_4^4=(gh)^2\frac{16\beta^8}{ch^4(2\beta z+8\alpha\beta \xi -\delta)}.
\]
We note that the solution of our equations obtained has no the oscillatory
character of the original Schr\"odinger 1-soliton solution, it is a
(3+1)-localized running wave and moves as a whole with the velocity
$4c\alpha$, so in the relativistic frame we must require $4c\alpha<c$.

\vskip 0.5cm
4.{\it Boomerons}. The system of differential equations, having soliton
solutions, known as {\it boomerons}, is defined by the following functions:
${\bf K}:{\cal R}^2\rightarrow {\cal R}^3$,
$H:{\cal R}^2\rightarrow {\cal R}$,
and besides, two {\it constant} 3-dimensional vectors
${\bf r}$ and ${\bf s}$, where ${\bf s}$ is a unit vector: $|{\bf s}|=1$.
The equations have the form
\[
H_\xi-{\bf s}.{\bf K}_z =0,\  {\bf K}_{z\xi}=H_{zz}{\bf s}+{\bf r}\times {\bf K}_z
-2{\bf K}_z\times({\bf K}\times{\bf s}).
\]
Now we have to define our functions $U(z,\xi),\ V(z,\xi)$ and $b^o(z,\xi)$.
The defining relations are:
\[
U=H_z=|{\bf s}|^2 H_z,\ V={\bf s}.{\bf K}_z=({\bf s.K})_z,\
b^o=-{\bf s}.\left[{\bf r}\times {\bf K}_z-2{\bf K}_z\times({\bf K}\times {\bf s})\right].
\]
Under these definitions our equations $U_\xi-V_z=0,\ V_\xi-U_z=-b^o$ look as
follows:
\[
\left[H_z-{\bf s}.{\bf K}_z\right]_z=0,\
{\bf s}.\left[{\bf K}_{z\xi}-H_{zz}{\bf s}-{\bf r}\times{\bf K}_z+
2{\bf K}_z\times({\bf K}\times {\bf s})\right]=0.
\]
It is clear that every solution of the "boomeron" system determines a
solution of our system of equations according to the above given rules and
with the multiplicative factors $g(x)$ and $h(y)$. The two our functions
$A(z,\xi)$ and $B(z,\xi)$ are easily then computed.

\vskip 0.5cm
Following this procedure we can generate a spatial soliton solution to our
system of equations by means of {\it every}  solution, in particular every
soliton solution, to {\it any} (1+1)-soliton equation, as well as to compute
the corresponding conserved quantities. It seems senseless to give here these
easily obtainable results.  The richness of this comparatively simple family
of solutions, as well as the availability of corresponding correctly defined
integral conserved quantities, are obvious and should not be neglected as a
generating procedure for obtaining (3+1) one-soliton, as well as, (3+1)
many-soliton solutions.
\vskip 1cm
\newpage

\section{Retrospect and Outlook}
In trying to understand our observational knowledge of the real world we must
be able to separate the {\it important} structural and behavioral
properties of the real objects from those, the changes of which
during time-evolution do not lead to annihilation of the objects
under consideration. One of the basic in our view lessons that we more or
less have been taught is that the physical objects are {\it spatially finite
entities}, and that for their detection and further study some
energy-momentum exchange is necessarily {\it required}. So, every physical
object necessarily carries energy-momentum and every interaction between two
physical objects has such an energy-momentum exchange aspect. The second
lesson concerning any interaction is that, beyond its {\it universality},
energy-momentum is a {\it conserved} quantity, so NO loss of it is allowed:
it may only pass from one object to another. This means that every {\it
annihilation} process causes {\it creation} process(es), and the full
energy-momentum that has been carried by the annihilated objects, is carried
away by the created ones. Energy-momentum always needs carriers, as well as
every physical object always carries energy-momentum.  Hence, the
energy-momentum exchange abilities of any physical object realize its
protection against external influence on one side, and reveal its intrinsic
nature, on the other side. Therefore, our knowledge about the entire
complex of properties of a physical object relies on getting information
about its abilities in this respect and finding corresponding quantities
describing quantitatively these abilities.

The spatially finite nature of a physical object implies spatial structure
and finite quantity of energy-momentum needed  for its creation, so NO
structureless and infinite objects may exist. The approximations for
"point object" and "infinite field", although useful in some respects, seem
theoretically inadequate and should not be considered as basic ones. More
reliable appears to be the approximation "finite continuous object", which we
tried to follow throughout our exposition. This last approximation suggests
that nonlinear partial differential equations should be the basic tool for
building mathematical models of local nature of such objects. The natural
physical sense of these equations is naturally supposed to be local
energy-momentum exchange.

Another useful observation is that physical objects are {\it many-aspect
entities}, they have complicated structure and their very existence is
connected with {\it internal} energy-momentum exchange/redistribution among
the various structural components. So, the mathematical model objects should
be many-component ones, and with appropriate mathematical structure.  Of
basic help in finding appropriate mathematical objects is having knowledge
of the internal symmetry properties of the physical object under
consideration.  This "step by step" process of getting and accumulating
important information about the physical properties of natural objects
reflects in the "step by step" process of refining the corresponding
mathematical models.

The greatest discovery at the very beginning of the last century was that the
notion of electromagnetic field as suggested by Maxwell equations is
inadequate: the time dependent electromagnetic field is not an infinite
smooth perturbation of the aether, on the contrary, it consists of many
individual time-stable objects, called later {\it photons}, which are
created/destroyed mainly during intra-atomic energy-transition processes.
Photons are finite objects, they carry energy-momentum and after they have
been radiated outside their atom-creator, they propagate as a whole
translationally by the speed of light.  Moreover, their propagation is not
just translational, it includes rotational component, which is of {\it
intrinsic} and {\it periodical} nature. The corresponding intrinsic action
for one period $T$ is $h=ET$, where $E$ is the full energy of the photon, and
all photons carry the same intrinsic action $h$. During the entire 20th
century physicists have tried to understand the dynamical structure/nature of
photons from various points of view, and this process is still going on
today.

Extended Electrodynamics, as presented in the preceding sections, is an
attempt in this direction. The basic starting observation for approaching
the problem is that the energy-momentum local quantities and relations
of Maxwell theory do agree with the experiment, but the free field equations
$\delta F=0,\ \ \delta *F=0$ give non-realistic free time-dependent
solutions: they are either {\it strongly time-unstable}, or {\it infinite}.
Hence, these solutions can not be used as mathematical models of photons,
since the latter are time-stable and finite objects. The basic idea of
writing down new nonlinear equations was to pass to local energy-momentum
relations, describing how the internal energy-momentum exchanges are carried
out. The mathematical structure of Maxwell equations and local conservation
relations allowed the extension procedure to be used, so the new
nonlinear equations contain all Maxwell solutions as exact solutions. This
feature is important from the point of view of applications.

In order to come to the new equations we made use of the dual symmetry of the
linear solutions, which suggested to consider the field as having two vector
components $(F,*F)$, and introduced the concept of $\mathbb{G}$-covariant
Lorentz force as a natural generalization of the known Lorentz force.
The $\mathbb{G}$-covariant Lorentz force has three vector components, and
each component has a clear physical sense of corresponding
local energy-momentum change, as it should be. The two components
$F_{\mu\nu}\delta F^\nu$ and $(*F)_{\mu\nu}(\delta *F)^\nu$ determine how
much energy-momentum the field is potentially able to give out to some other
physical object through any of the two components $F$ and $*F$, and the third
component $F_{\mu\nu}(\delta *F)^\nu+(*F)_{\mu\nu}\delta F^\nu$ characterizes
the internal energy-momentum exchange between the two components $F$ and
$*F$.  We concluded that in case of energy-momentum isolation, i.e. in
vacuum, these three components must be zero, so, the dynamics obtained
is of intrinsic nature.

In studying the vacuum nonlinear solutions, i.e. those with $\delta F\neq 0$
and $\delta *F\neq0$, we found their basic property: every nonlinear solution
has zero-invariants: $F_{\mu\nu}F^{\mu\nu}=(*F)_{\mu\nu}F^{\mu\nu}=0$. Then
the eigen properties of $F$, $*F$ and of the energy-momentum tensor
$Q_\mu^\nu$  determine unique isotropic direction $\zeta$ along which the
solution propagates translationally as a whole, which fits well with the
photons' way of propagation .  The obtained simple form of $F=A\wedge\zeta$
and $*F=A^*\wedge\zeta$ allowed a complete analysis of the nonlinear
solutions to be made. The whole set of nonlinear solutions consists of
nonoverlapping subsets, and each subset is characterized by the corresponding
isotropic eigen direction. Every solution of a given subclass is uniquely
determined by two functions: the amplitude function $\phi$ which is arbitrary
with respect to the spatial variables and is a running wave along $\zeta$;
the phase function $\varphi, |\varphi|\leq 1$, where $\varphi$ depends
arbitrarily on all space-time variables (we made use of the function
$\psi=arccos(\varphi)$). The components of $F$ have the form $\phi\,cos\psi$
and $\phi\,sin\psi$, so, finite solutions with photon-like behavior are
allowed.

A basic characteristic of the nonlinear solutions with $|\delta F|\neq 0$
turned out to be the {\it scale factor} $\mathcal{L}$ defined by the relation
$\mathcal{L}=|A|/|\delta F|$ and having physical dimension of [length].
Hence, every nonlinear solution with $|\delta F|\neq 0$ defines {\it its own
scale}. As it was shown further, the case $\mathcal{L}=const$ is allowed, so
corresponding intrinsically defined time-period $T=\mathcal{L}/c$ and
frequency $\nu=T^{-1}$ can be introduced, and the corresponding finite
solutions with integral energy $E$ admit the characteristic {\it intrinsic
action} given by $ET$.

The natural question "do there exist nonlinear finite solutions with
rotational component of propagation" was answered positively. Seven
equivalent conditions of quite different nature, determining when this is
possible, were found to exist. It is remarkable that the condition
$\mathcal{L}\neq 0$ is one of them. So, for finite solutions with
$\mathcal{L}=const$, we have a natural and intrinsically defined measure of
this rotational component of propagation, namely, the elementary action $ET$,
which recalls the corresponding invariant characteristic of photons $h$, the
Planck constant. Anyway, these solutions deserve to be called "photon-like".
It is also remarkable that completely integrable and nonintegrable
2-dimensional Pfaff systems can be associated with a nonlinear solution, and
a natural integral characteristic of the nonintegrable Pfaff systems is also
the quantity $ET$. This point shows that both, integrable and nonintegrable
features of a solution are important, the nonintegrability characteristics
give information about  the contacts between the system and the
physical environment.

It was very interesting to find that the energy-momentum tensor $Q$ of
a nonlinear solution defines a boundary operator in the tangent and cotangent
bundles of the Minkowski space-time. The corresponding homology/cohomology
spaces are 2-dimensional, the classes represented by
the electric and magnetic components of the field $F$ form a basis of the
corresponding homology/cohomology space. Moreover, since for each nonlinear
solution the corresponding image spaces of $Q$ are 1-dimensional, $Q$ is
extended to boundary operator in the whole algebra of exterior forms and
antisymmetric tensors.  The $Q$-image of any differential 2-form is collinear
to the nonlinear solution that generates $Q$, and the $Q$-image $Q(\alpha)$
of any $p$-form $\alpha$ is isotropic:  $[Q(\alpha)]^2=0$.

The natural representation of $\mathbb{G}$ in the space of 2-forms leaves
the scale factor $\mathcal{L}$ invariant. As a consequence we obtained
that $\mathcal{L}$ is Lorentz-invariant only if it is a constant, and in this
case the 3-form $\delta F\wedge F$ is closed, which generates a conserved
quantity through Stokes theorem, and this conserved quantity is proportional
to the elementary action $ET$. If the group parameters depend on the
space-time points, then the commutative group structure of $\mathbb{G}$
generates a group structure inside the subset of solutions with the same
$\zeta$. In such a case a "vacuum state" $F_o$ can be defined such, that every
solution of the subclass is defined by an action upon $F_o$ of a
point-dependent group element with determinant $(a^2+b^2)$ having running
wave character along $\zeta$.

The connection-curvature interpretations showed that rotational component of
propagation may have only those nonlinear solutions which generate non-zero
curvature in the three cases of general, principal amd linear connection, so,
roughly speaking, nonzero curvature means nonzero spin.

We gave five ways to compute the integral spin $ET$ of a nonlinear solution.
It worths noting that an appropriate representative of the
Godbillon-Vey class determined by the completely integrable
1-dimensional Pfaff system $\zeta$, or $f\zeta$, can be made equal to $\delta
F\wedge F$, and so it can be used to compute the elementary action $ET$.

Finally, in presence of external fields, we showed that our general system of
nonlinear equations together with the additional Pfaff equations is
consistent, and we found a large family of solutions. This family is
parametrized by one function of one-space (say $z$) and one-time independent
variables: $V(z,\xi)$, and two other functions $g(x)$ and $h(y)$. So,
choosing $V(z,\xi)$ to be any (one, or many) soliton solution of any soliton
equation, and $g(x)$and $h(y)$ to be finite, we obtain its (3+1) image as a
finite/concentrated solution of our nonlinear equations with well defined
energy-momentum quantities. We illustrated our approach with examples from
the 1-soliton solutions of the well known Sine-Gordon, KdV and NLS
equations.
\vskip 0.4cm
We would like to note also that the nonlinear vacuum equations obtained
follow the idea that was formulated in Sec.1.1: the admissible changes (in
our case $\delta *F$ and $\delta F$, or correspondingly, $\mathbf{d}F$ and
$\mathbf{d}*F$) are projected upon the field components $(F,*F)$, and these
projections are assumed to be zero, or intrinsically connected. This leads
directly to nonlinear equations with corresponding physical sense of
energy-momentum balance relations. From mathematical point of view this
suggests a natural generalization of the geometrical concept of parallelism,
and this idea is developed and illustrated with many examples from
differential geometry and theoretical physics in the Appendix. This
generalization, in turn, suggests natural ways to nonlinearization of
important physical linear and nonlinear equations.

\vskip 0.4cm
The development of theoretical physics during the last century, and especially
during the past 30 years, shows growing interest to nonlinearization of the
widely used linear equations. At the beginning of the century Einstein
declared [32] that the fundamental equation of Optics $\square \phi =0$ must
be replaced by nonlinear equation(s), and he had worked in trying to find
appropriate one(s). General Relativity made a decisive step along the road of
nonlinearization. Mie's [44] and Born-Infeld's [9] nonlinearizations of CED
are also steps in this direction. One of the greatest achievements was the
Yang-Mills approach, which dominates nowadays the various models of field
theory (the "standard model").  Unfortunately, spatially finite solutions of
the vacuum Yang- Mills equations with soliton behavior were not found.

The later studies with appropriate interaction term in the Yang-Mills
lagrangian brought the monopole solutions [33]. A somewhat curious example is
presented by the instanton solutions [34] because of their mathematical
significance on one side, and because of the interpretational difficulties
coming from the necessary zero value of their energy-momentum tensor, on the
other side.  Some development should be noted of the theory of the so called
"nontopological solitons", or Q-balls, [35,36,37,38], dyons [39], vortices
[40], sphalerons [41], skyrmeons [42], knots [43], which in most cases give
spherically symmetric static solutions of equations obtained from
correspondingly nonlinearized lagrangians. We have to mention that all these
examples follow the rule: "add nonlinear interaction term to the "free"
lagrangian giving linear equations and see what happens". We do not share the
view that this is the right way to pass to (3+1)-dimensional nonlinear
equations. We explained our view in the preceding sections and showed that
our approach works in the important case of the vacuum photon-like finite
solutions and in presence of external fields.  All free and not-free
time-stable objects in Nature are finite, so, the widely used "free" terms in
any lagrangian could play some misleading role, especially those "free" terms
that generate, or imply, D'Alembert-like wave equations.

Finally, we just mention (without any comments) the (super)string and brane
approaches, which also try to incorporate in theoretical physics spatially
finite solutions.

\newpage
{\bf \large References}
\vskip 0.5cm
[1]. {\bf S.J.Farlow}, "Partial Differential Equations for Scientists and
Engineers", John Wiley \& Sons, Inc., 1982.

[2]. {\bf G.N.Lewis}, Nature {\bf 118}, 874, 1926.

[3]. {\bf A.Einstein}, Sobranie Nauchnih Trudov, vols.2,3, Nauka,
Moskva, 1966.

[4]. {\bf M.Planck}, J.Franklin Institute, 1927 (July), p.13.

[5]. {\bf J.J.Thomson}, Philos.Mag.Ser. 6, 48, 737 (1924), and 50, 1181
(1925), and Nature, vol.137, 23 (1936).

[6] {\bf N.Rashevsky}, Philos.Mag. Ser.7, 4, 459 (1927).

[7] {\bf B. Lehnert, S. Roy}, {\it Extended Electromagnetic Theory}, World
Scientific, 1998.

[8] {\bf G.Hunter,R.Wadlinger}, Phys.Essays, vol.2, 158 (1989).

[9] {\bf M.Born, L.Infeld}, Proc.Roy.Soc., A 144 (425), 1934.

[10] {\bf A.Lees}, Phyl.Mag., 28 (385), 1939.

[11] {\bf N.Rosen}, Phys.Rev., 55 (94), 1939.

[12] {\bf D.Finkelstein, C.Misner}, Ann.Phys., 6 (230), 1959.

[13] {\bf D.Finkelstein}, Journ.Math.Phys., 7 (1218), 1966.

[14] {\bf J.P.Vigier}, Found. of Physics, vol.21 (1991), 125.

[15] {\bf S.Donev},  Compt. Rend. Bulg. Acad. Sci., vol.34, No.4, 1986.

[16] {\bf S.Donev},  Bulg.Journ.Phys., vol.13, (295), 1986.

[17] {\bf S.Donev},  Bulg.Journ.Phys., vol.15, (419), 1988.

[18] {\bf S.Donev},  Helvetica Physica Acta, vol.65, (910), 1992.

[19] {\bf S.Donev, M.Tashkova}, Proc.R.Soc.of Lond., A 443, (301), 1993.

[20] {\bf S.Donev, M.Tashkova}, Proc.R.Soc. of Lond., A 450, (281), 1995.

[21] {\bf A.Einstein}, J.Franklin Institute, 221 (349-382), 1936.

[22] {\bf J. Jackson}, {\it Classical Electrodynamics}, Wiley,
New York, 1962.

[23] {\bf H.K. Moffatt}, J. Fluid Mech., 35, 117-129 (1969).

[24] {\bf A.F. Ranada}, Eur.J.Phys, vol.13, 70 (1992);
{\bf J.L. Trueba, A.F. Ranada}, Eur.J.Phis, vol.17, 141 (1996).

[25] {\bf H.Cartan}, {\it Calcul Differentiel. Formes Differentielles},
Herman, Paris, 1967.

[26] {\bf C.Godbillon}, {\it Geometry Differentielle et Mecanique
Analitique},    Herman, Paris 1969.

[27] {\bf J.Synge}, {\it Relativity: The Special Theory}, North Holland,
Amsterdam, 1958.

[28] {\bf W. Greub}, {\it Multilinear Algebra}, Springer-Verlag, 1967.

[29] {\bf G.Rainich}, Trans. Am. Math. Soc., 27, 106 (1925).

[30] {\bf W. Greub, S. Halperin, R. Vanstone}, {\it Connections, Curvature
and Cohomology}, vol.2, Academic Press, 1973.

[31] {\bf C. Godbillon, J. Vey}, C. R. Acad. Sci., Paris, 273 (1971), 92-95.

[32] {\bf A. Einstein}, Phys. Zs., vol.9, p.662 (1908).

[33] {\bf G.'t Hooft}, Nucl. Phys. B79, 276 (1974); {\bf A. Polyakov},
JETP Lett., 20, 430 (1974).

[34] {\bf A. Belavin, A. Polyakov, A. Schwartz, Y. Tyupkin}, Phys. Lett,
v.59B, p. 85 (1975).

[35] {\bf R. Friedberg, T. Lee, A. Sirlin}, Phys. Rev. D, 13, 2739 (1976).

[36] {\bf S. Coleman}, Nucl. Phys., B262, 263-283 (1985).

[37] {\bf T. Lee, Y. Pang}, Phys. Rep., 221, No.5-6 (1992).

[38] {\bf M. Volkov, E. W\"ohnert}, Phys. Rev. D 66, 085003 (2002).

[39] {\bf B. Julia, A. Zee}, Phys. Rev. D 11, 2227 (1975).

[40] {\bf H. Nielsen, P. Olsen}, Nucl. Phys., B61, 45 (1973).

[41] {\bf F. Klinkhamer, N. Manton}, Phys. Rev. D30, 2212 (1984).

[42] {\bf T. Skyrme}, Nucl. Phys., 31, 556 (1962).

[43] {\bf L. Faddeev, A. Niemi}, Nature 387, 58 (1997).

[44] {\bf G. Mie}, Ann. der Phys. Bd.37, 511 (1912); Bd.39, 1 (1912); Bd.40,
1 (1913)

\newpage
\appendix


\section{\bf Appendix: Extended Parallelism. Applications in Physics}

\begin{abstract}
This appendix considers a generalization of the existing concept of parallel
(with respect to a given connection) geometric objects and its possible usage
as a suggesting rule in searching for adequate field equations and local
conservation laws in theoretical physics. The generalization tries to
represent mathematically the two-sided (or dual) nature of the physical
objects, the {\it change} and the {\it conservation}.  The physical objects
are presented mathematically by sections $\Psi$ of vector bundles, the
admissible changes $D\Psi$ are described as a result of the action of
appropriate differential operators $D$ on these sections, and the
conservation proprieties are accounted for by the requirement that suitable
projections of $D\Psi$ on $\Psi$ and on other appropriate sections must be
zero. It is shown that the most important equations of theoretical physics
obey this rule.  Extended forms of Maxwell and Yang-Mills equations are also
considered.
\end{abstract}

\subsection{The general Rule}

We are going to consider here a more general view  on the geometrical concept
of parallelism, more or less already implicitly used in some physical
theories. The concept of parallelism appropriately unifies two features:
{\it change} and {\it conservation}, through suitable {\it differential
operators} and suitable {\it "projections"}.  In some cases this concept may
be applied to produce dynamical equations, and in other cases the
corresponding physical interpretation is {\it conservation (balance)
equations} (mainly energy-momentum balance).  The examples presented further
show how it has been used so far, and how it could be used as a generating
tool for producing new useful relations and for finding natural ways to
extension of the existing equations in physics.  An important feature, that
deserves to be noted even at this moment, is that the corresponding equations
may become {\it nonlinear} in a natural way, so we might be fortunately
surprised by appearing of spatially finite (or soliton-like) solutions.
\vskip 0.5cm
We begin with the algebraic structure to be used further in the bundle
picture.  The basic concepts used are the {\it tensor product} $\otimes$ of
two linear spaces (we shall use the same term {\it linear space} for a vector
space over a field, and for a module over a ring, and from the context it
will be clear which case is considered) and {\it bilinear maps}. The bilinear
maps used are as a rule assumed to be nondegenerate [1]. Also, all linear
spaces, manifolds and bundles are assumed to be finite dimensional, and the
dimensions will be pointed out only if needed.

Let $(U_1,V_1)$, $(U_2,V_2)$ and $(U_3,V_3)$ be three couples of linear
spaces.  Let $\Phi:  U_1\times U_2\rightarrow U_3$ and $\varphi:  V_1\times
V_2\rightarrow V_3$ be two bilinear maps.
We consider the decomposable elements
\[
(u_1\otimes v_1)\in U_1\otimes V_1 \quad \text{and} \quad
(u_2\otimes v_2)\in U_2\otimes V_2.
\]
Now we make use of the given bilinear maps as follows:
$$
(\Phi,\varphi)(u_1\otimes v_1,u_2\otimes v_2)=\Phi(u_1,u_2)\otimes
\varphi(v_1,v_2),
$$
and on nondecomposable elements we extend this map by linearity:
\[
(\Phi,\varphi)((u_1)^i\otimes (v_1)_i,(u_2)^j\otimes
(v_2)_j)=\Phi((u_1)^i,(u_2)^j)\otimes \varphi((v_1)_i,(v_2)_j),
\]
where summation over the repeated indecies is assumed. The obtained elements
are in $U_3\otimes V_3$. Note that if $\{e_i\}$ is a basis of $V$ then for
every element $u\otimes v\in U\otimes V$ we have $u\otimes v=u\otimes v^i
e_i=v^i u\otimes e_i=K^i\otimes e_i$, where $K^i\equiv v^i u\in U$, so
further we may use such a representation.

We give now the corresponding bundle picture. Let $M$ be a smooth real
manifold. We assume that the following vector bundles over $M$ are
constructed:  $\xi_i, \eta_i$, with standard fibers $U_i,V_i$ and sets of
sections $Sec(\xi_i), Sec(\eta_i), i=1,2,3$.

Assume the two bundle maps are given: $(\Phi,id_M):
\xi_1\times\xi_2\rightarrow\xi_3$ and $(\varphi,id_M):
\eta_1\times\eta_2\rightarrow\eta_3$. Let $\{e_i\}$ and $\{\tau_j\}$ be local
bases of $Sec(\eta_1)$ and $Sec(\eta_2)$ respectively.
Then if $A^i\otimes e_i$ and
$B^j\otimes\tau_j$ are sections of $\xi_1\otimes\eta_1$ and
$\xi_2\otimes\eta_2$ respectively, we can form an element of
$Sec(\xi_3\otimes\eta_3)$ by the rule :
\begin{equation}
(\Phi,\varphi)(A^i\otimes e_i,B^j\otimes\tau_j)=           
\Phi(A^i,B^j)\otimes\varphi(e_i,\tau_j).
\end{equation}
Note that this rule assumes that the summation commutes with the
bilinear maps $\Phi$ and $\varphi$.

Let now $\tilde\xi$ be a new vector bundle on $M$ and $w \in Sec(\tilde\xi)$.
Assume that we have a differential operator
$D:Sec(\tilde\xi)\rightarrow Sec(\xi_2\otimes\eta_2)$
so we can form the section $Dw\in Sec(\xi_2\otimes \eta_2)$ and
represent it as $(Dw)^j\otimes\tau_j\in Sec(\xi_2\otimes\eta_2)$.
Applying now the two bilinear maps we obtain
$$
\Phi(A^i,(Dw)^j)\otimes\varphi(e_i,\tau_j)\in Sec(\xi_3\otimes\eta_3).
$$
We give now the following
\vskip 0.4cm
\noindent
{\bf Definition}:
The section $w\in Sec(\tilde{\xi})$, called further {\it the section of
interest}, will be called $(\Phi,\varphi;D)$-{\it parallel/conservative} with
respect to $A^i\otimes e_i\in Sec(\xi_1\otimes\eta_1)$ if

\begin{equation}
(\Phi,\varphi;D)(A^i\otimes e_i,w)
= (\Phi,\varphi)(A^i\otimes e_i,(Dw)^j\otimes\tau_j)=       
\Phi(A^i,(Dw)^j)\otimes\varphi(e_i,\tau_j)=0.
\end{equation}
\vskip 0.4cm
\noindent
The term {\it parallel} seems more appropriate when, making use of (85), the
relation obtained is interpreted as equation of motion (field equation), and
the term {\it conservative} is more appropriate when the relation obtained is
interpreted as local conservation law.

The above relation (85) we call the {\bf GENERAL RULE (GR)}. The bilinear maps
$\Phi,\varphi$ "project" the "changes" $Dw$ of the section $w$ on the section
$A^i\otimes e_i$ in the bases chosen. If the relation (85) holds we understand
it in the following sense: those characteristics of $w$ which {\it do feel}
the operator $D$ may be NOT {\it constant} with respect to $D$, but the
"projections" of the possible nonzero changes $Dw$ on an appropriate
section $A^i\otimes e_i$ {\it do vanish}. In particular, if
$A^i\otimes e_i=w$,
then we may say that $w$ is $(\Phi,\varphi;D)$-{\it autoparallel}. In
this sense the above definition (85) goes along with our concept of {\it
physical object} as given in Sec.1 of this paper.

As an example of a differential operator we note the
{\it exterior derivative} $\mathbf{d}:
\Lambda^p(M)\rightarrow\Lambda^{p+1}(M)$.  In the case of the physically
important example of Lie algebra $\mathfrak{g}$-valued differential forms,
with "$\Phi=$ exterior product" and "$\varphi=$ Lie bracket $[,]$",
$\xi_1=\Lambda^p(M)$, $\xi_2=\Lambda^{p+1}(M)$,
$\tilde\xi=\Lambda^p(M)\otimes\mathfrak{g}$,
$\eta_1=\eta_2=M\times\mathfrak{g}$, the {\bf GR} (85) looks as follows:
\[
(\wedge,[,];\mathbf{d}\times id_{\mathfrak{g}})
(\alpha^i\otimes E_i, \beta^j\otimes E_j)=
(\wedge,[,])(\alpha^i\otimes E_i, \mathbf{d}\beta^j\otimes E_j)=
\alpha^i\wedge\mathbf{d}\beta^j\otimes [E_i,E_j]=0,
\]
where $\{E_i\}$ is a basis of $\mathfrak{g}$, and a summation over the
repeated indexes is understood.  Further we are going to consider particular
cases of the {\bf (GR)} (85) with explicitly defined operators $D$
and additional operators whenever they participate in the definition of the
corresponding section of interest.


\subsection{The General Rule in Action}
\subsubsection{Examples from Geometry and Mechanics}
We begin with some simple examples from differential geometry and
classical mechanics.
\vskip 0.4cm
\noindent 1. {\bf Integral invariance relations}
\vskip 0.2cm
These relations have been introduced and studied from the point
of view of applications in mechanics by Lichnerowicz [2].

We specify the bundles over the real finite dimensional manifold $M$:

$\xi_1=TM;\  \xi_2=T^*(M);\  \
\eta_1=\eta_2=\xi_3=\eta_3=M\times\mathbb{R},\  \text{denote}\
Sec(M\times \mathbb{R})\equiv C^{\infty}(M)$

$\Phi$=substitution operator, denoted by\  $i(X), X\in Sec(TM)$;

$\varphi$=point-wise product of functions.

We denote by $1$ the function $f(x)=1, x\in M$. Consider the sections
\newline $X\otimes 1\in Sec(TM\otimes(M\times \mathbb{R}));\ \ \alpha\otimes
1\in Sec(T^*M\otimes(M\times\mathbb{R}))$.  Then the {\bf GR} leads to
\begin{equation}
(\Phi,\varphi)(X\otimes 1, \alpha\otimes 1)=i(X)\alpha\otimes 1      
=i(X)\alpha=0.
\end{equation}
We introduce now the differential operator $\mathbf{d}$: if $\alpha$ is an
exact 1-form, $\alpha=\mathbf{d}f$, so that $\tilde{\xi}=M\times\mathbb{R}$,
the {\bf GR} gives
\[
(\Phi,\varphi;\mathbf{d})(X\otimes 1, f)=
(\Phi,\varphi)(X\otimes 1, \mathbf{d}f\otimes 1)=
i(X)\mathbf{d}f\otimes (1\otimes 1) =i(X)\mathbf{d}f=X(f)=0.
\]
i.e. the derivative of $f$ along the vector field $X$ is equal to zero. So,
we obtain the well known relation, defining the first integrals  $f$ of the
dynamical system determined by the vector field $X$.  In this sense $f$ may
be called $(\Phi,\varphi,\mathbf{d})$-{\it conservative} with respect to $X$,
where $\Phi$ and $\varphi$ are defined above.
\vskip 0.3cm
\noindent
2.{\bf Absolute and relative integral invariants}
\vskip 0.2cm
These quantities have been introduced and studied in mechanics by Cartan
[3]. By definition, a $p$-form $\alpha$ is called an {\it absolute integral
invariant} of the vector field $X$ if $i(X)\alpha=0$ and
$i(X)\mathbf{d}\alpha=0$. And $\alpha$ is called a {\it relative integral
invariant} of the field $X$ if $i(X)\mathbf{d}\alpha=0$. So, in our
terminology (the same bundle picture as above), we can call the relative
integral invariants of $X$ $(\Phi,\varphi;\mathbf{d})$-{\it conservative} with
respect to $X$, and the absolute integral invariants of $X$ are additionally
$(\Phi,\varphi)$-{\it conservative} with respect to $X$, with $(\Phi,\varphi)$
as defined above.  A special case is when $p=n$, and $\omega\in\Lambda^n(M)$
is a volume form on $M$.
\vskip 0.5cm
\noindent 3. {\bf Symplectic geometry and mechanics}
\vskip 0.2cm
Symplectic manifolds are even dimensional and have a distinguished
nondegenerate closed $2$-form $\omega$, $\mathbf{d}\omega=0$.  This structure
may be defined in terms of the {\bf GR} in the following way.  Choose
$\xi_1=\eta_1=\eta_2=M\times\mathbb{R}$, $\xi_2\otimes
\eta_2=\Lambda^3(T^*M)\otimes \mathbb{R}$, and $\mathbf{d}$ as a differential
operator. Consider now the section $1\in Sec(M\times\mathbb{R})$ and
$\omega\in Sec(\Lambda^2(T^*M)$, with $\omega$ - nondegenerate.  The map
$\Phi$ is the product $f.\omega$ and the map $\varphi$ is the product of
functions. So, we have
\[
(\Phi,\varphi;\mathbf{d})(1\otimes1, \omega)=
\Phi(1,\mathbf{d}\omega)\otimes \varphi(1,1)=\mathbf{d}\omega=0.
\]
Hence, the relation $\mathbf{d}\omega=0$ is equivalent to the requirement
$\omega$ to be $(\Phi,\varphi;\mathbf{d})$-{\it parallel} with respect to the
section $1\otimes 1$.

The hamiltonian vector fields $X$ are defined by the condition
$L_X\omega=\mathbf{d}i(X)\omega=0$.  If $\Phi=\varphi$ is the point-wise
product of functions we have
\[
(\Phi,\varphi;\mathbf{d})(1\otimes 1,i(X)\omega)=
(\Phi,\varphi)(1\otimes 1,\mathbf{d}i(X)(\omega)\otimes 1)=
L_X\omega\otimes 1=L_X\omega=0.
\]
In terms of the {\bf GR} we can say that $X$ is hamiltonian if $i(X)\omega$ is
$(\Phi,\varphi;\mathbf{d})$-{\it parallel} with respect to $1\otimes 1$.

The induced Poisson structure $\{f,g\}$, is given in terms of the {\bf GR} by
setting $\Phi=\omega^{-1}$, where $\omega^{-1}.\omega=id_{TM}$,
$\varphi$=point-wise product of functions, and $1\in Sec(M\times\mathbb{R})$.
We get
\[
(\Phi,\varphi)(\mathbf{d}f\otimes 1, \mathbf{d}g\otimes 1)=
\omega^{-1}(\mathbf{d}f,\mathbf{d}g)\otimes 1.
\]
A closed 1-form $\alpha,\ \mathbf{d}\alpha=0$, is a first integral of the
hamiltonian system $Z$, $\mathbf{d}i(Z)\omega=0$, if $i(Z)\alpha=0$. In terms
of the {\bf GR} we can say that the first integrals $\alpha$ are
$(i,\varphi)$-parallel with respect to $Z$:  $(i,\varphi)(Z\otimes
1,\alpha\otimes 1)=i(Z)\alpha\otimes 1=0$.  From $L_Z\omega=0$ it follows
$L_Z\omega^{-1}=0$. The Poisson bracket $(\alpha,\beta)$ of two first
integrals $\alpha$ and $\beta$ is equal to
$[-\mathbf{d}(\omega^{-1}(\alpha,\beta))]$ [6]. The well known property that
the Poisson bracket of two first integrals of $Z$ is again a first integral of
$Z$ may be formulated as: the function $\omega^{-1}(\alpha,\beta)$ is
$(i,\varphi;\mathbf{d})$-parallel with respect to $Z$,
\[
(i,\varphi;\mathbf{d})(Z\otimes 1, \omega^{-1}(\alpha,\beta))=
i(Z)\mathbf{d}\omega^{-1}(\alpha,\beta)\otimes 1=0.
\]

\subsubsection{Frobenius integrability theorems and linear connections}
\vskip 0.4cm
\noindent
1.{\bf Frobenius integrability theorems [4]}
\vskip 0.2cm
Let $\Delta=(X_1,\dots,X_r)$ be a differential system on $M$, i.e. the vector
fields $X_i, i=1,\dots,r$ define a locally stable submodule of $Sec(TM)$ and
at every point $p\in M$ the subspace $\Delta_p^r\subset T_p(M)$ has dimension
$r$. Then $\Delta^r$ is called involutive
(or integrable) if $[X_i,X_j]\in \Delta^r,
i,j=1,\dots,r$. Denote by $\Delta^{n-r}_p\subset T_p(M)$ a complimentary
subspace: $\Delta_p^r\oplus\Delta^{n-r}_p=T_p(M)$, and let $\pi:
T_p(M)\rightarrow \Delta^{n-r}_p$ be the corresponding projection. So,
the corresponding Frobenius integrability condition means $\pi([X_i,X_j])=0,
i,j=1,\dots,r$.

In terms of the {\bf GR} we set $D(X_i)=\pi\circ L_{X_i}$, $\Phi$="product of
functions and vector fields",  and $\varphi$ again the product of functions.
The integrability condition now is
\[
\begin{split}
&(\Phi,\varphi;D(X_i))
(1\otimes 1, X_j)\\&=(\Phi,\varphi)
(1\otimes 1,\pi([X_i,X_j]\otimes 1))=1.\pi([X_i,X_j])\otimes 1
=0,
\quad i,j=1,\dots,r.
\end{split}
\]

In the dual formulation we have the Pfaff system $\Delta^*_{n-r}$, generated
by the linearly independent 1-forms $(\alpha_1,\dots,\alpha_{n-r})$, such
that $\alpha_m(X_i)=0, i=1,\dots r; m=1,\dots n-r$.  Then $\Delta^*_{n-r}$ is
integrable if $\mathbf{d}\alpha\wedge \alpha_1\wedge\dots \wedge
\alpha_{n-r}=0, \alpha\in \Delta^*_{n-r}$. In terms of {\bf GR} we set
$\varphi$ the same as above, $\Phi=\wedge$ and $\mathbf{d}$ as differential
operator and obtain
\[ (\Phi,\varphi;\mathbf{d})(\alpha_1\wedge\dots
\wedge\alpha_{n-r}\otimes 1, \alpha)=
(\wedge,\varphi)(\alpha_1\wedge\dots \wedge\alpha_{n-r}\otimes 1,
\mathbf{d}\alpha\otimes 1)
\]
\[
=\mathbf{d}\alpha\wedge\alpha_1\wedge\dots \wedge\alpha_{n-r}\otimes
\varphi(1,1)=\mathbf{d}\alpha\wedge\alpha_1\wedge\dots \wedge\alpha_{n-r}=0,
\]
i.e. $\alpha$ is $(\Phi,\varphi;\mathbf{d})$-parallel
with respect to $\alpha_1\wedge\dots \wedge\alpha_{n-r}\otimes 1$.

\vskip 0.4cm
\noindent
2. {\bf Linear connections}
\vskip 0.2cm
The concept of a linear connection in a vector bundle has proved to be of
great importance in geometry and physics. In fact, it allows to differentiate
sections of vector bundles along vector fields, which is a basic operation in
differential geometry, and in theoretical physics the physical fields are
represented mainly by sections of vector bundles. We recall now how one comes
to it.

Let $f:\mathbb{R}^n\rightarrow\mathbb{R}$ be a differentiable function. Then
we can find its differential $\mathbf{d}f$. The map $f\rightarrow\mathbf{d}f$
is $\mathbb{R}$-linear: $\mathbf{d}(\kappa.f)=\kappa.\mathbf{d}f$, $\kappa
\in \mathbb{R}$, and it has the derivative property
$\mathbf{d}(f.g)=f\mathbf{d}g+g\mathbf{d}f$. These two properties are
characteristic ones, and they are carried to the bundle situation as follows.

Let $\xi$ be a vector bundle over $M$. We always have the trivial bundle
$\xi_o=M\times\mathbb{R}$. Consider now $f\in C^{\infty}(M)$ as a section of
$\xi_o$. We note that $Sec(\xi_o)=C^{\infty}(M)$ is a module over itself, so
we can form $\mathbf{d}f$ with the above two characteristic  properties. The
new object $\mathbf{d}f$ lives in the space $\Lambda^1(M)$ of 1-forms on $M$,
so it defines a linear map $\mathbf{d}f: Sec(TM)\rightarrow Sec(\xi_o),
\mathbf{d}f(X)=X(f)$.  Hence, we have a map $\nabla$ from $Sec(\xi_o)$ to
the 1-forms with values in $Sec(\xi_o)$, and this map has the above two
characteristic properties.  We say that $\nabla$ defines a linear connection
in the vector bundle $\xi_o$.

In the general case the sections $Sec(\xi)$ of the vector bundle $\xi$ form a
module over $C^{\infty}(M)$. So, a linear connection $\nabla$ in $\xi$ is a
$\mathbb{R}$-linear map $\nabla: Sec(\xi)\rightarrow \Lambda^1(M,\xi)$. In
other words, $\nabla$ sends a section $\sigma\in Sec(\xi)$ to a 1-form
$\nabla \sigma$ valued in $Sec(\xi)$ in such a way, that
\begin{equation}
\nabla(k\,\sigma)=k\,\nabla(\sigma), \quad                     
\nabla(f\,\sigma)=df\otimes\sigma+f\,\nabla(\sigma),
\end{equation}
where $k\in \mathbb{R}$ and $f\in C^{\infty}(M)$. If $X\in Sec(TM)$ then we
have the composition $i(X)\circ\nabla$, so that
\[
i(X)\circ\nabla(f\,\sigma)=X(f)\,\sigma+f\,\nabla_X(\sigma),
\]
where $\nabla_X(\sigma)\in Sec(\xi)$.

In terms of the {\bf GR} we put $\xi_1=TM=\tilde\xi$ and
$\xi_2=\Lambda^1(M)\otimes\xi$,
and $\eta_1=\eta_2=\xi_o$. Also, $\Phi(X,\nabla\sigma)=\nabla_X \sigma$ and
$\varphi(f,g)=f.g$. Hence, we obtain
\begin{equation}
(\Phi,\varphi;\nabla)(X\otimes 1, \sigma)           
=(\Phi,\varphi)(X\otimes 1,(\nabla \sigma)\otimes 1)
=\nabla_X \sigma\otimes 1=\nabla_X \sigma,
\end{equation}
and the section $\sigma$ is called $\nabla$-{\it parallel} with respect to
$X$ if $\nabla_X\sigma=0$.

\vskip 0.4cm
\noindent
3. {\bf Covariant exterior derivative}
\vskip 0.2cm
The space of $\xi$-valued $p$-forms $\Lambda^p(M,\xi)$ on $M$ is isomorphic
to $\Lambda^p(M)\otimes Sec(\xi)$. So, if $(\sigma_1,\dots,\sigma_r)$ is a
local basis of $Sec(\xi)$, every $\Psi\in \Lambda^p(M,\xi)$ is represented by
$\psi^i\otimes \sigma_i, i=1,\dots,r$, where $\psi^i\in \Lambda^p(M)$.
Clearly the space $\Lambda(M,\xi)=\Sigma^n_{p=0}\Lambda^p(M,\xi)$, where
$\Lambda^o(M,\xi)=Sec(\xi)$, is a
$\Lambda(M)=\Sigma^n_{p=0}\Lambda^p(M)$-module:
$\alpha.\Psi=\alpha\wedge\Psi=(\alpha\wedge\psi^i)\otimes \sigma_i$.

A linear connection $\nabla$ in $\xi$ generates covariant exterior
derivative $\mathbf{D}: \Lambda^p(M,\xi)\rightarrow\Lambda^{p+1}(M,\xi)$ in
$\Lambda(M,\xi)$ according to the rule
\[
\begin{split}
\mathbf{D}\Psi&=\mathbf{D}(\psi^i\otimes \sigma_i)=
\mathbf{d}\psi^i\otimes \sigma_i+(-1)^p \psi^i\wedge\nabla(\sigma_i)\\
&=(\mathbf{d}\psi^i+(-1)^p \psi^j\wedge\Gamma_{\mu j}^i dx^\mu)\otimes\sigma_i
=(\mathbf{D}\Psi)^i\otimes\sigma_i.
\end{split}
\]
We may call now a $\xi$-valued $p$-form $\Psi$ $\nabla$-{\it parallel} if
$\mathbf{D}\Psi=0$, and $(X,\nabla)$-{\it parallel} if
$i(X)\mathbf{D}\Psi=0$. This definition extends in a natural way to
$q$-vectors with $q\le p$. Actually, the substitution operator $i(X)$ extends
to (decomposable) $q$-vectors $X_1\wedge X_2\wedge\dots\wedge X_q$ as
follows:
\[
i(X_1\wedge X_2\wedge\dots\wedge X_q)\Psi =i(X_q)\circ
i(X)_{q-1}\circ\dots\circ i(X_1)\Psi,
\]
and extends to nondecomposable $q$-vectors by linearity. Hence, if $\Theta$
is a section of $\Lambda^q(TM)$ we may call $\Psi$ $(\Theta,\nabla)$-{\it
parallel} if $i(\Theta)\mathbf{D}\Psi=0$.

Denote now by $L_\xi$ the vector bundle of (linear) homomorphisms $(\Pi,id):
\xi\rightarrow \xi$, and let $\Pi\in Sec(L_\xi)$. Let $\chi \in
Sec(\Lambda^q(TM)\otimes L_\xi)$ be represented as $\Theta\otimes\Pi$. The
map $\Phi$ will act as: $\Phi(\Theta,\Psi)= i(\Theta)\Psi$, and the map
$\varphi$ will act as: $\varphi(\Pi,\sigma_i)= \Pi(\sigma_i)$.  So, if
$\nabla(\sigma_k)=\Gamma^j_{\mu k}dx^\mu\otimes \sigma_j$, we may call $\Psi$
$(\nabla)$-{\it parallel} with respect to $\chi=\Theta\otimes\Pi$ if
\begin{equation}
(\Phi,\varphi;\mathbf{D})(\Theta\otimes\Pi,\Psi=\psi^i\otimes\sigma_i)=
(\Phi,\varphi)(\Theta\otimes\Pi,(\mathbf{D}\Psi)^i\otimes\sigma_i)=     
i(\Theta)(\mathbf{D}\Psi)^i\otimes\Pi(\sigma_i)=0.
\end{equation}
If we have isomorphisms $\otimes^p TM\backsim
\otimes^p T^*M, p=1,2,\dots$, defined in some natural way (e.g. through a
metric tensor field), then to any $p$-form $\alpha$ corresponds unique
$p$-vector $\tilde\alpha$. In this case we may talk about "$\backsim$"-
{\it autopaparallel} objects with respect a (point-wise) bilinear map
$\varphi:  (\xi\times\xi)\rightarrow \eta$, where $\eta$ is also a vector
bundle over $M$. So, $\Psi=\alpha^k\otimes\sigma_k\in \Lambda^p(M,\xi)$ may
be called $(i,\varphi;\nabla)$-{\it autoparallel} with respect to the
isomorphism "$\backsim$" if
\begin{equation}
\begin{split}
&(i,\varphi;\nabla)
(\tilde\alpha^k\otimes \sigma_k,\alpha^m\otimes\sigma_m)\\
&=i(\tilde\alpha^k)\mathbf{d}\alpha^m\otimes\varphi(\sigma_k,\sigma_m)+
(-1)^p i(\tilde\alpha^k)(\alpha^j\wedge \Gamma^m_{\mu j}dx^\mu)
\otimes\varphi(\sigma_k,\sigma_m)\\                                   
&=\big[i(\tilde\alpha^k)\mathbf{d}\alpha^m+
(-1)^p i(\tilde\alpha^k)(\alpha^j\wedge \Gamma^m_{\mu j}dx^\mu)\big]
\otimes\varphi(\sigma_k,\sigma_m)=0.
\end{split}
\end{equation}
\noindent
Although the above examples do not, of course, give a complete list of the
possible applications of the {\bf GR} (85), they will serve as a good basis
for the physical applications we are going to consider further.

\subsection{Physical applications of GR}

{\bf 1. Autoparallel vector fields and 1-forms}
\vskip 0.2cm
In nonrelativistic and relativistic mechanics the vector fields $X$ on a
manifold $M$ are the local representatives (velocity vectors) of the
evolution trajectories for point-like objects.  The condition that a particle
is {\it free} is mathematically represented by the requirement that the
corresponding vector field $X$ is autoparallel with respect to a given
connection $\nabla$ (covariant derivative) in $TM$:
\begin{equation}
i(X)\nabla X=0,\quad
\text{or in components},\quad
X^\sigma \nabla_\sigma X^\mu +\Gamma^\mu_{\sigma\nu}X^\sigma X^\nu=0.   
\end{equation}
In view of the physical interpretation of $X$ as velocity vector field the
usual latter used instead of $X$ is $u$. The above equation (91) presents a
system of nonlinear partial differential equations for the components
$X^\mu$, or $u^\mu$. When reduced to 1-dimensional submanifold which is
parametrized locally by the appropriately chosen parameter $s$, (91) gives a
system of ordinary differential equations:
\begin{equation}
\frac{d^2 x^\mu}{ds^2}+\Gamma^\mu_{\sigma\nu}\frac{dx^\nu}{ds}       
\frac{dx^\nu}{ds}=0,
\end{equation}
and (92) are known as ODE defining the geodesic (with respect to $\Gamma$)
lines in $M$. When $M$ is riemannian with metric tensor $g$ and $\Gamma$ the
corresponding Levi-Civita connection, i.e. $\nabla g=0$ and
$\Gamma^\mu_{\nu\sigma}=\Gamma^\mu_{\sigma\nu}$, then the solutions of (92)
give the extreme (shortest or longest) distance $\int^b_a ds$ between the two
points $a,b\in M$, so (92) are equivalent to
\[
\delta\left(\int^b_a ds\right)=
\delta\left(\int^b_a
\sqrt{g_{\mu\nu}\frac{dx^\mu}{ds}\frac{dx^\nu}{ds}}\right)=0.
\]
A system of particles that move along the solutions to (92) with $g$-the
Minkowski metric and $g_{\mu\nu}\frac{dx^\mu}{ds}\frac{dx^\nu}{ds}>0$,
is said to form an {\it inertial frame of reference}.

It is interesting to note that the system (91) has (3+1)-soliton-like (even
spatially finite) solutions on Minkowski space-time [5]. In fact, in
canonical coordinates $(x^1,x^2,x^3,x^4)=(x,y,z,\xi=ct)$ let $u^\mu=(0,0,\pm
\frac vc f,f)$ be the components of $u$, where $0<v=const<c$, and $c$ is the
velocity of light, so $\frac vc < 1$ and $u^\sigma u_\sigma =
\left(1-\frac{v^2}{c^2}\right)f^2>0$. Then every function $f$ of the kind
\[
f(x,y,z,\xi)=f\left(x,y,\alpha(z\mp\frac vc
\xi)\right),\  \alpha=const, \quad\text{for example}\quad
\alpha=\frac{1}{\sqrt{1-\frac{v^2}{c^2}}},
\]
defines a solution to (91).  If $u_\sigma u^\sigma =0$ then equations (91)
($X=u$), are equivalent to $u^\mu(\mathbf{d}u)_{\mu\nu}=0$, where
$\mathbf{d}$ is the exterior derivative. In fact, since the connection used
is riemannian, we have $0=\nabla_\mu\frac12(u^\nu u_\nu)=u^\nu\nabla_\mu
u_\nu$, so the relation $u^\nu\nabla_\nu u_\mu -u^\nu\nabla_\mu u_\nu=0$
holds and is obviously equal to $u^\mu(\mathbf{d}u)_{\mu\nu}=0$. The
soliton-like solution is defined by $u=(0,0,\pm f,f)$ where the function $f$
is of the form
\[
f(x,y,z,\xi)=f(x,y,z\mp \xi).
\]
Clearly, for every autoparallel vector field $u$ (or one-form $u$) there
exists a canonical coordinate system on the Minkowski space-time, in
which $u$ takes such a simple form: $u^\mu=(0,0,\alpha f,f), \alpha=const$.
The dependence of $f$ on the three spatial coordinates $(x,y,z)$ is arbitrary
, so it is allowed to be chosen {\it soliton-like} and, even, {\it finite}.
Let now $\rho$ be the mass-energy density function, so that
$\nabla_\sigma(\rho u^\sigma)=0$ gives the mass-energy conservation, i.e. the
function $\rho$ defines those properties of our physical system which
identify the system during its evolution. In this way the tensor conservation
law
\[
\nabla_\sigma(\rho u^\sigma u^\mu)=(\nabla_\sigma \rho u^\sigma)u^\mu+
\rho u^\sigma\nabla_\sigma u^\mu=0
\]
describes the two aspects of the physical system: its dynamics through
equations (91) and its mass-energy conservation properties.

The properties described give a connection between free point-like objects
and (3+1) soliton-like autoparallel vector fields on Minkowski space-time.
Moreover, they suggest that extended free objects with more complicated
space-time dynamical structure may be described by some appropriately
generalized concept of autoparallel mathematical objects.


\vskip 0.4cm
{\bf 2. Electrodynamics}
\vskip 0.2cm
{\bf 2.1 Maxwell equations}

The Maxwell equations $\mathbf{d}F=0, \mathbf{d}*F=0$ in their 4-dimensional
formulation on Minkowski space-time $(M,\eta), sign(\eta)=(-,-,-,+)$ and the
Hodge $*$ is defined by $\eta$, make use of the exterior derivative as a
differential operator.  The field has, in general, 2 components $(F,*F)$, so
the interesting bundle is $\Lambda^2(M)\otimes V$, where $V$ is a real
2-dimensional vector space. Hence the adequate mathematical field will look
like $\Omega=F\otimes e_1+*F\otimes e_2$, where $(e_1,e_2)$ is a basis of
$V$. The exterior derivative acts on $\Omega$ as:
$\mathbf{d}\Omega=\mathbf{d}F\otimes e_1+\mathbf{d}*F\otimes e_2$, and the
equation $\mathbf{d}\Omega=0$ gives the vacuum Maxwell equations.

In order to interpret in terms of the above given general view ({\bf GR})
on parallel objects with respect to given sections of vector bundles and
differential operators we consider the sections (see the above introduced
notation) $(1\otimes 1, \Omega\otimes 1)$ and the differential operator
$\mathbf{d}$. Hence, the {\bf GR} acts as follows:
\[
(\Phi,\varphi;\mathbf{d})(1\otimes 1, \Omega\otimes 1)=
(\Phi,\varphi)(1\otimes 1,\mathbf{d}\Omega\otimes 1)=
(1.\mathbf{d}\Omega\otimes 1.1)=\mathbf{d}\Omega.
\]
The corresponding $(\Phi,\varphi;\mathbf{d})$-parallelism leads to
$\mathbf{d}\Omega=0$. In presence of electric $\mathbf{j}$ and magnetic
$\mathbf{m}$ currents, considered as 3-forms, the parallelism condition does
not hold and on the right-hand side we'll have non-zero term, so the full
condition is
\begin{equation}
(\Phi,\varphi)[1\otimes 1, (\mathbf{d}F\otimes e_1+
\mathbf{d}*F\otimes e_2)\otimes 1]=
(\Phi,\varphi)[1\otimes1, (\mathbf{m}\otimes e_1+        
\mathbf{j}\otimes e_2)\otimes 1]
\end{equation}
The case $\mathbf{m}=0, F=\mathbf{d}A$ is, obviously a special case.

\vskip 0.2cm
{\bf 2.2 Extended Maxwell equations}

The extended Maxwell equations (on Minkowski space-time) in vacuum read
[6]-[7]:
\begin{equation}
F\wedge *\mathbf{d}F=0,\quad (*F)\wedge(*\mathbf{d}*F)=0,\quad      
F\wedge(*\mathbf{d}*F)+(*F)\wedge(*\mathbf{d}F)=0
\end {equation}
They may be expressed through the {\bf GR} in the following way.
On $(M,\eta)$ we have the bijection between $\Lambda^2(TM)$ and
$\Lambda^2(T^*M)$ defined by $\eta$, which we denote by
$\tilde F\leftrightarrow F$. So, equations (94) are equivalent to
\[
i(\tilde F)\mathbf{d}F=0,\quad i(\widetilde{*F})\mathbf{d}*F=0,\quad
i(\tilde F)\mathbf{d}*F+i(\widetilde{*F})\mathbf{d}F=0.
\]
We consider the sections $\tilde\Omega=\tilde F\otimes
e_1+\widetilde{*F}\otimes e_2$ and $\Omega=F\otimes e_1+*F\otimes e_2$ with
the differential operator $\mathbf{d}$. The maps $\Phi$ and $\varphi$ are
defined as:  $\Phi$ is the substitution operator $i$, and $\varphi=\vee$ is
the symmetrized tensor product in $V$. So we obtain
\begin{equation}
\begin{split} &(\Phi,\varphi;\mathbf{d})(\tilde F\otimes
e_1+\widetilde{*F}\otimes e_2, F\otimes e_1+*F\otimes e_2)\\ &=i(\tilde
F)\mathbf{d}F\otimes e_1\vee e_1+ i(\widetilde{*F})\mathbf{d}*F\otimes  
e_2\vee e_2+ (i(\tilde F)\mathbf{d}*F+i(\widetilde{*F})\mathbf{d}F)
\otimes e_1\vee e_2=0.
\end{split}
\end{equation}
Equations (95) may be written down also as
$\big((i,\vee)\tilde\Omega\big)\mathbf{d}\Omega=0$.

Equations (95) are physically interpreted as describing locally the intrinsic
energy-momentum exchange between the two components $F$ and $*F$ of $\Omega$:
the first two equations $i(\tilde F)\mathbf{d}F=0$ and $i(\widetilde
*F)\mathbf{d}*F=0$ say that every component keeps locally its
energy-momentum, and the third equation $i(\tilde F)\mathbf{d}*F+i(\widetilde
*F)\mathbf{d}F=0$ says (in accordance with the first two) that if $F$
transfers energy-momentum to $*F$, then $*F$ transfers the same quantity
of energy-momentum to $F$.

If the field exchanges (loses or gains) energy-momentum with some external
systems, Extended Electrodynamics describes the potential abilities of the
external systems to gain or lose energy-momentum from the field by means of 4
one-forms (currents) $J_a, a=1,2,3,4$, and explicitly the exchange is
given by [7]
\begin{equation}
i(\tilde F)\mathbf{d}F=i(\tilde J_1)F,\ \
i(\widetilde {*F})\mathbf{d}*F=i(\tilde J_2)F,\ \
i(\tilde F)\mathbf{d}*F+i(\widetilde {*F})\mathbf{d}F=             
i(\tilde J_3)F+i(\tilde J_4)*F.
\end{equation}
It is additionally assumed that every couple $(J_a,J_b)$ defines
a completely integrable Pfaff system, i.e. the following equations hold:
\begin{equation}
\mathbf{d}J_a\wedge J_a\wedge J_b=0,\ \ a,b=1,\dots,4.             
\end{equation}
The system (95) has (3+1)-localized photon-like (massless) solutions, and
the system (96)-(97) admits a large family of (3+1)-soliton solutions [7].

\vskip 0.4cm
{\bf 3. Yang-Mills theory}
\vskip 0.2cm
{\bf 3.1 Yang-Mills equations}

In this case the field is a connection, represented locally by its connection
form $\omega\in \Lambda^1(M)\otimes\mathfrak{g}$, where $\mathfrak{g}$ is the
Lie algebra of the corresponding Lie group $G$.  If $\mathbf{D}$ is the
corresponding covariant derivative, and $\Omega=\mathbf{D}\omega$ is the
curvature , then Yang-Mills equations read $\mathbf{D}*\Omega=0$. The formal
difference with the Maxwell case is that $G$ may NOT be commutative, and may
have, in general, arbitrary finite dimension. So, the two sections are
$1\otimes 1$ and $*\Omega\otimes 1$, the maps $\Phi$ and $\varphi$ are
product of functions and the differential operator is $\mathbf{D}$. So, we
may write
\begin{equation}
(\Phi,\varphi;\mathbf{D})(1\otimes 1, *\Omega\otimes 1)=      
\mathbf{D}*\Omega\otimes 1=0.
\end{equation}
Of course, equations (98) are always coupled to the Bianchi identity
$\mathbf{D}\Omega=0$.

\vskip 0.3cm
{\bf 3.2 Extended Yang-Mills equations}
\vskip 0.2cm
The extended Yang-Mills equations are written down in analogy with the
extended Maxwell equations.  The field of interest is an arbitrary 2-form
$\Psi$ on $(M,\eta)$ with values in a Lie algebra $\mathfrak{g}$,
$\dim(\mathfrak{g})=r$. If $\{E_i\}, i=1,2,\dots,r$ is a basis of
$\mathfrak{g}$ we have $\Psi=\psi^i\otimes E_i$ and
$\tilde\Psi=\tilde\psi^i\otimes E_i$. The map $\Phi$ is the substitution
operator, the map $\varphi$ is the corresponding Lie product $[,]$, and the
differential operator is the exterior covariant derivative with respect to a
given connection $\omega$:  $\mathbf{D}\Psi=\mathbf{d}\Psi+[\omega,\Psi]$. We
obtain
\begin{equation}
(\Phi,\varphi;\mathbf{D})(\tilde\psi^i\otimes E_i,\psi^j\otimes E_j)=
i(\tilde\psi^i)(\mathbf{d}\psi^m+\omega^j\wedge\psi^k\,C_{jk}^m)      
\otimes[E_m,E_i]=0,
\end{equation}
where $C_{jk}^m$ are the corresponding structure constants.  If the
connection is the trivial one, then $\omega=0$ and $\mathbf{D}\rightarrow
\mathbf{d}$, so, (99) gives
\begin{equation}
i(\tilde\psi^i)\mathbf{d}\psi^j\,C_{ij}^k\otimes E_k=0 .        
\end{equation}
If, in addition, instead of $[,]$ we assume for $\varphi$ some bilinear map
$f:\mathfrak{g}\times\mathfrak{g}\rightarrow\mathfrak{g}$, such that in the
basis $\{E_i\}$ $f$ is given by $f(E_i,E_i)=E_i$, and $f(E_i,E_j)=0$ for
$i\neq j$ the last relation reads
\begin{equation}
i(\tilde\psi^i)\mathbf{d}\psi^i\otimes E_i=0,\quad i=1,2,\dots,r.  
\end{equation}
The last equations (101) define the components $\psi^i$ as independent 2-forms
(of course $\psi^i$ may be arbitrary $p$-forms). If the bilinear map
$\varphi$ is chosen to be the symmetrized tensor product
$\vee:\mathfrak{g}\times \mathfrak{g}\rightarrow \mathfrak{g}\vee
\mathfrak{g}$, we obtain
\begin{equation}
i(\tilde\psi^i)\mathbf{d}\psi^j\otimes E_i\vee E_j=0,
\quad i\leqq j=1,\dots,r.                                         
\end{equation}
Equations (100) and (102) may be used to model bilinear interaction among the
components of $\Psi$. If the components of
$i(\tilde\psi^i)\mathbf{d}\psi^j\otimes E_i\vee E_j$
have the physical sense of energy-momentum exchange we may say
that every component $\psi^i$ gets locally as much energy-momentum from
$\psi^j$ as it gives to it. Since $C^k_{ij}=-C^k_{ji}$, equations (100)
consider only the case $i<j$, while equations (102) consider $i\leq j$, in
fact, for every $i,j=1,2,\dots,r$ we obtain from (102)
\[
i(\tilde\psi^i)\mathbf{d}\psi^i=0,\ \ \text{and}\ \
i(\tilde\psi^i)\mathbf{d}\psi^j+i(\tilde\psi^j)\mathbf{d}\psi^i=0.
\]
Clearly, these last equations may be considered as a natural generalization of
equations (94), so spatial soliton-like solutions are expectable.

\vskip 0.4cm
{\bf 4. General Relativity}
\vskip 0.2cm
In General Relativity the field function of interest is in a definite sense
identified with a Lorentz pseudometric $g$ on a 4-dimensional manifold, and
only those $g$ are considered as appropriate to describe the real
gravitational fields in vacuum which satisfy the equations $R_{\mu\nu}=0$,
where $R_{\mu\nu}$ are the components of the Ricci tensor.  The main
mathematical object which detects possible gravity is the Riemann curvature
tensor $R_{\alpha\mu,\beta\nu}$, which is a second order nonlinear
differential operator $R:g\rightarrow R(g)$. We define the map $\Phi$ to be
the contraction, or taking a {\it trace}:
$$
\Phi:(g_{\alpha\beta},R_{\alpha\mu,\beta\nu})
=g^{\alpha\beta}R_{\alpha\mu,\beta\nu}=R_{\mu\nu},
$$
so it is obviously bilinear. The map $\varphi$ is a product of functions, so
the {\bf GR} gives
\begin{equation}
(\Phi,\varphi;R)(g\otimes 1,g)=\Phi[g\otimes 1,R(g)\otimes 1]    
=\Phi[g,R(g)]\otimes\varphi(1,1)=Ric[R(g)]=0.
\end{equation}
In presence of matter fields $\Psi^a, a=1,2,\dots,r$, the system of equations
is
$$
R_{\mu\nu}-\kappa\left(T_{\mu\nu}-\frac12 g_{\mu\nu}T\right)=0.
$$
It is easily obtained through the {\bf GR} if we modify the differential
operator $R_{\alpha\mu,\beta\nu}$ to
$$
R_{\alpha\mu,\beta\nu}-\frac{\kappa}{2}\left(T_{\alpha\beta}g_{\mu\nu}+
T_{\mu\nu}g_{\alpha\beta}-T_{\alpha\nu}g_{\mu\beta}-
T_{\mu\beta}g_{\alpha\nu}\right)+
\frac{\kappa}{3}\left(g_{\alpha\beta}g_{\mu\nu}-
g_{\alpha\nu}g_{\mu\beta}\right)T,
$$
where $\kappa$ is the gravitational constant, $T_{\mu\nu}(\Psi^a)=
T_{\nu\mu}(\Psi^a)$ is the corresponding stress energy momentum tensor, and
$T=g^{\mu\nu}T_{\mu\nu}$.
\vskip 0.4cm
{\bf 5. Schr\"odinger equation}
\vskip 0.3cm
The object of interest in this case is a map $\Psi: \mathbb{R}^4\rightarrow
\mathbb{C}$, and $\mathbb{R}^4=\mathbb{R}^3\times\mathbb{R}$ is
parametrized by the canonical coordinates $(x,y,z;t)$, where $t$ is the
(absolute) time "coordinate". The operator $D$ used here is
$$
D=i\hbar\frac{\partial }{\partial t}-\mathbf{H},
$$
where $\mathbf{H}$ is the corresponding {\it hamiltonian}. The maps $\Phi$
and $\varphi$ are products of functions, so the {\bf GR} gives
\begin{equation}
(\Phi,\varphi;\mathbf{D})(1\otimes 1,\Psi)=
\left(1\otimes\left(i\hbar\frac{\partial \Psi}                  
{\partial t}-\mathbf{H}\Psi\right)\right)\otimes 1=0.
\end{equation}

\vskip 0.4cm
{\bf 6. Dirac equation}
\vskip 0.2cm
The original free Dirac
equation on the Minkowski space-time $(M,\eta)$ makes use of the following
objects:
$\mathbb{C}^4$ - the canonical 4-dimensional complex vector space,
$L_{\mathbb{C}^4}$-the space of $\mathbb{C}$-linear maps
$\mathbb{C}^4\rightarrow\mathbb{C}^4$, $\Psi\in Sec(M\times \mathbb{C}^4)$,
$\gamma\in Sec(T^*M\otimes L_{\mathbb{C}^4})$, and the usual differential
$\mathbf{d}: \psi^i\otimes e_i\rightarrow \mathbf{d}\psi^i\otimes e_i$, where
$\{e_i\}, i=1,2,3,4$, is a basis of $\mathbb{C}^4$. We identify further
$L_{\mathbb{C}^4}$ with $(\mathbb{C}^4)^*\otimes \mathbb{C}^4$ and if
$\{\varepsilon^i\}$ is a basis of $(\mathbb{C}^4)^*$, dual to $\{e_i\}$, we
have the basis $\varepsilon^i\otimes e_j$ of $L_{\mathbb{C}^4}$. Hence, we
may write
\[
\gamma=\gamma_{\mu i}^j dx^\mu\otimes(\varepsilon^i\otimes e_j),
\]
and
\[
\begin{split}
\gamma(\Psi)&=
\gamma_{\mu i}^j dx^\mu\otimes(\varepsilon^i\otimes e_j)(\psi^k\otimes e_k)\\
&=\gamma_{\mu i}^j dx^\mu\otimes \psi^k<\varepsilon^i,e_k>e_j=
\gamma_{\mu i}^j dx^\mu\otimes \psi^k\delta^i_k e_j=
\gamma_{\mu i}^j \psi^i dx^\mu\otimes e_j.
\end{split}
\]
The 4 matrices $\gamma_\mu$ satisfy
$\gamma_\mu\gamma_\nu+\gamma_\nu\gamma_\mu=2\eta_{\mu\nu}id_{\mathbb{C}^4}$,
so they are nondegenerate: $det(\gamma_\mu)\neq 0, \mu=1,2,3,4$, and we can
find $(\gamma_\mu)^{-1}$ and introduce $\gamma^{-1}$ by
\[
\gamma^{-1}=((\gamma_\mu)^{-1})_i^jdx^\mu\otimes(\varepsilon^i\otimes e_j)
\]
We introduce now the differential operators
$\mathcal{D}^{\pm}:Sec(M\times\mathbb{C}^4)\rightarrow Sec(T^*M\otimes
\mathbb{C}^4)$ through the formula:
$\mathcal{D}^{\pm}=i\mathbf{d}\pm\frac 12 m\gamma^{-1}, i=\sqrt{-1},
m\in \mathbb{R}$.
The corresponding maps are: $\Phi=\eta$, $\varphi:
L_{\mathbb{C}^4}\times\mathbb{C}^4\rightarrow\mathbb{C}^4$ given by
$\varphi(\alpha^*\otimes\beta,\rho)=<\alpha^*,\rho>\beta$.  We obtain

\begin{equation}
\begin{split}
&(\Phi,\varphi;\mathcal{D}^{\pm})(\gamma,\Psi)=
(\Phi,\varphi)(\gamma^j_{\mu i}dx^\mu\otimes(\varepsilon^i\otimes e_j),
i\frac{\partial \psi^k}{\partial x^\nu}dx^\nu\otimes e_k\pm\frac 12 m
(\gamma_\nu^{-1})^s_r dx^\nu\otimes(\varepsilon^r\otimes e_s)\psi^m e_m)\\
&=i\gamma^j_{\mu i}\frac{\partial \psi^k}{\partial x^\nu}\eta(dx^\mu,dx^\nu)
<\varepsilon^i,e_k>e_j\pm\frac 12 m\gamma^j_{\mu i}(\gamma_\nu^{-1})^s_r
\psi^r\eta(dx^\mu,dx^\nu)<\varepsilon^i,e_s>e_j\\
&=i\eta^{\mu\nu}\gamma^j_{\mu i}\frac{\partial \psi^k}
{\partial x^\nu}\delta^i_k e_j\pm                                   
\frac 12 m\eta^{\mu\nu}\gamma^j_{\mu i}(\gamma_\nu^{-1})^s_r\psi^r
\delta^i_s e_j\\
&=i\gamma^{\mu j}_i\frac{\partial \psi^i}{\partial x^\mu} e_j\pm
\frac 12 m(-2\delta^j_r\psi^r)e_j
=\left(i\gamma^{\mu j}_i \frac{\partial \psi^i}{\partial
x^\mu}\mp m\psi^j\right)e_j=0
\end{split}
\end{equation}
In terms of extended parallelism we can say that the Dirac equation is
equivalent to the requirement the section $\Psi\in Sec(M\times \mathbb{C}^4)$
to be ($\eta,\varphi;\mathcal{D}^{\pm}$)-parallel with respect to the given
$\gamma\in Sec(M\times L_{\mathbb{C}^4})$. Finally, in presence of external
electromagnetic field $\mathbf{A}=A_\mu dx^\mu$ the differential operators
$\mathcal{D}^{\pm}$ modify to
$\mathfrak{D}^{\pm}=(i\mathbf{d}-
e\mathbf{A})\pm \frac 12 m \gamma^{-1}$,
where $e$ is the electron charge.

\subsection{Conclusion}
It was shown that the {\bf GR}, defined by relation (85), naturally
generalizes the geometrical concept of parallelism,  and that it may
be successfully used as a unified tool to represent formally important
equations in geometry and theoretical physics. If $\Psi$ is the object of
interest then the {\bf GR} specifies mainly the following things: the change
$\mathbf{D}\Psi$ of $\Psi$, the object $\Psi_1$ with respect to which we
consider the change, the "projection" of the change $\mathbf{D}\Psi$ on
$\Psi_1$ through the bilinear map $\Phi$, and the bilinear map $\varphi$
determines the space where the final object takes values.  When $\Psi=\Psi_1$
we may speak about {\it autoparallel} objects in the corresponding sense, and
in this case, as well as when the differential operator $\mathbf{D}$ depends
on $\Psi$ and its derivatives, we obtain {\it nonlinear} equation(s).  In
most of the examples considered the main differential operator used was the
exterior derivative {\bf d} and its covariant generalization.

In the case of vector fields and one-forms on the Minkowski space-time we
recalled our previous result that among the corresponding autoparallel vector
fields there are finite (3+1) soliton-like ones, time-like, as well as
isotropic.  This is due to the fact that the trajectories of these fields
define straight lines, so their "transverse" components should be zero if the
spatially finite (or localized) configuration must move along these
trajectories.  Moreover, the corresponding equation can be easily modified to
be interpreted as local energy-momentum conservation relation.

It was further shown that Maxwell vacuum equations appear as
$\mathbf{d}$-parallel with somewhat trivial projection
procedure.  This determines their {\it linear} nature and leads to the lack
of spatial soliton-like solutions. The vacuum extended Maxwell equations
are naturally cast through the {\bf GR} in the form of autoparallel
(nonlinear) equations, and, as it was shown in our former works and in this
review, they admit photon-like (3+1) {\it spatially finite}
 solutions, and some of them admit naturally defined spin
properties [8]. The general extended Maxwell equations (96) may also be
given such a form if we replace the operator $\mathbf{d}$ with
$(\mathbf{d}-i(J^a))$, and (3+1) soliton solutions of this system were
also found [7].

The Yang-Mills equations were also described in this way.
The corresponding extended Yang-Mills equations  are expected to give
spatial soliton solutions. The Einstein equations of General
Relativity also admit such a formulation.

In quantum physics the Schr\"odinger equation admits the "parallel"
formulation with trivial projection. A bit more complicated was to put the
Dirac equation in  this formulation, and this is due to the bit more
complicated mathematical structure of $\gamma=\gamma_{\mu i}^j
dx^\mu\otimes(\varepsilon^i\otimes e_j)$.

These important examples make us think that the introduced here
extended concept for $(\Phi,\varphi;\mathbf{D})$-{\it parallel/conservative
objects} as a natural generalization of the existing geometrical concept for
$\nabla$-parallel objects, may be successfully used in various directions, in
particular, in searching for appropriate nonlinearizations of the existing
linear equations in theoretical and mathematical physics. It may also turn
out to find it useful in looking for appropriate lagrangians in some cases.
In our view, this is due to the fact that it expresses in a unified manner
the dual {\it change-conservation} nature of the physical objects.

\vskip 1.5cm
{\bf References to the Appendix}
\vskip 0.6cm

[1]. {\bf Greub, W}. {\it Multilinear Algebra}, Springer, 1967

[2]. {\bf Lichn$\acute{\mathrm{e}}$rowicz, A}., {\it Les Relations
Int\'egrales d'Invariance et leurs Applications $\grave{\mathrm{a}}$ la
Dinamique}, Bull. Sc. Math., 70 (1946), 82-95

[3]. {\bf Cartan, E}., {\it Le\c{c}ons sur les Invariants int\'egraux},
Hermann \& Fils, Paris, 1922

[4]. {\bf Godbillon, C.}, {\it G\'eom\'etrie Differentielle et M\'echanique
Analytique}, Hermann, Paris, 1969

[5]. {\bf Donev, S}., {\it Autoclosed Differential Forms and (3+1)-Solitary
Waves}, Bulg. J. Phys., 15 (1988), 419-426

[6]. {\bf Donev, S., Tashkova, M}., {\it Energy-momentum Directed
Nonlinearization of Maxwell's Pure Field Equations}, Proc. R. Soc. Lond. A,
443 (1993), 301-312

[7]. {\bf Donev, S., Tashkova, M}., {\it Energy-momentum Directed
Nonlinearization of Maxwell's Equations in the Case of a Continuous Madia},
Proc. R. Soc.  Lond.  A, 443 (1995), 281-291

[8]. {\bf Donev, S}., {\it Screw Photon-like (3+1)-Solitons in Extended
Electrodynamics}, The European Physical Journal "B", vol.29, No.2, 233-237
(2002).

\end{document}